\documentclass[onecolumn,draftcls]{IEEEtran}
\usepackage[utf8]{inputenc}
\usepackage[mathscr]{eucal}
\usepackage[cmex10]{amsmath}
\usepackage{amssymb,amsmath,amsfonts,latexsym}
\usepackage{graphicx,bm,xcolor,url}
\usepackage[caption=false]{subfig} 
\usepackage{array}
\usepackage{verbatim}
\usepackage{bm}
\usepackage{cite}
\usepackage{algpseudocode}
\usepackage{algorithm}
\usepackage{verbatim}
\usepackage{textcomp}
\usepackage{mathrsfs}
\usepackage{multirow}
\usepackage{hyperref}
\usepackage{epstopdf}
\usepackage{centernot}
\usepackage{dsfont}
 \usepackage{color,pgfplots,tikz}
\usepackage[nolist]{acronym}

\usepackage{thmtools}

\usepackage{xpatch}

\makeatletter
\newif\ifthmt@listswap
\def\thmt@TRUE{true}
\def\thmt@FALSE{false}
\define@key{thmt-listof}{swapnumber}[true]{%
  \def\thmt@tmp{#1}%
  \ifx\thmt@tmp\thmt@TRUE
    \thmt@listswaptrue
  \else\ifx\thmt@tmp\thmt@FALSE
    \thmt@listswapfalse
  \else
    \PackageError{thmtools}{Unknown value `#1' to key swapnumber}{}%
  \fi\fi
}

\xpatchcmd\thmt@mklistcmd
  {\protect\numberline{\protect\let\protect\autodot\protect\@empty}}
  {%
    \protect\ifthmt@listswap
    \protect\else
      \protect\numberline{\protect\let\protect\autodot\protect\@empty}%
    \protect\fi
  }
  {}{\fail}

\xpatchcmd\thmt@mklistcmd
  {%
    \protect\numberline{\csname the\thmt@envname\endcsname}%
    \thmt@thmname
  }
  {%
    \protect\ifthmt@listswap
      \thmt@thmname~\csname the\thmt@envname\endcsname
    \protect\else
      \protect\numberline{\csname the\thmt@envname\endcsname}%
      \thmt@thmname
    \protect\fi
  }
  {}{\fail}
\makeatother

\catcode`~=11 \def\UrlSpecials{\do\~{\kern -.15em\lower .7ex\hbox{~}\kern .04em}} \catcode`~=13 

\allowdisplaybreaks[1]
 
\newcommand{\nn}{\nonumber}

\newcommand{\x}{\boldsymbol{x}}

\begin{acronym}
\acro{DMC}{discrete memoryless channel}
\acro{BSC}{binary symmetric channel}
\acro{CC}{constant composition}
\acro{FSC}{finite-state channels}
\acro{i.i.d.}{independently and identically distributed}
\acro{JSCC}{joint source-channel coding}
\acro{ML}{maximum likelihood}
\acro{RCU}{Random Coding Union}
\acro{RGV}{Random Gilbert-Varshamov}
\acro{TRC}{typical random coding}
\acro{RCE}{random coding exponent}
\end{acronym}

\newcommand{\Pecsnub}{P_{\rme}^{\rm ub}(\mathcal{C}_n)}
\newcommand{\Eexr}{E_{\rm ex}(R)}
\newcommand{\Peubn}{P_{\rme}^{\rm ub}(\Cn)}

\newcommand{\calB}{\mathcal{B}}
\newcommand{\calH}{\mathcal{H}}
\newcommand{\calI}{\mathcal{I}}

\newcommand{\calL}{\mathcal{L}}
\newcommand{\calN}{\mathcal{N}}

\newcommand{\calP}{\mathcal{P}}
\newcommand{\calT}{\mathcal{T}}

\newcommand{\calY}{\mathcal{Y}}

\newcommand{\Pro}{\mathbb{P}}

\newcommand{\rmb}{\mathrm{b}}
\newcommand{\rmB}{\mathrm{B}}

\newcommand{\rme}{\mathrm{e}}

\DeclareMathAlphabet{\mathbsf}{OT1}{cmss}{bx}{n}
\DeclareMathAlphabet{\mathssf}{OT1}{cmss}{m}{sl}

\DeclareSymbolFont{bsfletters}{OT1}{cmss}{bx}{n}  
\DeclareSymbolFont{ssfletters}{OT1}{cmss}{m}{n}
\DeclareMathSymbol{\bsfGamma}{0}{bsfletters}{'000}
\DeclareMathSymbol{\ssfGamma}{0}{ssfletters}{'000}
\DeclareMathSymbol{\bsfDelta}{0}{bsfletters}{'001}
\DeclareMathSymbol{\ssfDelta}{0}{ssfletters}{'001}
\DeclareMathSymbol{\bsfTheta}{0}{bsfletters}{'002}
\DeclareMathSymbol{\ssfTheta}{0}{ssfletters}{'002}
\DeclareMathSymbol{\bsfLambda}{0}{bsfletters}{'003}
\DeclareMathSymbol{\ssfLambda}{0}{ssfletters}{'003}
\DeclareMathSymbol{\bsfXi}{0}{bsfletters}{'004}
\DeclareMathSymbol{\ssfXi}{0}{ssfletters}{'004}
\DeclareMathSymbol{\bsfPi}{0}{bsfletters}{'005}
\DeclareMathSymbol{\ssfPi}{0}{ssfletters}{'005}
\DeclareMathSymbol{\bsfSigma}{0}{bsfletters}{'006}
\DeclareMathSymbol{\ssfSigma}{0}{ssfletters}{'006}
\DeclareMathSymbol{\bsfUpsilon}{0}{bsfletters}{'007}
\DeclareMathSymbol{\ssfUpsilon}{0}{ssfletters}{'007}
\DeclareMathSymbol{\bsfPhi}{0}{bsfletters}{'010}
\DeclareMathSymbol{\ssfPhi}{0}{ssfletters}{'010}
\DeclareMathSymbol{\bsfPsi}{0}{bsfletters}{'011}
\DeclareMathSymbol{\ssfPsi}{0}{ssfletters}{'011}
\DeclareMathSymbol{\bsfOmega}{0}{bsfletters}{'012}
\DeclareMathSymbol{\ssfOmega}{0}{ssfletters}{'012}

\newcommand{\tild}{\tilde{d}}
\newcommand{\tilD}{\tilde{D}}

\newcommand{\tilh}{\tilde{h}}

\newcommand{\hatP}{\hat{P}}

\newcommand{\tilT}{\tilde{T}}

\newcommand{\tilV}{\tilde{V}}

\newcommand{\tilY}{\tilde{Y}}

\newcommand{\bard}{\bar{d}}

\newcommand{\barE}{\bar{E}}

\newcommand{\barZ}{\bar{Z}}

\newcommand{\eps}{\varepsilon}

\newcommand{\dotleq}{\stackrel{.}{\leq}}

\newcommand{\dotgeq}{\stackrel{.}{\geq}}

\DeclareMathOperator*{\argmax}{arg\,max}
\DeclareMathOperator*{\argmin}{arg\,min}

\DeclareMathOperator{\var}{\mathrm{Var}}

\newcommand{\bone}{\mathbf{1}}

\newtheorem{theorem}{Theorem} 
\newtheorem{lemma}{Lemma}
\newtheorem{corollary}{Corollary}
\newtheorem{definition}{Definition} 

\newcommand{\qednew}{\nobreak \ifvmode \relax \else
      \ifdim\lastskip<1.5em \hskip-\lastskip
      \hskip1.5em plus0em minus0.5em \fi \nobreak
      \vrule height0.75em width0.5em depth0.25em\fi}

\usepackage{ifthen}
\usepackage{xargs}
\newcommandx{\yaHelper}[2][1=\empty]{%
\ifthenelse{\equal{#1}{\empty}}%
  { \ensuremath{ \scriptstyle{ #2 } } } 
  { \raisebox{ #1 }[0pt][0pt]{ \ensuremath{ \scriptstyle{ #2 } } } }  
}
\newcommandx{\yrightarrow}[4][1=\empty, 2=\empty, 4=\empty, usedefault=@]{%
  \ifthenelse{\equal{#2}{\empty}}
  { \xrightarrow{ \protect{ \yaHelper[ #4 ]{ #3 } } } } 
  { \xrightarrow[ \protect{ \yaHelper[ #2 ]{ #1 } } ]{ \protect{ \yaHelper[ #4 ]{ #3 } } } } 
}
\newcommand{\pto}{\smash{\stackrel{({\rm p})}{ \,\longrightarrow\,}}}
\newcommand{\dto}{\smash{\stackrel{({\rm d})}{\, \longrightarrow\,}}}
\newcommand{\asto}{\xrightarrow{(\rm a.s.)}}

\newcommand{\notdto}{\smash{\stackrel{({\rm d})}{\,\centernot \longrightarrow\,}}}

\usepackage{xspace}

\DeclareFontFamily{U}{mathc}{}
\DeclareFontShape{U}{mathc}{m}{it}%
{<->s*[1.03] mathc10}{}

\DeclareMathAlphabet{\mathscr}{U}{mathc}{m}{it}

\newcommand{\bx}{\boldsymbol{x}}
\newcommand{\bX}{\boldsymbol{X}}
\newcommand{\by}{\boldsymbol{y}}

\newcommand{\calA}{\mathcal{A}}
\newcommand{\calC}{\mathcal{C}}
\newcommand{\calE}{\mathcal{E}}
\newcommand{\calX}{\mathcal{X}}

\newcommand{\calF}{\mathcal{F}}

\newcommand{\calV}{\mathcal{V}}

\newcommand{\Cn}{\calC_n}

\newcommand{\cn}{\mathscr{c}_n}

\newcommand{\mb}{{\bar m}}

\newcommand{\indicator}{\mathds{1}}

\newcommand{\PeCn}{P_{\rm e}(\Cn)}
\newcommand{\PeCnQ}{P_{\rm e}(\Cn,Q)}
\newcommand{\Pecn}{P_{\rm e}(\cn)}

\newcommand{\ECn}{E_n(\Cn)}
\newcommand{\Ecn}{E_n(\cn)}

\newcommand{\Erce}{E_{\rm rce}(R)}
\newcommand{\Etrc}{E_{\rm trc}(R)}
\newcommand{\Etrcbar}{\bar E_{\rm{trc}}}
\newcommand{\Etrcub}{E_{\rm trc}^{\rm ub}}

\newcommand{\ErceQ}{E_{\rm rce}(R,Q)}
\newcommand{\Erceiid}{E_{\rm rce}^{\rm iid}(R,Q)}
\newcommand{\Ercecc}{E_{\rm rce}^{\rm cc}(R,Q)}
\newcommand{\EtrcQ}{E_{\rm trc}(R,Q)}
\newcommand{\EtrcubQ}{E_{\rm trc}^{\rm ub}(R,Q)}

\newcommand{\Eoiid}{E_0^{\rm iid}}
\newcommand{\Eocc}{E_0^{\rm cc}}

\newcommand{\Qiid}{Q^{\rm iid}}
\newcommand{\Qcc}{Q^{\rm cc}}

\newcommand{\tilQ}{\tilde{Q}} 

\newcommand{\bbR}{\mathbb{R}} 
\newcommand{\bbZ}{\mathbb{Z}} 
\newcommand{\bbE}{\mathbb{E}} 
\newcommand{\bbP}{\mathbb{P}} 

\newcommand{\EE}{\mathbb{E}} 
\newcommand{\PP}{\mathbb{P}} 

\IEEEoverridecommandlockouts

\begin{document}

\title{Concentration Properties of Random Codes}
	
\author{Lan V.~Truong, Giuseppe Cocco, Josep Font-Segura and Albert Guill\'en i F\`abregas
		\thanks{L. V. Truong is with the Department of Engineering, 
			University
			of Cambridge, Cambridge CB2 1PZ, U.K. (e-mail: lt407@cam.ac.uk).
			G. Cocco and J. Font-Segura are with Department of Information and 
			Communication
			Technologies, Universitat Pompeu Fabra, Barcelona 08018, Spain (e-mail: giuseppe.cocco@upf.edu, josep.font@upf.edu).
			A.~Guill\'en i F\`abregas is with the  Department of Engineering, 
			University
			of Cambridge, Cambridge CB2 1PZ, U.K. and the Department of Information and 
			Communication
			Technologies, Universitat Pompeu Fabra, Barcelona 08018, Spain (e-mail: guillen@ieee.org). 
			
			This work has been funded in part by the European Research Council under ERC grant agreement 725411, by the Secretary of Universities and Research (Catalan Government) under a Beatriu de Pin\'{o}s postdoctoral fellowship, and by the European Union's Horizon 2020 research and innovation programme under the Marie Sk{\l{}}odowska-Curie grant agreement 801370.
		}
		\thanks{This work has been presented in part at the 2021 IEEE Information Theory Workshop, Kanazawa, Japan.}
	}

\maketitle

\begin{abstract}
This paper studies concentration properties of random codes. Specifically, we show that, for discrete memoryless channels, the error exponent of a randomly generated code with pairwise-independent codewords converges in probability to its expectation---the typical error exponent. For high rates, the result is a consequence of the fact that the random-coding error exponent and the sphere-packing error exponent coincide. For low rates, instead, the convergence is based on the fact that the union bound accurately characterizes the probability of error. The paper also zooms into the behavior at asymptotically low rates, and shows that the error exponent converges in distribution to a Gaussian-like distribution. Finally, we present several results on the convergence of the error probability and error exponent for generic ensembles and channels.
\end{abstract}

\section{Introduction}
The \ac{DMC} has been devoted a lot of interest in information theory ever since in \cite{Shannon48} Shannon  showed that for \ac{DMC} there exist codes whose probability of error vanishes with the codewords length for rates below the channel capacity. Since then, one of the most active areas of research in Information Theory has been the study of properties of the probability of error. For rates below capacity, Fano \cite{Fano} characterized the exponential decay of the error probability defining the error exponent as the negative normalized logarithm of the ensemble-average error probability, i.e., the  \ac{RCE}. In \cite{Gallager1965a}, Gallager derived the \ac{RCE} in a simpler way and introduced the idea of expurgation in order to obtain an improved exponent the at low rates. A lower bound on the  error probability in the \ac{DMC}, called sphere-packing bound, was first introduced in \cite{sgb} and it was shown to coincide with the \ac{RCE} for rates higher than a certain critical rate. Nakibo{\u{g}}lu in~\cite{Nakiboglu2020} recently derived sphere-packing bounds for some stationary memoryless channels using Augustin's method~\cite{nakibouglu2019augustin}.

In \cite{Barg2002a}, Barg and Forney  studied the random-coding ensemble over the \ac{BSC} with maximum likelihood decoding and showed that the error exponent of most random codes in the i.i.d. ensemble is close to the so-called \ac{TRC} exponent, strictly larger than the \ac{RCE} at low rates. Upper and lower bounds on the \ac{TRC} for constant-composition codes and general \ac{DMC}s were provided in \cite{Nazari}. For the same type of codes and channels, Merhav \cite{Merhav2018a} determined the exact \ac{TRC} error exponent and a wide class of stochastic decoders called generalized likelihood decoder (GLD), of which maximum-likelihood is a special case. Merhav derived the \ac{TRC} exponent for spherical codes over coloured Gaussian channels~\cite{merhav2019error} and for random convolutional code ensembles \cite{merhav2019error2}. The error exponent of a random pairwise-independent constant-composition code with GLD was shown to converge in probability to the TRC in \cite{Tamir2020a}. The convergence is non-symmetric: the lower tail decays exponentially while the upper tail decays doubly-exponentially. The latter was first established for a limited range of rates in \cite{Ahlswede1982}. The TRC was shown to be universally achievable with the likelihood mutual information decoder in \cite{tamir2021universal}. For pairwise-independent ensembles and arbitrary channels, Cocco~\emph{et al.} showed in~\cite{Giusseppe2021e} that the probability that the exponent of a given code in the ensemble is smaller than a lower bound on the \ac{TRC} exponent is vanishingly small.

The main motivation of our work is the fact that the aforementioned results highlight the importance of the statistical properties of the error probability and the error exponent across the random-coding ensemble.  After describing the main performance metrics of random codes for reliable communication in the next section, namely the error probability and the error exponent, we use the notion of convergence in probability and convergence in distribution to obtain a number of concentration results of such performance metric, seen as sequences of random variables, as the blocklength tends to infinity. Since neither the error probability or the error exponent are sums of i.i.d.~terms, our results are based on probability results beyond the central limit theorem, such as the Stein's method and a novel, modified Wasserstein metric. We anticipate here some of our main results in~Sec.~\ref{sec:dmc}, valid for the \ac{DMC} and the i.i.d.~and constant-composition ensembles.
\begin{itemize}
	\item We show in Theorem~\ref{thm:main} that the error exponent converges in probability the \ac{TRC} exponent.
	\item In Theorems~\ref{asko}--\ref{theo:double_exp_0R} we provide bounds on the rate of such convergence.
	\item For codes with a constant number of codewords, we obtain in Theorem~\ref{lem:aux2} that the error exponent converges in distribution to a quasi-Gaussian, while for codes with a sub-exponential number of codewords it does converge to a Gaussian, as shown in Theorem~\ref{lem:ct2}.
\end{itemize}
For general channels under some additional conditions, we obtain in Sec.~\ref{sec:general} the following results.
\begin{itemize}
	\item For any channel and capacity-achieving ensemble, Theorem~\ref{theo:1} states that the error probability converges to the ensemble average.
	\item Theorems~\ref{thm:abet}--\ref{corox2} discuss several convergence results relating random properties of the error probability, the random-coding error exponent and the \ac{TRC} exponent.
	\item Sufficient conditions for the union bound on the error probability and any general function of the error probability to converge to a Gaussian are respectively described in Theorems~\ref{thm:ub}--\ref{mainthm4}.
	\end{itemize}
The proofs of our main results, including our modified Wasserstein metric, are reported in Sec.~\ref{sec:proof}.
\section{Preliminaries} 
We consider the problem of transmitting $M_n$ equiprobable messages over a \ac{DMC} with transition probability $W$ and finite input and output alphabets $\calX$ and $\calY$, respectively. We employ a codebook $\mathscr{c}_n=\{\bx_1,\bx_2,\cdots, \bx_{M_n}\}$ with $\bx_m \in \calX^n$, for $m=1,\ldots,M_n$. The conditional distribution of a channel output $\by\in\calY^n$ given a transmitted codeword $\bx$ is given by $W^n(\by|\bx) = \prod_{i=1}^nW(y_i|x_i)$.
 We consider maximum-likelihood decoding, that is, the decoder produces an estimate of the transmitted codeword as $
\hat\bx = \argmax_{\bx\in\cn} \,W^n(\by|\bx)$. The error probability of such code is
\begin{align}
\Pecn=\frac{1}{M_n}\sum_{m=1}^{M_n} \bbP\bigg[\bigcup_{\mb\neq m} \{\bx_m \to \bx_\mb\}\bigg],
\label{eq:Pe}
\end{align} 
where $\{\bx_m\to\bx_\mb\}=\{\by\in\calY: W^n(\by|\bx_{\mb})\geq W^n(\by|\bx_m)\}$ is the maximum-likelihood pairwise error event, i.e., the event of deciding in favor of codeword $\bx_\mb$ when codeword $\bx_m$ was transmitted. 
The error exponent of code $\mathscr{c}_n$ is defined  as 
\begin{align}
\Ecn = -\frac{1}{n}\log P_{\rme}(\mathscr{c}_n).
\label{eq:en}
\end{align}
Let $R=\lim_{n\to \infty} \frac{1}{n}\log M_n$ be the rate of the code in bits per channel use. An error exponent $E(R)$ is said to be achievable when there exists a sequence of codes $\{\mathscr{c}_n\}_{n=1}^{\infty}$ such that $\liminf_{n\to \infty} \Ecn\geq E(R)$. The channel capacity $C$ is the supremum of the code rates $R$ such that $E(R)>0$. 

We next consider the random generation of the codebook. Similarly to random variables, $\Cn$ denotes a random code, and $\mathscr{c}_n$ denotes a specific code in the ensemble. In particular, we consider the pairwise-independent random-coding ensemble, i.e., the set of random codes $\Cn$ whose codewords $\bX_1,\bX_2,\cdots, \bX_{M_n}$ are pairwise-independently generated. We consider the i.i.d. ensemble, in which each codeword is generated according to the distribution
\begin{align}
\Qiid(\bx) = \prod_{i=1}^nQ(x_i),
\end{align}
$Q$ being the single-letter distribution and the constant-composition ensemble, in which each codeword is generated according to the distribution
\begin{align}
\Qcc(\bx) = \frac{1}{|\calT_n(Q_n)|} \indicator\{\bx\in\calT_n(Q_n)\}
\end{align}
where $\calT_n(Q_n)$ is the type class of composition $Q_n\in\calP_n(\calX)$, i.e., all $n$-length sequences whose empirical distribution is $Q_n$ such that $\max_x|Q_n(x) - Q(x)|\leq \frac1n$ for a given distribution $Q$.
 For a given distribution or composition $Q$, we define the random-coding error exponent $\ErceQ$ as
\begin{align}
\ErceQ = \lim_{n\to \infty}- \frac{1}{n}\log \bbE[ \PeCnQ],
\label{eq:Erce}
\end{align} 
where $\PeCnQ$ denotes the error probability of the random code ensemble $\Cn$ parametrized by the distribution or composition $Q$ and where the expectation is taken over the code ensemble. Eq.~\eqref{eq:Erce} suggests that $\ErceQ$ is the asymptotic exponent of the ensemble-average probability of error. 
For the i.i.d. ensemble, it is known that \cite[Th.~1]{Gallager1965a}
\begin{align}
\Erceiid = \max_{0\leq \rho \leq 1} \big\{\Eoiid(\rho,Q)-\rho R\big\}
\label{eq:rcedef}
\end{align} 
with
\begin{align}
\Eoiid(\rho,Q) = -\log\sum_y\bigg(\sum_x Q(x)W(y|x)^{\frac{1}{1+\rho}}\bigg)^{1+\rho};
\label{eq:e0def}
\end{align}
while for the constant-composition ensemble, we have that \eqref{eq:rcedef} remains valid, but $\Eoiid(\rho,Q)$ is replaced by \cite[Eq.~(53)]{Scarlett13}
\begin{align}
\Eocc(\rho,Q) =  \sup_{a(x)}-\log\sum_y\biggl(\sum_x Q(x)W(y|x)^{\frac{1}{1+\rho}}e^{a(x) - \phi_a}\biggr)^{\!\!1+\rho}
\end{align}
where $a(x)$ is an auxiliary function and $\phi_a  =  \sum_x Q(x) a(x)$. It is known that for any given $Q$, $\Erceiid\leq\Ercecc$ (see e.g. \cite{gallager_fcc_notes}).


While $\ErceQ$ in~\eqref{eq:Erce} is the limiting exponential rate of decay of the ensemble average probability of error, the typical random-coding exponent $\EtrcQ$ is instead defined as the limiting expected error exponent over the ensemble, that is,
\begin{align}
\EtrcQ = \lim_{n\to \infty}- \frac{1}{n}\bbE\big[\log \PeCnQ\big] \label{detrc}.
\end{align}  
We observe that Jensen's inequality implies that the random-coding error exponent in~\eqref{eq:Erce} and the typical random-coding error exponent in~\eqref{detrc} satisfy $\ErceQ\leq \EtrcQ$. The proofs of our results exploit the idea that there are two rate regimes; one where $\ErceQ< \EtrcQ$ and the other where $\ErceQ= \EtrcQ$. For DMCs with constant-composition codes \cite{Merhav2018a} and for i.i.d. codes over the BSC \cite{Barg2002a}, the typical error exponent can be further expressed in terms of the expurgated error exponent $E_{\rm ex}(R,Q)$ as
\begin{align}
\EtrcQ = \max\{E_{\rm ex}(2R,Q) + R, \ErceQ\}.
\end{align}

In the next sections, we derive concentration results of the error probability~\eqref{eq:Pe} and the error exponent~\eqref{eq:en} of sequences of random codes ${\cal C}_n$ in the asymptotic regime as $n\to\infty$. As sequence of random variables, we assume throughout the paper that $\PeCn$ and $\ECn$ do not diverge as $n\to\infty$, leaving particular cases such as channels with positive zero-error capacity beyond the scope of this paper. We will use the notion of convergence in probability and convergence in distribution. A sequence of random variables $\{A_n\}_{n=1}^{\infty}$ converges to $A$ in probability, denoted as $A_n \pto A$ if for all $\delta>0$ \cite[Sec.~2.2]{Durrett},
\begin{align}
\lim_{n\to\infty}\bbP[|A_n-A|>\delta]= 0.
\label{eq:convp}
\end{align}
If $A_n = \frac1n\sum_{i=1}X_i$, where $X_i, i=1,\dotsc,n$ are i.i.d. random variables, then $A=\bbE[X_1]$ and~\eqref{eq:convp} reduces to the weak law of large numbers \cite[Th.~2.2.3]{Durrett}. The weak law of large numbers is at the core of the asymptotic equipartition property, a widely used tool in information theory to establish the achievable rates using random coding \cite{coverThomas}; it is well known that the asymptotic equipartition property is not sufficient to show the achievability of error exponents. Alternatively to~\eqref{eq:convp}, we say that a sequence of random variables $\{A_n\}_{n=1}^{\infty}$ converges to $A$ in distribution, denoted as $A_n \dto A $ if~\cite[Sec.~3.2]{Durrett}
\begin{equation}
	\lim_{n\to\infty} \sup_{x \in \bbR}\big|\bbP[A_n \leq x]-\bbP[A \leq x]\big| = 0
\end{equation}
for all continuous points $x$ of $\bbP[A \leq x]$.

We first state in Sec.~\ref{sec:dmc} our main results for the relevant case of i.i.d.~and constant-composition ensembles over generic \ac{DMC}s. Additional results are shown in Sec.~\ref{sec:general} for general channels with few additional assumptions on the ensemble or conditions on the statistical behavior of the error probability as $n\to\infty$. The proofs of our theorems are included in Sec.~\ref{sec:proof}, while most lemmas thereby used are proved in the Appendix.


\section{Discrete Memoryless Channels}
\label{sec:dmc}

In this section, we introduce our main concentration results for \ac{DMC}s.
Our first result states the convergence in probability of the error exponent $\ECn$ to the TRC exponent $\Etrc$. Since the exponent of the probability of error is not a sum of i.i.d. terms, the weak law of large numbers cannot be applied. This result holds for i.i.d. and constant-composition ensembles over \ac{DMC}s with input distribution $Q$.

\begin{theorem}
\label{thm:main} 
For a general DMC channel, i.i.d. and constant-composition ensembles and rates $0\leq R < C$, it holds that
	\begin{align}
	\ECn \pto \EtrcQ \label{biscuitge}.
	\end{align}
\end{theorem}
\begin{IEEEproof}
Sec.~\ref{sub:p1}.
\end{IEEEproof}


Theorem \ref{thm:main} not only proves the achievability of the TRC exponent, but also shows that the probability of finding a code in the ensemble with higher or lower exponent than the TRC exponent tends to zero. The above concentration property gives more information about the error exponent behaviour of the ensemble than the traditional derivation of the random coding error exponent, which computes the exponent of the expected error probability. This way, the TRC emerges as the most likely error exponent for pairwise-independent random-coding ensembles as the block length $n$ tends to infinity ---if one wishes to improve the error exponent, one must improve the ensemble. The \ac{TRC} exponent is lower than or equal to the expurgated exponent and can in some case be strictly smaller. This implies that the codes in the pairwise independent ensemble that achieve the expurgated exponent are not typical codes and are unlikely to be found by random generation.

The proof of Theorem~\ref{thm:main} requires different techniques for the rate regimes  $\EtrcQ=\ErceQ$ and $\EtrcQ>\ErceQ$. For the first regime, corresponding to high rates such that $R_{\rm{crit}}(R,Q)\leq R\leq C$, where $R_{\rm{crit}}(R,Q)$ is the critical rate, we exploit the fact that $\EtrcQ=\ErceQ$ and the Levy's continuity theorem \cite[Sec.~XIII.1]{feller} to obtain that the moment-generating function of the random variable $E_n({\cal C}_n)$ converges to the moment-generating function of a deterministic variable with value $\EtrcQ$. In other words, $E_n({\cal C}_n)$ converges in distribution to a constant $\EtrcQ$, implying the convergence in probability in~\eqref{biscuitge} for both the i.i.d.~and the constant-composition ensembles.



For the second regime where $\EtrcQ>\ErceQ$, corresponding to low rates, we shift the convergence analysis to the union upper bound on the error probability $P_{\rm e}^{\rm ub}({\cal C}_n)$ in~\eqref{defRBE}, a bound that is tight enough at low rates. Then, we use De Caen's inequality~\cite{cohen2004lower}, a bound proved to be tight for the random-coding ensemble average error probability in~\cite{ScarlettMF2013aler}, to argue that such convergence also happens for the exact error probability $\PeCn$. In particular, we set $\delta=3\varepsilon$ in~\eqref{eq:convp} and write
\begin{equation}
	\bbP\big[|\ECn -\Etrc|>3 \eps\big]\leq \alpha_n + \beta_n + \gamma_n,
	\label{eq:abc}
\end{equation}
where $\alpha_n$ accounts for the convergence of the exact error exponent to that of the union bound
\begin{equation}
	\alpha_n = \bbP\Big[\big|\ECn  - E_n^{\rm ub}(\Cn)\big|>\eps\Big],
\end{equation}
the second term studies the convergence of the error exponent of the union bound, $E_n^{\rm ub}(\Cn)$ to its ensemble average,
\begin{equation}
	\beta_n = \bbP\bigg[\bigg| E_n^{\rm ub}(\Cn) - \left(-\frac1n\bbE\big[\log P_{\rme}^{\rm ub}(\Cn)\big]\right)\bigg|>\eps\bigg],
\end{equation}
and the last term deals with the convergence of such ensemble average to the TRC, namely
\begin{equation}
	\gamma_n = \bbP\bigg[\bigg|\left(-\frac1n\bbE\big[\log P_{\rme}^{\rm ub}(\Cn)\big]\right)-\EtrcQ\bigg|>\eps\bigg].
\end{equation}

For the i.i.d.~ensemble, we exploit the symbol-wise independence to show that the three terms $\alpha_n$, $\beta_n$ and $\gamma_n$ vanish as $n\to\infty$ in~\eqref{eq:abc}, hence obtaining our result in~\eqref{biscuitge}. For constant-composition codes, we are also able to obtain a vanishing  $\alpha_n$, $\beta_n$ and $\gamma_n$ exploiting the independence of the joint type between two pairs of codewords and consider the expression of the limiting ensemble average of $E_n^{\rm ub}({\cal C}_n)$ for constant-composition codes reported in~\cite{Merhav2018a}.

Theorem~\ref{thm:main} shows the converge of sequences of random variables to the statistical mean. A refined analysis to that of Theorem~\ref{thm:main} consists of studying, separately, the probability tails involved in the definition of convergence in probability in~\eqref{eq:convp}. The work in \cite{Tamir2020a}, addressed this issue for the constant-composition ensemble over \ac{DMC}s. Specifically,   \cite{Tamir2020a} showed an interesting asymmetry: the probability  $\bbP[\ECn<\EtrcQ]$ decays exponentially, while $\bbP[\ECn>\EtrcQ]$ decays double-exponentially. This implies that, beyond the concentration property, it is significantly more difficult to find a code in the ensemble with exponent higher than $\EtrcQ$. 

We next derive some results on the convergence rate of the  error exponent $E_n({\cal C}_n)$ to the typical random-coding exponent $E_{\rm trc}(R,Q)$.

\begin{theorem} \label{asko} 
	For the i.i.d. or constant-composition ensembles with rate $0\leq R\leq C$ and any $\eps>0$, it holds that
	\begin{align}
	\bbP\big[E_n({\cal C}_n)  <\EtrcQ-\eps\big]\dotleq 2^{-n\eps} \label{espa},
	\end{align} 
	that is an exponential decay in the coding  blocklength. In addition, for any $0< R\leq C$ and $\eps>0$, it holds that
	\begin{align}
	\bbP\big[E_n({\cal C}_n)  >\EtrcQ+\eps\big]= O\bigg(\frac{1}{\sqrt{n}}\bigg) \label{astamoto}.
	\end{align}
\end{theorem}
\begin{IEEEproof}
Sec.~\ref{sub:p2}.
\end{IEEEproof}

Theorem \ref{asko}, that strengthens Theorem \ref{thm:main}, implicitly assumes that $\EtrcubQ=\EtrcQ$ for all rates below capacity to obtain~\eqref{espa} and uses the Berry-Esseen theorem \cite{Billingsley} to obtain~\eqref{astamoto}. For the union bound to the error probability,
\begin{align}
P_{\rme}^{\rm ub}(\mathscr{c}_n) = \frac{1}{M_n}\sum_{i=1}^{M_n}\sum_{j\neq i}\bbP[\bx_i \to \bx_j] \label{defRBE},
\end{align}
we are able to refine the upper tail in~\eqref{astamoto} as follows.
\begin{theorem}\label{theo:double_exp}
	For all rates satisfying $\EtrcQ>\ErceQ$, there exists some $\epsilon>0$ such that the following holds:
	\begin{align}\label{eqn:double_exp1}
	\Pro\left[E_n^{\rm ub}({\cal C}_n) \geq \Eexr + \epsilon\right]\leq 2^{-2^{n\epsilon}},
	\end{align}
	where $E_{\rm ex}(R)$ is the expurgated error exponent~\cite{gallagerBook}.
\end{theorem}
\begin{IEEEproof}
	Appendix~\ref{sec:double_exp}.
\end{IEEEproof}

For strictly zero rate, that is $R=0$, the expurgated is tight. Therefore, we have the following result.
\begin{theorem}\label{theo:double_exp_0R}
	For the i.i.d.~or constant-composition ensembles with rate $R=0$ and any $\varepsilon>0$, we have that
	\begin{align}\label{eqn:double_exp_0R_1}
	\Pro\left[E_n({\cal C}_n) \geq E_{\rm trc}(0,Q) + \epsilon\right]\leq 2^{-2^{n\epsilon}}.
	\end{align}
\end{theorem}
\begin{IEEEproof}
Sec.~\ref{sub:p3}.
\end{IEEEproof}
So far, we have introduced results related to the convergence in probability of the error exponent for pairwise-independent random codes. In the remaining of the section, we discuss the concentration in distribution of the error exponent $E_n(\calC_n)$ to its ensemble-average $\mathbb{E}[E_n({\cal C}_n)]$ as $n\to\infty$ at the low rate regime. While Theorem~\ref{lem:aux2}, valid for an exactly constant number of messages, states that the random-coding error exponent converges to a Gaussian-like distribution, we let the number of messages $M_n$ in Theorem~\ref{lem:ct2} to grow sub-exponentially with $n$, yet at a minimum rate $M_n \gg \sqrt{n}$, and show that it converges to a Gaussian. As in Theorem~\ref{lem:ct2} and Theorem~\ref{mainthm4}, most of the following results use the Stein's method \cite{Nathan2011a}. 

\begin{theorem} 
\label{lem:aux2} 
Let $M_n=M$ be a constant number of messages, fixed for every $n$, and let $U_{ij}\sim{\cal N}(0,1)$, for $i=1,\ldots,M$ and $j=1,\ldots,M$ such that $i\neq j$, be a set of independent standard normal random variables. Then, the error exponent for both i.i.d. and constant-composition random-coding ensembles satisfies
	\begin{align}
	\frac{\ECn-\bbE[\ECn]}{\sqrt{\var(\ECn)}} \dto \frac{\min_{i\neq j}U_{ij}-\bbE[\min_{i\neq j}U_{ij}]}{\sqrt{\var(\min_{i\neq j}U_{ij})}} \label{eheldb}.
	\end{align}	 
\end{theorem}
\begin{IEEEproof}
Sec.~\ref{sub:p4}.
\end{IEEEproof}

The proof of Theorem~\ref{lem:aux2} is based on the fact that $\ECn$ is a minimization of a constant number of $M(M-1)$ terms where each term is a sum of independent random variables in the i.i.d.~ensemble, and a sum of dependent random variables with an additional vanishing term in the constant-composition ensemble. Hence, the central limit theorm~\cite[Ch.~VIII]{feller} and the Levy's continuity theorem \cite{Billingsley} can be applied. In fact, it is easy to see that~\eqref{eheldb} holds when the exponent of the pairwise error probability of two different codewords $\bX_i$ and $\bX_j$, that is $-\frac 1n \log \bbP[\bX_i \to \bX_j]$ forms a set of independent random variables. However, even for the i.i.d.~ensemble, such variables are only pairwise-independent. To give an example, for the binary symmetric channel with only three codewords $\bX_1$, $\bX_2$ and $\bX_3$, given the Hamming distance between two pairs, the Hamming distance between the third pair is not independent on the previous ones. Theorem~\ref{lem:aux2} argues that, despite such dependence, this becomes negligible as $n\to \infty$ when the number of codewords is constant.

\begin{figure*}[t!]
	\begin{tabular}{cc}
	~\hspace{-1em}\begin{tikzpicture}
\footnotesize
\begin{axis}[%
width=0.425*\textwidth,
height=0.425*\textwidth,
xmin=0.225,
xmax=0.2375,
ymin=0,
ymax = 300,
scale only axis,
xlabel = (a) i.i.d.,
legend style={legend cell align=left,align=left,draw=black,at={(0.983,0.98)},anchor=north east},
xtick= {0.225, 0.2275, 0.23, 0.2325, 0.235, 0.2375},
xticklabels={0.225, 0.2275, 0.23, 0.2325, 0.235, 0.2375}
]

\addplot[ybar,solid,fill,opacity=0.2,bar width=3pt]
  table{iid_n10000_hist.txt};

\addplot[line width = 1pt, dashed]  table {iid_n10000_pdf1.txt};

\addplot[line width = 1pt, blue] table {iid_n10000_pdf2.txt};

\end{axis}

\end{tikzpicture}%
 & 
	\begin{tikzpicture}
\footnotesize
\begin{axis}[%
width=0.425*\textwidth,
height=0.425*\textwidth,
xmin=0.225,
xmax=0.2375,
ymin=0,
ymax = 300,
scale only axis,
xlabel = (c) constant-composition,
legend style={legend cell align=left,align=left,draw=black,at={(0.983,0.98)},anchor=north east},
xtick={0.225, 0.2275, 0.23, 0.2325, 0.235, 0.2375},
xticklabels={0.225, 0.2275, 0.23, 0.2325, 0.235, 0.2375}
]

\addplot[ybar,solid,fill,opacity=0.2,bar width=3pt]
  table{cc_n10000_hist.txt};

\addplot[line width = 1pt, dashed]  table {cc_n10000_pdf1.txt};

\addplot[line width = 1pt, blue] table {cc_n10000_pdf2.txt};

\end{axis}

\end{tikzpicture}%
	\end{tabular}
	\caption{Distribution of the error exponent of the (a) i.i.d. and (b) constant-composition codes over the BSC with $M=4$, $n=10,000$, symmetric input distribution and composition, and $p=0.11$. Histograms of $\ECn$ with $10^7$ trials, dashed black lines are normal distributions, and solid blue lines are the distributions of $\min_{i\neq j}U_{ij}$.}
	\label{fig:2}
\end{figure*}
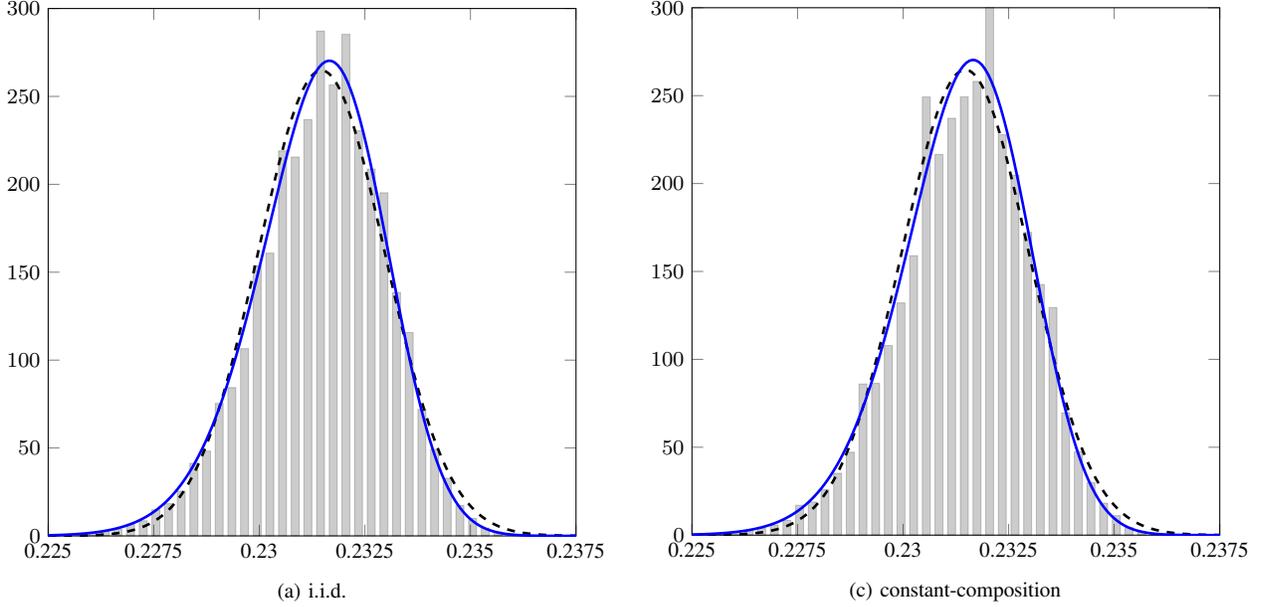

We illustrate in Fig.~\ref{fig:2} the histogram of the error exponent $E_n({\cal C}_n)$  used over a binary symmetric channel (BSC) with bit-flipping probability $p=0.11$, equiprobable bits and $M=4$ codewords for a blocklength of $n=10,000$. The histograms are obtained for the i.i.d. and constant-composition ensembles using the Monte Carlo method after $10^7$ trials. For the sake of comparison, we also depict the asymptotic distribution of the random variable $\min_{i\neq j}U_{ij}$ in the right-hand side of~\eqref{eheldb} (solid), and a normal approximation with the same mean and variance (dashed). We observe that the histogram matches the Gaussian-like distribution predicted by Theorem \ref{lem:aux2}, with a slightly asymmetric tail tilting.

We observe that the cumulative distribution function of the random variable in the right-hand side of~\eqref{eheldb} can be easily obtained using that, for any sequence of random variables of length $L>1$, for example $X_1,\ldots,X_L$, we have that
\begin{align}
\bbP[\min\{X_1,X_2,\cdots,X_L\}\geq t] =\bbP[X_1\geq t] \cdots \bbP[X_L\geq t].
\end{align}
If such sequence is i.i.d.~standard-normally distributed, then
\begin{equation}
	\bbP[\min\{X_1,X_2,\cdots,X_L\}\geq t] = Q(t)^L,
	\label{eq:37}
\end{equation}
where $Q(t)=\frac{1}{\sqrt{2\pi}}\int_{t}^{\infty} e^{-x^2/2}{\rm d}x$ is the Gaussian tail function. It is obvious that the right-hand side of~\eqref{eq:37} does \emph{not} correspond to a Gaussian cumulative distribution function.
\begin{theorem} \label{lem:ct2}  Let $M_n$ be a subexponential number of messages, namely $\lim_{n\to\infty}\frac{1}{n}\log M_n=0$, satisfying the condition
	\begin{align}
	\sum_{n=1}^{\infty} \frac{1}{M_n (M_n-1) }< \infty \label{condbat}.
	\end{align}
Then, the error exponent of codes in the i.i.d. and constant-composition random-coding ensembles satisfies
	\begin{align}
	\frac{\ECn-\bbE[\ECn]}{\sqrt{\var(\ECn)}}\dto \calN(0,1) \label{cota}.
	\end{align}
\end{theorem}
\begin{IEEEproof}
Sec.~\ref{sub:p5}.
\end{IEEEproof}

The proof of Theorem~\ref{lem:ct2} is based on the fact that, for both i.i.d. and constant-composition, the error exponent $\ECn$ is the minimum of an infinite number of terms, where each term converges to a Gaussian distribution.

For a constant number of messages $M_n=M$, the condition in Theorem~\ref{lem:ct2} is not satisfied, and therefore the error exponent does not concentrate according to~\eqref{cota} but to~\eqref{eheldb} instead. The fact that $M_n$ does not grow with $n$ implies that the dependence between the codewords vanishes as $n\to\infty$, and therefore the independence of $U_{ij}$ is preserved. On the contrary, for a (sub-exponentially) growing number of messages $M_n$, the dependence among the codewords, and therefore the correlation among $U_{ij}$, increases such that the random variables $U_{ij}$ can be represented by a common Gaussian random variable $U$. One example of sub-exponential growth of the number of messages satisfying~\eqref{condbat} is the polynomial function of $n$ given, for some $\delta>0$, by $
	 M_n=\Omega\bigl(n^{\frac{1+\delta}{2}}\bigr)$.

\section{General Channels}
\label{sec:general}

In this section, we introduce a number of results related to the concentration of the error probability and error exponent for relatively general channels and ensembles.
 The first result is a direct consequence of elementary probability results such as Chebyshev's inequality or Jensen's inequality.

\begin{theorem}\label{theo:1}
For a general channel and general ensemble such that $\EE[\PeCn]\to  0$ for $0\leq R<C$ and $\EE[\PeCn]\to  1$ for $R>C$, we have 
\begin{align}\label{eqn:cep_1}
\PeCn -\EE[\PeCn] \pto 0.
\end{align}
\end{theorem}
\begin{IEEEproof}
Sec.~\ref{sub:p6}.
\end{IEEEproof}

The above result holds for general channels and general ensembles as long as the strong converse property is satisfied. For channels and ensembles only satisfying a weak converse, namely that $\lim_{n\to\infty}\EE[\PeCn]>\epsilon$, then~\eqref{eqn:cep_1} is valid if $
\EE[\PeCn^2]/\EE[\PeCn]^2\to  1$ as $n\to\infty$. In fact, such condition also guarantees the convergence in probability of the error exponent $E_n({\cal C}_n)$ to the TRC, obtained by a direct application of Markov's inequality for general channels and pairwise-independent ensembles. Since the next results are very general and only assume pairwise-independent codewords, we drop the single-letter input distribution $Q$ in the notation of $\Etrc$ and $\Erce$ for the rest of the section.

\begin{theorem}  \label{thm:abet}  For a general channel and a pairwise-independent ensemble, under the condition that 
	\begin{align}
	\frac{\bbE[P_\rme(\Cn)^2]}{\bbE[P_\rme(\Cn)]^2}\to 1 \label{asspx},
	\end{align}
	we have
	\begin{align}
	E_n(\Cn) \pto \Etrc \label{mainreskq}.
	\end{align}
\end{theorem}
\begin{IEEEproof}
Sec.~\ref{sub:p7}.
\end{IEEEproof}

As a remark, the condition \eqref{asspx} might hold for all the rate less than or equal to the critical rate, but not too small. This idea is made more precise in the following result, based on~\cite[Th.~1]{Giusseppe2021e} and the Paley-Zygmund inequality \cite{paley_zyg1932}, valid for general channels and pairwise-independent ensembles.

\begin{theorem}\label{theo:paley_zygmund}
For a general channel and pariwise-independent ensemble with rate such that $\Etrc > \Erce$, we have
\begin{align}
\frac{\EE[P_{\rm e}(\Cn)]^2}{\EE[P_{\rm e}(\Cn)^2]}\to 0.
\end{align} 
\end{theorem}
\begin{IEEEproof}
Sec.~\ref{sub:p8}.
\end{IEEEproof}

In words, at low rates where the typical random-coding error exponent is strictly larger than the random-coding error exponent, the second-order moment of the error probability vanishes slower than the squared first-order moment. This implies that  $\var\left(P_{\rm e}(\Cn)\right)$ vanishes slower than the squared ensemble average $\bbE[P_\rme(\Cn)]^2$, suggesting that the error probability cannot converge to a Gaussian distribution this rate regime. Such intuition is formalized in the next result, based on Theorem~\ref{theo:paley_zygmund} and  Slutsky's theorem~\cite{Billingsley}.
\begin{theorem}\label{lem:d_nonconv} 
	For any code ensemble and channel such that $\Etrc>\Erce$, it holds that
	\begin{align}
	\frac{P_{\rme} (\Cn)-\bbE[\PeCn]}{\sqrt{\var(\PeCn)}} \notdto \calN(0,1).
	\label{eq:notg}
	\end{align}
\end{theorem}
\begin{IEEEproof}
Sec.~\ref{sub:p9}.
\end{IEEEproof}


\begin{theorem} \label{corox2}  For a general channel and pairwise-independent ensemble such that the normalized error probability converges in distribution to the standard normal distribution, that is
	\begin{align}
	\frac{P_\rme(\Cn)-\bbE[P_\rme(\Cn)]}{\sqrt{\var(P_\rme(\Cn))}} \dto \calN(0,1) \label{tacon},
	\end{align}
	we have that
	\begin{align}
	E_n(\Cn)\pto \Etrc \label{reverd}.
	\end{align}
\end{theorem}
\begin{IEEEproof}
Sec.~\ref{sub:p10}.
\end{IEEEproof}

We remark that condition~\eqref{tacon} is sufficient, but not necessary. To show~\eqref{reverd}, we bound the tail probabilities of the error probability $P_\rme(\Cn)$ around $2^{-n\Etrc}$ and relate such bounds with the standard normal distribution.

In the remaining of the section, we state two auxiliary results related to the convergence in distribution of the union bound to the error probability of a code $\mathscr{c}_n$ in~\eqref{defRBE}, and the convergence in distribution of an arbitrary function of the error probability.

By applying~ \cite[Th.~3.6]{Nathan2011a}, a result for the sum of random variables with local dependence, to the union bound~\ref{defRBE}, we obtain the following result.

\begin{theorem} \label{thm:ub}
Let $Y_{12}$ and $\gamma^2$ be two parameters respectively given by $Y_{12}  =  \bbP\big[ \{\bX_1 \to \bX_2\}\big]- \bbE\big[\bbP[ \{\bX_1 \to \bX_2\}]\big]$ and $\gamma^2 =   \var \bigl(\bbP[ \{\bX_1 \to \bX_2\}]\bigr)$. For general channels and i.i.d.~ensembles such that
	\begin{gather}
	\frac{M_n}{\gamma^3}\bbE[|Y_{12}|^3] \to 0 \label{c:cond1}\\
	\frac{M_n}{\gamma^4}\bbE[|Y_{12}|^4] \to 0 \label{c:cond2}
	\end{gather}
as $n\to\infty$, we have that
	\begin{align}
	\frac{P_{\rme}^{\rm ub}(\Cn)-\bbE[P_{\rme}^{\rm ub}(\Cn)]}{\sqrt{\var(P_{\rme}^{\rm ub}(\Cn))}} \dto \calN(0,1) \label{rever1b}.
	\end{align}
\end{theorem}
\begin{IEEEproof}
Sec.~\ref{sub:p11}.
\end{IEEEproof}

Despite the result is about an upper bound on the error probability, Theorem~\ref{thm:ub} gives sufficient conditions for convergence in probability of the probability of error, while Theorem~\ref{lem:d_nonconv} gives a sufficient condition that prevents this to happen. This implies that for all codes and channels such that the two conditions \eqref{c:cond1}, \eqref{c:cond2} hold, the condition $\EtrcQ>\ErceQ$ cannot be satisfied.

To obtain a similar result for the exact error probability $\PeCn$, that is a sum of all-dependent random variables, we note that a simple application of H\"older's inequality suggests that~ \cite[Th.~3.6]{Nathan2011a} is too loose. While other Stein method-based approaches such as exchangeable pairs could be applied, it is actually very challenging to find a partner for the error probability. In addition, the error exponent $E_n({\cal C}_n)$ is \emph{not} even a sum of random variables.

In the last result, we develop a general condition for the convergence in distribution of a random variable sequence to the standard normal random variable based on Stein's method. Our method is based on a modification of the Wasserstein metric as in the proof of Theorem \ref{lem:ct2}, and requires channels and decoding rules to satisfy certain conditions. It is open to find which specific channels and (random) codebook ensembles such that these conditions hold.
\begin{theorem} \label{mainthm4} Let $g_n: [0,1]\to \bbR$ be an arbitrary sequence of functions. For general channels and random codebook ensembles, under the condition that 
	\begin{align}
	\bbE\bigg[\bigg|\frac{g_n(\PeCn)-\bbE[g_n(\PeCn)]}{\sqrt{\var(g_n(\PeCn))}}\bigg|\bigg] \to 0 \label{ass1},
	\end{align}  
	and
	\begin{align}
 \bbE\bigg[\bigg|\bigg(\frac{g_n(\PeCn)-\bbE[g_n(\PeCn)]}{\sqrt{\var(g_n(\PeCn))}}\bigg)^2-1\bigg|\bigg] \to 0 \label{ass2},
	\end{align}
	the following holds:
	\begin{align}
	\frac{g_n(\PeCn)-\bbE[g_n(\PeCn)]}{\sqrt{\var(g_n(\PeCn))}} \dto \calN(0,1) \label{rever}.
	\end{align}
\end{theorem}
\begin{IEEEproof}
Sec.~\ref{sub:p12}.
\end{IEEEproof}
Summarizing, in Sec.~\ref{sec:dmc} and Sec.~\ref{sec:general} we discussed a number of convergence results of the error probability $P_{\rm e}({\cal C}_n)$ and the error exponent $E_n({\cal C}_n)$, important performance metrics of random codes used in the study of reliable communication. In the next section, we present the proofs for such results.
\section{Proofs of Theorems} \label{sec:proof}
We begin by introducing some definitions used in the Stein's method \cite{Nathan2011a}. We also introduce a novel, modified Wasserstein metric that is used throughout the section.
\begin{definition} \label{def1} Define
	\begin{align}
	\calV =  \bigg\{h:\bbR\to \bbR_+: h(u)=\begin{cases}c,& x\leq a\\ 0,&x\geq a+c\\ \mbox{linear between $a$ and $a+c$},&\text{otherwise} \end{cases},\quad \mbox{for some} \quad c>0, a \in \bbR \bigg\}.
	\end{align}
\end{definition}
\begin{definition}[Probability metrics] \label{steindef} For two random variables $X$ and $Y$, the probability metrics have the following form:
	\begin{gather}
	d_{\calH}(X,Y)=\sup_{h \in \calH}\big|\bbE[h(X)]-\bbE[h(Y)]\big|,\\
	\bard_{\calH}(X,Y)=\sup_{h \in \calH}\min\bigg\{\big|\bbE[h(X)]-\bbE[h(Y)]\big|,\big|\bbE[h(-X)]-\bbE[h(Y)]\big|\bigg\}  \label{eqsec2:1},
	\end{gather} where $\calH$ is some family of ``test" functions on $\bbR$.
\end{definition} 
We now details examples of metrics of this form along with some useful properties and relations:
\begin{itemize}
	\item By taking $\calH=\{\bone\{ \cdot \leq u\}\: u \in \bbR\}$ in \eqref{eqsec2:1} and the probability metric $d_{\calH}(X,Y)$, we obtain the Kolmogorov metric, which denote by $d_K$. By definition, the convergence in the Kolmogorov metric means the convergence in distribution.
	\item By taking $\calH=\{h:\bbR\to \bbR:|h(u)-h(v)|\leq |u-v| \}$ and the probability metric $d_{\calH}(X,Y)$, we obtain the Wasserstein metric, which we denote $d_{W}$.
	\item By taking $\calH=\{h\in \calV: c\leq 4 \sqrt{2\pi} \}$ and the probability metric $d_{\calH}(X,Y)$, we obtain a slightly modified Wasserstein metric $\tild_{W,\rm{mod}}$.
	\item By taking $\calH=\{h\in \calV: c\leq 4 \sqrt{2\pi} \}$ and the probability metric $\bard_{\calH}(X,Y)$, we obtain a modified Wasserstein metric\footnote{This definition of Wasserstein metric is a variant of the definition in \cite{Nathan2011a}, where we constraint the set $\calH$ to achieve a tighter bound.}, which we denote $d_{W,\rm{mod}}$.
\end{itemize}

The following auxiliary lemma is also very important in deriving the convergence in distributions in most of lemmas and theorems in this paper, whose proof can be found in the Appendix~\ref{ap:proof24}.
\begin{lemma} \label{lem:aux2021} Assume that $U_n \dto U$ as $n\to \infty$ for some random variable $U$, and $\bbE[U_n]=0$ and $\var(U_n)=1$. Then, it holds that
	\begin{align}
	\bbE[U]=0, \qquad \var(U)=1.
	\end{align}
\end{lemma}
\begin{IEEEproof}
	Appendix~\ref{ap:proof24}.	
\end{IEEEproof}
\subsection{Proof of Theorem~\ref{thm:main}}
\label{sub:p1}
We start by stating De Caen's inequality.

\begin{lemma}[De Caen \cite{Caenlem}]  \label{DECAENLB}  Let $\{A_i\}_{i \in \calI}$ be finite family of events in a probability space $(\Omega,\bbP)$. Then\footnote{We make the convention $\frac{0}{0}=0$, so that events of probability zero are not counted in \eqref{caenineq}.} 
	\begin{align}
	\bbP\bigg(\bigcup_{i \in \calI}A_i\bigg)\geq \sum_{i \in \calI} \frac{\bbP^2(A_i)}{\sum_{j \in \calI}\bbP[A_i \cap A_j]} \label{caenineq}
	\end{align}
\end{lemma}

\begin{lemma} \label{lem:bestcha} Under the condition that $\EtrcQ=\ErceQ$ and for $\lambda>0$, the following holds for i.i.d. and constant-composition ensembles.
	\begin{align}
	\lim_{n\to \infty}\bbE\big[\PeCnQ^{\frac{\lambda}{n}}\big]=2^{-\lambda \EtrcQ}.
	\end{align}
\end{lemma}
\begin{IEEEproof} 
Appendix \ref{app:proof_rce_trc}.
\end{IEEEproof}

\subsubsection{Proof of Theorem \ref{thm:main} for $\EtrcQ=\ErceQ$} \label{sub:sec1}

Let
\begin{align}
\varphi(\lambda) = 2^{-\lambda \EtrcQ}
\end{align} for all $\lambda>0$, which is the Laplace transform of the constant random variable $-\EtrcQ$. Let
$\varphi_n(\lambda)$ be the Laplace transform of the distribution of $\frac1n\log \PeCnQ$. Then, we have
\begin{align}
\varphi_n(\lambda)&=\bbE\bigg[2^{\lambda \frac1n\log \PeCnQ }\bigg]\\
&=\bbE\big[\PeCnQ^{\frac{\lambda}{n}}\big] \label{F1b}.
\end{align}
Then, by Lemma \ref{lem:bestcha}, the following holds:
\begin{align}
\lim_{n\to \infty }\varphi_n(\lambda)= 2^{-\lambda \EtrcQ}\label{T1b}.
\end{align} 
Applying the Levy's continuity theorem \cite[Sec.~XIII.1]{feller}, we obtain from \eqref{T1b} that
\begin{align}
-\frac1n\log \PeCnQ \dto \EtrcQ.
\end{align}
However, we know that the convergence in distribution to a constant implies convergence in probability, i.\,e.
\begin{align}
-\frac1n\log \PeCnQ \pto \EtrcQ \label{phub}.
\end{align}	

\subsubsection{Proof of Theorem \ref{thm:main} for $\EtrcQ>\ErceQ$}

This section is devoted to the proof \eqref{biscuitge} for the range of rates for which $\EtrcQ>\ErceQ$. We first need some definitions and lemmas. For this range of rates, the proof uses the union bound to the error probability~\eqref{eq:Pe} and shows that it gives a good estimate of the probability of error. The union bound is given by, 
\begin{equation}
P_{\rme}(\mathscr{c}_n)\leq P_{\rme}^{\rm ub}(\mathscr{c}_n),
\label{eq:UB}
\end{equation}
where $P_{\rme}^{\rm ub}(\mathscr{c}_n)$	
is defined in~\eqref{defRBE}, and we define its finite-length error exponent as
\begin{equation}
E_n^{\rm ub}(\mathscr{c}_n)  =  -\frac{1}{n}\log P_{\rme}^{\rm ub}(\mathscr{c}_n).
\label{eq:Etrcn}
\end{equation}
We denote by $\EtrcQ$ and $\ErceQ$ respectively the typical error and the random coding error exponents for the fixed underlying distribution $Q$, and we define
\begin{align}
d_{\rm B}(x,x')=-\log\bigg(\sum_{y} \sqrt{W(y|x)W(y|x')}\bigg)
\end{align}
to be the Bhattacharyya distance between symbols $x,x'\in\calX$. 

We assume that the DMC is such that
\begin{align}
0<D_{\rmb} =   \max_{x,x'}d_{\rm B}(x,x') <\infty \label{cond0},
\end{align}
that is, we leave the cases where $W(y|x)W(y|x')=0$ for for some $x$ and $x'$ and all $y$ beyond the scope of the paper. This case would correspond to a positive zero-error capacity, where some symbols cannot be confused at the decoder.

First, we introduce some auxiliary results about the exponential decay f the pairwise error probability between two codewords, using the method of types. We let $\calP_n(\calX \times \calX)$ be the set of all joint types on $\calX \times \calX$, and $\calP(\calX \times \calX)$ be the set of all possible probability distributions on $\calX \times \calX$.

\begin{lemma} \label{lem:GLem} 
For $R<R_{\rm{crit}}$, the pairwise codeword error probability between two codewords $\bx_i,\bx_j$ given their joint type $P_{XX'}$ satisfies
	\begin{align}
	\bbP[\bx_i \to \bx_j | P_{XX'}]=g_n(P_{XX'})\doteq 2^{-n \sum_{x,x'} d_{\rm B}(x,x') P_{XX'}(x,x')} \label{tfact1}
	\end{align} for some function $g_n: \calP_n(\calX \times \calX) \to [0,1]$.
\end{lemma}
\begin{IEEEproof}
Appendix~\ref{ap:proof5}	
\end{IEEEproof}

\begin{lemma} \label{lem:Vtypi} Given an i.i.d. random codebook ensemble. For each $P_{XX'} \in \calP_n(\calX \times \calX)$, let $\calN(P_{XX'})$ be the number of codeword pairs in a specific code such that their joint type is $P_{XX'}$. Let $Q_X=Q_X'=Q$. Define
	\begin{align}
	\calV_n=\bigg\{\calN(P_{XX'})=0,\quad \forall P_{XX'}\in \calP_n(\calX \times \calX): D(P_{XX'}\|Q_X Q_X')> 2R\bigg\}
	\end{align} 
which is the event that the (random) number of pairs $(i,j) \in [M_n] \times [M_n]$ such that $i\neq j$ and $(\bX_i,\bX_j) \in \calT(P_{XX'})$ is equal to zero for each $n$-joint type $P_{XX'}$ with $D(P_{XX'}\|Q_X Q_X')> 2R$. Then, we have
\begin{align}
	\bbP\big[\calV_n^c\big]\leq 2^{-n \alpha(R)}
	\end{align} for some $\alpha(R)>0$ for all $R\geq 0$.
\end{lemma}
\begin{IEEEproof}
Appendix~\ref{ap:proof6}	
\end{IEEEproof}

\begin{lemma} \label{lem:aux} 
Assume that $R>0$. Take an arbitrary $\nu\geq 0$ such that $\nu\leq 2R$. Let
	$Q_X=Q_{X'}=Q$ and
	\begin{align}
	\calP =  \bigg\{P_{XX'} \in \calP(\calX\times \calX): D(P_{XX'}\|Q_XQ_X')\leq 2R-\nu\bigg\}
	\end{align}
	and
	\begin{align}
	D_n& =  \frac{1}{M}\sum_{P_{XX'}\in \calP} \calN(P_{XX'}) g_n(P_{XX'}),\\
	\Etrcbar(\nu,R,Q)& =  \begin{cases} R+\sum_{x,x'} d_{\rmB}(x,x') P_{XX'}^*(x,x'), \qquad \mbox{if} \qquad D(P_{XX'}^*\|Q_X Q_X')=2R-\nu,\\
	\ErceQ, \qquad \mbox{otherwise}\end{cases}.\label{eq:etrcbar}
	\end{align} where $P_{XX'}^*$ is an optimizer of $\min_{P_{XX'}\in \calP} D(P_{XX'}\|Q_X Q_{X'})
	+\sum_{x,x'}d_{\rmB}(x,x') P_{XX'}(x,x')-R$. Then,
	the following holds:
	\begin{align}
	\bbE[D_n]\doteq  2^{-n \barE_{\rm{trc}}(\nu,R)}
	\label{eq167a}
	\end{align}
	and
	\begin{align}
	\frac{\var(D_n)}{\big(\bbE[D_n]\big)^2} \dotleq 2^{-n\zeta(\nu,R)}
	\end{align}  for some constant $\zeta(\nu,R)$ such that $\zeta(\nu,R)>0$ if $0<\nu\leq 2R$.
\end{lemma}
\begin{IEEEproof}
Appendix~\ref{ap:proof7}	
\end{IEEEproof}

\begin{lemma} \label{lem:tmc} 
Let
	\begin{align}
	\Etrcub(R,Q) =  \lim_{n\to \infty}-\frac1n \bbE[\log P_{\rme}^{\rm ub}(\Cn)].
	\end{align} Then, for $0< R< R_{\rm{crit}}(Q)$, the following holds:
	\begin{align}
	\Etrcub(R,Q)&=\min_{P_{XX'} \in \calP(\calX \times \calX): D(P_{XX'}\|Q_XQ_X')\leq 2R} D(P_{XX'}\|Q_X Q_X')+\sum_{x,x'} d_{\rmB}(x,x') P_{XX'}(x,x')-R \label{eqn:key0}\\
	&=\Etrcbar(0,R,Q) \label{eqn:key1},
	\end{align} where $\Etrcbar$ is defined in \eqref{eq:etrcbar}, Lemma \ref{lem:aux}.
\end{lemma}
\begin{IEEEproof}
Appendix~\ref{ap:proof8}	
\end{IEEEproof}

\begin{lemma} \label{lem:sup} For the range of rates  $0\leq R< R_{\rm{crit}}(Q)$, any $\eps>0$ and for some $\kappa>0$, it holds that
	\begin{align}
	&\bbP\bigg[P_{\rme}^{\rm ub}(\Cn)>\frac{1}{2}2^{-n(\Etrcub(R,Q)-\eps)}\bigg] + \bbP\Big[P_{\rme}^{\rm ub}(\Cn)< 2^{-n(\Etrcub(R,Q)+\eps)}\Big]\leq \frac{1}{n^{1+\kappa}}.
	\end{align} 
\end{lemma}
\begin{IEEEproof}
Appendix~\ref{ap:proof9}	
\end{IEEEproof}

\begin{lemma} \label{caenlem} 
	For all rate $R$ such that $0< R< R_{\rm{crit}}(Q)$ and for some $\delta(R)>0$, it holds that
	\begin{align}
	0\leq \frac{\bbE[P_{\rme}^{\rm ub}(\Cn)]}{\bbE[\PeCn]}- 1 \leq  2^{-n\big(\delta(R)+\Etrcub(R,Q)- \ErceQ\big)}. \label{keyfoafix}
	\end{align}
\end{lemma}
\begin{IEEEproof}
Appendix~\ref{ap:proof10}	
\end{IEEEproof}

We are now equipped to prove Theorem \ref{thm:main} by observing that for any $\varepsilon>0$, the convergence in probability of $\ECn$ to $E_{\rm trc}(R,Q)$ can be written and upper bounded as
\begin{align}
\bbP\big[|\ECn -\Etrc|>3 \eps\big]&\leq \underbrace{\bbP\Big[\big|\ECn  - E_n^{\rm ub}(\Cn)\big|>\eps\Big]}_{\alpha_n} \nn\\
&\quad\quad+ \underbrace{\bbP\bigg[\bigg| E_n^{\rm ub}(\Cn) - \left(-\frac1n\bbE\big[\log P_{\rme}^{\rm ub}(\Cn)\big]\right)\bigg|>\eps\bigg]}_{\beta_n}\nn\\
&\quad\quad + \underbrace{\bbP\bigg[\bigg|\left(-\frac1n\bbE\big[\log P_{\rme}^{\rm ub}(\Cn)\big]\right)-\EtrcQ\bigg|>\eps\bigg]}_{\gamma_n} \label{eq46eq}.
\end{align}
We next show that the terms $\alpha_n$, $\beta_n$ and $\gamma_n$  in \eqref{eq46eq} tend to zero as $n\to \infty$, implying the concentration result in~\eqref{biscuitge}.

\subsubsection{First term of \eqref{eq46eq}}
\label{sec:step1}
The term $\alpha_n$ quantifies the deviation of the error exponent of the error probability~\eqref{eq:en} with that of the union bound~\eqref{defRBE}. By the symmetry of the pairwise-independent i.i.d. random-coding ensemble, for any pair of codewords $\bX_i$ and $\bX_j$ with $i\neq j$  we have that
\begin{align}
\bbE\big[\bbP[\bX_i \to \bX_j]\big]=\bbE\big[\bbP[\bX_1 \to \bX_2]\big].\label{keycondition}
\end{align}
Similarly, for any triplet of codewords $\bX_i$, $\bX_j$ and $\bX_k$ with $j, k \neq i$ and $j\neq k$, it holds that
\begin{align}
\bbE &\Big[\bbP\big[\{\bX_i \to \bX_j\}\cap \{\bX_i \to \bX_k\}\big]\Big]= \bbE \Big[\bbP\big[\{\bX_1 \to \bX_2\}\cap \{\bX_1 \to \bX_3\}\big]\Big]
\label{eq3a}
\end{align}
where in both~\eqref{keycondition} and~\eqref{eq3a}, the expectations are calculated with respect to the i.i.d. ensemble codeword distribution $Q^n(\bx) = \prod_{k=1}^nQ(x_k)$, where $Q(x)$ is the single-letter input distribution. We next provide separate convergence of $\alpha_n$ for $R=0$ and for $0<R<R_{\rm crit}(Q)$. 

For the case of $R=0$, we first observe that the union bound~\eqref{defRBE} can be bounded from above as
\begin{align}
P_{\rme}^{\rm ub}(\Cn)&=\frac{1}{M_n}\sum_{i=1}^{M_n}\sum_{j\neq i} \bbP[\bx_i \to \bx_j]\\
&\leq (M_n-1)\max_{i\neq j} \bbP[\bx_i \to \bx_j] \label{gfact1},
\end{align}
while the probability of error~\eqref{eq:Pe} can be lower bounded by
\begin{align}
\PeCn&=\frac{1}{M_n}\sum_{i=1}^{M_n}\bbP\bigg[\bigcup_{j\neq i} \{\bx_i \to \bx_j\}\bigg]\\
&\geq \frac{1}{M_n}\max_{i\neq j} \bbP[\bx_i \to \bx_j] \label{gfact2}.
\end{align}
From \eqref{gfact1} and \eqref{gfact2}, we have that the first term in the r.h.s.~of~\eqref{eq46eq} satisfies
\begin{align}
\alpha_n & =\bbP\bigg[P_{\rme}^{\rm ub}(\Cn)> 2^{n\eps }\PeCn   \bigg]\\
& \leq \bbP\bigg[(M_n-1)\max_{i\neq j} \bbP[\bX_i \to \bX_j]> 2^{n\eps }\frac{1}{M_n}\max_{i\neq j} \bbP[\bX_i \to \bX_j]  \bigg]\\
&= \bbP\bigg[(M_n-1)> 2^{n\eps }\frac{1}{M_n} \bigg].\label{eq:22}
\end{align}
Since $M_n$ is any sub-exponential sequence in $n$, the probability in~\eqref{eq:22}  vanishes as $n\to\infty$ for $\eps>0$. 

We now consider the case of $0<R<R_{\rm crit}(Q)$. We define the sequence $a_n$ as
\begin{align}
a_n = 2^{-n (\Etrcub(R,Q)+\frac\eps2)} \label{defAN}.
\end{align} 
Then, we have
		\begin{align}
		\bbP\bigg[\bigg|\ECn -E_n^{\rm ub}(\Cn)\bigg|>\eps\bigg]&=\bbP\bigg[P_{\rme}^{\rm ub}(\Cn)> 2^{\eps n}\PeCn   \bigg]\\
		&=\bbP\bigg[ P_{\rme}^{\rm ub}(\Cn)- a_n-2^{\eps n}\big(\PeCn-a_n\big)>(2^{\eps n}-1) a_n \bigg]\\
		& \leq \bbP\bigg[ P_{\rme}^{\rm ub}(\Cn)-a_n>\frac{1}{2}(2^{\eps n}-1)a_n \bigg]\nn\\
		&\qquad \qquad +\bbP\bigg[ -2^{\eps n}\big(\PeCn-a_n\big)>\frac{1}{2}(2^{\eps n}-1)a_n \bigg] \label{L1},
		\end{align} where \eqref{L1} follows from the fact that $\bbP[A+B>2C]=\bbP[\{A>C\}\cup \{B>C\}]\leq \bbP[A>C]+\bbP[B>C]$.
		
		Now, observe that
		\begin{align}
		\bbP\bigg[ P_{\rme}^{\rm ub}(\Cn)-a_n>\frac{1}{2}(2^{\eps n}-1)a_n \bigg]& =\bbP\bigg[ P_{\rme}^{\rm ub }(\Cn)>\frac{1}{2}(2^{\eps n}+1)a_n \bigg]
		\label{fromV}\\
		&=\bbP\bigg[ P_{\rme}^{\rm ub}(\Cn)>\frac{1}{2}(2^{\eps n}+1)2^{-n (\Etrcub(R,Q)+\eps/2)} \bigg]   \\
		& \leq \bbP\bigg[P_{\rme}^{\rm ub}(\Cn)> \frac{1}{2} 2^{-n (\Etrcub(R,Q)-\eps/2)} \bigg]  \label{L2}.
		\end{align} 
		
		On the other hand, we also have
		\begin{align}
		&\bbP\bigg[ -2^{\eps n}\big(\PeCn-a_n\big)>\frac{1}{2}(2^{\eps n}-1) a_n\bigg]\nn\\
		&\qquad =\bbP\bigg[2^{\eps n}(P_{\rme}^{\rm ub}(\Cn)-\PeCn) -2^{\eps n}\bigg(\PeCn-a_n\bigg)  >\frac{1}{2}(2^{\eps n}-1) a_n\bigg]\\
		&\qquad \leq \bbP\bigg[2^{\eps n}(P_{\rme}^{\rm ub}(\Cn)-\PeCn)>\frac{1}{4}(2^{\eps n}-1) a_n\bigg]\nn\\
		&\qquad \quad+ \bbP\bigg[ -2^{\eps n}\bigg(P_{\rme}^{\rm ub}(\Cn)-a_n\bigg)>\frac{1}{4}(2^{\eps n}-1)a_n\bigg].
		\label{L3}
		\end{align}
Now, we know that
		\begin{align}
		\bbP\bigg[ -2^{\eps n}\bigg(P_{\rme}^{\rm ub}(\Cn)- a_n\bigg)>\frac{1}{4}(2^{\eps n}-1) a_n\bigg]&=\bbP\bigg[P_{\rme}^{\rm ub}(\Cn)<\bigg(1-\frac{1}{4}\bigg(\frac{2^{\eps n}-1}{2^{\eps n}}\bigg)\bigg)a_n \bigg]\\
		&=\bbP\bigg[P_{\rme}^{\rm ub}(\Cn)<\bigg(1-\frac{1}{4}\bigg(\frac{2^{\eps n}-1}{2^{\eps n}}\bigg)\bigg)2^{-n (\Etrcub(R,Q)+\eps/2)}\bigg]\\
		& \leq \bbP\bigg[P_{\rme}^{\rm ub}(\Cn)<2^{-n (\Etrcub(R,Q)+\eps/2)}\bigg]\label{L3a}.
		\end{align} 	
In addition, we also have
		\begin{align}
		\bbP\bigg[2^{\eps n}(P_{\rme}^{\rm ub}(\Cn)-\PeCn)>\frac{1}{4}(2^{\eps n}-1) a_n\bigg]& \dotleq a_n^{-1}\bbE[P_{\rme}^{\rm ub}(\Cn)-\PeCn] \label{layka}\\
		& = 2^{(\Etrcub(R,Q)+\eps/2) n} \bbE[P_{\rme}^{\rm ub}(\Cn)-\PeCn] \label{layka2},
		\end{align} where \eqref{layka} follows from
		$P_{\rme}^{\rm ub}(\Cn)\geq \PeCn$ and Markov's inequality, and \eqref{layka2} follows from \eqref{defAN}.
		
		Now, for $R>0$ and $\Etrcub(R,Q)>\ErceQ$, from Lemma \ref{caenlem}, we have
		\begin{align}
		\bbE[P_{\rme}^{\rm ub}(\Cn)-\PeCn]
		&=\bbE[\PeCn]\bigg(\frac{\bbE[P_{\rme}^{\rm ub}(\Cn)]}{\bbE[\PeCn]}-1\bigg)\\
		& \dotleq 2^{-n\ErceQ }  \bigg(2^{-n\big(\delta(R)+ \Etrcub(R,Q)-\ErceQ\big)}\bigg) \label{cute2}.
		\end{align}
		
From \eqref{layka2} and \eqref{cute2}, we obtain
		\begin{align}
		\bbP\bigg[2^{\eps n}(P_{\rme}^{\rm ub}(\Cn)-\PeCn)>\frac{1}{4}(2^{\eps n}-1) a_n\bigg]& \dotleq 2^{ \big(\Etrcub(R,Q)+\eps/2\big) n} 2^{-n\ErceQ  }  \bigg[2^{-n\big(\delta(R)+ \Etrcub(R,Q)-\ErceQ\big)}\bigg]\\
		& \dotleq 2^{-n \big(\delta(R)-\eps/2\big)} \label{L3b}.
		\end{align} 
Hence, from \eqref{L3}, \eqref{L3a}, and \eqref{L3b}, we have
		\begin{align}
	\bbP\bigg[ -2^{\eps n}\big(\PeCn-a_n\big)>\frac{1}{2}(2^{\eps n}-1) a_n\bigg]& \leq \bbP\bigg[P_{\rme}^{\rm ub}(\Cn)<2^{-N (\Etrcub(R,Q)+\eps/2)}\bigg]+   2^{-n \big(\delta(R)-\eps/2\big)} \label{champa1}.
		\end{align}
From \eqref{L1}, \eqref{L2}, and \eqref{champa1}, we have
		\begin{align}
		&\bbP\bigg[\bigg|-\frac1n\log \PeCn -\frac{-\log P_{\rme}^{\rm ub} (\Cn)}{n}\bigg|>\eps\bigg]\nn\\
		&\qquad \leq  \bbP\bigg[ P_{\rme}^{\rm ub}(\Cn)> \frac{1}{2} 2^{-n (\Etrcub(R,Q)-\eps/2)} \bigg] +  \bbP\bigg[P_{\rme}^{\rm ub}(\Cn)<2^{-n (\Etrcub(R,Q)+\eps/2)}\bigg]+  2^{-n \big(\delta(R)-\eps/2\big)} \\
		&\qquad \leq  \frac{1}{n^{1+\beta}} +  2^{-n \big(\delta(R)-\eps/2\big)}                  \label{suple}\\ 
		&\qquad \to 0 \label{champa2},
		\end{align} for any $0<\eps<2 \delta(R)$, where \eqref{suple} follows from Lemma \ref{lem:sup} with $\beta$ being a positive constant. Since $\bbP\big[\big|\ECn -E_n^{\rm ub}(\Cn)\big|>\eps\big]$ is a non-increasing function in $\eps$, \eqref{champa2} must hold for all $\eps>0$.
		
		Furthermore, since $\bbP\big[\big|\ECn -E_n^{\rm ub}(\Cn)\big|>\eps\big]$ is a non-increasing function in $\eps$, for any $\eps> 0$, there exists an  $\eps_0 \in (0,2 \delta(R))$ such that 
		\begin{align}
		\bbP\big[\big|\ECn -E_n^{\rm ub}(\Cn)\big|>\eps\big] \leq \frac{1}{n^{1+\beta}} +  2^{-n \big(\delta(R)-\eps_0/2\big)}  \label{cuchi}
		\end{align} for some $\eps_0 \in (0,2 \delta(R))$. It follows from \eqref{cuchi} that
		\begin{align}
		\sum_{n=1}^{\infty }\bbP\big[\big|\ECn -E_n^{\rm ub}(\Cn)\big|>\eps\big]<\infty.
		\end{align}
		Hence, by Borel-Cantelli's lemma \cite{Billingsley}, we have
		\begin{align}
		\ECn -E_n^{\rm ub}(\Cn) \asto 0,
		\label{tesg}
		\end{align}
		where $\asto$ denotes almost sure convergence as $n\to\infty$, that is, a sequence of random variables $\{A_n\}_{n=1}^{\infty}$ converge almost surely to $A$ if
		\begin{equation}
			\mathbb{P}\left[ \lim_{n\to\infty} A_n = A  \right] = 1.
		\end{equation}
		On the other hand, observe that
		\begin{align}
		\big|\ECn -E_n^{\rm ub}(\Cn)\big|&\leq - \frac{2 \log \PeCn}{n} \\
		&\leq 2 E_{\rm{sp}}(R),
		\label{spac}
		\end{align} where \eqref{spac} follows from the fact that the error exponent of any sufficiently long code is upper bounded by the sphere-packing bound. 
				
		Hence, from \eqref{tesg} and \eqref{spac}, by the bounded convergence theorem, it holds that
		\begin{align}
		\lim_{n\to \infty} \bbE\big[\ECn -E_n^{\rm ub}(\Cn)\big]=0. 
		\end{align}
	This means that
		\begin{align}
		\EtrcQ&=\lim_{n\to \infty}\ECn\\
		&=\lim_{n\to \infty} E_n^{\rm ub}(\Cn)\\
		&=\Etrcub(R,Q) \label{keypo}.
		\end{align}
	
\subsubsection{Second term of \eqref{eq46eq}}
Using Chebyshev's inequality, we have
		\begin{align}
		&\bbP\bigg[\bigg| E_n^{\rm ub}(\Cn) - \left(-\frac1n\bbE\big[\log P_{\rme}^{\rm ub}(\Cn)\big]\right)\bigg|>\eps\bigg]\nn\\
		&\qquad \leq \frac{1}{\eps^2}\var\bigg(-\frac{\log P_{\rme}^{\rm ub}(\Cn)}{n}\bigg)\\
		&\qquad =\frac{1}{n^2 \eps^2} \var\bigg(-\log P_{\rme}^{\rm ub}(\Cn)\bigg)\\
		& \qquad =\frac{1}{n^2 \eps^2}\bbE\bigg[\bigg(-\log(M_n-1)-\log\bigg(\frac{P_{\rme}^{\rm ub}(\Cn)}{M_n-1} \bigg)\bigg)^2\bigg]-\frac{1}{\eps^2}\bigg(\frac{\bbE\big[-\log P_{\rme}^{\rm ub}(\Cn)\big]}{n}\bigg)^2 \label{culi2}.
		\end{align}
		
		Now, define
		\begin{align}
		\xi(p,n,R) =  2^{-n(\Etrcub(R,Q)+R)}  \label{defzeta}.
		\end{align}
		
		From \eqref{culi2}, we obtain
		\begin{align}
		&\bbP\bigg[\bigg| E_n^{\rm ub}(\Cn) - \left(-\frac1n\bbE\big[\log P_{\rme}^{\rm ub}(\Cn)\big]\right)\bigg|>\eps\bigg]\nn\\
		&\qquad \leq \frac{1}{n^2 \eps^2}\bbE\bigg[\bigg(-\log(M_n-1)-\log\xi(p,n,R)-\log\bigg(\frac{P_{\rme}^{\rm ub}(\Cn)}{(M_n-1)\xi(p,n,R)} \bigg)\bigg)^2\bigg]\nn\\
		&\qquad \qquad -\frac{1}{\eps^2}\bigg(\frac{\bbE\big[-\log P_{\rme}^{\rm ub}(\Cn)\big]}{n}\bigg)^2
		\label{culi4}.
		\end{align}
		By Lemma \ref{lem:tmc}, we know that
		\begin{align}
		\lim_{n\to \infty} \frac{\bbE\big[-\log P_{\rme}^{\rm ub}(\Cn)\big]}{n} =\Etrcub(R,Q),
		\end{align}
		hence, it holds that
		\begin{align}
		&\limsup_{n\to \infty} \bbP\bigg[\bigg| E_n^{\rm ub}(\Cn) - \left(-\frac1n\bbE\big[\log P_{\rme}^{\rm ub}(\Cn)\big]\right)\bigg|>\eps\bigg]\nn\\ &\qquad =\limsup_{n\to \infty}\frac{1}{\eps^2}\bbE\bigg[\bigg(\Etrcub(R,Q)-\frac{1}{n} \log \bigg(\frac{P_{\rme}^{\rm ub}(\Cn)}{(M_n-1)\xi(p,n,R)}\bigg) \bigg)^2\bigg]-\frac{\big(\Etrcub(R,Q)\big)^2}{\eps^2}\\
		&\qquad \leq \frac{1}{\eps^2} \bigg((\Etrcub(R,Q))^2- 2\Etrcub(R,Q)\liminf_{n\to \infty} \bbE\bigg[\frac{1}{n}\log \bigg(\frac{P_{\rme}^{\rm ub} (\Cn)}{(M_n-1)\xi(p,n,R)}\bigg)\bigg]\nn\\
		&\qquad \qquad+\limsup_{n\to \infty}\bbE\bigg[\bigg(\frac{1}{n}\log \bigg(\frac{P_{\rme}^{\rm ub}(\Cn)}{(M_n-1)\xi(p,n,R)}\bigg)\bigg)^2\bigg]\bigg)-\frac{\big(\Etrcub(R,Q)\big)^2 }{\eps^2} \label{keypair},
		\end{align}
where \eqref{keypair} follows from the sub-additivity of $\limsup$.
Now, we need to estimate
		\begin{align*}
		\liminf_{n\to \infty} \bbE\bigg[\frac{1}{n}\log \bigg(\frac{P_{\rme}^{\rm ub}(\Cn)}{(M_n-1)\xi(p,n,R)}\bigg)\bigg]
		\end{align*}
		and
		\begin{align*}
		\limsup_{n\to \infty}\bbE\bigg[\bigg(\frac{1}{n}\log \bigg(\frac{P_{\rme}^{\rm ub}(\Cn)}{(M_n-1)\xi(p,n,R)}\bigg)\bigg)^2\bigg].
		\end{align*}
		
First, we show that
		\begin{align}
		\frac{1}{n}\log \bigg(\frac{P_{\rme}^{\rm ub}(\Cn)}{(M-1)\xi(p,n,R)}\bigg) \asto 0  \label{keyfact}
		\end{align}
		
Indeed, take an arbitrary $\nu>0$ and observe that
		\begin{align}
		&\bbP\bigg[\bigg|\frac{1}{n}\log \bigg(\frac{P_{\rme}^{\rm ub} (\Cn)}{(M_n-1)\xi(p,n,R)}\bigg)\bigg| >\nu\bigg]\nn\\
		& \qquad =\bbP\bigg[\bigg|\log \bigg(\frac{P_{\rme}^{\rm ub} (\Cn)}{(M_n-1)\xi(p,n,R)}\bigg)\bigg| >n\nu\bigg]\\
		&\qquad =\bbP\bigg[\bigg|-\log \xi(p,n,R)+ \log\bigg(\frac{P_{\rme}^{\rm ub} (\Cn)}{M_n-1} \bigg)\bigg| >n\nu\bigg]\\
		&\qquad =\bbP\bigg[\bigg|n(\Etrcub(R,Q)+R)+ \log\bigg(\frac{P_{\rme}^{\rm ub}(\Cn)}{M_n-1} \bigg)\bigg| >n\nu\bigg]\\
		&\qquad = \bbP\bigg[\frac{P_{\rme}^{\rm ub}(\Cn)}{M_n-1}> 2^{-n(\Etrcub(R,Q)+R-\nu)}\bigg] +\bbP\bigg[\frac{P_{\rme}^{\rm ub}(\Cn)}{M_n-1}< 2^{-N(\Etrcub(R,Q)+R+\nu)}\bigg]\\
		&\qquad \leq \bbP\bigg[P_{\rme}^{\rm ub}(\Cn)>  \frac{1}{2} 2^{-n(\Etrcub(R,Q)-\nu)}\bigg] + \bbP\bigg[P_{\rme}^{\rm ub}(\Cn)< 2^{-n(\Etrcub(R,Q)+\nu)}\bigg]\label{key2x}\\
		&\qquad \leq   \frac{1}{n^{1+\beta}} \label{key2},
		\end{align} for some constants $\beta>0$, where \eqref{key2} follows from Lemma \ref{lem:sup}.
		
		From \eqref{key2}, we obtain
		\begin{align}
		\sum_{n=1}^{\infty}\bbP\bigg[\bigg|\frac{1}{n}\log \bigg(\frac{P_{\rme}^{\rm ub}(\Cn)}{(M_n-1)\xi(p,n,R)}\bigg)\bigg| >\nu\bigg]
		< \sum_{n=1}^{\infty}\frac{1}{n^{1+\beta}} <\infty \label{eq101}
		\end{align}  by using D'Alembert criterion. 
		
		This means that \eqref{keyfact} holds, or
		\begin{align}
		\frac{1}{n}\log \bigg(\frac{P_{\rme}^{\rm ub}(\Cn)}{(M_n-1)\xi(p,n,R)}\bigg) \asto 0  \label{bunhi1}
		\end{align} by Borel-Cantelli lemma \cite{Billingsley}.
		
		Now, since $0\leq \bbP(\bX_i \to \bX_j)\leq 1$ for all $i,j \in [M]: i\neq j$, it holds that
		\begin{align}
		\frac{1}{n}\log \bigg(\frac{P_{\rme}^{\rm ub}(\Cn)}{(M_n-1)\xi(p,n,R)}\bigg)
		&=\frac{1}{n}\log\bigg(\frac{1}{M_n(M_n-1)\xi(p,n,R)}\sum_{i\neq j} \bbP(\bX_i\to \bX_j) \bigg)\\
		&\leq \frac{1}{n}\log\bigg(\frac{1}{\xi(p,n,R)}\bigg)\\
		&\leq \Etrcub(R,Q)+R \label{bes1},
		\end{align} where \eqref{bes1} follows from \eqref{defzeta}.
On the other hand, from the sphere-packing bound \footnote{In case that the sphere packing bound diverges, we can use $E_{\rm ex}(R=0)$ as an upper bound, which is finite at $R=0$ unless the zero error capacity $C_0>0$.}, it holds almost surely that
		\begin{align}
		\frac{1}{n}\log \bigg(\frac{P_{\rme}^{\rm ub} (\Cn)}{(M_n-1)\xi(p,n,R)}\bigg)&\geq\frac{1}{n}\log \bigg(\frac{\PeCn}{(M_n-1)\xi(p,n,R)}\bigg)\\
		& \dotgeq \frac{1}{n}\log \bigg(\frac{2^{-n E_{\rm{sp}}(R)}}{(M_n-1)\xi(p,n,R)}\bigg) \label{tax10}\\
		&=-E_{\rm{sp}}(R)-R+ \big(\Etrcub(R,Q)+R\big) \label{tax11} \\
		&= \Etrcub(R,Q)-E_{\rm{sp}}(R) \label{bes10},
		\end{align}
		where \eqref{tax10} follows from the sphere-packing bound \cite{MoserBook}, 
 and \eqref{tax11} follows from \eqref{defzeta} and $M_n\doteq 2^{nR}$.
		
From \eqref{bes1} and \eqref{bes10}, $\frac{1}{n}\log \bigg(\frac{P_{\rme}^{\rm ub}(\Cn)}{(M_n-1)\xi(p,n,R)}\bigg)$ is bounded (both below and above).
		
Hence, by the bounded convergence theorem and the continuous mapping theorem \cite{Billingsley},  it holds that
		\begin{align}
		\bbE\bigg[\frac{1}{n}\log \bigg(\frac{P_{\rme}^{\rm ub}(\Cn)}{(M_n-1)\xi(p,n,R)}\bigg)\bigg] &\to 0  \label{facto2a},\\
		\bbE\bigg[\bigg(\frac{1}{n}\log \bigg(\frac{P_{\rme}^{\rm ub}(\Cn)}{(M_n-1)\xi(p,n,R)}\bigg)\bigg)^2\bigg] &\to 0 \label{facto3a}.
		\end{align}
		From \eqref{keypair}, \eqref{facto2a}, and \eqref{facto3a}, we finally have
		\begin{align}
		\limsup_{n\to \infty} \bbP\bigg[\bigg| E_n^{\rm ub}(\Cn) - \left(-\frac1n\bbE\big[\log P_{\rme}^{\rm ub}(\Cn)\big]\right)\bigg|>\eps\bigg] = 0 \label{facfinal}
		\end{align} for any arbitrary $\eps>0$. This leads to 
		\begin{align}
		\lim_{n\to \infty} \bbP\bigg[\bigg| E_n^{\rm ub}(\Cn) - \left(-\frac1n\bbE\big[\log P_{\rme}^{\rm ub}(\Cn)\big]\right)\bigg|>\eps\bigg] = 0
		\end{align} 
by the fact that the probability measure is bounded from below by zero. 
\subsubsection{Third term of \eqref{eq46eq}}
By Lemma \ref{lem:tmc}, it is known that
		\begin{align}
		\bbE\bigg[\frac{-\log P_{\rme}^{\rm ub}(\Cn)}{n}\bigg]\to \Etrcub(R,Q) \label{bess1}.
		\end{align}
On the other hand, from \eqref{keypo} in Step 1, we know that
		\begin{align}
		\Etrcub(R,Q)=\EtrcQ \label{bess2}.
		\end{align}
		It follows from \eqref{bess1} and \eqref{bess2} that 
	\begin{align}
		\bbP\bigg[\bigg|\bbE\bigg[\frac{-\log P_{\rme}^{\rm ub} (\Cn)}{n}\bigg]-\EtrcQ\bigg|>\eps\bigg] \to 0.
		\end{align}
In conclusion, as anticipated, the three terms of \eqref{eq46eq} tend to zero as $n\to \infty$, showing~\eqref{biscuitge} for rates below the critical rate. Together with Subsection \ref{sub:sec1}, we proved Theorem~\ref{thm:main}, which states the convergence in probability of the error exponent of the codes in the ensemble to the typical random-coding error exponent.

\subsubsection{Extension to Constant Composition Codes} \label{sec:extend}


For the constant-composition code, for all the rate $R_{\rm{crit}}(R,Q)\leq R\leq C$, the proof of Theorem \ref{thm:main} holds by using the Levy's continuity theorem since it is not hard to see that $\ErceQ=\EtrcQ$ for this case.  At all the rate $0\leq R\leq \EtrcQ$, Lemma \ref{lem:GLem} - Lemma \ref{lem:aux} still hold since  $\bone\{(\bX_i,\bX_j) \in \calT(Q_{XX'})\}$ and  $\bone\{(\bX_k,\bX_l) \in \calT(\tilQ_{XX'})\}$ are still pairwise-independent for the constant-composition code for all $\{i,j,k, l \in [M]: i\neq j, k\neq l\}$. In Lemma \ref{lem:tmc}, the typical error exponent of the union bound should be replaced by $E_{\rm{trc}}^{\rm{ub}}$ for the constant-composition code in \cite{Merhav2018a}. To show that Theorem \ref{thm:main} still holds for the constant-composition code, we need to prove that the mapping from the error probability and the union bound in Lemma \ref{lem:sup} and Lemma \ref{caenlem} still work. It is not hard to see that the proof of Lemma \ref{lem:sup} still holds for the constant-composition code since its correctness depends on  Lemma~\ref{lem:GLem}, Lemma~ \ref{lem:aux} and the fact that $\tilV_{ij}$'s are pairwise-independent where $\tilV_{ij}$ is defined \eqref{deftilVij}. Lemma \ref{caenlem} still holds for the constant-composition code, i.e.,
\begin{lemma} \label{caenlemfixed} 
For any constant-composition code with type $Q$ and for all the rate $R$ such that $0< R< R_{\rm{crit}}$, it holds that
	\begin{align}
	0\leq \frac{\bbE[P_{\rme}^{\rm ub}(\Cn)]}{\bbE[ P_{\rme}(\Cn)]}- 1 \leq  2^{-N\big(\delta(R)+\EtrcubQ- \ErceQ\big)}  \label{keyfoafixx}
	\end{align}  for some $\delta(R)>0$.
\end{lemma}
\begin{IEEEproof}
To prove Lemma \ref{caenlemfixed}, we use the same proof as Lemma \ref{caenlem} in Appendix~\ref{ap:proof10}. It is easy to check that \eqref{eq69} still holds for the constant-composition code. In addition, the pairwise error probability only depends on the joint-type of the two codewords as in the i.i.d~.case.
\end{IEEEproof}

\subsection{Proof of Theorem~\ref{asko}}
\label{sub:p2}

\emph{First, we prove \eqref{espa}.} Under the condition that $\ErceQ=\EtrcQ$, observe that
		\begin{align}
		\bbP\bigg[-\frac{1}{n} \log \PeCn <\EtrcQ-\eps\bigg]&=\bbP\bigg[ \PeCn >2^{-n(\EtrcQ-\eps)} \bigg]\\
		&\leq 2^{n(\EtrcQ-\eps)}\bbE[\PeCn] \label{balat}\\
		&\doteq 2^{n(\EtrcQ-\eps)}2^{-n\ErceQ} \label{balat2}\\
		&=2^{-n \eps} \label{balat3},
		\end{align} where \eqref{balat} follows from the Markov inequality, \eqref{balat2} follows from $\bbE\big[\PeCn\big]\doteq 2^{-n\ErceQ}$, \eqref{balat3} follows from $\ErceQ=\EtrcQ$. 
		
		Now, for any $s>0$, observe that
		\begin{align}
		\bbP\bigg[-\frac{1}{n} \log \PeCn <\EtrcQ-\eps\bigg]&=\bbP\bigg[\frac{s}{n} \log \PeCn >-s(\Etrc-\eps)\bigg]\\
		&=\bbP\bigg[2^{\frac{s}{n} \log \PeCn}> 2^{-s(\EtrcQ-\eps)}\bigg]\\
		&\leq 2^{s(\EtrcQ-\eps)}\bbE\bigg[2^{\frac{s}{n} \log \PeCn}\bigg] \\
		&=2^{s(\EtrcQ-\eps)}\bbE\big[\big(\PeCn\big)^{s/n}\big]\\
		&\leq 2^{s(\EtrcQ-\eps)}\bbE\big[\big(P_{\rme}^{\rm ub}(\Cn)\big)^{s/n}\big] \label{wor1}.
		\end{align}
		On the other hand, for any $0\leq s\leq n$ and $\lambda>0$, we have
		\begin{align}
		\bbE\big[\big(P_{\rme}^{\rm ub}(\Cn)\big)^{s/n}\big]&=\bbE\bigg[\bigg(\frac{1}{M}\sum_{i\neq j} \bbP(\bX_i \to \bX_j)\bigg)^{s/n}\bigg]\\
		&\leq \bbE\bigg[\sum_{i\neq j}  \bigg(\frac{1}{M}\sum_{i\neq j} \bbP(\bX_i \to \bX_j)\bigg)^{s/n}\bigg] \label{wor2}\\
		&=  \frac{1}{M^{s/n}}\sum_{i\neq j} \bbE\bigg[\bigg(\bbP(\bX_i \to \bX_j)\bigg)^{s/n}\bigg]\\
		&=\frac{M(M-1)}{M^{s/n}}\bbE\bigg[\bigg(\bbP(\bX_1 \to \bX_2)\bigg)^{s/n}\bigg]\\
		&\leq M^{2-\frac{s}{n}} \bbE\bigg[\bigg(\bbP(\bX_1 \to \bX_2)\bigg)^{s/n}\bigg]\\
		&= M^{2-\frac{s}{n}} \bbE\bigg[\bigg(\big[\bbP(\bX_1 \to \bX_2)\big]^{1+\lambda}\bigg)^{\frac{s}{n(1+\lambda)}}\bigg]\label{wor3},
		\end{align} where \eqref{wor2} follows from
		\begin{align}
		(x_1+x_2+\cdots +x_n)^{\alpha}\leq x_1^{\alpha}+x_2^{\alpha}+\cdots + x_n^{\alpha}
		\end{align} for any $x_1,x_2,\cdots, x_n\geq 0$ and $\alpha \in [0,1]$, and \eqref{wor3} follows from $0\leq \bbP(\bX_1 \to \bX_2)\leq 1$ and $\lambda>0$, so $\big(\bbP(\bX_1 \to \bX_2)\big)^{s/n}\leq \big(\bbP(\bX_1 \to \bX_2)\big)^{\frac{s}{n(1+\lambda)}}$.
		
		On the other hand, by Lemma \ref{lem:GLem}, the pairwise codeword error probability $\Pro(\bX_1 \to \bX_2)$ given their joint type $Q_{XX'}$ satisfies
		\begin{align}
		\bbP(\bX_1 \to \bX_2) \doteq 2^{-n \sum_{k=1}^n \sum_{x,x'} d_{\rmB}(x,x')\bone\{(X_{1k},X_{2k})=(x,x')\}} \label{eq1modex},
		\end{align} where
		\begin{align}
		d_{\rmB}(x,x')=-\log\bigg(\sum_{y} \sqrt{W(y|x)W(y|x')}\bigg).
		\end{align}
		Hence, for any $0\leq s\leq n$, we have
		\begin{align}
		\bbE\Big[\big(\bbP(\bX_1 \to \bX_2)\big)^{\frac{s}{n}}\Big]&=\bbE\bigg[2^{\frac{s}{n}\log \bbP(\bX_1 \to \bX_2)}\bigg]\\
		&\doteq \bbE\bigg[2^{-\frac{s}{n} \sum_{k=1}^n \sum_{x,x'} d_{\rmB}(x,x')\bone\{(X_{1k},X_{2k})=(x,x')\}}\bigg]\label{mode1}.
		\end{align}
Now, since $\{\sum_{x,x'} d_{\rmB}(x,x')\bone\{(X_{1k},X_{2k})=(x,x')\}\}_{k=1}^n$ are i.i.d., hence by the SLLN, we have
\begin{align}
\frac{1}{n} \sum_{k=1}^n \sum_{x,x'} d_{\rmB}(x,x')\bone\{(X_{1k},X_{2k})=(x,x')\} \asto  \sum_{x,x'} Q(x)Q(x')d_{\rmB}(x,x').
\end{align}	
On the other hand, we have
\begin{align}
0\leq \frac{1}{n} \sum_{k=1}^n \sum_{x,x'} d_{\rmB}(x,x')\bone\{(X_{1k},X_{2k})=(x,x')\}\leq \max_{x,x'} d_{\rmB}(x,x')<\infty.
\end{align}
Hence, by the bounded convergence theorem \cite{Billingsley}, we have
\begin{align}
\bbE\bigg[2^{-\frac{s}{n} \sum_{k=1}^n \sum_{x,x'} d_{\rmB}(x,x')\bone\{(X_{1k},X_{2k})=(x,x')\}}\bigg]\to 2^{-s\sum_{x,x'} Q(x)Q(x')d_{\rmB}(x,x')} \label{mode2}.
\end{align}
Similarly, for any fixed constant $\lambda\geq 0$, we have
\begin{align}
\bbE\bigg[2^{-\frac{s}{n(1+\lambda)} \sum_{k=1}^n \sum_{x,x'} d_{\rmB}(x,x')\bone\{(X_{1k},X_{2k})=(x,x')\}}\bigg]\to 2^{-\frac{s}{1+\lambda} \sum_{x,x'} Q(x)Q(x')d_{\rmB}(x,x')} \label{mode3}.
\end{align}
Hence, from \eqref{mode2} and \eqref{mode3}, for any fixed constant $\lambda\geq 0$, it holds that
\begin{align}
&\bbE\bigg[2^{-\frac{s}{n} \sum_{k=1}^n \sum_{x,x'} d_{\rmB}(x,x')\bone\{(X_{1k},X_{2k})=(x,x')\}}\bigg]\nn\\
&\qquad =(1+o(1)) \bigg(\bbE\bigg[2^{-\frac{s}{n(1+\lambda)} \sum_{k=1}^n \sum_{x,x'} d_{\rmB}(x,x')\bone\{(X_{1k},X_{2k})=(x,x')\}}\bigg]\bigg)^{1+\lambda} \label{mode4}\\
&\qquad =(1+o(1))\bigg( \bbE\bigg[2^{-\frac{s}{n(1+\lambda)}  \sum_{x,x'} d_{\rmB}(x,x')\bone\{(X_{11},X_{21})=(x,x')\}}\bigg]\bigg)^{n(1+\lambda)} \label{ask6}.
		\end{align}
From \eqref{wor3} and \eqref{ask6}, we obtain
		\begin{align}
		\bbE\big[\big(P_{\rme}^{\rm ub}(\Cn)\big)^{s/n}\big]&\leq (1+o(1)) M^{2-\frac{s}{n}}\bigg( \bbE\bigg[2^{-\frac{s}{n(1+\lambda)}  \sum_{x,x'} d_{\rmB}(x,x')\bone\{(X_{11},X_{21})=(x,x')\}}\bigg]\bigg)^{n(1+\lambda)} \label{ask7}.
		\end{align}
Now, observe that
		\begin{align}
		&\bbE\bigg[2^{-\frac{s}{n(1+\lambda)}  \sum_{x,x'} d_{\rmB}(x,x')\bone\{(X_{11},X_{21})=(x,x')\}}\bigg]\nn\\
		&\qquad = \sum_{x,x'} \bbP((X_{11},X_{21})=(x,x')) \bbE\bigg[2^{-\frac{s}{n(1+\lambda)}  \sum_{x,x'} d_{\rmB}(x,x')\bone\{(X_{11},X_{21})=(x,x')\}}\bigg|(X_{11},X_{21})=(x,x')\bigg]\\
		&\qquad = \sum_{x,x'} Q(x)Q(x') 2^{-\frac{s}{n(1+\lambda)} d_{\rmB}(x,x')} \label{wor5}.
		\end{align}
From \eqref{ask7} and \eqref{wor5}, we obtain
		\begin{align}
		\bbE\big[\big(P_{\rme}^{\rm ub}(\Cn)\big)^{\frac{s}{n}}\big]&\leq (1+o(1)) M^{2-\frac{s}{n}}\bigg(\sum_{x,x'} Q(x)Q(x') 2^{-\frac{s}{n(1+\lambda)} d_{\rmB}(x,x')}\bigg)^{n(1+\lambda)} \label{ask8}.
		\end{align}
		From \eqref{wor1} and \eqref{ask8}, for any $s$ such that $0\leq s\leq n$ and any fixed constant $\lambda>0$, we have
		\begin{align}
		&\bbP\bigg[-\frac{1}{n} \log \PeCn <\EtrcQ-\eps\bigg]\nn\\
		&\qquad =\bbP\bigg[\frac{s}{n} \log \PeCn >-s(\EtrcQ-\eps)\bigg]\\
		&\qquad =\bbP\bigg[2^{\frac{s}{n} \log \PeCn}> 2^{-s (\EtrcQ-\eps)}\bigg]\\
		&\qquad \leq 2^{s (\EtrcQ-\eps)}\bbE\bigg[2^{\frac{s}{n} \log \PeCn}\bigg] \\
		&\qquad =2^{s (\EtrcQ-\eps)}\bbE\big[\big(\PeCn\big)^{\frac{s}{n}}\big]\\
		&\qquad \leq (1+o(1)) 2^{s (\EtrcQ-\eps)}M^{2-\frac{s}{n}}\bigg(\sum_{x,x'} Q(x)Q(x') 2^{-\frac{s}{n(1+\lambda)} d_{\rmB}(x,x')}\bigg)^{n(1+\lambda)} \label{wor11}.
		\end{align}
From \eqref{wor11}, by choosing $s=n$ and using $M=2^{nR}$, we have
		\begin{align}
		\bbP\bigg[-\frac{1}{n} \log \PeCn <\EtrcQ-\eps\bigg]&\leq (1+o(1)) 2^{n\big( \EtrcQ+\big(2-\frac{s}{n}\big)R-\eps\big)}\bigg(\sum_{x,x'} Q(x)Q(x') 2^{- \frac{d_{\rmB}(x,x')}{1+\lambda}}\bigg)^{n(1+\lambda)}\\
		&=(1+o(1))2^{n(\EtrcQ+R-\eps)}2^{n(1+\lambda)\log \big(\sum_{x,x'} Q(x)Q(x') 2^{-\frac{d_{\rmB}(x,x')}{1+\lambda}}\big)}\\ 
		&=(1+o(1)) 2^{n \big[\EtrcQ+R+ (1+\lambda)\log \big(\sum_{x,x'} Q(x)Q(x') 2^{- \frac{d_{\rmB}(x,x')}{1+\lambda}}\big)\big]}2^{-n\eps} \label{wor12}.
		\end{align}
Now, for $\EtrcQ \neq \ErceQ$, from \eqref{keypo} and Lemma \ref{lem:tmc}, observe that
\begin{align}
\EtrcQ=\min_{P_{XX'}: D(P_{XX'}\|Q_X Q_{X'})\leq 2R} D(P_{XX'}\|Q_X Q_{X'})+ \sum_{x,x'} d_{\rmB}(x,x') P_{XX'}(x,x')-R \label{eq914}.
\end{align}
Given the distribution $Q$ and $Q_X=Q_{X'}=Q$, the optimization problem in \eqref{eq914} is convex in $\{P_{XX'}(x,x')\}_{x,x'}$ since the KL divergence is convex. By using standard KKT conditions,  it is easy to see that \eqref{eq914} has as optimal solution:
\begin{align}
P_{XX'} \in \{P_{XX'}^{0}, P_{XX'}^*\},
\end{align}
where
\begin{align}
P_{XX'}^{0}(x,x')=\frac{Q(x)Q(x') 2^{-d_{\rmB}(x,x')}}{\sum_{x,x'}Q(x)Q(x') 2^{-d_{\rmB}(x,x')}},
\end{align}
and
\begin{align}
P_{XX'}^*(x,x')=\frac{Q(x)Q(x') 2^{-\frac{d_{\rmB}(x,x')}{1+\lambda_*}}}{\sum_{x,x'}Q(x)Q(x')2^{-\frac{d_{\rmB}(x,x')}{1+\lambda_*}}}.
\end{align}
Here, $\lambda^*$ is the unique positive solution of the following equation:
\begin{align}
2R=D(P_{XX'}^*(x,x')\|Q_X Q_{X'}).
\end{align}
Now, if $P_{XX'}=P_{XX'}^{0}$, then we have
\begin{align}
&\EtrcQ+R+ \log \bigg(\sum_{x,x'} Q(x)Q(x') 2^{-d_{\rmB}(x,x')}\bigg)\nn\\
&\qquad = D(P_{XX'}^{0}\|Q_X Q_{X'})+ \sum_{x,x'} d_{\rmB}(x,x') P_{XX'}^{0}(x,x')+ \log \bigg(\sum_{x,x'} Q(x)Q(x') 2^{-d_{\rmB}(x,x')}\bigg)\\
&\qquad= \sum_{x,x'} \frac{Q(x)Q(x') 2^{-d_{\rmB}(x,x')}}{\sum_{x,x'}Q(x)Q(x') 2^{-d_{\rmB}(x,x')}}\log \frac{ 2^{-d_{\rmB}(x,x')}}{\sum_{x,x'}Q(x)Q(x') 2^{-d_{\rmB}(x,x')}}\nn\\
&\qquad \qquad +  \sum_{x,x'} d_{\rmB}(x,x') P_{XX'}^{0}(x,x')+ \log \bigg(\sum_{x,x'} Q(x)Q(x') 2^{-d_{\rmB}(x,x')}\bigg)\\
&\qquad= -\sum_{x,x'} \frac{Q(x)Q(x') 2^{-d_{\rmB}(x,x')}}{\sum_{x,x'}Q(x)Q(x') 2^{-d_{\rmB}(x,x')}}d_{\rmB}(x,x')- \log \bigg(\sum_{x,x'}Q(x)Q(x') 2^{-d_{\rmB}(x,x')}\bigg)\nn\\
&\qquad \qquad  + \sum_{x,x'} d_{\rmB}(x,x') P_{XX'}^{0}(x,x')+ \log \bigg(\sum_{x,x'} Q(x)Q(x') 2^{-d_{\rmB}(x,x')}\bigg)\\
&\qquad= -\sum_{x,x'} P_{XX'}^{0}(x,x')d_{\rmB}(x,x')- \log \bigg(\sum_{x,x'}Q(x)Q(x') 2^{-d_{\rmB}(x,x')}\bigg)\nn\\
&\qquad \qquad + \sum_{x,x'} d_{\rmB}(x,x') P_{XX'}^{0}(x,x')+ \log \bigg(\sum_{x,x'} Q(x)Q(x') 2^{-d_{\rmB}(x,x')}\bigg)\\
&\qquad =0 \label{mamat1}.
\end{align}
Hence, by choosing $\lambda=0$, from \eqref{wor12} and \eqref{mamat1}, we obtain
\begin{align}
\bbP\bigg[-\frac{1}{n} \log \PeCn <\EtrcQ-\eps\bigg]\dotleq  2^{-n\eps} \label{ipad1}.
\end{align}
Similarly, for the case $P_{XX'}=P_{XX'}^*$, we have
\begin{align}
&\EtrcQ+R+ (1+\lambda^*)\log \bigg(\sum_{x,x'} Q(x)Q(x') 2^{-\frac{d_{\rmB}(x,x')}{1+\lambda^*}}\bigg)\nn\\
&\qquad = D(P_{XX'}^{*}\|Q_X Q_{X'})+ \sum_{x,x'} d_{\rmB}(x,x') P_{XX'}^{*}(x,x')+ (1+\lambda^*) \log \bigg(\sum_{x,x'} Q(x)Q(x') 2^{-\frac{d_{\rmB}(x,x')}{1+\lambda^*}}\bigg)\\
&\qquad \leq (1+\lambda^*) D(P_{XX'}^{*}\|Q_X Q_{X'})+ \sum_{x,x'} d_{\rmB}(x,x') P_{XX'}^{*}(x,x')+  (1+\lambda^*)\log \bigg(\sum_{x,x'} Q(x)Q(x') 2^{-\frac{d_{\rmB}(x,x')}{1+\lambda^*}}\bigg)\\
&\qquad=  (1+\lambda^*)\sum_{x,x'} \frac{Q(x)Q(x') 2^{-\frac{d_{\rmB}(x,x')}{1+\lambda^*}}}{\sum_{x,x'}Q(x)Q(x') 2^{-\frac{d_{\rmB}(x,x')}{1+\lambda^*}}}\log \frac{ 2^{-\frac{d_{\rmB}(x,x')}{1+\lambda^*}}}{\sum_{x,x'}Q(x)Q(x') 2^{-\frac{d_{\rmB}(x,x')}{1+\lambda^*}}}\nn\\
&\qquad \qquad +  \sum_{x,x'} d_{\rmB}(x,x') P_{XX'}^{*}(x,x')+  (1+\lambda^*)\log \bigg(\sum_{x,x'} Q(x)Q(x') 2^{-\frac{d_{\rmB}(x,x')}{1+\lambda^*}}\bigg)\\
&\qquad= - (1+\lambda^*)\sum_{x,x'} \frac{Q(x)Q(x') 2^{-\frac{d_{\rmB}(x,x')}{1+\lambda^*}}}{\sum_{x,x'}Q(x)Q(x') 2^{-\frac{d_{\rmB}(x,x')}{1+\lambda^*}}}\frac{d_{\rmB}(x,x')}{1+\lambda^*}-  (1+\lambda^*)\log \bigg(\sum_{x,x'}Q(x)Q(x') 2^{-\frac{d_{\rmB}(x,x')}{1+\lambda^*}}\bigg)\nn\\
&\qquad \qquad  + \sum_{x,x'} d_{\rmB}(x,x') P_{XX'}^{*}(x,x')+ \log \bigg(\sum_{x,x'} Q(x)Q(x') 2^{-\frac{d_{\rmB}(x,x')}{1+\lambda^*}}\bigg)\\
&\qquad= - \sum_{x,x'} P_{XX'}^{*}(x,x')d_{\rmB}(x,x')-  (1+\lambda^*)\log \bigg(\sum_{x,x'}Q(x)Q(x') 2^{-\frac{d_{\rmB}(x,x')}{1+\lambda^*}}\bigg)\nn\\
&\qquad \qquad + \sum_{x,x'} d_{\rmB}(x,x') P_{XX'}^{*}(x,x')+ \log \bigg(\sum_{x,x'} Q(x)Q(x') 2^{-\frac{d_{\rmB}(x,x')}{1+\lambda^*}}\bigg)\\
&\qquad =0 \label{mamat2}.
\end{align}
From \eqref{wor12},\eqref{mamat1}, and \eqref{mamat2}, where we set $\lambda=0$ for the first case and $\lambda=\lambda^*$ for the second one, we have
\begin{align}
\bbP\bigg[-\frac{1}{n} \log \PeCn <\EtrcQ-\eps\bigg]&\leq (1+o(1)) 2^{-n\eps}\\
&\doteq  2^{-n\eps} \label{wor14}.
\end{align}
Finally, from \eqref{balat3} and \eqref{wor14}, we obtain \eqref{espa}. This concludes our proof of  \eqref{espa} for the i.i.d. random codebook ensemble.

For the constant-composition codebook ensemble, to prove \eqref{espa}, we first prove the following lemma, which is somewhat similar to Lemma \ref{steinlem}.
\begin{lemma} \label{steinlemmu} Let $X_1,X_2,\cdots,X_n$ be Bernoulli random variables on $\bbR$. In addition, there exists a set $\calV \subset \bbR^n$ with cardinality $|\calV|$ such that for all $x_1,x_2,\cdots,x_n \in \calV$,
	\begin{align}
	\Pro[X_1=x_1,X_2=x_2,\cdots,X_n=x_n]\leq \big(1+o(1)\big) \prod_{k=1}^n P(x_k)
	\end{align} for some distribution $P$ on $\{0,1\}$, and 
	\begin{align}
	\Pro[X_1=x_1,X_2=x_2,\cdots,X_n=x_n]&\leq 2^{-n \zeta}, \quad \forall (x_1,x_2,\cdots,x_n) \in \calV^c 
	\end{align} for some $\zeta>0$.
	In addition, 
	\begin{align}
	2^{-n \zeta}|\calV^c|\to 0.
	\end{align} 
	Let $S_n=X_1+X_2+\cdots+X_n$. Then, for any $t>0$ it holds that 
	\begin{align}
	\bbE_{\prod_{i=1}^n P(x_i)}\big[2^{-tS_n/n}\big]+ 2^{-n \zeta} |\calV^c|.
	\end{align}
\end{lemma}
\begin{IEEEproof}
Appendix~\ref{prooflemma10}.
\end{IEEEproof}
It is known that $\{\bone\{(\bX_1,\bX_2) \in \calT_{Q_{XX'}}\}\}_{Q_{XX'}}$ satisfies all the conditions of Lemma \ref{steinlemmu}. Hence, for any $0\leq s\leq n$, from \eqref{mode1}, we have
\begin{align}
&\bbE\bigg[\bigg(\bbP(\bX_1 \to \bX_2)\bigg)^{\frac{s}{n}}\bigg]\nn\\
&\qquad =\bbE\bigg[2^{\frac{s}{n}\log \bbP(\bX_1 \to \bX_2)}\bigg]\\
&\qquad \doteq \bbE_{}\bigg[2^{-\frac{s}{n} \sum_{k=1}^n \sum_{x,x'} d_{\rmB}(x,x')\bone\{(X_{1k},X_{2k})=(x,x')\}}\bigg]\\
&\qquad \doteq \bbE_{\bone\{(\bX_1,\bX_2)\in \calT_{Q_{XX'}}\} \enspace \mbox{is independent}}\bigg[2^{-s \sum_{Q_{XX'}} \sum_{x,x'} d_{\rmB}(x,x')Q_{XX'}(x,x') \bone\{(\bX_1,\bX_2)\in \calT_{Q_{XX'}}\}}\bigg]+2^{-n \zeta} |\calV^c|
\label{mode1mod}.
\end{align}
The rest follows the same as the proof of \eqref{espa} for the i.i.d. random codebook ensemble.

\emph{Now, we prove \eqref{astamoto}.} For any i.i.d. and constant-composition random codebooks, it is easy to see that
	\begin{align}
	\PeCn\geq \max_{i\neq j} \bbP(\bX_i \to \bX_j) \label{gjmod}.
	\end{align}
	Recall the definition of $V_n$ in \eqref{defVn}. It follows from \eqref{gjmod} that 
	\begin{align}
	&\bbP\bigg[-\frac{1}{n} \log \PeCn >\EtrcQ+\eps\bigg]\nn\\
	&\qquad \leq \bbP\bigg[-\frac{1}{n} \log \max_{i\neq j} \bbP(\bX_i \to \bX_j) >\EtrcQ+\eps\bigg]\\
	&\qquad =\bbP\bigg[V_n >n \big(\EtrcQ+\eps\big)\bigg]\\
	&\qquad = \bbP\bigg[ \frac{V_n-\bbE[V_n]}{\sqrt{\var(V_n)}}> \frac{n \big(\EtrcQ+\eps-\frac{\bbE[V_n]}{n}\big)}{\sqrt{\var(V_n)}}\bigg]\\
	&\qquad= Q\bigg( \frac{n \big(\EtrcQ+\eps-\frac{\bbE[V_n]}{n}\big)}{\sqrt{\var(V_n)}}\bigg)+O\bigg(\frac{1}{\sqrt{n}}\bigg) \label{eq315} \\
	&\qquad = Q\Bigg( \frac{\sqrt{n} \big(\EtrcQ+\eps-E_{\rm{trc}}(Q,0)\big)}{\sqrt{ \sum_{x,x'} d_{\rmB}^2(x,x') Q(x)Q(x')-\big(\sum_{x,x'} d_{\rmB}(x,x') Q(x)Q(x')\big)^2}}\Bigg)+O\bigg(\frac{1}{\sqrt{n}}\bigg) \label{eq316}
	\end{align} as $n\to \infty$ since $\EtrcQ \geq E_{\rm{trc}}(Q,0)$, where \eqref{eq315} follows from Theorem \ref{lem:ct2}, and \eqref{eq316} follows from \eqref{factTc} and the Berry–Esseen theorem \cite{Billingsley}.

\subsection{Proof of Theorem~\ref{theo:double_exp}}
\label{sec:double_exp}

We start with some accessory results, then prove the main part of Theorem \ref{theo:double_exp}.

In the lemma below we use the simplified notation: $\mathcal{I}\{i,j\} =  \mathcal{I}\{(\mathbf{x}_i,\mathbf{x}_j)\in \mathcal{T}(P_{XX'})\}$, $\mathcal{I}\{.\}$ being the indicator function, i.e., the two considered codewords have joint type $P_{XX'}$.
\begin{lemma}\label{lemma:mean_joint_type_prod}
	\begin{align}
	2^{-n2D(P_{XX'}||Q_X Q_X')}\leq\mathbb{E}[\mathcal{I}\{i,j\}\mathcal{I}\{i,k\}]\leq 2^{-n[D(P_{XX'}||Q_X Q_X')+\eta]}
	\end{align}
	where $0\leq\eta\leq D(P_{XX'}||Q_X Q_X')$.
\end{lemma}
\begin{IEEEproof}
Appendix~\ref{prooflemma11}.
\end{IEEEproof}

\begin{lemma}\label{lemma:type_enum_double}
	For any $\epsilon>0$ and for any joint type $P_{XX'}$ such that $D(P_{XX'}||Q_X Q_X')\leq R-\epsilon$, $\forall\epsilon>0$, the following holds:
	\begin{align}\label{eqn:type_enum_double1}
	\bbP\left[\calN(P_{XX'})\leq 2^{-n\epsilon}\bbE[\calN(P_{XX'})]\right]\dot{\leq} 2^{-2^{n\epsilon}}
	\end{align}
\end{lemma}

\begin{IEEEproof}
Appendix~\ref{prooflemma12}.
\end{IEEEproof}
Using Lemma \ref{lemma:mean_joint_type_prod} and Lemma \ref{lemma:type_enum_double} we prove the following theorem, which states that the probability of finding a code for which the exponent of $\Pecsnub$ is larger than the expurgated exponent $\Eexr$ is double exponentially decaying in $n$.

Now we can prove the main part of theorem \ref{theo:double_exp}. We have that
	\begin{align}\label{eqn:pub_doteq}
	\Pecsnub&\dot{=}\max_{P_{XX'}}\calN(P_{XX'})e^{-n[\sum_{x,x'}d_{\rmB}(x,x')P_{XX'}(x,x')+R] }
	\end{align}
	Let us refer to the maximizing joint type  of \eqref{eqn:pub_doteq} as $P_{XX'}^*$. We define the following complementary events:
	\begin{align}
	A&=\{P_{XX'}^*\in \calP\}\\
	\overline{A}&=\{P_{XX'}^*\in \overline{\mathcal{P}}\}
	\end{align}
	where
	$$\calP  =  \{P_{XX'} | D(P_{XX'}\|Q_X Q_X')\leq 2R\}$$
	while
	$$\overline{\mathcal{P}} =  \{P_{XX'} | D(P_{XX'}\|Q_X Q_X')> 2R\},$$
	$Q_X Q_X'$ being the theoretical joint type.
	Consider a positive real number $E_2 > \EtrcQ$. We have:
	\begin{align}\label{eqn:proof_double_exp1}
	\Pro\left[-\frac{1}{n}\log \Pecsnub \geq E_{2}\right]=\Pro\left[-\frac{1}{n}\log \Pecsnub \geq E_{2}, A\right]+\Pro\left[-\frac{1}{n}\log \Pecsnub \geq E_{2}, \overline{A}\right] 
	\end{align}
	Now we proceed to bound from above both terms at the right hand side of \eqref{eqn:proof_double_exp1}. 
	
	\subsubsection{First Term}
	\begin{align}\label{eqn:proof_double_exp3}
	\Pro\left[-\frac{1}{n}\log \Pecsnub \geq E_{2}, A\right]&=\Pro\left[ \Pecsnub \leq 2^{-n E_{2}}, A\right]
	\\&=\Pro\left[ \frac{1}{M_n}\sum_{P_{XX'}}\calN(P_{XX'})2^{-n\sum_{x,x'}d_{\rmB}(x,x')P_{XX'}(x,x')} \leq 2^{-n E_{2}}, A\right]\label{eqn:proof_double_exp4}
	\\&=\Pro\left[\sum_{P_{XX'}}\calN(P_{XX'})e^{-n\sum_{x,x'}d_{\rmB}(x,x')P_{XX'}(x,x')} \leq 2^{-n (E_{2}-R) }, A\right]\label{eqn:proof_double_exp5}
	\\&\dot{=}\Pro\left[\max_{P_{XX'}}\calN(P_{XX'})2^{-n\sum_{x,x'}d_{\rmB}(x,x')P_{XX'}(x,x')} \leq 2^{-n (E_{2}-R) }, A\right]\label{eqn:proof_double_exp6}
	\\&{=}\Pro\left[\max_{P_{XX'}\in \calP}\calN(P_{XX'})2^{-n\sum_{x,x'}d_{\rmB}(x,x')P_{XX'}(x,x')} \leq 2^{-n (E_{2}-R) }, A\right]\label{eqn:proof_double_exp7}
	\\&\leq\Pro\left[\max_{P_{XX'}\in \calP}\calN(P_{XX'})2^{-n\sum_{x,x'}d_{\rmB}(x,x')P_{XX'}(x,x')} \leq 2^{-n (E_{2}-R) }\right]\label{eqn:proof_double_exp8}
	\\&{=}\Pro\left[\bigcap_{P_{XX'}\in \mathcal{P}}\left[\mathcal{N}(P_{XX'})e^{-n\sum_{x,x'}d_{\rmB}(x,x')P_{XX'}(x,x')} \leq 2^{-n (E_{2}-R) }\right]\right\}\label{eqn:proof_double_exp9}
	\\&{=}\Pro\left[\bigcap_{P_{XX'}\in \calP}\left[\calN(P_{XX'}) \leq 2^{-n (E_{2}-R - \sum_{x,x'}d_{\rmB}(x,x')P_{XX'}(x,x')) }\right]\right\}\label{eqn:proof_double_exp10}
	\end{align}
	where \eqref{eqn:proof_double_exp7} comes from the definition of $A$ while	 \eqref{eqn:proof_double_exp8} comes from removing the event $A$. 
	Let us now define $\mathcal{P}'$:
	\begin{align}\label{eqn:subsetP}
	\mathcal{P}' =  \{P_{XX'} | D(P_{XX'}\|Q_X Q_X')\leq R\},
	\end{align}
	and note that $\mathcal{P}'\subset \mathcal{P}$. Let us consider the term $2^{-n (E_{2}-R - \sum_{x,x'}d_{\rmB}(x,x')P_{XX'}(x,x')) }$.
	We now look for a $P_{XX'}\in \mathcal{P}'$ such that this is smaller than the mean of the enumerator function, i.e, a $P_{XX'}\in \mathcal{P}'$ such that the following holds:
	\begin{align}\label{eqn:smaller_mean}
	&2^{-n (E_2-R -\sum_{x,x'}d_{\rmB}(x,x')P_{XX'}(x,x')) }\leq 2^{n[2R-D(P_{XX'}||Q_X Q_X')-\epsilon]}\\
	&E_2-R -\sum_{x,x'}d_{\rmB}(x,x')P_{XX'}(x,x') \geq -[2R-D(P_{XX'}||Q_X Q_X')-\epsilon]\\
	&E_2-R -\sum_{x,x'}d_{\rmB}(x,x')P_{XX'}(x,x') \geq -2R+D(P_{XX'}||Q_X Q_X')+\epsilon\\
	&E_2 \geq -R+D(P_{XX'}||Q_X Q_X') +\sum_{x,x'}d_{\rmB}(x,x')P_{XX'}(x,x') +\epsilon.\label{eqn:smaller_mean1}
	\end{align}
	Let us indicate the $P_{XX'}$ that minimizes \eqref{eqn:smaller_mean1} with $P_{XX'}'$.
	Minimizing the term at the right hand side of \eqref{eqn:smaller_mean1} we can set the value of $E_2$ to:
	\begin{align}\label{eqn:smaller_mean12}
	E_{2}  = \min_{P_{XX'}\in \calP'}-R+D(P_{XX'}||Q_X Q_X') +\sum_{x,x'}d_{\rmB}(x,x')P_{XX'}(x,x')) +\epsilon.
	\end{align}
	The right hand side of \eqref{eqn:smaller_mean12} is strictly larger than $\EtrcQ$. To see this note the following:
	\begin{align}
	\min_{P_{XX'}\in \calP'}-R+D(P_{XX'}||Q_X Q_X') &+\sum_{x,x'}d_{\rmB}(x,x')P_{XX'}(x,x') +\epsilon \\&= \min_{P_{XX'}\in \calP'}+R-2R+D(P_{XX'}||Q_X Q_X') +\sum_{x,x'}d_{\rmB}(x,x')P_{XX'}(x,x') +\epsilon \label{eqn:smaller_mean3}\\
	&> \min_{P_{XX'}\in \calP}+R-2R+D(P_{XX'}||Q_X Q_X') +\sum_{x,x'}d_{\rmB}(x,x')P_{XX'}(x,x') +\epsilon \label{eqn:smaller_mean4}\\
	&= \min_{P_{XX'}\in \mathcal{Z}_{GGV}}+ R-2R +D(P_{XX'}||Q_X Q_X')+\sum_{x,x'}d_{\rmB}(x,x')P_{XX'}(x,x') +\epsilon \label{eqn:smaller_mean5}\\
	&= \min_{P_{XX'}\in \mathcal{Z}_{GGV}} +R +\sum_{x,x'}d_{\rmB}(x,x')P_{XX'}(x,x') +\epsilon \label{eqn:smaller_mean6a}\\
	&= \EtrcQ +\epsilon \label{eqn:smaller_mean6}
	\end{align}
	where \eqref{eqn:smaller_mean4} follows from the fact that $\mathcal{P}'\subset\mathcal{P}$, \eqref{eqn:smaller_mean5} follows from the concavity of the objective function (minimum is on the border) while \eqref{eqn:smaller_mean6} follows from the definition of $\mathcal{Z}_{GGV}$.
	With this definition of $E_{2}$ we ensure that for at least one joint type the conditions for applying Lemma \ref{lemma:type_enum_double} (i.e., \eqref{eqn:smaller_mean}) hold. Using the definition in \eqref{eqn:proof_double_exp10} together with the statement of Lemma \ref{lemma:type_enum_double} we have:
	\begin{align}\label{eqn:proof_double_exp11}
	\Pro\left[-\frac{1}{n}\log \Pecsnub \geq E_{2}, A\right]&\dotleq \Pro\left[\bigcap_{P_{XX'}\in \calP}\left[\calN(P_{XX'}) \leq 2^{-n (E_{2}-R + \sum_{x,x'}d_{\rmB}(x,x')P_{XX'}(x,x')) }\right]\right\}\\
	&\dotleq 2^{-2^{n[R-D(P_{XX'}'||Q_X Q_X')]}}\\
	&\leq 2^{-2^{n\epsilon'}}
	\end{align}
	with $\epsilon'>0$.

	\subsubsection{Second Term}
	\begin{align}\label{eqn:proof_double_exp12}
	\Pro\left[-\frac{1}{n}\log \Pecsnub \geq E_2, \overline{A}\right]&=\Pro\left[ \Pecsnub \leq 2^{-n E_2}, \overline{A}\right]
	\\&=\Pro\left[ \frac{1}{M_n}\sum_{P_{XX'}}\mathcal{N}(P_{XX'})2^{n\sum_{x,x'}d_{\rmB}(x,x')P_{XX'}(x,x')} \leq 2^{-n E_{2}}, \overline{A}\right]\label{eqn:proof_double_exp13}
	\\&=\Pro\left[\sum_{P_{XX'}}\calN(P_{XX'})2^{n\sum_{x,x'}d_{\rmB}(x,x')P_{XX'}(x,x')} \leq 2^{-n (E_2-R) }, A\right]\label{eqn:proof_double_exp14}
	\\&\dot{=}\Pro\left[\max_{P_{XX'}}\calN(P_{XX'})2^{n\sum_{x,x'}d_{\rmB}(x,x')P_{XX'}(x,x')} \leq 2^{-n (E_2-R) }, \overline{A}\right]\label{eqn:proof_double_exp15}
	\end{align}
	Consider \eqref{eqn:proof_double_exp15}. The event $\overline{A}$ implies that the joint type maximizing the expression at the left hand side lays outside $\calP$. This implies that any $P_{XX'}$ which lies inside $\calP$ leads to a value which is no greater than the maximum. Since this is an implication of the events within brackets, its probability is larger than or equal to the one of \eqref{eqn:proof_double_exp15}.
	Thus we have:
	\begin{align}\label{eqn:proof_double_exp16}
	\Pro\left[-\frac{1}{n}\log \Pecsnub \geq E_2, \overline{A}\right]&\dot{=}\Pro\left[\max_{P_{XX'}}\calN(P_{XX'})2^{n\sum_{x,x'}d_{\rmB}(x,x')P_{XX'}(x,x')} \leq 2^{-n (E_2-R) }, \overline{A}\right]\\\label{eqn:proof_double_exp17}
	&\leq \Pro\left[\max_{P_{XX'}\in\calP}\calN(P_{XX'})2^{n\sum_{x,x'}d_{\rmB}(x,x')P_{XX'}(x,x')} \leq 2^{-n (E_2-R) }\right]\\\label{eqn:proof_double_exp18}
	&\leq 2^{-2^{n\epsilon'}}
	\end{align}
	where \eqref{eqn:proof_double_exp18} is because \eqref{eqn:proof_double_exp17} has the same form as \eqref{eqn:proof_double_exp8} and thus the same inequalities as for the first term hold. 
	
	Finally, we note that from \eqref{eqn:smaller_mean12} we can further state the following:
	\begin{align}\label{eqn:smaller_mean12_1}
	E_{2}  &= \min_{P_{XX'}\in \calP'}-R+D(P_{XX'}||Q_X Q_X') +\sum_{x,x'}d_{\rmB}(x,x')P_{XX'}(x,x') +\epsilon\\
	&= \min_{P_{XX'}\in \calP'} -\sum_{x,x'}d_{\rmB}(x,x')P_{XX'}(x,x') +\epsilon \label{eqn:smaller_mean12_2}\\
	&=\Eexr +\epsilon \label{eqn:smaller_mean12_3}
	\end{align}
	where \eqref{eqn:smaller_mean12_2} follows from the concavity of the objective function, which implies that the minimum is on the border of the region $ \calP'$, and from the definition of  $\calP'$ while \eqref{eqn:smaller_mean12_3} is found by calculating the derivative of \cite[Eq. (5.7.11)]{gallagerBook} with respect to the optimization variable $\rho$ and, after some change of variable, equating to zero. 

\subsection{Proof of Theorem~\ref{theo:double_exp_0R}}
\label{sub:p3}
Now let us consider the following inequality
	\begin{equation}\label{eqn:bound_Pe}
	\Pecsnub\leq M_n\PeCn
	\end{equation}
	which follows from upper-bounding the probability $\bbP[\bx_i \to \bx_j]$ in \eqref{defRBE} by $\bbP\bigg[\bigcup_{j\neq i} \{\bx_i \to \bx_j\}\bigg]$ in \eqref{eq:Pe}.
	From Theorem \ref{theo:double_exp} and using \eqref{eqn:bound_Pe} we have
	\begin{align}\label{eqn:double_exp_0R_2}
	\Pro\left[-\frac{1}{n}\log \Pecsnub \geq \Eexr +R+ \epsilon\right]\leq 2^{-2^{n\epsilon}}
	\end{align}
	and finally
	\begin{align}\label{eqn:double_exp_0R_3}
	\Pro\left[-\frac{1}{n}\log \Pecsnub \geq E_{\rm{ex}}(0)+ \epsilon\right]\leq 2^{-2^{n\epsilon}}.
	\end{align}

\subsection{Proof of Theorem~\ref{lem:aux2}}
\label{sub:p4}
\subsubsection{i.i.d.~ensemble}

Observe that
	\begin{align}
	\max_{i\neq j} \bbP(\bX_i \to \bX_j) \leq \PeCn\leq \sum_{i=1}^{M_n} \sum_{j\neq i} \bbP(\bX_i \to \bX_j) \leq M_n(M_n-1) \max_{i\neq j}\bbP(\bX_i \to \bX_j) \label{stat1}.
	\end{align}
On the other hand, by Lemma \ref{lem:GLem}, the pairwise codeword error probability $\Pro(\bX_i \to \bX_j)$ given their joint type $P_{XX'}$ satisfies
\begin{align}
\bbP(\bX_i \to \bX_j) \doteq 2^{-n \sum_{x,x'} d_{\rmB}(x,x') \hatP_{\bX_i\bX_j} (x,x')} \label{eq1mod},
\end{align} where $\hatP_{\bX_i\bX_j}$ is the $n$-joint type of $(\bX_1,\bX_2)$, and  
\begin{align}
d_{\rmB}(x,x')=-\log\bigg(\sum_{y} \sqrt{W(y|x)W(y|x')}\bigg).
\end{align}
Observe that
\begin{align}
\hatP_{\bX_i\bX_j}(x,x')=\frac{1}{n}\sum_{k=1}^n \bone\{(X_{ik},X_{jk})=(x,x')\} \label{eq2mod}.
\end{align}
It follows from \eqref{eq1mod} and \eqref{eq2mod} that
\begin{align}
\bbP(\bX_i \to \bX_j) \doteq 2^{-\sum_{k=1}^n \sum_{x,x'} d_{\rmB}(x,x') \bone\{(X_{ik},X_{jk})=(x,x')\}} \label{stat3}
\end{align} for all $i,j \in [M_n], i\neq j$.
Since $M_n$ sub-exponential in $n$, from \eqref{stat1} and \eqref{stat3}, we obtain
	\begin{align}
	-\frac1n\log \PeCn \asto \frac{V}{n},
	\label{cota2mod}
	\end{align} where
	\begin{align}
	V =  \min_{i\neq j} Z_{ij}
	\end{align} with
	\begin{align}
	Z_{ij} =   -\sum_{k=1}^n \sum_{x,x'}  d_{\rmB}(x,x') \bone\{(X_{ik},X_{jk})=(x,x')\} \label{defZij},
	\end{align} for all $i,j \in [M_n]$ and $i\neq j$.\\
Now, observe that
\begin{align}
\bbE[Z_{ij}]&=-\sum_{k=1}^n \sum_{x,x'}  d_{\rmB}(x,x')\bbP\{(X_{ik},X_{jk})=(x,x')\}\\
&=-\sum_{k=1}^n \sum_{x,x'}  d_{\rmB}(x,x')\Pro(X_{ik}=x)\Pro(X_{jk}=x')\\
&=-\sum_{k=1}^n \sum_{x,x'}  d_{\rmB}(x,x')Q(x)Q(x')	 \label{fact1mod}.
\end{align}
In addition, we have
\begin{align}
\var(Z_{ij})&=\sum_{k=1}^n \var\bigg(\sum_{x,x'}  d_{\rmB}(x,x')\bone\{(X_{ik},X_{jk})=(x,x')\}\bigg)\\
&=n\bigg(\bbE\bigg[\bigg(\sum_{x,x'}  d_{\rmB}(x,x')\bone\{(X_{i1},X_{j1})=(x,x')\}\bigg)^2\bigg]-\bigg(\sum_{x,x'}  d_{\rmB}(x,x')Q(x)Q(x')	\bigg)^2\bigg)\\
&=n\bigg(\bbE\bigg[\sum_{x,x'}  d_{\rmB}^2(x,x') \bone\{(X_{1i},X_{2i})=(x,x')\}\bigg]-\bigg(\sum_{x,x'}  d_{\rmB}(x,x')Q(x)Q(x')	\bigg)^2\bigg)\\
&= n\bigg(\sum_{x,x'}  d_{\rmB}^2(x,x') \bbP\{(X_{1i},X_{2i})=(x,x')\}-\bigg(\sum_{x,x'}  d_{\rmB}(x,x')Q(x)Q(x')	\bigg)^2\bigg)\\
&= n\bigg(\sum_{x,x'}  d_{\rmB}^2(x,x') Q(x)Q(x')-\bigg(\sum_{x,x'}  d_{\rmB}(x,x')Q(x)Q(x')	\bigg)^2\bigg) \label{cswmod}.
\end{align} for all $i\neq j$.

Now, define
	\begin{align}
	T_{ij}:&=\frac{Z_{ij}-\bbE[Z_{ij}]}{\sqrt{\var(Z_{ij})}}\\
	&=\frac{Z_{ij}-\bbE[Z_{12}]}{\sqrt{\var(Z_{12})}} \label{cubi},
	\end{align} where \eqref{cubi} follows from the fact that $Z_{ij}$'s are identically distributed.
	
Then, by CLT, it holds that
	\begin{align}
	T_{ij} \dto \calN(0,1), \qquad \forall i\neq j.
	\end{align}
On the other hand, for any fixed tuple $(\{\alpha_{ij}\}: i,j \in [M],i\neq j)$, we have
\begin{align}
\sum_{i\neq j} \alpha_{ij} T_{ij}&=\frac{\sum_{i\neq j}\alpha_{ij} \sum_{k=1}^N \sum_{x,x'}  d_{\rmB}(x,x') \big( Q(x)Q(x')-\bone\{(X_{ik},X_{jk})=(x,x')\}\big)}{\sqrt{\var(Z_{12})}} \\
&=\sum_{k=1}^N\frac{ \sum_{i\neq j}\alpha_{ij} \sum_{x,x'}  d_{\rmB}(x,x') \big( Q(x)Q(x')-\bone\{(X_{ik},X_{jk})=(x,x')\}\big)}{\sqrt{\var(Z_{12})}} \label{cubi0}.
\end{align}
Now, by the i.i.d. random codebook generation, it holds that $\{V_k\}_{k=1}^n$ are i.i.d. random variables, where
\begin{align}
V_k =  \frac{\sum_{i\neq j} \alpha_{ij}\sum_{x,x'}  d_{\rmB}(x,x')  \big( Q(x)Q(x')-\bone\{(X_{ik},X_{jk})=(x,x')\}\big)}{\sqrt{\var(Z_{12})}}.
\end{align} 
In addition, since $(X_{i1},X_{j1})_{i\neq j}$'s are pairwise independent, we have
\begin{align}
\var(V_1)&=\frac{\sum_{i\neq j} \alpha_{ij}^2 \sum_{x,x'}  d_{\rmB}^2(x,x') \var\big(\bone\{(X_{ik},X_{jk})=(x,x')\}- Q(x)Q(x')\big))}{\var(Z_{12})}\\
&=\frac{\sum_{i\neq j} \alpha_{ij}^2 \big[\sum_{x,x'}  d_{\rmB}^2(x,x') Q(x)Q(x')-\big(\sum_{x,x'}  d_{\rmB}(x,x')Q(x)Q(x')	\big)^2\big]}{n\big[\sum_{x,x'}  d_{\rmB}^2(x,x') Q(x)Q(x')-\big(\sum_{x,x'}  d_{\rmB}(x,x')Q(x)Q(x')	\big)^2\big]} \\
&=\frac{\sum_{i\neq j} \alpha_{ij}^2}{n}\label{cubi2}.
\end{align}
Hence, it holds from \eqref{cubi0} and \eqref{cubi2} that
\begin{align}
&\sum_{i\neq j} \alpha_{ij} T_{ij}=\sqrt{\sum_{i\neq j} \alpha_{ij}^2 }\bigg(\frac{\sum_{k=1}^n V_k}{\sqrt{n\var(V_1)}}\bigg) \nn\\
&\qquad \dto \calN\bigg(0, \sum_{i\neq j} \alpha_{ij}^2\bigg) \label{lastb},
\end{align} where \eqref{lastb} follows from the CLT. Hence, the distribution of the vector $\{T_{ij}:i,j \in [M],i\neq j\}$ goes to the distribution of a jointly Gaussian random vector by the Levy's continuity theorem \cite{Billingsley}. 

Now, it is known that the distribution of any Gaussian random vector (both p.d.f and c.d.f.) is defined by its mean and covariance matrix. Since the covariance matrix of the vector $\{T_{ij}:i,j \in [M],i\neq j\}$ is the identity matrix by the pairwise independence of $T_{ij}$, which originates from the pairwise independence of $\Pro(\bX_i\to \bX_j)'s$, hence, the limit distribution is the standard normal Gaussian vector with dimension $M(M-1)$. This distribution is equal to the joint distribution of $M(M-1)$ independent standard normal variables $\{U_{ij}\}_{i\neq j}$. Hence, by the continuous mapping theorem \cite{Billingsley}, it follows that 
\begin{align}
\min_{i\neq j}T_{ij}\dto \min_{i\neq j} U_{ij} \label{kingsu}.
\end{align}
Now, let
\begin{align}
T=\frac{V-\bbE[V]}{\sqrt{\var(V)}} \label{defT},
\end{align}
then, we have
\begin{align}
T=\min_{i\neq j} \tilT_{ij},
\end{align}
where
\begin{align}
\tilT_{ij} =  \frac{Z_{ij}-\bbE[V]}{\sqrt{\var(V)}},
\end{align}
Hence, it is easy to see that
\begin{align}
T&= \min_{i\neq j} \frac{\sqrt{\var(Z_{ij})}}{\sqrt{\var(V)}}\tilT_{ij}+ \frac{\bbE[Z_{ij}]-\bbE[V]}{\sqrt{\var(V)}}\\
&= \min_{i\neq j} \frac{\sqrt{\var(Z_{12})}}{\sqrt{\var(V)}}\tilT_{ij}+ \frac{\bbE[Z_{22}]-\bbE[V]}{\sqrt{\var(V)}} \label{cofact}\\
&=\frac{\sqrt{\var(Z_{12})}}{\sqrt{\var(V)}}\min_{i\neq j}\tilT_{ij}+ \frac{\bbE[Z_{22}]-\bbE[V]}{\sqrt{\var(V)}} \label{cofact2b},
\end{align} where \eqref{cofact} follows from $Z_{ij}$ are identically distributed by the i.i.d. random codebook generation.

Now, assume that 
\begin{align}
\lim_{n\to \infty}\sqrt{\frac{\var(Z_{12})}{\var(V)}}&=\zeta,\\
\lim_{n\to \infty}\frac{\bbE[Z_{12}]-\bbE[V]}{\sqrt{\var(V)}}&=\beta,
\end{align} for some $\zeta, \beta \in \bbR$ \footnote{The existence of these limits can be proved easily.}. Then, by applying Slutsky's theorem, from \eqref{cofact2b} and \eqref{kingsu}, we have
\begin{align}
T \dto \zeta \min_{i\neq j}U_{ij} + \beta 
\end{align} where $U_{ij} \sim \calN(0,1)$ and $U_{ij}$'s are independent.

Since $\bbE[T]=0$ and $\var(T)=1$, it follows from Lemma \ref{lem:aux2021} that
\begin{align}
T   \dto \frac{\min_{i\neq j}U_{ij}-\bbE[\min_{i\neq j}U_{ij}]}{\sqrt{\var(\min_{i\neq j}U_{ij})}} \label{factT}.
\end{align} 

Finally, from \eqref{cota2mod}, \eqref{defT} and \eqref{factT}, by appying Slutsky's theorem, we have
\begin{align}
\frac{\frac{-\log \PeCn}{n}-\bbE[\frac{-\log \PeCn}{n}]}{\sqrt{\var{\big(\frac{-\log \PeCn}{n}}\big)}}\dto \frac{\min_{i\neq j}U_{ij}-\bbE[\min_{i\neq j}U_{ij}]}{\sqrt{\var(\min_{i\neq j}U_{ij})}}.
\end{align}

\subsubsection{Constant-composition ensemble}

In this part, we use Stein's method to derive some criteria that provide sufficient conditions for the convergence in distribution to the normal random variable of the error probabilities and error exponents for general random coding ensemble over general channels, including the zero rate where $M_n \to \infty$ as we mentioned. This includes other random codebooks than i.i.d. random codebook ensembles.

We start by showing that Theorem \ref{lem:aux2} also holds for the constant-composition codes.
In order to do this, we need some extra lemmas.
\begin{lemma}\label{lem:extra1} Let $X$ be a random variable on some finite set $\calX$. Assume that for a certain $\beta>0$
	\begin{align}
	\Pro\big[\{X\in \calA_1\} \cap \{X\in \calA_2\}\big]\leq \beta \Pro[X\in \calA_1]\Pro[X\in \calA_2]
	\label{ta1}
	\end{align} holds for any $A_1, A_2 \subset \calX$. Then, for any sequence of sets $\{A_k\}_{k=1}^{n-1}$ such that $A_k \subset \calX$ and there exists $A_i \cap A_j =\emptyset$ for some $i\neq j$ and $i,j \in [n]$, it holds that
	\begin{align}
	\Pro\bigg[\bigcap_{k=1}^n \big\{X \in \calA_k\big\}\bigg]\leq \beta \prod_{k=1}^n \Pro\big[X \in \calA_k\big]  \label{ta0}.
	\end{align}
\end{lemma}
\begin{IEEEproof}
Appendix~\ref{prooflemma14}
\end{IEEEproof}

Next, we have the following lemma.
\begin{lemma} \label{lem:extra2} Let $\calT_n(\calX \times \calX)$ be the set of all $n$-joint-types in $\calX \times \calX$ and let $Q_{XX'},\tilQ_{XX'} \in \calT_N(\calX \times \calX)$ such that $Q_{XX'}\neq \tilQ_{XX'}$. Let $\{\bX_1,\bX_2,\cdots,\bX_M\}$ be codewords of a constant-composition code with type $Q$. For a fixed pair $(i,j)$ with $i\neq j$ and $i,j \in [n]$, define $Z_{Q_{XX'}} = \bone \{(\bX_i,\bX_j) \in \calT(Q_{XX'})\}$ and $Z_{\tilQ_{XX'}} = \bone \{(\bX_i,\bX_j) \in \calT(\tilQ_{XX'})\}$. Then, it holds that
	\begin{align}
	&\Pro\big[\{Z_{Q_{XX'}}=a\} \cap \{Z_{\tilQ_{XX'}}=b\} \big]\leq \frac{1}{1-2^{-n I_{\min}}(Q)}\Pro\big[ Z_{Q_{XX'}}=a\big]\Pro\big[ Z_{\tilQ_{XX'}}=b\big]\bigg| \label{bacha1},
	\end{align} for any $(a,b) \in \{0,1\} \times \{0,1\}$. Here,
	\begin{align}
	I_{\min}(Q) = \min_{Q_{XX'} \in \calP(\calX \times \calX):I_{Q_{XX'}}(X;X')>0} I_{Q_{XX'}}(X;X') >0 \label{defimin}. 
	\end{align}
\end{lemma}
\begin{IEEEproof}
Appendix~\ref{prooflemma15}
\end{IEEEproof}

\begin{corollary} \label{showcor} Let $\calT_n(\calX \times \calX)$ be the set of all $n$-joint types in $\calX\times \calX$. For a constant-composition code and $i\neq j$ and $i,j \in [M]$, let
	\begin{align}
	Z_{\tilQ_{XX'}} = \bone\big\{(\bX_i,\bX_j) \in \calT(Q_{XX'})\big\}
	\end{align} for all $Q_{XX'} \in \calT_n(\calX \times \calX)$. Then, for any vector $\{z_{Q_{XX'}}\}_{Q_{XX'} \in \calT_n(\calX\times \calX)}$ such that there are at least two joint types $Q_{XX'}$ and $\tilQ_{XX'}$ such that $\tilQ_{XX'}\neq Q_{XX'}$ and $z_{Q_{XX'}}=z_{\tilQ_{XX'}}=1$, it holds that
	\begin{align}
	\Pro\bigg[\bigcap_{Q_{XX'} \in \calT_n(\calX \times \calX)} \{Z_{Q_{XX'}}=z_{Q_{XX'}}\} \bigg]\leq \frac{1}{1-2^{-n I_{\min}}(Q)} \prod_{Q_{XX'} \in \calT_n(\calX \times \calX)}\Pro\big[Z_{Q_{XX'}}=z_{Q_{XX'}}\big] 
	\end{align}
\end{corollary}
\begin{IEEEproof} Since the vector $\{z_{Q_{XX'}}\}_{Q_{XX'} \in \calT_n(\calX\times \calX)}$ has at least two $n$-joint types $Q_{XX'}$ and $\tilQ_{XX'}$ such that $\tilQ_{XX'}\neq Q_{XX'}$ and $z_{Q_{XX'}}=z_{\tilQ_{XX'}}=1$, it holds that
	\begin{align}
	\Pro\bigg[\bigcap_{Q_{XX'} \in \calT_n(\calX \times \calX)} \{Z_{\tilQ_{XX'}}=z_{Q_{XX'}}\} \bigg]&\leq \Pro\bigg[ \{Z_{\tilQ_{XX'}}=z_{Q_{XX'}}\}\cap \{Z_{\tilQ_{XX'}}=z_{\tilQ_{XX'}}\}\bigg]\\
	&=\Pro\bigg[ \{Z_{\tilQ_{XX'}}=1\}\cap \{Z_{\tilQ_{XX'}}=1\}\bigg]\\
	&=\Pro\bigg[\{(\bX_i,\bX_j) \in \calT(Q_{XX'})\bigg\} \cap \bigg\{ (\bX_i,\bX_j) \in \calT(\tilQ_{XX'})\bigg\}\bigg]\\
	&=0. 
	\end{align}
	This concludes our proof of this corollary.	
\end{IEEEproof}
Now, we show the following fact which is based on Stein's method.
\begin{lemma} \label{steinlem} Let $f$ be a bounded function with bounded first and second derivative.  Let $X_1,X_2,\cdots,X_n$ be zero-mean random variables on $\bbR$ such that $\bbE|X_k^4|<\infty$ for all $k \in [n]$.   
	In addition, assume there exists a function $f:\bbZ_+ \to \bbR^+$ such that $f(n)\to \infty$ as $n\to \infty$ and a set $\calV \subset \bbR^n$ with cardinality $|\calV|$ such that 
	\begin{align}
	\Pro[X_1=x_1,X_2=x_2,\cdots,X_n=x_n]\leq \big(1+o(1)\big) \prod_{k=1}^n \Pro[X_k=x_k]
	\end{align} for all $x_1,x_2,\cdots,x_n \in \calV$ and
	\begin{align}
	\Pro[X_1=x_1,X_2=x_2,\cdots,X_n=x_n]&\leq 2^{-n \zeta},\\
	\sum_{k=1}^n |x_k|^4 &\leq g(n),\quad \forall (x_1,x_2,\cdots,x_n) \in \calV^c 
	\end{align} for some $\zeta>0$.
	Assume also that:
	\begin{align}
	2^{-n \zeta}|\calV^c|g(n)\to 0 \quad \mbox{as} \quad n\to \infty,\\
	ng(n)\to \infty \qquad \mbox{as} \quad n\to \infty.
	\end{align} 
	Let $S_n=X_1+X_2+\cdots+X_n$ and
	\begin{align}
	\tilT = \frac{S_n}{\sqrt{\var(S_n)}}. 
	\end{align}
	Then, under the condition that 
	\begin{align}
	\frac{1}{n^{3/2}}\sum_{i=1}^n \bbE[|X_i^3|] \to 0 \label{conbat1},\\
	\frac{1}{n^2}\sum_{i=1}^n \bbE[X_i^4] \to 0 \label{conbat2},
	\end{align}
	we have
	\begin{align}
	\tilT \dto  \calN(0,1).
	\end{align}
\end{lemma}
\begin{IEEEproof}
Appendix~\ref{prooflemma16}.
\end{IEEEproof}

Now, we return to proof Theorem \ref{lem:aux2}. As in the i.i.d.~case, we have
	\begin{align}
	\frac{-\log \PeCn}{n} \asto \frac{V}{n},
	\label{cota2mod2}
	\end{align} where
	\begin{align}
	V = \min_{i\neq j} Z_{ij}
	\end{align} with
	\begin{align}
	Z_{ij} =  -\sum_{k=1}^n \sum_{x,x'} d_{\rmB}(x,x') \bone\{(X_{ik},X_{jk})=(x,x')\} \label{defZij2},
	\end{align} for all $i,j \in [M]$ and $i\neq j$.
	
	For the constant-composition code, we have
	\begin{align}
	Z_{ij}&= -\sum_{k=1}^n \sum_{x,x'} d_{\rmB}(x,x') \bone\{(X_{ik},X_{jk})=(x,x')\}\\
	&= -\sum_{x,x'} d_{\rmB}(x,x') \sum_{k=1}^n \bone\{(X_{ik},X_{jk})=(x,x')\}   \sum_{Q_{XX'}} \bone\{(\bX_i,\bX_j)\in \calT(Q_{XX'}) \}\\
	&= -n \sum_{Q_{XX'}} \sum_{x,x'}  Q_{XX'}(x,x') d_{\rmB}(x,x') \bone\{(\bX_i,\bX_j)\in \calT(Q_{XX'}) \}.
	\end{align} 
	Let 
	\begin{align}
	U_{Q_{XX'}}& = \frac{\barZ_{ij}-\bbE[\barZ_{ij}]}{\sqrt{\var(\barZ_{ij})}}
	\end{align} 
	where
	\begin{align}
	\barZ_{ij} = \frac{Z_{ij}}{n}=-\sum_{Q_{XX'}} \sum_{x,x'}  Q_{XX'}(x,x') d_{\rmB}(x,x') \bone\{(\bX_i,\bX_j)\in \calT(Q_{XX'}) \}.
	\end{align}
	Then, we have
	\begin{align}
	\frac{Z_{ij}-\bbE[Z_{ij}]}{\sqrt{\var(Z_ij)}}&=\sum_{Q_{XX'}} U_{Q_{XX'}},
	\end{align}
	and
	\begin{align}
	U_{Q_{XX'}}(Z_{Q_{XX'}})& = \frac{-\sum_{x,x'}  Q_{XX'}(x,x') d_{\rmB}(x,x') Z_{Q_{XX'}}+\sum_{x,x'}  Q_{XX'}(x,x') d_{\rmB}(x,x') \Pro\big(Z_{Q_{XX'}}=1\big) }{\sqrt{\var(\barZ_{ij})}}
	\end{align}
	where 
	\begin{align}
	Z_{Q_{XX'}} = \bone\{(\bX_i,\bX_j)\in \calT(Q_{XX'}) \}.
	\end{align}
	Let 
	\begin{align}
	\calV = \calV_1 \cup \calV_2 \cup \calV_3,
	\end{align}
	where 
	\begin{align}
	\calV_1& = \bigg\{ \{z_{Q_{XX'}}\}_{Q_{XX'} \in \calT_n(\calX \times \calX)}: \mbox{there are at least two different $n$-joint types $Q_{XX'}$ and $\tilQ_{XX'}$}\nn\\
	\qquad & \qquad \qquad  \mbox{such that $z_{Q_{XX'}}=z_{\tilQ_{XX'}}=1$} \bigg\},
	\end{align}
	and
	\begin{align}
	&\calV_2 =  \bigg\{ \{z_{Q_{XX'}}\}_{Q_{XX'} \in \calT_n(\calX \times \calX)}: \mbox{there is exactly one $n$-joint types $Q_{XX'}^*$}\nn\\
	\qquad & \qquad \qquad  \mbox{such that $z_{Q_{XX'}}^*=1$ and $Q_{XX'}^*=Q_X^* Q_{X'}^*$}\bigg\},
	\end{align}
	and
	\begin{align}
	&\calV_3 = \bigg\{ \{z_{Q_{XX'}}\}_{Q_{XX'} \in \calT_n(\calX \times \calX)}:\mbox{such that $z_{Q_{XX'}}=1$ for all $Q_{XX'} \in \calT_n(\calX  \times \calX)$} \bigg\}.
	\end{align}
	Now, for any $\{z_{Q_{XX'}}\}_{Q_{XX'}\in \calV}$, there are three subcases:
	\begin{itemize}
		\item $\{z_{Q_{XX'}}\}_{Q_{XX'}} \in \calV_3$. Then, we have
		\begin{align}
		\Pro\bigg[\bigcap_{Q_{XX'} \in \calT_n(\calX \times \calX)} \{Z_{Q_{XX'}}=z_{Q_{XX'}}\} \bigg]&=\Pro\bigg[\bigcap_{Q_{XX'} \in \calT_n(\calX \times \calX)} \{(\bX_i,\bX_j)\in \calT(Q_{XX'})\} \bigg]\\
		&=0\\
		&\leq \prod_{Q_{XX'} \in \calT_n(\calX \times \calX)}\Pro\big[Z_{Q_{XX'}}=z_{Q_{XX'}}\big].
		\end{align}
		\item  $\{z_{Q_{XX'}}\}_{Q_{XX'}} \in \calV_2$. Then, we have
		\begin{align}
		\Pro\big[(\bX_i,\bX_j)\in \calT(Q_{XX'}^*)\big]&\doteq 1 \label{betto1}.
		\end{align}
		\begin{align}
		&\Pro\bigg[\bigcap_{Q_{XX'} \in \calT_n(\calX \times \calX)} \{Z_{Q_{XX'}}=z_{Q_{XX'}}\} \bigg]\nn\\
		&\qquad =\Pro\bigg[\bigcap_{Q_{XX'} \in \calT_n(\calX \times \calX)} \{(\bX_i,\bX_j)\in \calT(Q_{XX'})\} \bigg]\\
		&\qquad = \Pro\bigg[\{Z_{Q_{XX'}^*}=1\} \cap \bigcap_{Q_{XX'} \in \calT_n(\calX \times \calX): Q_{XX'}\neq Q_{XX'}^*} \{Z_{Q_{XX'}}=0\}\bigg]  \\
		&\qquad =\Pro\bigg[\{(\bX_i,\bX_j)\in \calT(Q_{XX'}^*)\} \cap \bigcap_{Q_{XX'} \in \calT_n(\calX \times \calX): Q_{XX'}\neq Q_{XX'}^*} \{(\bX_i,\bX_j)\notin \calT(Q_{XX'})\}\bigg]  \\
		&\qquad= \Pro\bigg[(\bX_i,\bX_j)\in \calT(Q_{XX'}^*)\bigg] \label{betto2}.
		\end{align}
		On the other hand, it is known that
		\begin{align}
		&\prod_{Q_{XX'} \in \calT_n(\calX \times \calX): Q_{XX'}\neq Q_{XX'}^*} \Pro\big[Z_{Q_{XX'}}=z_{Q_{XX'}}\big]\nn\\
		&\qquad =\prod_{Q_{XX'} \in \calT_n(\calX \times \calX): Q_{XX'}\neq Q_{XX'}^*} \Pro\big[(\bX_i,\bX_j)\notin \calT(Q_{XX'})\big]\\
		&\qquad \doteq \prod_{Q_{XX'} \in \calT_n(\calX \times \calX): Q_{XX'}\neq Q_{XX'}^*} \big(1-2^{-n I_{Q_{XX'}}(X;X')}\big)\\
		&\qquad \dotgeq \prod_{Q_{XX'} \in \calT_n(\calX \times \calX): Q_{XX'}\neq Q_{XX'}^*} \big(1-2^{-n I_{\min}(Q)}\big)\\
		&\qquad =\big(1-2^{-n I_{\min}(Q)}\big)^{|\calT_n(\calX\times \calX)|-1}\\
		&\qquad \geq 1-\big(|\calT_n(\calX\times \calX)|-1\big)2^{-n I_{\min}(Q)}  \label{stree}\\
		&\qquad =1+o(1) \label{stree2},
		\end{align}  where \eqref{stree} follows from $(1-x)^n \geq 1-nx$ for all $x \in (0,1)$, and \eqref{stree2} follows from $|\calT_n(\calX\times \calX)|\leq (n+1)^{|\calX|}$.
		
		From \eqref{betto2} and \eqref{stree}, we have
		\begin{align}
		\Pro\bigg[\bigcap_{Q_{XX'} \in \calT_n(\calX \times \calX)} \{Z_{Q_{XX'}}=z_{Q_{XX'}}\} \bigg] \leq (1+o(1))\prod_{Q_{XX'} \in \calT_n(\calX \times \calX)}\Pro\big[Z_{Q_{XX'}}=z_{Q_{XX'}}\big].
		\end{align}
		\item $\{z_{Q_{XX'}}\}_{Q_{XX'}} \in \calV_1$. Then, from Lemma \ref{lem:extra1} and Lemma \ref{lem:extra2}, we obtain
		\begin{align}
		\Pro\bigg[\bigcap_{Q_{XX'} \in \calT_n(\calX \times \calX)} \{Z_{Q_{XX'}}=z_{Q_{XX'}}\} \bigg] \leq (1+o(1))\prod_{Q_{XX'} \in \calT_n(\calX \times \calX)}\Pro\big[Z_{Q_{XX'}}=z_{Q_{XX'}}\big].
		\end{align}
	\end{itemize}
	From case 1, case 2, and case 3, on $\calV$, it holds
	\begin{align}
	\Pro\bigg[\bigcap_{Q_{XX'} \in \calT_N(\calX \times \calX)} \{Z_{Q_{XX'}}=z_{Q_{XX'}}\} \bigg] \leq (1+o(1)) \prod_{Q_{XX'} \in \calT_N(\calX \times \calX)}  \Pro\big[Z_{Q_{XX'}}=z_{Q_{XX'}}\big].
	\end{align} 
	
	Now, for  $\{z_{Q_{XX'}}\}_{Q_{XX'}} \in \calV^c$, it holds that $z_{Q_{XX'}}=1$ for exact one $N$-joint type $Q_{XX'}^* \in \calT_n(\calX\times \calX)$ and $I_{Q_{XX'}^*}(X;X')\geq I_{\min}(Q)+o_N(1)$. Note that there are at most $|\calT_n(\calX\times \calX)|\leq (n+1)^{|\calX|}$ such sequences.  Hence, on this subset,
	we have
	\begin{align}
	&\Pro\bigg[\bigcap_{Q_{XX'} \in \calT_n(\calX \times \calX)} \{Z_{Q_{XX'}}=z_{Q_{XX'}}\} \bigg]\nn\\
	&\qquad =\Pro\bigg[\bigcap_{Q_{XX'} \in \calT_n(\calX \times \calX)} \{(\bX_i,\bX_j)\in \calT(Q_{XX'})\} \bigg]\\
	&\qquad = \Pro\bigg[\{Z_{Q_{XX'}^*}=1\} \cap \bigcap_{Q_{XX'} \in \calT_n(\calX \times \calX): Q_{XX'}\neq Q_{XX'}^*} \{Z_{Q_{XX'}}=0\}\bigg]  \\
	&\qquad =\Pro\bigg[\{(\bX_i,\bX_j)\in \calT(Q_{XX'}^*)\} \cap \bigcap_{Q_{XX'} \in \calT_n(\calX \times \calX): Q_{XX'}\neq Q_{XX'}^*} \{(\bX_i,\bX_j)\notin \calT(Q_{XX'})\}\bigg]  \\
	&\qquad= \Pro\bigg[(\bX_i,\bX_j)\in \calT(Q_{XX'}^*)\bigg]\\
	&\qquad \dotleq 2^{-n I_{\min}(Q)} \label{betto4}.
	\end{align}

	In addition, on $\calV^c$, we have
	\begin{align}
	\sum_{Q_{XX'}} \frac{U_{Q_{XX'}}^4} {\var(\barZ_{ij})} \leq \bigg(\var(\barZ_{ij})\bigg)^{-1}\bigg(\sum_{Q_{XX'}} U_{Q_{XX'}}^4\bigg) \label{mote}.
	\end{align}
	Now, observe that
	\begin{align}
	&\var(\barZ_{ij})= \bbE\bigg[\bigg(\sum_{Q_{XX'}} \sum_{x,x'}  Q_{XX'}(x,x') d_{\rmB}(x,x') \bone\{(\bX_i,\bX_j)\in \calT(Q_{XX'}) \} \bigg)^2 \bigg]\nn\\
	& \qquad-\bigg(\bbE\bigg[\sum_{Q_{XX'}} \sum_{x,x'}  Q_{XX'}(x,x')d_{\rmB}(x,x')  \bone\{(\bX_i,\bX_j)\in \calT(Q_{XX'}) \}  \bigg]\bigg)^2\\
	&= \bbE\bigg[\sum_{\tilQ_{XX'}} \sum_{Q_{XX'}} \sum_{x,x'}\sum_{y,y'} \tilQ_{XX'}(y,y') Q_{XX'}(x,x') d_{\rmB}(x,x') d_{\rmB}(y,y')\nn\\
	&\qquad \qquad \times  \bone\{(\bX_i,\bX_j)\in \calT(Q_{XX'}) \} \bone\{(\bX_i,\bX_j)\in \calT(\tilQ_{XX'}) \}  \bigg]\nn\\
	& \qquad-\bigg(\bbE\bigg[\sum_{Q_{XX'}} \sum_{x,x'}  Q_{XX'}(x,x') d_{\rmB}(x,x')  \bone\{(\bX_i,\bX_j)\in \calT(Q_{XX'}) \}  \bigg]\bigg)^2\\
	&= \bbE\bigg[\sum_{Q_{XX'}} \sum_{x,x'}\sum_{y,y'} Q_{XX'}(y,y') Q_{XX'}(x,x') d_{\rmB}(x,x') d_{\rmB}(y,y') \bone\{(\bX_i,\bX_j)\in \calT(Q_{XX'}) \} \bigg]\nn\\
	& \qquad-\bigg(\bbE\bigg[\sum_{Q_{XX'}} \sum_{x,x'}  Q_{XX'}(x,x') d_{\rmB}(x,x')  \bone\{(\bX_i,\bX_j)\in \calT(Q_{XX'}) \}  \bigg]\bigg)^2\\
	&=\sum_{Q_{XX'}} \Pro\big[(\bX_i,\bX_j)\in \calT(Q_{XX'})\big]\sum_{x,x'}\sum_{y,y'} Q_{XX'}(y,y') Q_{XX'}(x,x')d_{\rmB}(x,x') d_{\rmB}(y,y') \nn\\
	& \qquad-\bigg(\bbE\bigg[\sum_{Q_{XX'}} \sum_{x,x'}  Q_{XX'}(x,x') d_{\rmB}(x,x') \bone\{(\bX_i,\bX_j)\in \calT(Q_{XX'}) \}  \bigg]\bigg)^2\\
	&= \sum_{Q_{XX'}} \Pro\big[(\bX_i,\bX_j)\in \calT(Q_{XX'})\big]\bigg(\sum_{x,x'} Q_{XX'}(x,x') d_{\rmB}(x,x')\bigg)^2\nn\\
	&\qquad  -\bigg(\sum_{Q_{XX'}} \Pro\big[(\bX_i,\bX_j)\in \calT(Q_{XX'})\big] \sum_{x,x'}  Q_{XX'}(x,x') d_{\rmB}(x,x')   \bigg)^2 \label{cuchipu}\\
	&= \sum_{Q_{XX'}} \Pro\big[(\bX_i,\bX_j)\in \calT(Q_{XX'})\big]\bigg(\sum_{x,x'} Q_{XX'}(x,x') d_{\rmB}(x,x')\bigg)^2\nn\\
	&\qquad- \sum_{Q_{XX'}} \big(\Pro\big[(\bX_i,\bX_j)\in \calT(Q_{XX'})\big]\big)^2 \bigg(\sum_{x,x'} Q_{XX'}(x,x') d_{\rmB}(x,x')\bigg)^2\nn\\
	&\qquad \qquad - 2\sum_{Q_{XX'}}\sum_{\tilQ_{XX'}}\sum_{x,x'}\sum_{y,y'} Q_{XX'}(x,x') \tilQ_{XX'}(y,y')d_{\rmB}(x,x') d_{\rmB}(y,y')\nn\\
	&\qquad \qquad \qquad  \times  \Pro\big[(\bX_i,\bX_j)\in \calT(Q_{XX'})\big]\Pro\big[(\bX_i,\bX_j)\in \calT(\tilQ_{XX'})\big] \label{poly0}.
	\end{align}
	Since $Q_{XX'}\neq \tilQ_{XX'}$, we have
	\begin{align}
	\Pro\big[(\bX_i,\bX_j)\in \calT(Q_{XX'})\big]\Pro\big[(\bX_i,\bX_j)\in \calT(\tilQ_{XX'})\big]&\doteq 2^{-n I_{Q_{XX'}}(X;X')}2^{-n I_{\tilQ_{XX'}}(X;X')}\\
	&\leq 2^{-n I_{\min}(Q)}.
	\end{align}
	Hence, we have
	\begin{align}
	&\sum_{Q_{XX'}}\sum_{\tilQ_{XX'}}\sum_{x,x'}\sum_{y,y'} Q_{XX'}(x,x') \tilQ_{XX'}(y,y')d_{\rmB}(x,x') d_{\rmB}(y,y')\nn\\
	&\qquad \qquad \qquad  \times  \Pro\big[(\bX_i,\bX_j)\in \calT(Q_{XX'})\big]\Pro\big[(\bX_i,\bX_j)\in \calT(\tilQ_{XX'})\big]\nn\\
	&\qquad \leq 2^{-n I_{\min}(Q)} \sum_{Q_{XX'}}\sum_{\tilQ_{XX'}}\sum_{x,x'}\sum_{y,y'} Q_{XX'}(x,x') \tilQ_{XX'}(y,y')d_{\rmB}(x,x')d_{\rmB}(y,y')\\
	&\qquad \leq 2^{-n I_{\min}(Q)}\bigg( \sum_{Q_{XX'}} \sum_{x,x'}Q_{XX'}(x,x')d_{\rmB}(x,x') \bigg )^2\\
	&\qquad \leq 2^{-n I_{\min}(Q)} d_{\max}^2 \bigg( \sum_{Q_{XX'}} \sum_{x,x'}Q_{XX'}(x,x')\bigg )^2 \\
	&\qquad \leq 2^{-n I_{\min}(Q)} d_{\max}^2 |\calT_n(\calX \times \calX)|^2\\
	&\qquad \leq 2^{-n I_{\min}(Q)} d_{\max}^2 (n+1)^{2|\calX|} \label{poly1},
	\end{align} where
	\begin{align}
	d_{\max} = \max_{x,x'} \bigg[-\log \bigg(\sum_{y} \sqrt{W(y|x)W(y|x')}\bigg)\bigg].
	\end{align}
	From \eqref{poly0} and \eqref{poly1}, we obtain
	\begin{align}
	&\var(\barZ_{ij})\nn\\
	& \geq  \sum_{Q_{XX'}} \Pro\big[(\bX_i,\bX_j)\in \calT(Q_{XX'})\big]\bigg(\sum_{x,x'} Q_{XX'}(x,x') d_{\rmB}(x,x')\bigg)^2\nn\\
	&\qquad  - \sum_{Q_{XX'}} \big(\Pro\big[(\bX_i,\bX_j)\in \calT(Q_{XX'})\big]\big)^2 \bigg(\sum_{x,x'} Q_{XX'}(x,x') d_{\rmB}(x,x')\bigg)^2 -  2^{-n I_{\min}(Q)} d_{\max}^2 (n+1)^{2|\calX|} \\
	&= \sum_{Q_{XX'}} \Pro\big[(\bX_i,\bX_j)\in \calT(Q_{XX'})\big]\bigg(1-\Pro\big[(\bX_i,\bX_j)\in \calT(Q_{XX'})\big]\bigg) \bigg(\sum_{x,x'} Q_{XX'}(x,x') d_{\rmB}(x,x')\bigg)^2\nn\\
	&\qquad   -  2^{-n I_{\min}(Q)} d_{\max}^2 (n+1)^{2|\calX|}\\
	& \geq \sum_{Q_{XX'}: Q_{XX'}\neq Q_X Q_{X'}} \Pro\big[(\bX_i,\bX_j)\in \calT(Q_{XX'})\big]\bigg(1-\Pro\big[(\bX_i,\bX_j)\in \calT(Q_{XX'})\big]\bigg) \bigg(\sum_{x,x'} Q_{XX'}(x,x') d_{\rmB}(x,x')\bigg)^2\nn\\
	& \qquad   -  2^{-n I_{\min}(Q)} d_{\max}^2 (n+1)^{2|\calX|}\\
	& \geq \sum_{Q_{XX'}: Q_{XX'}\neq Q_X Q_{X'}} \Pro\big[(\bX_i,\bX_j)\in \calT(Q_{XX'})\big]\bigg(1-2^{-n I_{\min}(Q)}\bigg) \bigg(\sum_{x,x'} Q_{XX'}(x,x') d_{\rmB}(x,x')\bigg)^2\nn\\
	& \qquad   -  2^{-n I_{\min}(Q)} d_{\max}^2 (n+1)^{2|\calX|}\\
	& \geq \bigg(1-2^{-n I_{\min}(Q)}\bigg)d_{\min}^2 \sum_{Q_{XX'}: Q_{XX'}\neq Q_X Q_{X'}} \Pro\big[(\bX_i,\bX_j)\in \calT(Q_{XX'})\big]  -  2^{-n I_{\min}(Q)} d_{\max}^2 (n+1)^{2|\calX|} \label{moly}
	\end{align} where
	\begin{align}
	d_{\min} = \min_{x,x'} \bigg[-\log \bigg(\sum_{y} \sqrt{W(y|x)W(y|x')}\bigg)\bigg]>0.
	\end{align}
	Now, observe that
	\begin{align}
	&\sum_{Q_{XX'}: Q_{XX'}\neq Q_X Q_{X'}} \Pro\big[(\bX_i,\bX_j)\in \calT(Q_{XX'})\big]\nn\\
	&\qquad\qquad  =1-\Pro\big[(\bX_i,\bX_j)\in \calT(Q_X Q_{X'})\big]\\
	&\qquad \qquad =1-\sum_{(\bx,\bx') \in \calT(Q_X Q_{X'})} 2^{-n H(Q_X)}2^{-n H(Q_X')} \\
	&\qquad \qquad =1- \big|\calT(Q_X Q_{X'})\big| 2^{-n H(Q_X)}2^{-n H(Q_X')} \label{moly2}.
	\end{align} 
	Now, by \cite{Csis00}, it holds that
	\begin{align}
	2^{nH(Q_X Q_{X'})} (n+1)^{-|\calX|}\leq \big|\calT(Q_X Q_{X'})\big|\leq 2^{n H(Q_X Q_{X'})}.
	\end{align}
	We can assume that $ \big|\calT(Q_X Q_{X'})\big|\neq 2^{n H(Q_X Q_{X'})}$, then 
	\begin{align}
	\big|\calT(Q_X Q_{X'})\big|\leq (\mbox{poly}(n))^{-1}2^{nH(Q_X Q_{X'})}
	\end{align} 
	where $\mbox{poly}(n)\neq 1$ where $\mbox{poly}(n)$ is some polynomial in $n$. Then, from \eqref{moly2} and $Q_X=Q_{X'}=Q$, we obtain
	\begin{align}
	\sum_{Q_{XX'}: Q_{XX'}\neq Q_X Q_{X'}} \Pro\big[(\bX_i,\bX_j)\in \calT(Q_{XX'})\big]\geq 1-(\mbox{poly}(n))^{-1}\geq 1-\alpha \label{moly3}
	\end{align} for some $\alpha\in (0,1)$ and $n$ sufficiently large.
	
	Hence, from \eqref{moly}, \eqref{moly2}, and \eqref{moly3}, we have
	\begin{align}
	\var(\barZ_{ij})&\geq \bigg(1-2^{-n I_{\min}(Q)}\bigg)d_{\min}^2 (1-\alpha)  -  2^{-n I_{\min}(Q)} d_{\max}^2 (n+1)^{2|\calX|}\\
	&=\Omega(1) \label{keypoint}.
	\end{align}
	From \eqref{mote} and \eqref{keypoint},
	\begin{align}
	\sum_{Q_{XX'}} \frac{V^4_{Q_{XX'}}(Z_{Q_{XX'}})} {\var(\barZ_{ij})} &\leq \bigg(\var(\barZ_{ij})\bigg)^{-1}\bigg(\sum_{Q_{XX'}} V_{Q_{XX'}}^4(Z_{Q_{XX'}})\bigg) \label{moteb}\\
	&\qquad = O(1) \sum_{Q_{XX'}}V_{Q_{XX'}}^4(Z_{Q_{XX'}}) \label{eq786}
	\end{align}
	where 
	\begin{align}
	V_{Q_{XX'}}(Z_{Q_{XX'}})& = -\sum_{x,x'}  Q_{XX'}(x,x') d_{\rmB}(x,x') Z_{Q_{XX'}}+\sum_{x,x'}  Q_{XX'}(x,x') d_{\rmB}(x,x') \Pro\big(Z_{Q_{XX'}}=1\big).
	\end{align}
	Note that
	\begin{align}
	\big|V_{Q_{XX'}}(Z_{Q_{XX'}})\big|^4 &\leq 2\bigg(\bigg|\sum_{x,x'}  Q_{XX'}(x,x') d_{\rmB}(x,x')  Z_{Q_{XX'}}\bigg|^4 +\bigg|\sum_{x,x'}  Q_{XX'}(x,x') d_{\rmB}(x,x') \Pro\big(Z_{Q_{XX'}}=1\big)\bigg|^4\bigg)\\
	&\leq  4 d_{\max}^4  \label{eq787}
	\end{align} for all $Q_{XX'} \in \calT_n(\calX \times \calX)$, and
	\begin{align}
	|V_{Q_{XX'}}(Z_{Q_{XX'}})|^3 &\leq 2\bigg(\bigg|\sum_{x,x'}  Q_{XX'}(x,x') d_{\rmB}(x,x')  Z_{Q_{XX'}}\bigg|^3 +\bigg|\sum_{x,x'}  Q_{XX'}(x,x') d_{\rmB}(x,x') \Pro\big(Z_{Q_{XX'}}=1\big)\bigg|^4\bigg)\\
	&\leq  4 d_{\max}^3  \label{eq788}
	\end{align}
	
	Hence, from \eqref{eq786} and \eqref{eq787}, we have
	\begin{align}
	\sum_{Q_{XX'}} \frac{V^4_{Q_{XX'}}(Z_{Q_{XX'}})} {\var(\barZ_{ij})}&\leq 4 |\calT_n(\calX\times \calX)|d_{\max}^4 \label{alo0} \\
	&\leq 4 (n+1)^{|\calX|} d_{\max}^4 \label{alo}.
	\end{align}
	From \eqref{betto4} and \eqref{alo}, we have
	\begin{align}
	2^{-nI_{\min}(Q)}|\calV^c| \sum_{Q_{XX'}} \frac{V^4_{Q_{XX'}}(Z_{Q_{XX'}})} {\var(\barZ_{ij})}&\leq 2^{-nI_{\min}(Q)}(n+1)^{|\calX|} 4 (n+1)^{|\calX|} d_{\max}^4 \to 0 \label{check1}
	\end{align} as $n\to \infty$.
	
	Now, we have
	\begin{align}
	\frac{1}{|\calT_n(\calX \times \calX)|^2}\sum_{Q_{XX'}} V^4_{Q_{XX'}}(Z_{Q_{XX'}}) &\leq  \frac{1}{|\calT_n(\calX \times \calX)|} 4 d_{\max}^4  \label{alo2}\\
	&\doteq \frac{1}{2^{nH(Q_{XX'})}} 4 d_{\max}^4\\
	&\to 0
	\end{align} as $n\to \infty$, where \eqref{alo2} follows from \eqref{alo0}.
	
	Similarly, we have
	\begin{align}
	\frac{1}{|\calT_n(\calX \times \calX)|^{3/2}} \sum_{Q_{XX'}} |V_{Q_{XX'}}(Z_{Q_{XX'}})|^3 &\leq \frac{1}{|\calT_n(\calX \times \calX)|^{3/2}} \sum_{Q_{XX'}}4 d_{\max}^3\\
	&\to 0.
	\end{align}
	From the above facts and Lemma \ref{steinlem}, we conclude that 
	\begin{align}
	T_{ij} = \frac{Z_{ij}-\bbE[Z_{ij}]}{\sqrt{\var(Z_{ij})}} \dto \calN(0,1).
	\end{align}
	Similarly, we can prove that if $M$ is a constant, we have
	\begin{align}
	\sum_{i\neq j} \alpha_{ij} T_{ij} \dto  \calN(0,1).
	\end{align} for any sequence $\{\alpha_{ij}\}_{i, j \in [M], i\neq j}$. Then, by using the same arguments as the proof of Lemma \ref{lem:aux2} for the rest, we obtain our result in \eqref{eheldb}.
\subsection{Proof of Theorem~\ref{lem:ct2}}
\label{sub:p5}
Our proof of this theorem is based on a modification of the Wasserstein metric, inspired by the classical Kolmogorov and Wasserstein metrics, that measures the distance between the distribution of the error exponent and that of the standard Gaussian. Such modification is needed to deal with an infinite number of terms as $n\to\infty$, a case where the classical Wasserstein metric upper bound fails to work~\cite[Prop. 2.4]{Nathan2011a}. 

Recall the definitions of probability metrics in Definition \ref{steindef}. First, we prove the following fundamental lemma.
\begin{lemma} \label{lem:boundmea} If $Z \sim \calN(0,1)$, then for any random variable $T$, it holds that
	\begin{align}
	\big|\Pro(T\leq x)-\Pro(Z\leq x)\big| \leq 2 (8\pi)^{-1/4}\sqrt{d_{W,\rm{mod}}(T,Z)} + \big|\Pro(T\leq x)-\Pro(T\geq -x)\big|\label{cachua}
	\end{align} for all $x \in \bbR$. In addition, if the distribution of $T$ is tight\footnote{A distribution on $(\bbR, \calB(\bbR))$ is tight if for any fixed $\eps>0$, there exists $u,v \in \bbR$ such that $\Pro(u<T\leq v)>1-\eps$ \cite{Billingsley}.}, for any $x\to 0$, which is a continuous point of the limit distribution of $T$, as $N\to \infty$, we have
	\begin{align}
	\limsup_{N\to \infty} \big|\Pro(T\leq x)-\Pro(Z\leq x)\big|\leq 2 (8\pi)^{-1/4}\limsup_{N\to \infty} \sqrt{d_{W,\rm{mod}}(T,Z)} \label{cachuab}.
	\end{align}
\end{lemma}
\begin{IEEEproof}
Appendix~\ref{ap:proof43}.
\end{IEEEproof}

By using the definition of $d_{W,\rm{mod}}$ and setting $T=X$ and $Y=Z$ where $Z \sim \calN(0,1)$, we obtain the following result, which is tighter than (or at least equal to) the upper bound of $d_K(T,Z)$ in \cite[Prop. 2.4]{Nathan2011a}. However, we note that the probability metric here is the modified Wasserstein metric. See the same arguments to achieve a similar result in \cite[Prop.~2.4]{Nathan2011a}.
\begin{lemma} \label{lem:nathan1} For $h \in \calH$, let $f_h$ solve 
	\begin{align}
	f_h'(w)-wf_h(w)=h(w)-\bbE[h(Z)] \label{eqkey}.
	\end{align}	If $T$ is a random variable and $Z$ has the standard normal distribution, then
	\begin{align}
	d_{W,\rm{mod}}(T,Z)\leq \sup_{h \in \calH} \min\bigg\{\big|\bbE\big[f_h'(T)-Tf_h(T)\big]\big|,\big|\bbE\big[f_h'(-T)+Tf_h(-T)\big]\big|\bigg\}.
	\end{align}
\end{lemma}
\begin{IEEEproof}
Left as exercise.
\end{IEEEproof}
Now, we prove the following lemma.
\begin{lemma} \label{lem:co1} Assume that $T=\min\{T_1,T_2,\cdots,T_L\}$ for some $L \in \bbZ^+$ and $T_1,T_2,\cdots,T_L$ are identically distributed random variables. Then, it holds that
	\begin{align}
	d_{W,\rm{mod}}(T,Z)&\leq \max\bigg\{\sup_{h\in \calH} \big|\bbE[f_h'(T_1)-T_1 f_h(T_1)]\big|,\sup_{h\in \calH} \big|\bbE[f_h'(-T_1)+T_1 f_h(-T_1)]\big|\bigg\}\nn\\ &\qquad +\sup_{h\in \calH} \min\bigg\{\bbE[h(T)-h(T_1)],\bbE[h(-T_1)-h(-T)]\bigg\} . \label{ba1}
	\end{align}
\end{lemma}
\begin{IEEEproof}
Appendix~\ref{ap:proof45}.
\end{IEEEproof}

\begin{lemma} \cite[Th.~3.2]{Nathan2011a} \label{lem:nathanstan} Let $X_1,X_2,\cdots,X_n$ be independent mean zero random variables such that $\bbE[|X_i|^4]<\infty$ and $\bbE[X_i^2]=1$. If $T=\sum_{i=1}^n X_i/\sqrt{n}$ and $Z$ has the standard normal distribution, then
	\begin{align}
	\max\bigg\{\sup_{h \in \calH} \big|\bbE\big[f'_h(T)-T f_h(T)\big]\big|, \sup_{h \in \calH} \big|\bbE\big[f'_h(-T)+T f_h(-T)\big]\big|\bigg\}  \leq \frac{1}{n^{3/2}}\sum_{i=1}^n \bbE[|X_i|^3]+\frac{\sqrt{2}}{n \sqrt{\pi}}\sqrt{\sum_{i=1}^n \bbE[X_i^4]} \label{cut}.
	\end{align}
\end{lemma}
We can observe the fact \eqref{cut} since $T$ and $-T$ are both the sums of independent random variables. Now, we are ready to prove Theorem \ref{lem:ct2}. Observe that
	\begin{align}
	\max_{i\neq j} \bbP(\bX_i \to \bX_j) \leq \PeCn\leq \sum_{i=1}^{M_n} \sum_{j\neq i} \bbP(\bX_i \to \bX_j) \leq M_n(M_n-1) \max_{i\neq j}\bbP(\bX_i \to \bX_j) \label{gj}.
	\end{align}
	Hence, for $M_n$ sub-exponential in $n$, it holds that
	\begin{align}
	-\frac1n\log P_\rme(\Cn) \asto \frac{V_n}{n}, \label{cota2modx}
	\end{align} where
	\begin{align}
	V_n =  \min_{i\neq j} Z_{ij}(n) \label{defVn}
	\end{align} with
	\begin{align}
	Z_{ij}(n) =   -\sum_{k=1}^n \sum_{x,x'} d_{\rmB}(x,x') \bone\{(X_{ik},X_{jk})=(x,x')\},
	\end{align} for all $i,j \in [M_n]$ and $i\neq j$ (See the proof for this fact from the proof of Theorem \ref{lem:aux2}).
	
	Define
	\begin{align}
	T_{ij}(n) =  \frac{Z_{ij}(n)-\bbE[Z_{ij}(n)]}{\sqrt{\var(Z_{ij}(n))}},
	\end{align}
	we have
	\begin{align}
	\min_{i\neq j} T_{ij}(n)&=\min_{i\neq j} \frac{Z_{ij}(n)-\bbE[Z_{ij}(n)]}{\sqrt{\var(Z_{ij}(n))}} \label{eqsh}.
	\end{align} 
	Now, for any $\eps>0$, let the event
	\begin{align}
	\calE_n =  \bigg\{\frac{1}{M_n(M_n-1)}\bigg|\sum_{i\neq j}\frac{Z_{ij}(n)-\bbE[Z_{ij}(n)]}{\sqrt{\var(Z_{ij}(n))}}\bigg|\geq \eps \bigg\}
	\end{align} for all $n \in \bbZ^+$. Then, we have
	\begin{align}
	\sum_{n=1}^{\infty}\Pro(\calE_n)
	&\leq \sum_{n=1}^{\infty}\bbP\bigg[ \frac{1}{(M_n-1)M_n}\bigg| \sum_{i\neq j} \frac{Z_{ij}(n)-\bbE[Z_{ij}(n)]}{\sqrt{\var(Z_{ij}(n))}}\bigg|\geq \eps\bigg]\\
    &\leq \sum_{n=1}^{\infty}\frac{1}{\eps^2 M_n^2(M_n-1)^2}\var\bigg(\sum_{i\neq j} \frac{Z_{ij}(n)-\bbE[Z_{ij}(n)]}{\sqrt{\var(Z_{ij}(n))}}\bigg) \label{cheby}\\
    &=\sum_{n=1}^{\infty} \frac{1}{\eps^2 M_n^2(M_n-1)^2} \sum_{i\neq j} \var\bigg(\frac{Z_{ij}(n)-\bbE[Z_{ij}(n)]}{\sqrt{\var(Z_{ij}(N))}}\bigg) \label{cu}\\
    &= \sum_{n=1}^{\infty}\frac{1}{\eps^2 M_n(M_n-1)}\\
    &<\infty \label{final},
    \end{align} where \eqref{cheby} follows from Chebyshev's inequality, \eqref{cu} follows from the pairwise independence of $Z_{ij}$'s, and \eqref{final} follows from the condition \eqref{condbat}.
    
    Hence, by Borel–Cantelli lemma, from \eqref{final}, we have
    \begin{align}
    \bbP\bigg[\bigcup_{n=1}^{\infty} \bigcap_{k=n}^{\infty}\calE_k^c\bigg]=1 \label{ohat}.
    \end{align} 
    However, we have
    \begin{align}
    \bbP\bigg[\bigcup_{n=1}^{\infty} \bigcap_{k=n}^{\infty}\calE_k^c\bigg]&=\bbP\bigg[\bigcup_{n=1}^{\infty} \bigcap_{k=n}^{\infty}\bigg\{\frac{1}{M_n(M_n-1)}\bigg|\sum_{i\neq j}\frac{Z_{ij}(k)-\bbE[Z_{ij}(k)]}{\sqrt{\var(Z_{ij}(k))}}\bigg|<\eps \bigg\}\bigg]    \label{ohat2}.
    \end{align}
    It follows from \eqref{ohat} and \eqref{ohat2} that
    \begin{align}
    \bbP\bigg[\bigcup_{n=1}^{\infty} \bigcap_{k=n}^{\infty}\bigg\{\frac{1}{M_n(M_n-1)}\bigg|\sum_{i\neq j} \frac{Z_{ij}(k)-\bbE[Z_{ij}(k)]}{\sqrt{\var(Z_{ij}(k))}}\bigg|<\eps \bigg\}\bigg]=1,
    \end{align}
    or 
    \begin{align}
    \frac{1}{M_n(M_n-1)}\sum_{i\neq j} \frac{Z_{ij}(n)-\bbE[Z_{12}(n)]}{\sqrt{\var(Z_{12}(n))}}\asto 0,\label{eq611}
    \end{align} as $n\to \infty$. Hence, there exists a subset $\calA$ such that $\Pro(\calA)=1$ and 
    \begin{align}
    \frac{1}{M_n(M_n-1)}\sum_{i\neq j} \frac{Z_{ij}(n)-\bbE[Z_{12}(n)]}{\sqrt{\var(Z_{12}(n))}}\to 0
    \end{align} on $\calA$. 
    
    Now, from Theorem \ref{lem:aux2}, we have $T_{ij}(n),T_{i'j'}(n),T_{12}(n)$ are independent as $n\to \infty$ if $(i,j)\neq (i',j')\neq (1,2)$. Then, for any $B_1,B_2 \in \calB(R)$ (Borel sets in $\bbR$), as $n\to \infty$, we have
    \begin{align}
    &\bbP\bigg[\bigg\{T_{ij}(n)-T_{12}(n)\in B_1\bigg\}\cap \bigg\{T_{ij}(n)-T_{12}(n)\in B_2\bigg\} \bigg]\nn\\
    &\qquad =\int_{\bbR} \bbP\bigg[\bigg\{T_{ij}(n)-T_{12}(n)\in B_1\bigg\}\cap \bigg\{T_{i'j'}(n)-T_{12}(n)\in B_2\bigg\}\bigg|T_{12}(n)=\alpha \bigg]f_{T_{12}(n)}(\alpha) d\alpha\\
    &\qquad=\int_{\bbR} \bbP\bigg[\bigg\{T_{ij}(n)\in \alpha+B_1\bigg\}\cap \bigg\{T_{i'j'}(n)\in \alpha+ B_2\bigg\}\bigg|T_{12}(n)=\alpha \bigg]f_{T_{12}(n)}(\alpha) d\alpha\\ 
    &\qquad=\int_{\bbR} \bbP\bigg[\bigg\{T_{ij}(n)\in \alpha+B_1\bigg\}\cap \bigg\{T_{i'j'}(n)\in \alpha+ B_2\bigg\} \bigg]f_{T_{12}(n)}(\alpha) d\alpha \label{ra1}\\
    &\qquad=\int_{\bbR} \bbP\big[T_{ij}(n)\in \alpha+B_1\big] \bbP\big[T_{i'j'}(n)\in \alpha+ B_2 \big]f_{T_{12}}(\alpha) d\alpha\\ 
    &\qquad=\bbP\big[T_{ij}(n)-T_{12}(n)\in B_1\big]\bbP\big[T_{i'j'}(n)-T_{12}(n)\in B_2\big]+o(1),
    \end{align} i.e., $T_{ij}(n)-T_{12}(n)$ and $T_{i'j'}(n)-T_{12}(n)$ are asymptotically independent. This means that $\{T_{ij}(n)-T_{12}(n)\}$ are asymptotically pairwise independent. Hence, by using the same arguments to achieve \eqref{eq611}, we have
    \begin{align}
    \frac{1}{M_n(M_n-1)}\sum_{i\neq j} T_{ij}(n)-T_{12}(n)\asto 0,\label{eq611b}
    \end{align} as $n\to \infty$ (point-wise convergence). Then, there exist a subset $\calB$ such that $\Pro(\calB)=1$, and
    \begin{align}
    \frac{1}{M_n(M_n-1)}\sum_{i\neq j} T_{ij}(n)-T_{12}(n) \to 0 \label{calt}
    \end{align} on $\calB$ as $n\to \infty$.
    
    Hence, we have $P_\rme(\Cn)=1$, where $\calC =  \calA \cap \calB$. It follows that, for any $\eps>0$, on the set $\calC$, as $n$ sufficiently large (which depends on each realization of $\{T_{ij}(n)\}_{n=1}^{\infty}$'s), we have
    \begin{align}
    -\eps &<\frac{1}{M_n(M_n-1)}\sum_{i\neq j} T_{ij}(n)<\eps \label{han1},\\
    -\eps &< \frac{1}{M_n(M_n-1)}\sum_{i\neq j} T_{ij}(n)-T_{12}(n)<\eps \label{han2}.
    \end{align} 
    
The first step consists of showing that $\min\big\{\bbE[h(\min_{i\neq j}T_{ij}(n))-h(T_{12}(n))],\bbE[h(-T_{12}(n))-h(-\min_{i\neq j}T_{ij}(n))]\big\}\to 0$ as $n\to \infty$. We carry out with two sub-steps, step 1a and step 1b.

\subsubsection{Step 1a}

To begin with, we prove that $\bbE[h(\min_{i\neq j}T_{ij}(n))-h(T_{12}(n))]\to 0$ as $n\to \infty$ for all $h\in \{\calH:a\geq 0\}$. We divide into different cases based on the value of $a$ as following:
    \begin{itemize}
    \item Case 1: $\liminf_{n\to \infty} a>0$. \\
    Now, take an arbitrary small $\eps>0$ such that $\eps<a$ as $n\to \infty$. From \eqref{han1}, we have $\min_{i\neq j} T_{ij}(n)<\eps<a$. It follows that
    \begin{align}
    h(\min_{i\neq j} T_{ij}(n))=h\bigg(\frac{1}{M(M-1)}\sum_{i\neq j} T_{ij}(n)\bigg)=c
    \end{align} by the definition of $\calH$. 
    
    Then, we have
    \begin{align}
    h(\min_{i\neq j}T_{ij}(n))-h(T_{12}(n))&=\bigg[h(\min_{i\neq j}T_{ij}(n))-h\bigg(\frac{1}{M_n(M_n-1)}\sum_{i\neq j} T_{ij}(n)\bigg)\bigg] \nn\\
    &\qquad \qquad + \bigg[h\bigg(\frac{1}{M_n(M_n-1)}\sum_{i\neq j} T_{ij}(n)\bigg)-h(T_{12}(n))\bigg] \\
    &=h\bigg(\frac{1}{M_n(M_n-1)}\sum_{i\neq j} T_{ij}(n)\bigg)-h(T_{12}(n))\\
    &\leq \bigg|h\bigg(\frac{1}{M_n(M_n-1)}\sum_{i\neq j} T_{ij}(n)\bigg)-h(T_{12}(n))\bigg| \label{cas1}.
    \end{align}
    \item Case 2: $a\geq 0$ and $\lim_{n\to \infty} a = 0$. \\
    Then, if $\min_{i\neq j} T_{ij}(n)\leq a$ as $n\to \infty$, we have
    \begin{align}
    h(\min_{i\neq j} T_{ij}(n))-h\bigg(\frac{1}{M_n(M_n-1)}\sum_{i\neq j} T_{ij}(n)\bigg)&=h(a)-h\bigg(\frac{1}{M_n(M_n-1)}\sum_{i\neq j} T_{ij}(n)\bigg)\\
    &\leq \bigg|a- \frac{1}{M_n(M_n-1)}\sum_{i\neq j} T_{ij}(n)\bigg| \label{covas}\\
    &\leq \max\bigg\{a,\frac{1}{M_n(M_n-1)}\sum_{i\neq j} T_{ij}(n) \bigg\}\\
    &\leq 2\eps \label{cas2a},
    \end{align} where \eqref{covas} follows from $1$-Lipschitz property of $h$ for all $h\in \calV$.
    
   On the other hand, if $a<\min_{i\neq j} T_{ij}(n)\leq \frac{1}{M_n(M_n-1)}\sum_{i\neq j} T_{ij}(n)\leq \eps$ and $\liminf_{n\to \infty} c>0$, we have
   \begin{align}
   h(\min_{i\neq j} T_{ij}(n))-h\bigg(\frac{1}{M_n(M_n-1)}\sum_{i\neq j} T_{ij}(n)\bigg)&= \frac{1}{M_n(M_n-1)}\sum_{i\neq j} T_{ij}(n)-\min_{i\neq j} T_{ij}(n)\\
   &\leq \eps-a\\
   &\leq \eps \label{cas2b}.
   \end{align}
   In addition, if $a<\min_{i\neq j} T_{ij}(n)\leq \frac{1}{M_n(M_n-1)}\sum_{i\neq j} T_{ij}(n)\leq \eps$ and $\lim_{n\to \infty} c=0$, we have
   \begin{align}
   h(\min_{i\neq j} T_{ij}(n))-h\bigg(\frac{1}{M_n(M_n-1)}\sum_{i\neq j} T_{ij}(n)\bigg)&\leq c\\
   &\leq \eps \label{cas2c}.
   \end{align}
   From \eqref{cas2a}, \eqref{cas2b}, and \eqref{cas2c}, it holds that
   \begin{align}
   h(\min_{i\neq j} T_{ij}(n))-h\bigg(\frac{1}{M_n(M_n-1)}\sum_{i\neq j} T_{ij}(n)\bigg)\leq 2\eps \label{cas2}
   \end{align} for this case.
\end{itemize}
   From \eqref{cas1} and \eqref{cas2}, we have
   \begin{align}
   h(\min_{i\neq j}T_{ij}(n))-h(T_{12}(n))&\leq 2\eps+ \bigg|h\bigg(\frac{1}{M_n(M_n-1)}\sum_{i\neq j} T_{ij}(n)\bigg)-h(T_{12}(n))\bigg| \label{stres}\\
    &\leq 2\eps+ \bigg|\frac{1}{M_n(M_n-1)}\sum_{i\neq j} T_{ij}(n)-T_{12}(n)\bigg| \label{bus}
   \end{align} on $\calC$.
   
 From \eqref{bus}, on $\calC$, we have
   \begin{align}
   \bbE\bigg[h(\min_{i\neq j}T_{ij}(n))-h(T_{12}(n))\bigg]&=\bbE\bigg[\big(h(\min_{i\neq j}T_{ij}(n))-h(T_{12}(n))\big)\bone\{\calC\}]+\bbE\bigg[\big(h(\min_{i\neq j}T_{ij})-h(T_{12})\big)\bone\{\calC^c\}]\\
   &\leq \bbE\bigg[\big(h(\min_{i\neq j}T_{ij}(n))-h(T_{12}(n))\big)\bone\{\calC\}]+ c \Pro(\calC^c)\\
   &=\bbE\bigg[\big(h(\min_{i\neq j}T_{ij}(n))-h(T_{12}(n))\big)\bone\{\calC\}]
    \label{chivas}.
   \end{align}
   Now, since $|\big(h(\min_{i\neq j}T_{ij}(n))-h(T_{12}(n))\big)\bone\{\calC\}|\leq c$, hence by the bounded convergence theorem \cite{Billingsley}, we have
   \begin{align}
   &\limsup_{n\to \infty} \bbE\bigg[h(\min_{i\neq j}T_{ij}(n))-h(T_{12}(n))\bigg]\nn\\
   &\leq \bbE\bigg[\lim_{n\to \infty}  h(\min_{i\neq j}T_{ij}(n))-h(T_{12}(n))\bigg]\\
   &=\bbE\bigg[\bigg(2\eps+ \bigg|\frac{1}{M_n(M_n-1)}\sum_{i\neq j} T_{ij}(n)-T_{12}(n)\bigg|\bigg)\bone\{\calC\}\bigg]+ c \Pro(\calC^c) \\
   &\leq 2\eps+ \eps \label{han3} \\
   &=3\eps \label{han4}
   \end{align} for any $\eps>0$, where \eqref{han3} follows from \eqref{han2}.
   
   From \eqref{han4}, by taking $\eps\to 0$, we obtain
   \begin{align}
   \limsup_{n\to \infty} \bbE\big[h(\min_{i\neq j}T_{ij}(n))-h(T_{12}(n))\big]\leq 0 \label{eqsh2a}.
   \end{align} 
   Since $h(\min_{i\neq j}T_{ij}(n))-h(T_{12}(n))\geq 0$, by the fact that $h$ is non-increasing for all $h\in \calV$, from \eqref{eqsh2a}, we obtain
   \begin{align}
   \lim_{n\to \infty} \bbE\big[h(\min_{i\neq j}T_{ij}(n))-h(T_{12}(n))\big]= 0 \label{eqsh2}.
   \end{align}
   
   \subsubsection{Step 1b}
   
Next, we prove that $\bbE[h(-\min_{i\neq j}T_{ij}(n))-h(-T_{12}(n))]\to 0$ as $n\to \infty$ for all $h\in \{\calH:a< 0\}$.

For all $h\in \calH$, let $\tilh(x) =  h(-x)$ for all $x \in \bbR$. Then, we have
   \begin{align}
   h(\min_{i\neq j}T_{ij}(n))-h(T_{12}(n))=
   \tilh(T_{12}(n))-\tilh(\min_{i\neq j}\{T_{ij}\}(n)) \label{keypu}.
   \end{align}
   Take an arbitrary $\eps>0$, under the condition \eqref{condbat}, from \eqref{han1} and \eqref{han2}, as $n\to \infty$, we obtain that
   \begin{align}
   -\eps&<\frac{1}{M_n(M_n-1)}\sum_{i\neq j} T_{ij}(n)<\eps,  \label{han1mod},\\
   -\eps&<\frac{1}{M(M-1)}\sum_{i\neq j} T_{ij}(n)-T_{12}(n) <\eps \label{hand2mod}
   \end{align} on $\calC$ where $P_\rme(\Cn)=1$.  

Now, we show that $\bbE[\tilh(T_{12}(n))-\tilh(\min_{i\neq j}T_{ij}(n))]\to 0$ as $n\to \infty$. 

Similar to Step 1a, we divide into different cases based on $a+c$ as following:
   \begin{itemize}
   	\item Case 1: $\limsup_{n\to \infty} (a+c)<0$. \\
   	Then, by taking $\eps>0$ small enough such that $\eps<-\limsup_{n\to \infty} (a+c)$, from \eqref{han1mod}, as $n\to \infty$, we have
   	\begin{align}
   	\min_{i\neq j}T_{ij}(n)\leq \frac{1}{M_n(M_n-1)}\sum_{i\neq j} T_{ij}(n)\leq \eps<-(a+c).
   	\end{align}
   	Hence, it holds that
   	\begin{align}
   	\tilh\big(\max_{i\neq j}\tilT_{ij}(n)\big)=\tilh\bigg(\frac{1}{M_n(M_n-1)}\sum_{i\neq j} T_{ij}(n) \bigg)=0.
   	\end{align}
    It follows that on $\calC$, as $n\to \infty$, we have
    \begin{align}
    \tilh\bigg(\frac{1}{M_n(M_n-1)}\sum_{i\neq j} T_{ij}(n) \bigg)-\tilh(\min_{i\neq j}T_{ij}(n))= 0\label{bu1mod}.
    \end{align}
	\item Case 2: $\lim_{n\to \infty} a+c=0$.\\
Then, if $\min_{i\neq j} T_{ij}(n)\leq -(a+c)$, as $n\to \infty$, we have
\begin{align}
\tilh\bigg(\frac{1}{M_n(M_n-1)}\sum_{i\neq j} T_{ij}(n) \bigg)-\tilh(\min_{i\neq j}\{T_{ij}(n)\})&=\tilh\bigg(\frac{1}{M_n(M_n-1)}\sum_{i\neq j} T_{ij}(n) \bigg)-\tilh(-(a+c))\\
&\leq \bigg|\frac{1}{M_n(M_n-1)}\sum_{i\neq j} T_{ij}(n) +(a+c)\bigg|\\
&\leq \bigg|\frac{1}{M_n(M_n-1)}\sum_{i\neq j} T_{ij}(n)\bigg| +|a+c|\\
&< 2\eps \label{cas1mod}.
\end{align}
In addition, if $\min_{i\neq j} T_{ij}(n)\geq -(a+c)$ and $\limsup_{n\to \infty} a<0$, we have $\lim_{n\to \infty} c=0$. Hence,
\begin{align}
\tilh\bigg(\frac{1}{M_n(M_n-1)}\sum_{i\neq j} T_{ij}(n) \bigg)-\tilh(\min_{i\neq j}T_{ij}(n))&\leq \bigg|\frac{1}{M_n(M_n-1)}\sum_{i\neq j} T_{ij}(n)-\min_{i\neq j}T_{ij}(n)\bigg|\\
&\leq \eps+(a+c)\\
&<2\eps \label{cas2mod}.
\end{align}
Finally, if if $\min_{i\neq j} T_{ij}(n)\geq -(a+c)$ and $\lim_{n\to \infty} a=0$, we have
\begin{align}
\tilh\bigg(\frac{1}{M_n(M_n-1)}\sum_{i\neq j} T_{ij}(n) \bigg)-\tilh(\min_{i\neq j}T_{ij}(n))&=\tilh(-a)-\tilh(\min_{i\neq j}T_{ij}(n))\\
&\leq \bigg|-a-\min_{i\neq j}T_{ij}(n)\bigg|\\
&\leq -a-(a+c)\\
&<2\eps \label{cas3mod}.
\end{align}
From \eqref{cas1mod}, \eqref{cas2mod}, and \eqref{cas3mod}, as $n\to \infty$, we have
\begin{align}
\tilh\bigg(\frac{1}{M_n(M_n-1)}\sum_{i\neq j} T_{ij}(n) \bigg)-\tilh(\min_{i\neq j}\{T_{ij}(n)\})< 2\eps+ \bigg|\tilh(T_{12}(n))-\tilh\bigg(\frac{1}{M_n(M_n-1)}\sum_{i\neq j} T_{ij}(n) \bigg)\bigg| \label{bu2mod}
\end{align} on $\calC$.
\end{itemize}
It follows from \eqref{bu1mod} and \eqref{bu2mod} that
\begin{align}
&\tilh(T_{12}(n))-\tilh(\min_{i\neq j}T_{ij}(n))\nn\\
&\qquad=\bigg[\tilh(T_{12}(n))-\tilh\bigg(\frac{1}{M_n(M_n-1)}\sum_{i\neq j} T_{ij}(n) \bigg)\bigg]+ \bigg[\tilh\bigg(\frac{1}{M_n(M_n-1)}\sum_{i\neq j} T_{ij}(n) \bigg)-\tilh(\min_{i\neq j}T_{ij}(n))\bigg]\\
&\qquad<\bigg|\tilh(T_{12}(n))-\tilh\bigg(\frac{1}{M_n(M_n-1)}\sum_{i\neq j} T_{ij}(n) \bigg)\bigg|+2\eps \label{bu3mod}\\
&\qquad \leq 2\eps + \bigg|T_{12}(n)-\frac{1}{M_n(M_n-1)}\sum_{i\neq j} T_{ij}(n)\bigg| \label{cimod}.
\end{align}   
   From \eqref{cimod}, on $\calC$, we have
  \begin{align}
  \bbE\bigg[\tilh(T_{12}(n))-\tilh(\min_{i\neq j}T_{ij}(n))\bigg]&=\bbE\bigg[\big(\tilh(T_{12}(n))-\tilh(\min_{i\neq j}T_{ij}(n)))\big)\bone\{\calC\}]+\bbE\bigg[\big(\tilh(T_{12}(n)-\tilh(\min_{i\neq j}T_{ij}(n)))\big)\bone\{\calC^c\}]\\
  &\leq \bbE\bigg[\big(\tilh(T_{12}(n))-\tilh(\min_{i\neq j}T_{ij}(n))\big)\bone\{\calC\}]+ c \Pro(\calC^c)\\
  &=\bbE\bigg[\big(\tilh(T_{12}(n)-\tilh(\min_{i\neq j}T_{ij}(n)))\big)\bone\{\calC\}]
  \label{chivasmod}.
  \end{align}
  Now, since $|\big(\tilh(\min_{i\neq j}T_{ij}(n))-\tilh(T_{12}(n))\big)\bone\{\calC\}|\leq c$, hence by the bounded convergence theorem \cite{Billingsley}, we have
  \begin{align}
  &\limsup_{N\to \infty} \bbE\bigg[\tilh(T_{12}(n))-\tilh(\min_{i\neq j}T_{ij}(n))-\bigg]\nn\\
  &\leq \bbE\bigg[\lim_{n\to \infty} \tilh(T_{12}(n))-\tilh(\min_{i\neq j}T_{ij}(n))\bigg]\\
  &=\bbE\bigg[\bigg(2\eps+ \bigg|\frac{1}{M_n(M_n-1)}\sum_{i\neq j} T_{ij}(n)-T_{12}(n)\bigg|\bigg)\bone\{\calC\}\bigg]+ c \Pro(\calC^c) \\
  &\leq 2\eps+ \eps \label{han3x} \\
  &=3\eps \label{han4x}
  \end{align} for any $\eps>0$, where \eqref{han3x} follows from \eqref{hand2mod}.
  
  From \eqref{han4x}, by taking $\eps\to 0$, we obtain
  \begin{align}
  \limsup_{N\to \infty} \bbE\big[\tilh(T_{12}(n))-\tilh(\min_{i\neq j}T_{ij}(n))\big]\leq 0 \label{eqsh2amod}.
  \end{align} 
  Since $h(\min_{i\neq j}T_{ij}(n))-h(T_{12}(n))\geq 0$, from \eqref{eqsh2amod}, we obtain
  \begin{align}
  \lim_{n\to \infty} \bbE\big[\tilh(T_{12}(n))-\tilh(\min_{i\neq j}T_{ij}(n))\big]= 0 \label{eqsh2mod},
  \end{align}
  or
  \begin{align}
  \lim_{n\to \infty} \bbE\big[h(-T_{12}(n))-h(-\min_{i\neq j}T_{ij}(n))\big]= 0 \label{cusure}.
  \end{align}
From \eqref{eqsh2} and \eqref{cusure}, we finally have 
\begin{align}	
	\lim_{n\to \infty} \min\bigg\{\bbE\big[h(\min_{i\neq j}T_{ij}(n))-h(T_{12}(n))\big], \bbE\big[h(-T_{12}(n))-h(-\min_{i\neq j}T_{ij}(n))\big]\bigg\}= 0 \label{cao1} 
\end{align} for all $h\in \calH$.\\

\subsubsection{Step 2}

In this step, we show that $\lim_{n\to \infty} d_{W,\rm{mod}}(\min_{i\neq j} T_{ij},Z)=0$. Indeed, from Lemma \ref{lem:nathanstan}, we have
	\begin{align}
	\sup_{h \in \calH} \big|\bbE\big[f'_h(T_{ij}(n))-T_{ij}(n) f_h(T_{ij}(n))\big]\big|&\leq \frac{1}{n^{3/2}}\sum_{k=1}^n \bbE[|X_k|^3]+\frac{\sqrt{2}}{n \sqrt{\pi}}\sqrt{\sum_{k=1}^n \bbE[X_k^4]}\\
	&= \frac{1}{\sqrt{n}} \bbE[|X_1|^3] + \frac{\sqrt{2}}{\sqrt{\pi n}} \sqrt{\bbE[X_1^4]} \label{key1}
	\end{align}
	where
	\begin{align}
	X_k:&=\frac{-\sum_{x,x'} d_{\rmB}(x,x') \big(\bone\{(X_{ik},X_{jk})=(x,x')\}-\bbP\big[(X_{ik},X_{jk})=(x,x')\big]\big)}{\sqrt{\var \big(-\sum_{x,x'} d_{\rmB}(x,x') \bone\{(X_{ik},X_{jk})=(x,x')\}\big)}},\qquad  \forall k \in [N].
	\end{align}
	Now, observe that
	\begin{align}
	\var(Z_{ij}(n))&=\var\bigg(-\sum_{k=1}^n \sum_{x,x'} d_{\rmB}(x,x') \bone\{(X_{ik},X_{jk})=(x,x')\}\bigg)\\
	&=\sum_{k=1}^n \var \bigg(\sum_{x,x'} d_{\rmB}(x,x') \bone\{(X_{ik},X_{jk})=(x,x')\}\bigg) \label{gh}\\
	&= n \var \bigg(\sum_{x,x'} d_{\rmB}(x,x') \bone\{(X_{ik},X_{jk})=(x,x')\}\bigg), \quad \forall k \in [n] \label{str},
	\end{align} where \eqref{gh} and \eqref{str} follow from the fact that $(X_{ik},X_{jk})$ are i.i.d. given $i,j$.
	
	Hence, we have
	\begin{align}
	&\var \bigg(-\sum_{x,x'} d_{\rmB}(x,x') \bone\{(X_{ik},X_{jk})=(x,x')\}\bigg)\nn\\
	&\qquad = \bbE\bigg[\bigg(\sum_{x,x'} d_{\rmB}(x,x') \bone\{(X_{ik},X_{jk})=(x,x')\}\bigg)^2\bigg]-\bigg(\bbE\bigg[\sum_{x,x'} d_{\rmB}(x,x') \bone\{(X_{ik},X_{jk})=(x,x')\}\bigg]\bigg)^2\\
	&\qquad= \sum_{x,x'} d_{\rmB}^2(x,x') \bbP\big[(X_{ik},X_{jk})=(x,x')\big]- \bigg( \sum_{x,x'} d_{\rmB}(x,x') \bbP\big[(X_{ik},X_{jk})=(x,x')\big] \bigg)^2\\
	&\qquad = \sum_{x,x'} d_{\rmB}^2(x,x') Q(x)Q(x')- \bigg( \sum_{x,x'} d_{\rmB}(x,x') Q(x)Q(x') \bigg)^2 =  L_2 \label{lanx1}.
	\end{align}
	In addition, we have
	\begin{align}
	&\bbE\bigg[\bigg|\sum_{x,x'} d_{\rmB}(x,x') \big(\bone\{(X_{ik},X_{jk})=(x,x')\}-\bbP\big[(X_{ik},X_{jk})=(x,x')\big]\big)\bigg|^3\bigg]\\
	&\qquad \leq 4 \bigg(\bbE\bigg[\bigg|\sum_{x,x'} d_{\rmB}(x,x') \bone\{(X_{ik},X_{jk})=(x,x')\}\bigg|^3\bigg]+\bigg|\sum_{x,x'} d_{\rmB}(x,x') \bbP\big[(X_{ik},X_{jk})=(x,x')\big]\bigg|^3\bigg) \label{lanx2}\\
	&\qquad =4 \bigg[\sum_{x,x'} d_{\rmB}^3(x,x') Q(x) Q(x')+ \bigg(\sum_{x,x'} d_{\rmB}(x,x') Q(x)Q(x')\bigg)^3\bigg] =  L_3 \label{lanx3},
	\end{align} where \eqref{lanx2} follows from $(a+b)^3 \leq 4(|a|^3+|b^3|)$.
	
	Similarly, we have
	\begin{align}
	&\bbE\bigg[\bigg|\sum_{x,x'} d_{\rmB}(x,x') \big(\bone\{(X_{ik},X_{jk})=(x,x')\}-\bbP\big[(X_{ik},X_{jk})=(x,x')\big]\big)\bigg|^4\bigg]\\
	&\qquad \leq 8 \bigg[\sum_{x,x'} d_{\rmB}^4(x,x') Q(x) Q(x')+ \bigg(\sum_{x,x'} d_{\rmB}(x,x') Q(x)Q(x')\bigg)^4 \bigg] =  L_4 \label{lanx4},
	\end{align} where we use $(a+b)^4 \leq 8(a^4+b^4)$ in \eqref{lanx4}.
	
	Hence, from \eqref{key1}, \eqref{lanx1}, \eqref{lanx3}, and \eqref{lanx4}, we obtain
	\begin{align}
\sup_{h \in \calH} \big|\bbE\big[f'_h(T_{ij}(n))-T_{ij}(n) f_h(T_{ij}(n))\big]\big| \leq \frac{1}{\sqrt{n}} \bigg(\frac{L_3}{L_2^{3/2}}\bigg)+ \sqrt{\frac{2}{\pi n}} \frac{L_4}{L_2^2}, \qquad \forall i\neq j \label{keyx1}.
	\end{align}
	
Similarly, we also have
\begin{align}
\sup_{h \in \calH} \big|\bbE\big[f'_h(-T_{ij}(n))+T_{ij}(n) f_h(-T_{ij}(n))\big]\big| \leq \frac{1}{\sqrt{n}} \bigg(\frac{L_3}{L_2^{3/2}}\bigg)+ \sqrt{\frac{2}{\pi n}} \frac{L_4}{L_2^2}, \qquad \forall i\neq j \label{keyx1mod}.
\end{align}
Since $T_{ij}(n)$'s (for $i\neq j$) are identically distributed by the random codebook generation, it follows from Lemma \ref{lem:co1} and \eqref{keyx1} that for any $x \in \bbR$,
	\begin{align}
	&d_{W,\rm{mod}}(\min_{i\neq j} T_{ij},Z)\nn\\
	&\qquad \leq \max\bigg\{\sup_{h \in \calH} \big|\bbE\big[f'_h(T_{12}(n))-T_{12}(n) f_h(T_{12}(n))\big]\big|,\sup_{h \in \calH} \big|\bbE\big[f'_h(-T_{12}(n))+T_{12}(n) f_h(-T_{12}(n))\big]\big|\bigg\} \nn\\
	&\qquad \qquad +\sup_{h\in \calH} \min\bigg\{\bbE[h(\min_{i\neq j} T_{ij}(n))- h(T_{12}(n))], \bbE[ h(-T_{12}(n))-h(-\min_{i\neq j} T_{ij}(n))]\bigg\}\\
	&\qquad \leq \frac{1}{\sqrt{n}} \bigg(\frac{L_3}{L_2^{3/2}}\bigg)+ \sqrt{\frac{2}{\pi n}} \frac{L_4}{L_2^2}\nn\\
	&\qquad \qquad +\sup_{h\in \calH} \min\bigg\{\bbE[h(\min_{i\neq j} T_{ij}(n))- h(T_{12}(n))], \bbE[ h(-T_{12}(n))-h(-\min_{i\neq j} T_{ij}(n))]\bigg\} \to 0 \label{keyx2},
	\end{align} where \eqref{keyx2} follows from \eqref{cao1}. 

\subsubsection{Step 3}

In the third step, we whow that $\lim_{n\to \infty} |\bbP(\min_{i\neq j} T_{ij}(n)\leq x)-\bbP(\min_{i\neq j} T_{ij}(n) \geq -x)\big|=0$ for all $x \in \bbR$ and $x$ is a continuous point of the limiting distribution of $\min_{i\neq j} T_{ij}(n)$.

First, by the i.i.d. random codebook generation, observe that $T_{12}(n), T_{23}(n),\cdots, T_{(M-1)M}(n)$ are i.i.d.. For any $x\in \bbR$, as $n,M_n\to \infty$, we have
\begin{align}
\bbP\big[\min_{i\neq j} T_{ij}(n)\geq x\big]&\leq  \bbP\big[\min\{T_{12}(n),T_{34}(n),\cdots, T_{(M-1)M}\}(n)\geq x\big]\\
&=\bbP[T_{12}(n)\geq x] \bbP[T_{34}(n)\geq x]\cdots \bbP[T_{(M-1)M}(n)\geq x]\\
&=\big(\bbP[T_{12}(n)\geq x]\bigg)^{\lfloor M/2\rfloor}\label{luck}\\
&\leq \bigg(Q(x)+o(1)\bigg)^{\lfloor M/2\rfloor} \label{luck2}\\
&\leq \bigg(\frac{Q(\limsup_{N\to \infty} x)}{2}\bigg)^{\lfloor M/2\rfloor} \label{lic2}\\
&\to 0 \label{baota},
\end{align} where \eqref{luck2} follows from the fact that $T_{12}(n)$ is the sum of $n$ i.i.d. terms, so we can apply the CLT, and \eqref{lic2} follows from the fact that as $n\to \infty$, $Q(x)\geq Q(\limsup_{n\to \infty}x )> Q(\limsup_{n\to \infty}x )/2$ as $n$ sufficiently large, and \eqref{baota} follows from $M_n\to \infty$ under the condition \eqref{condbat}.

Now, we divide into different cases depending on $x$:
\begin{itemize}
	\item Case 1: $x \in \bbR$ and $\liminf_{n\to \infty}x<0.$\\
	Then, there exists an $\eps>0$ such that $x<-\eps$ as $n$ sufficiently large. Hence, under the condition \eqref{condbat}, we have
	\begin{align}
	\bbP(\min_{i\neq j} T_{ij}(n) \geq -x)&\leq \bbP\bigg[\frac{1}{M_n(M_n-1)}\sum_{i\neq j} T_{ij}(n)  \geq -x\bigg]\\
	&\leq \bbP\bigg[\frac{1}{M_n(M_n-1)}\sum_{i\neq j} T_{ij}(n)  \geq \eps \bigg]\\
	&\to 0 \label{ila},
	\end{align} where \eqref{ila} follows from \eqref{eq611}.
	It follows from \eqref{baota} and \eqref{ila} that
	\begin{align}
	\lim_{n\to \infty} \bbP\big[\min_{i\neq j} T_{ij}(n)\geq x\big]-\bbP[\min_{i\neq j} T_{ij}(n) \geq -x]=0,
	\end{align} which leads to
	\begin{align}
	\lim_{n\to \infty} \big|\bbP\big[\min_{i\neq j} T_{ij}(n)\geq x\big]-\bbP[\min_{i\neq j} T_{ij}(n) \geq -x]\big|=0.
	\end{align}
	\item Case $\lim_{n\to \infty} x= 0$.\\
	From \eqref{eqsh2}, we have
	\begin{align}
	\lim_{n\to \infty} \bbE\big[h(\min_{i\neq j}T_{ij}(n))-h(T_{12}(n))\big]= 0, \label{eqsh2roi}
	\end{align} 
	for all $h\in \calH: \liminf_{n\to \infty} a\geq 0$.\\
	Now, for any fixed constants $(u,v)$ such that $0<u<v$, define $h_u(t) =  \bone\{t\leq x\}$ and $h_v(t) =  \bone\{t\leq x\}$ for all $t\in R$. Then, we have
	\begin{align}
	&\bbP\bigg[\min_{i\neq j} T_{ij}(n)\notin (u,v]\bigg)\nn\\
	&\qquad =\bbP\big[\min_{i\neq j} T_{ij}(n)> v\big]+\bbP\big[\min_{i\neq j} T_{ij}(n)\leq u\big]\\
	&\qquad \leq \bigg|\bbP\big[\min_{i\neq j} T_{ij}(n)> v\big]-\bbP(T_{12}(n)> v)\bigg]+\bigg[\bbP\big[\min_{i\neq j} T_{ij}(n)\leq u\big]-\bbP(T_{12}(n)\leq u)\bigg]\nn\\
	&\qquad \qquad + \bbP(T_{12}(n)>v)+ \bbP(T_{12}(n)\leq u)\\
	&\qquad = \bigg|\bbP\big[\min_{i\neq j} T_{ij}(n)\leq v\big]-\bbP(T_{12}(n)\leq v)\bigg]+\bigg[\bbP\big[\min_{i\neq j} T_{ij}(n)\leq u\big]-\bbP(T_{12}(n)\leq u)\bigg]\nn\\
	&\qquad \qquad + \bbP(T_{12}(n)>v)+ \bbP(T_{12}(n)\leq u)\\
	&\qquad = \bigg|\bbE\big[h_v(\min_{i\neq j}T_{ij}(n))-h_v(T_{12}(n))\big]+\bbE\big[h_u(\min_{i\neq j}T_{ij}(n))-h_u(T_{12}(n))\big]\bigg|\nn\\
	&\qquad \qquad + \bbP(T_{12}(n)>v)+ \bbP(T_{12}(n)\leq u)\\
	&\qquad =o(1)+ \bbP(T_{12}(n)>v)+ \bbP(T_{12}(n)\leq u) \label{khu3} \\
	&\qquad=o(1)+ Q(v)+ Q(u) \label{khu4},
	\end{align} where \eqref{khu3} follows from \eqref{eqsh2roi}, and \eqref{khu4} follows from $T_{12}(n) \dto \calN(0,1)$ by the CLT (it is a sum of i.i.d. random variables with finite variance). Hence, for any $\eps>0$, by letting $u$ and $v$ sufficiently large constants such that $Q(u)+Q(v)< \eps/2$. Then, for $N$ sufficiently large, we have
	\begin{align}
	\bbP\big[\min_{i\neq j} T_{ij}(n)\in (u,v]\big)>1-\eps.
	\end{align} This means that the probability distribution of $\min_{i\neq j} T_{ij}(n)$ is tight. Then, \cite{Billingsley}, the distribution of $\min_{i\neq j} T_{ij}(n)$ is convergent to some distribution of a random variable $Y$. Hence, by Lemma \ref{lem:boundmea}, we have
	\begin{align}
	\limsup_{n\to \infty} |\bbP(\min_{i\neq j} T_{ij}(n)\leq x)-\bbP(\min_{i\neq j} T_{ij}(n) \geq -x)\big| &\leq 2 (8\pi)^{-1/4}\limsup_{n\to \infty} \sqrt{d_{W,\rm{mod}}(\min_{i\neq j} T_{ij}(n),Z)}=0 \label{ask},
	\end{align} where \eqref{ask} follows from Step 3.
	\item For any $x\geq 0$, then $-x\leq 0$. Hence, from two cases above, we have
	\begin{align}
\lim_{n\to \infty} \big|\bbP(\min_{i\neq j} T_{ij}(n)\leq -x)-\bbP(\min_{i\neq j} T_{ij}(n) \geq x)\big|=0 \label{Q20}.
	\end{align}
However, for any continuous point $x$ of the limiting distribution of $\min_{i\neq j} T_{ij}(n)$, we have
\begin{align}
&\lim_{n\to \infty} \big|\bbP(\min_{i\neq j} T_{ij}(n)\leq -x)-\bbP(\min_{i\neq j} T_{ij}(n) \geq x)\big|\\
&\qquad=\lim_{n\to \infty}\big|(1-\bbP(\min_{i\neq j} T_{ij}(n)\geq -x))-(1-\bbP(\min_{i\neq j} T_{ij}(n) \leq x))\big|\\
&\qquad =\lim_{n\to \infty} |\bbP(\min_{i\neq j} T_{ij}(n)\leq x)-\bbP(\min_{i\neq j} T_{ij}(n) \geq -x)\big| \label{Q25}.
\end{align}
From \eqref{Q20} and \eqref{Q25}, we obtain
\begin{align}
\lim_{n\to \infty} |\bbP(\min_{i\neq j} T_{ij}(n)\leq x)-\bbP(\min_{i\neq j} T_{ij}(n) \geq -x)\big|=0 \label{Q21}.
\end{align}
\end{itemize} 
	
\subsubsection{Step 4}

The last step proves that $T_n =  \frac{V_n-\bbE[V_n]}{\sqrt{\var(V_n)}} \dto \calN(0,1)$. From Lemma \eqref{keyx2}, Lemma \ref{lem:boundmea}, and Step 3, we have 
\begin{align}
\bbP\big[\min_{i\neq j} T_{ij}(n)\leq x\big]-\bbP\big[Z\leq x\big] \to 0\label{songkey}
\end{align} as $n\to \infty$ for any continuous point  $x \in \bbR$ of the limiting distribution of $\min_{i\neq j} T_{ij}(n)$ (point-wise convergence),
or
\begin{align}
\min_{i\neq j} T_{ij}(n) \dto Z=\calN(0,1) \label{supfact}.
\end{align} 
Now, observe that
	\begin{align}
	T_n=\min_{i\neq j} \tilT_{ij}(n),
	\end{align}
	where
	\begin{align}
	\tilT_{ij}(n) =  \frac{Z_{ij}(n)-\bbE[V_n]}{\sqrt{\var(V_n)}}.
	\end{align}
Hence, by using Lemma \ref{lem:aux2021} and (by using the same arguments to achieve \eqref{factT} from \eqref{supfact} as in the proof of Theorem \ref{lem:aux2}), we obtain
	\begin{align}
	T_n   \dto \calN(0,1) \label{factTc}.
	\end{align} 
Finally, from \eqref{cota2modx} and \eqref{factTc}, by applying Slutsky's theorem, we have
	\begin{align}
	\frac{\frac{-\log \PeCn}{n}-\bbE[\frac{-\log \PeCn}{n}]}{\sqrt{\var{\big(\frac{-\log \PeCn}{n}}\big)}}\dto \calN(0,1).
	\end{align}

\subsection{Proof of Theorem~\ref{theo:1}}
\label{sub:p6}

Consider first the case $0\leq R<C$. Since the random variable $\PeCn$ takes values in $[0,1]$, we have that
\begin{align}\label{eqn:cep_2}
\var[\PeCn] &= \EE[\PeCn^2] -\EE[\PeCn]^2\\\label{eqn:cep_3}
&\leq \EE[\PeCn^2]\\   \label{eqn:cep_4}
&\leq \EE[\PeCn]\to 0
\end{align}
where \eqref{eqn:cep_4} follows from the assumption that $\EE[\PeCn]\to  0$ for $0\leq R<C$. Applying Chebyshev's inequality we have that
\begin{align}\label{eqn:cep_5}
\PP\Bigl [ \left|\PeCn - \EE[\PeCn] \right|\geq \delta\Bigr ]&\leq \frac{\var[\PeCn]}{\delta^2}\\\label{eqn:cep_6}
&\leq \frac{\EE[\PeCn]}{\delta^2}\to 0
\end{align}
where \eqref{eqn:cep_6} follows from \eqref{eqn:cep_4} and is valid for any given $\delta>0$.

Now let us consider the case $R>C$. The following hold:
\begin{gather}\label{eqn:cep_5_1}
\EE[\PeCn] \to  1\\
\EE[\PeCn]^2 \to  1 \label{eqn:cep_5_2}\\
\EE[\PeCn^2] \geq \EE[\PeCn]^2 \label{eqn:cep_5_3}\\
\EE[\PeCn^2] \to  1 \label{eqn:cep_5_4}\\
\var[\PeCn] \to  0 \label{eqn:cep_5_5}
\end{gather}
where \eqref{eqn:cep_5_1} follows from the theorem assumption, \eqref{eqn:cep_5_2} follows from \eqref{eqn:cep_5_1}, \eqref{eqn:cep_5_3} follows from Jensen's inequality, \eqref{eqn:cep_5_4} follows from \eqref{eqn:cep_5_2} and \eqref{eqn:cep_5_3} and the fact that $\EE[\PeCn^2]\leq 1$, while \eqref{eqn:cep_5_5} follows from \eqref{eqn:cep_5_4} and \eqref{eqn:cep_5_2} and the additivity of limits.

Finally, using Chebyshev's inequality again we find that
\begin{align}\label{eqn:cep_5}
\PP\Bigl [ \left|\PeCn - \EE[\PeCn] \right|\geq \delta\Bigr ] \leq \frac{\var[\PeCn]}{\delta^2}\to 0
\end{align}
for any $\delta>0$.

\subsection{Proof of Theorem~\ref{thm:abet}}
\label{sub:p7}

	First, by the condition \eqref{asspx}, we observe that
	\begin{align}
	\frac{\var(P_\rme(\Cn))}{\big(\bbE[P_\rme(\Cn)]\big)^2} 
	 &=\frac{\bbE[P_\rme^2(\Cn)]-\big(\bbE[P_\rme(\Cn)]\big)^2}{\big(\bbE[P_\rme(\Cn)]\big)^2}\\
	&= \frac{\bbE[P_\rme^2(\Cn)]}{\big(\bbE[P_\rme(\Cn)]\big)^2}-1\\
	& \to 0 \label{M6}.
	\end{align}
	
	On the other hand, we know that 
$
   \frac{\var(P_\rme(\Cn))}{\big(\bbE[P_\rme(\Cn)]\big)^2} \to \infty
$ if $\Erce<\Etrc$. Hence, from \eqref{M6}, we must have 
\begin{align}
\Etrc=\Erce \label{ladyfact}.
\end{align}

Now, for any $\eps>0$, we have
	\begin{align}
	\bbP\bigg[\bigg|-\frac{\log P_\rme(\Cn)}{N}-\Etrc\bigg|>\eps  \bigg] & =\bbP\bigg[\bigg\{P_\rme(\Cn)< 2^{-n(\Etrc+\eps)}\bigg\}\cup \bigg\{P_\rme(\Cn)> 2^{-n(\Etrc-\eps)}\bigg\} \bigg]\\
	& = \bbP\bigg[P_\rme(\Cn)< 2^{-n(\Etrc+\eps)}\bigg]+ \bbP\bigg[P_\rme(\Cn)> 2^{-n(\Etrc-\eps)} \bigg] \label{tas11}\\
	& \leq \bbP\bigg[P_\rme(\Cn)< 2^{-n(\Etrc+\eps)}\bigg]+ 2^{n(\Etrc-\eps)}\bbE[P_\rme(\Cn)] \label{mark}\\
	& \leq \bbP\bigg[P_\rme(\Cn)< 2^{-n(\Etrc+\eps)}\bigg]+ 2^{n(\Etrc-\eps)}2^{-n(\Erce-\eps/2\big)} \label{mark2}\\
	&=  \bbP\bigg[P_\rme(\Cn)< 2^{-n(\Etrc+\eps)}\bigg]+2^{-n\eps/2} \label{M0}
	\end{align}	 for $n$ sufficiently large, where \eqref{mark} follows from Markov's inequality, and \eqref{mark2} follows from $\bbE[P_\rme(\Cn)]\doteq 2^{-n\Erce} $, so $\bbE[P_\rme(\Cn)]\leq 2^{-n(\Erce-\eps/2)} $ for $n$ sufficiently large.
	
	Now, observe that
	\begin{align}
	\bbP\bigg[P_\rme(\Cn)< 2^{-n(\Etrc+\eps)}\bigg]
	& =\bbP\bigg[P_\rme(\Cn)-\bbE[P_\rme(\Cn)]< 2^{-n(\Etrc+\eps)}-\bbE[P_\rme(\Cn)]\bigg]\\
	& =\bbP\bigg[-\big(P_\rme(\Cn)-\bbE[P_\rme(\Cn)]\big)> \bbE[P_\rme(\Cn)]-2^{-n(\Etrc+\eps)}\bigg] \label{M1}.
	\end{align}
	Now, since $\bbE[P_\rme(\Cn)]\doteq 2^{-n\Erce}=2^{-n\Etrc}$ by \eqref{ladyfact}, so $\bbE[P_\rme(\Cn)]-2^{-n(\Etrc+\eps)}>0$ for $n$ sufficiently large. It follows from \eqref{M1} that
	\begin{align}
	\bbP\bigg[P_\rme(\Cn)< 2^{-n(\Etrc+\eps)}\bigg]
	& \leq \bbP\bigg[\big|P_\rme(\Cn)-\bbE[P_\rme(\Cn)]\big|> \bbE[P_\rme(\Cn)]-2^{-n(\Etrc+\eps)}\bigg]\\
	& \approx \frac{\var(P_\rme(\Cn))}{\big(\bbE[P_\rme(\Cn)]\big)^2} \label{M3}\\
	& \to 0 \label{M50},
	\end{align} where \eqref{M3} follows Markov's inequality and the fact that $\bbE[P_\rme(\Cn)]-2^{-n(\Etrc+\eps)}=\Theta\big(\bbE[P_\rme(\Cn)]\big)$, and \eqref{M50} follows from \eqref{M6}.

From \eqref{M0} and \eqref{M50}, we obtain 
\begin{align}
\bbP\bigg[\bigg|-\frac1n\log P_\rme(\Cn)-\Etrc\bigg|>\eps  \bigg]\to 0,
\end{align} or equivalently,
\begin{align}
-\frac1n\log P_\rme(\Cn) \pto \Etrc.
\end{align}

\subsection{Proof of Theorem~\ref{theo:paley_zygmund}}
\label{sub:p8}

From~\cite[Th.~1]{Giusseppe2021e} and from \eqref{keypo}, for $n$ sufficiently large we have:
\begin{align}\label{eqn:pz_proof_1}
\Pro\left[P_{\rm e}(\mathcal{C}_n) \geq \gamma_n^{\rho}\min_{\rho\in[1,\infty)}E[P_e(\mathcal{C}_n)^{\frac{1}{\rho}}]^{\rho}\right]&=
\Pro\left[P_{\rm e}(\Cn) \geq 2^{-n(\Etrc-\epsilon_n)}\right]\\
&\leq \frac{1}{\gamma_n}
\end{align}
where $\gamma_n\to \infty$, $\frac{\log\gamma_n}{n}\to 0$ and $\epsilon_n\to  0$.
On the other hand, the Paley-Zygmund inequality \cite{paley_zyg1932} implies that, for large enough $n$:
\begin{align}\label{eqn:pz_proof_2}
\Pro\left[P_{\rm e}(\Cn) \geq \delta_n\EE[P_{\rm e}(\Cn)]\right]&=\Pro\left[P_{\rm e}(\Cn) \geq 2^{-n(\Erce+\epsilon'_n)}\right]
\\&\geq (1-\delta_n)^2\frac{\EE[P_{\rm e}(\Cn)]^2}{E[P_{\rm e}(\Cn)^2]}
\end{align}
where we choose a sequence $\delta_n$ that goes to zero subexponentially, i.e., $\epsilon'_n\to  0$ and $0<\delta_n<1$ $\forall n$.
Let $n_0$ be such that $\Delta E>\epsilon_{n}$, $\forall n>n_0$. Note that such an $n_0$ must exist from the definition of limit for $\epsilon_n$. Now consider the following chain of inequalities for a large enough $n$, $n>n_0$:
\begin{align}\label{eqn:pz_proof_3a}
2^{-n(\Etrc-\epsilon_n)} &= 2^{-n(\Erce+\Delta E-\epsilon_n)} \\\label{eqn:pz_proof_3c}
& \leq 2^{-n(\Erce+\Delta E-\epsilon_{n_0})}\\\label{eqn:pz_proof_3d}
& < 2^{-n(\Erce+\epsilon'_{n})}
\end{align}
where \eqref{eqn:pz_proof_3a} is from the theorem statement, \eqref{eqn:pz_proof_3c} is valid from a certain $n$ onwards from the definition of limit for $\epsilon_n$, while \eqref{eqn:pz_proof_3d} is because $\Delta E-\epsilon_{n_0}$ is a positive constant and, for large enough $n$, $\epsilon'_{n}<\Delta E-\epsilon_{n_0}$.
Now, using \eqref{eqn:pz_proof_3d}, \eqref{eqn:pz_proof_1} and \eqref{eqn:pz_proof_2} we have:
\begin{align}
(1-\delta_n)\frac{\EE[P_{\rm e}(\Cn)]^2}{\EE[P_{\rm e}(\Cn)^2]}&\leq\Pro\left[P_{\rm e}(\Cn) \geq \delta_n\EE[P_{\rm e}(\Cn)]\right]\\ 
&= \Pro\left[P_{\rm e}(\Cn) \geq 2^{-n(\Erce+\epsilon'_n)}\right]\\	
\label{eqn:pz_proof_4c}
&\leq \Pro\left[P_{\rm e}(\Cn) \geq 2^{-n(\Etrc-\epsilon_n)}\right]\\
&=\Pro\left[P_{\rm e}(\Cn) \geq \gamma_n^{\rho}\min_{\rho\in[1,\infty)}\EE[P_{\rm e}(\Cn)^{\frac{1}{\rho}}]^{\rho}\right]
\\&\leq \frac{1}{\gamma_n}
\end{align}
where\eqref{eqn:pz_proof_4c} follows from \eqref{eqn:pz_proof_3d}. Finally, notice that, by definition, the following inequalities hold:
$$(1-\delta_n)\to 1$$
$$\frac{1}{\gamma_n}\to 0$$
that imply:
$$\frac{\EE[P_{\rm e}(\Cn)]^2}{\EE[P_{\rm e}(\Cn)^2]}\to 0.$$

\subsection{Proof of Theorem~\ref{lem:d_nonconv}}
\label{sub:p9}

Under the condition $\Etrc>\Erce$, it holds by Theorem \ref{theo:paley_zygmund}
	\begin{align}
	\frac{\bbE[\PeCn]}{\sqrt{\var{\PeCn}}} \to 0 \label{be1}.
	\end{align}
	Now, assume that 
	\begin{align}
	\frac{\PeCn-\bbE[\PeCn]}{\sqrt{\var(\PeCn)}}\dto \calN(0,1) \label{be2}.
	\end{align}
	Then, from \eqref{be1} and \eqref{be2} and Slutsky's theorem \cite{Billingsley}, it holds that
	\begin{align}
	\frac{\PeCn}{\sqrt{\var(\PeCn)}}\dto \calN(0,1) \label{tag},
	\end{align} which is a contradiction since the LHS of \eqref{tag} is a non-negative random variable.

\subsection{Proof of Theorem~\ref{corox2}}
\label{sub:p10}

 First, if $\liminf_{n\to \infty} \frac{\bbE[P_\rme^2(\Cn)]}{\big(\bbE[P_\rme(\Cn)]\big)^2}>1$, then it holds that
	\begin{align}
	\nu = \limsup_{n\to \infty} \frac{\bbE[P_\rme(\Cn)]}{\sqrt{\var(P_\rme(\Cn))}}< \infty.
	\end{align}
	Then, for $n$ sufficiently large, we have
	\begin{align}
	\frac{P_\rme(\Cn)-\bbE[P_\rme(\Cn)]}{\sqrt{\var(P_\rme(\Cn))}}&\geq \frac{P_\rme(\Cn)}{\sqrt{\var(P_\rme(\Cn))}}-\nu \\
	&\geq -\nu,
	\end{align} hence 
	\begin{align}
	\frac{P_\rme(\Cn)-\bbE[P_\rme(\Cn)]}{\sqrt{\var(P_\rme(\Cn))}} \notdto \calN(0,1).
	\end{align}
Hence, the condition \eqref{tacon}  implies that 
	\begin{align}
	\frac{\bbE[P_\rme^2(\Cn)]}{\big(\bbE[P_\rme(\Cn)]\big)^2} \to 1 \label{convex1},
	\end{align}
which leads to
	\begin{align}
	\frac{\bbE[P_\rme(\Cn)]}{\sqrt{\var(P_\rme(\Cn))}}\to \infty \label{afact2}.
	\end{align}
Now, for any $\eps>0$, we have
	\begin{align}
	&\bbP\bigg[\bigg|-\frac1n\log P_\rme(\Cn)-\Etrc\bigg|>\eps  \bigg]\nn\\
	&\qquad =\bbP\bigg[\bigg\{P_\rme(\Cn)< 2^{-n(\Etrc+\eps)}\bigg\}\cup \bigg\{P_\rme(\Cn)> 2^{-n(\Etrc-\eps)}\bigg\} \bigg]\\
	&\qquad \leq \bbP\bigg[P_\rme(\Cn)< 2^{-n(\Etrc+\eps)}\bigg]+ \bbP\bigg[P_\rme(\Cn)> 2^{-n(\Etrc-\eps)} \bigg] \label{tas1}.
	\end{align}	
	Furthermore, under the condition $\frac{\bbE[P_\rme(\Cn)]}{\sqrt{\var(P_\rme(\Cn))}} \to \infty$, we must have
	\begin{align}
	\Etrc=\Erce.
	\end{align}
	Observe that
	\begin{align}
	\bbP\bigg[P_\rme(\Cn)> 2^{-n(\Etrc-\eps)} \bigg]& \leq 2^{n(\Etrc-\eps)}\bbE\big[P_\rme(\Cn)\big]\\
	& \doteq 2^{n(\Etrc-\eps)}2^{-n \Erce}\\
	&= 2^{-n \eps} \label{bug10}.
	\end{align}
	On the other hand, from \eqref{tacon}, for  $Z \sim \calN(0,1)$, we have
	\begin{align}
	\bbP\bigg[P_\rme(\Cn)< 2^{-n(\Etrc+\eps)}\bigg]
	& \leq \bbP\bigg[\frac{P_\rme(\Cn)-\bbE[P_\rme(\Cn)]}{\sqrt{\var(P_\rme(\Cn))}}\leq  \frac{2^{-n(\Etrc+\eps)}-\bbE[P_\rme(\Cn)]}{\sqrt{\var(P_\rme(\Cn))}}\bigg]\\
	&=\bbP\bigg[Z \leq  \frac{2^{-n(\Etrc+\eps)}-\bbE[P_\rme(\Cn)]}{\sqrt{\var(P_\rme(\Cn))}}\bigg]+o_n(1)\label{K100b},
	\end{align} where \eqref{K100b} follows from the condition \eqref{tacon}.
	
	Now, since $2^{-n(\Etrc+\eps)}-\bbE[P_\rme(\Cn)]<0$ for $n$ sufficiently large since $\bbE[P_\rme(\Cn)]\doteq 2^{-n \Erce}$ and $\Etrc\geq \Erce$, it holds that
	\begin{align}
	\bbP\bigg[Z \leq  \frac{2^{-n(\Etrc+\eps)}-\bbE[P_\rme(\Cn)]}{\sqrt{\var(P_\rme(\Cn))}}\bigg]=Q\bigg(\frac{\bbE[P_\rme(\Cn)]-2^{-n(\Etrc+\eps)}}{\sqrt{\var(P_\rme(\Cn))}}\bigg) \to 0 \label{bun1}
	\end{align} as $n \to \infty$ since
	\begin{align}
	\frac{\bbE[P_\rme(\Cn)]-2^{-n(\Etrc+\eps)}}{\sqrt{\var(P_\rme(\Cn))}} \to \infty \label{bun2}.
	\end{align} 	
	From \eqref{tas1}, \eqref{bug10}, and \eqref{bun2}, we obtain
	\begin{align}
	\bbP\bigg[\bigg|-\frac1n\log P_\rme(\Cn)-\Etrc\bigg|>\eps  \bigg] \to 0
	\end{align} or \eqref{reverd} holds.

\subsection{Proof of Theorem~\ref{thm:ub}}
\label{sub:p11}

We first introduce a result which is developed in \cite{Nathan2011a} for sum of random variables with local dependence.
\begin{definition} We say that a collection of random variables $(Y_1,Y_2,\cdots,Y_n)$ has dependency neighbourhoods $\calN_i \subset \{1,2,\cdots,n\}, i=1,2,\cdots,n$, if $i \in \calN_i$ and $Y_i$ is independent of $\{Y_j\}_{j\notin \calN_i}$. 
\end{definition}
\begin{lemma} \cite[Th.~3.6]{Nathan2011a}  \label{thm:nathan} Let $Y_1,Y_2,\cdots, Y_n$ be random variables such that $\bbE[Y_i^4]<\infty$, $\bbE[Y_i=0]$, $\sigma^2=\var(\sum_{i=1}^n Y_i)$, and define $T=\sum_{i=1}^n Y_i/\sigma$. Let the collection $(Y_1,Y_2,\cdots, Y_n)$ have dependency neighborhoods $\calN_i, i=1,2,\cdots,n$, and also define  $D =  \max_{1\leq i\leq n} |\calN_i|$. Then for $Z$ a standard normal random variable,
	\begin{align}
	d_W(T,Z)\leq \frac{D^2}{\sigma^3}\sum_{i=1}^n \bbE|Y_i|^3+ \frac{\sqrt{28} D^{3/2}}{\sqrt{\pi}\sigma^2}\sqrt{\sum_{i=1}^n \bbE[Y_i^4]}.
	\end{align}	
\end{lemma}

Observe that
	\begin{align}
	P_{\rme}^{\rm ub}(\Cn)-\bbE[P_{\rme}^{\rm ub}(\Cn)]&=\frac{1}{M_n}\sum_{i=1}^{M_n} \sum_{j\neq i} Y_{ij}\\
	&=\frac{2}{M_n}\sum_{i=1}^{M_n} \sum_{i<j\leq M_n} Y_{ij}, 
	\end{align}
	where
	\begin{align}
	Y_{ij} =  \bbP\big[\{\bX_i \to \bX_j\}\big]- \bbE\big[\bbP\big[ \{\bX_i \to \bX_j\}\big]\big], \qquad \forall i,j \in [M_n]\times [M_n].
	\end{align}
	For i.i.d. random coding ensembles, $\{Y_{ij}\}_{1\leq i<j\leq M_n}$ are pairwise independent and identically distributed by the symmetry of the random codebook ensemble. Hence, we have
	\begin{align}
	\sigma^2& =  \var\bigg(\sum_{i=1}^{M_n} \sum_{j\neq i} Y_{ij}\bigg)\\
	&=4 \var\bigg(\sum_{1\leq i<j\leq M_n} Y_{ij}\bigg)\\
	&= 4 \sum_{1\leq i<j\leq M_n} \var(Y_{ij})\\
	&=2 M_n(M_n-1)\gamma^2.
	\end{align}
	In addition, it is easy to see that $D\leq 2(M_n-1)$. Hence, by Lemma \ref{thm:nathan}, we have
	\begin{align}
	d_W\bigg(\frac{P_{\rme}^{\rm ub}(\Cn)-\bbE[P_{\rme}^{\rm ub}(\Cn)]}{\sqrt{\var(P_{\rme}^{\rm ub}(\Cn)}},Z\bigg)&\leq \frac{4(M_n-1)^2}{(2M_n(M_n-1)\gamma^2)^{3/2}} \bigg(\frac{M_n(M_n-1)}{2}\bigg)\bbE\big[|Y_{12}|^3\big]\nn\\
	&\qquad + \frac{\sqrt{28} (2(M_n-1))^{3/2}}{\sqrt{\pi} (2M_n(M_n-1)\gamma^2)} \sqrt{\frac{M_n(M_n-1)}{2} \bbE[|Y_{12}|^4]}\\
	&\leq \frac{M_n}{\gamma^3}\bbE[|Y_{12}|^3]+ \sqrt{\frac{28}{\pi}} \sqrt{M_n\bbE[|Y_{12}|^4]}
   \label{hol},
	\end{align} which tends to zero if both \eqref{c:cond1} and \eqref{c:cond2} happen simultaneously.

\subsection{Proof of Theorem~\ref{mainthm4}}
\label{sub:p12}

We first state two auxiliary lemmas.

\begin{lemma} \label{lem:boundmeaext} If $Z \sim \calN(0,1)$, then for any random variable $T$, it holds that
	\begin{align}
	d_K(T,Z) \leq 2 (8\pi)^{-1/4}\sqrt{\tild_{W,\rm{mod}}(T,Z)} \label{cachuaext}.
	\end{align}
\end{lemma}
\begin{IEEEproof}
The proof is similar to the first part of the proof of Lemma~\ref{lem:boundmea} in Appendix~\ref{ap:proof43},  so we omit this proof.	
\end{IEEEproof}
\begin{lemma} \label{Wasserthm2} If $T$ is a random variable such that $\bbE[T]=0$ and $\var(T)=1$, and $Z$ has the standard normal distribution, then
	\begin{align}
	d_K(T,Z)<  14(8\pi)^{-1/4}\sqrt{\bbE[|T|]+ \bbE \big[|T^2-1|\big]} \label{keyfactox}.
	\end{align}
\end{lemma}
\begin{IEEEproof}
Appendix~\ref{Wasserthm2:proof}.	
\end{IEEEproof}

Theorem~\ref{mainthm4} is a direct application of Lemma \ref{Wasserthm2}  by setting $T=g_n(P_\rme(\Cn))$, gives a criterion for the convergence in distribution of the error exponent and any function of the error probability, in general.

\appendix

\section{Proofs of Lemmas}

\subsection{Proof of Lemma \ref{lem:aux2021}} 

\label{ap:proof24}

Since $U_n \dto U$, by Skorokhod's representation theorem \cite{Billingsley}, there exists a probability space $(\Omega,\calF,P)$ and two random variables $V_n$ and $V$ such that $V_n \sim U_n$ and $V \sim U$ such that $V_n \asto  V$ on $(\Omega,\calF,P)$. Now, for any fixed $\eps\in (0,1)$, observe that
	\begin{align}
	\bbE_P[|V_n|^{1+\eps}]&\leq \bigg(\bbE_P[|V_n|^2]\bigg)^{(1+\eps)/2}
	\label{conv}\\
	&=1<\infty \label{conv2},
	\end{align} where \eqref{conv} follows from the concavity of the function $f(x) =  x^{(1+\eps)/2}$ for any $\eps \in (0,1)$, and \eqref{conv2} follows from $\bbE_P[V_n^2]=\bbE[U_n^2]=1$. From \eqref{conv2}, it follows that $V_n$ is uniformly integrable on $(\Omega,\calF,P)$ \cite{Billingsley}. Hence, we have
	\begin{align}
	\bbE[U]&=\bbE_P[V]\\
	&=\lim_{n\to \infty} \bbE_P[V_n]\\
	&=\lim_{n\to \infty} \bbE[U_n]\\
	&=0.
	\end{align} 
	On the other hand, for any fixed $n$, we have
	\begin{align}
	|V_n|^2\bone\{|V_n|>\alpha\}\leq |V_n|^2, 
	\end{align} which satisfies
	\begin{align}
	\bbE_P[|V_n|^2]=\bbE[|U_n|^2]=1 \label{star}.
	\end{align}
	Hence, by the dominated convergence theorem \cite{Royden}, we have
	\begin{align}
	\lim_{\alpha\to \infty} \bbE_P\big[|V_n|^2\bone\{|V_n|>\alpha\}\big]&=\bbE_P\big[\lim_{\alpha\to \infty}|V_n|^2\bone\{|V_n|>\alpha\}\big]\\
	&=0 \label{cau0}
	\end{align} uniformly in $n$.
	
	On the other hand, for each fixed $\alpha$, we have
	\begin{align}
	\lim_{n\to \infty}\bbE_P\big[|V_n|^2\bone\{|V_n|>\alpha\}\big]&=\lim_{n\to \infty} \bbE_P[V_n^2] -\bbE_P\big[|V_n|^2\bone\{|V_n|\leq \alpha\}\big]\\
	&= 1 -\lim_{n\to \infty}\bbE_P\big[|V_n|^2\bone\{|V_n|\leq \alpha\}\big] \label{cau1},
	\end{align} where \eqref{cau1} follows from \eqref{star}. Note that
	\begin{align}
	|V_n|^2\bone\{|V_n|\leq \alpha\}\leq \alpha^2,
	\end{align}  and
	\begin{align}
	|V_n|^2\bone\{|V_n|\leq \alpha\} \asto  V^2 \bone\{|V|\leq \alpha\} 
	\end{align}
	by the continuous mapping theorem \cite{Billingsley} and $V_n \asto  V$. Hence, by the dominated convergence theorem \cite{Billingsley}, we also have
	\begin{align}
	\lim_{n\to \infty}\bbE_P\big[|V_n|^2\bone\{|V_N|\leq \alpha\}\big]= \bbE_P[|V|^2 \bone\{|V|\leq \alpha\}] \label{cau2}
	\end{align} point-wise in $\alpha$. Hence, from \eqref{cau1} and \eqref{cau2}, we have
	\begin{align}
	\lim_{n\to \infty}\bbE_P\big[|V_n|^2\bone\{|V_n|>\alpha\}\big]=1-\bbE_P[|V|^2 \bone\{|V|\leq \alpha\}] \label{cau3}
	\end{align}  point-wise in $\alpha$.
	
	From \eqref{cau0} and \eqref{cau3}, by Moore-Osgood theorem \cite{Stewart} on the interchange of limits, it holds that
	\begin{align}
	\lim_{\alpha \to \infty} \lim_{n\to \infty}\bbE_P\big[|V_n|^2\bone\{|V_n|\leq \alpha\}\big]&= \lim_{n\to \infty}\lim_{\alpha \to \infty}\bbE_P\big[|V_n|^2\bone\{|V_n|\leq \alpha\}\big]\\
	&=0,
	\end{align} which leads to $\{V_n\}_{n=1}^{\infty}$ be uniformly integrable. Hence, it holds that
	\begin{align}
	\lim_{n\to \infty} \bbE_P[V_n^2]&=\bbE_P[V^2]\\
	&=\bbE[U^2] \label{co10}.
	\end{align}  From \eqref{star} and \eqref{co10}, we obtain
	\begin{align}
	\bbE[U^2]=1.
	\end{align}

\subsection{Proof of Lemma \ref{lem:bestcha}}
\label{app:proof_rce_trc}

First, we prove that for any $\alpha>1$ and $\lambda>0$, the following holds:
	\begin{align}
	\bbE\big[\PeCnQ^{\frac{\lambda}{\alpha n}}\big]^{\frac{\alpha n}{\lambda}}\leq \bbE\big[\PeCnQ^{\frac{\lambda}{ n}}\big]^{\frac{ n}{\lambda}} \label{T11}.
	\end{align}
Indeed, let
\begin{align}
r=\frac{\lambda}{\alpha n}, \qquad p=\frac{\lambda}{n}, \qquad 	q=\frac{\lambda}{(\alpha-1)n}.
\end{align}
Then, it holds that
\begin{align}
\frac{1}{r}=\frac{1}{p}+\frac{1}{q}
\end{align} and $p,q,r \in (0,\infty)$ if $\alpha>1$.
By applying the generalized H\"{o}lder's inequality \cite{Royden,Gallager1965a}, we have
\begin{align}
\big(\bbE[\PeCnQ^r]\big)^{\frac{1}{r}}&\leq \big(\bbE[\PeCnQ^p]\big)^{\frac{1}{p}} \big(\bbE[1^q]\big)^{\frac{1}{q}}\\
&=\big(\bbE[\PeCnQ^p]\big)^{\frac{1}{p}},
\end{align} implying that \eqref{T11} holds.

Since \eqref{T11} holds for any $\alpha>1$, we have 
\begin{align}
\bbE\big[\PeCnQ^{\frac{\lambda}{ n}}\big]^{\frac{ n}{\lambda}}&\geq \lim_{\alpha \to \infty} \bbE\big[\PeCnQ^{\frac{\lambda}{\alpha n}}\big]^{\frac{\alpha n}{\lambda}}\\
&=2^{\bbE[\log \PeCnQ]} \label{T21}
\end{align} where \eqref{T21} follows from the identity $\bbE[\log X] = \lim_{x\to\infty} \log \EE[X^{\frac 1x}]^x$ for any given $X>0$. From the definition of $\EtrcQ$ in \eqref{detrc} and the definition of limit, we have that for every $\epsilon>0$ there exists an $n_0$ such that for $n>n_0$,
\begin{align}
\Big|-\frac1n\bbE[\log \PeCnQ] - \EtrcQ\Big|<\epsilon.
\end{align}
Therefore, from \eqref{T21} we have that
\begin{align}
\bbE\big[\PeCnQ^{\frac{\lambda}{ n}}\big]^{\frac{ n}{\lambda}} \geq 2^{-n(1-\eps)\EtrcQ} 
\label{LH2}.
\end{align} 

Thus, from \eqref{LH2} and \eqref{T11}, by letting $\eps \to 0$, it holds that
\begin{align}
\liminf_{n\to \infty} \bbE\big[\PeCnQ^{\frac{\lambda}{ n}}\big]\geq 2^{-\lambda\EtrcQ} \label{LH3}.
\end{align}

Now, by the concavity of the function $x^{\frac{\lambda}{n}}$ on $(0,\infty)$, we have by Jensen's inequality that
	\begin{align}
	\limsup_{n\to \infty} \bbE\big[\PeCnQ^{\frac{\lambda}{ n}}\big]&\leq \limsup_{n\to \infty} \bbE\big[\PeCnQ\big]^{\frac{\lambda}{ n}} \label{estat}\\
	&= 2^{-\lambda\ErceQ} \label{estat2}.
	\end{align} where \eqref{estat2} follows from the fact that $\bbE[\PeCnQ]\leq  2^{-n\ErceQ}$ \cite{Gallager1965a, MoserBook}.

From \eqref{LH3} and \eqref{estat2}, under the condition that $\EtrcQ=\ErceQ$, it holds that
\begin{align}
\lim_{n\to \infty} \bbE\big[\PeCnQ^{\frac{\lambda}{ n}}\big]= 2^{-\lambda\ErceQ}.
\end{align}	
This concludes the proof of this lemma.

\subsection{Proof of Lemma \ref{lem:GLem}}
\label{ap:proof5}
The upper bound follows from Bhattacharyya bound. Now, by \cite{Gallager1973a}, it holds that
	\begin{align}
	\bbE\big[\PeCn\big]\doteq 2^{-n \ErceQ} \label{gk1}
	\end{align} for all $R<R_{\rm{crit}}(Q)$. In addition, at this range of rate, the Bhattacharyya bound achieves the Gallager's random coding bound $\ErceQ$. Hence, from \eqref{gk1}, we have
	\begin{align}
	\bbE\big[P_{\rme}^{\rm ub}(\Cn)\big]\doteq 2^{-n \ErceQ} \label{gk2}
	\end{align} for all $R<R_{\rm{crit}}(Q)$, where $P_{\rme}^{\rm ub}(\Cn)$ is the union bound on $\PeCn$.
	
	Now, for all rate $R<R_{\rm{crit}}(Q)$, $\ErceQ=R_0(Q)-R$, where $R_0$ is the cut-off rate corresponding to the underlying distribution $Q$, i.e.,
	\begin{align}
	R_0(Q) =  -\log \bigg(\sum_y \bigg(\sum_x Q(x)\sqrt{W(y|x)}\bigg)^2\bigg) \label{gk4}.
	\end{align}
	
	Let $Q_X=Q_X'=Q$. By using standard KKT conditions for convex optimization, it is not hard to prove that
	\begin{align}
	R_0(Q)=\min_{P_{XX'} \in \calP(\calX\times \calX)} D(P_{XX'}\|Q_X Q_X') +\sum_{x,x'} P_{XX'}(x,x')d_{\rmB}(x,x') \label{gk6}.
	\end{align}	
	Hence, from \eqref{gk6}, we obtain
	\begin{align}
	\bbE\big[P_{\rme}^{\rm ub}(\Cn)\big]&\doteq 2^{nR} 2^{-n \min_{P_{XX'} \in \calP(\calX\times \calX)} \big(D(P_{XX'}\|Q_X Q_X') +\sum_{x,x'} P_{XX'}(x,x')d_{\rmB}(x,x') \big)}\\
	&\doteq M_n \sum_{P_{XX'} \in \calP(\calX\times \calX)} 2^{-n D(P_{XX'}\|Q_{XX'})} 2^{-n\sum_{x,x'} P_{XX'}(x,x')d_{\rmB}(x,x') } \label{gk7}.
	\end{align}
	Now, let $\calN(P_{XX'})$ be the number of codeword pairs which have the same join type $P_{XX'}$. Then, it holds that
	\begin{align}
	\calN(P_{XX'})=\sum_{i=1}^{M_n} \sum_{j \neq i} \bone\{(\bX_i,\bX_j) \in \calT(P_{XX'})\},
	\end{align} which leads to
	\begin{align}
	\bbE[\calN(P_{XX'})]&=\sum_{i=1}^{M_n} \sum_{j \neq i} \bbP\big[(\bX_i,\bX_j) \in \calT(P_{XX'})\big]\\
	&= M_n(M_n-1)2^{-n D(P_{XX'}\|Q_{XX'})} \label{gk8}.
	\end{align}
	From \eqref{gk7} and \eqref{gk8}, we have
	\begin{align}
	(M_n-1) \bbE\big[P_{\rme}^{\rm ub}(\Cn)\big]\doteq \sum_{P_{XX'} \in \calP(\calX\times \calX)} \bbE[\calN(P_{XX'})] 2^{-n \sum_{x,x'} P_{XX'}(x,x')d_{\rmB}(x,x') }  \label{gk9}.
	\end{align}
	On the other hand, observe that
	\begin{align}
	(M_n-1) \bbE\big[P_{\rme}^{\rm ub}(\Cn)\big]&= \bbE\bigg[\sum_{i=1}^{M_n} \sum_{j\neq i} \bbP\big(\bX_i \to \bX_j\big)\bigg] \label{gk10}.
	\end{align}
	From \eqref{gk9} and \eqref{gk10}, we obtain
	\begin{align}
	\bbE\bigg[\sum_{i=1}^{M_n} \sum_{j\neq i} \bbP\big(\bX_i \to \bX_j\big)\bigg]=\sum_{P_{XX'} \in \calP(\calX\times \calX)} 2^{-n D(P_{XX'}\|Q_{XX'})} 2^{-n\sum_{x,x'} P_{XX'}(x,x')d_{\rmB}(x,x')} \label{gk11}.
	\end{align}
	Since \eqref{gk11} holds for all random i.i.d. codebook ensembles, hence for any fixed type $P_{XX'} \in \calP(\calX \times \calX)$, by choosing a sub-random codebook ensemble which contains all the codewords with the same joint type $P_{XX'}$, we obtain
	\begin{align}
	\bbP\bigg[\bX_i \to \bX_j\big|(\bX_i,\bX_j) \in \calT(P_{XX'})\bigg] \doteq 2^{-n\sum_{x,x'} P_{XX'}(x,x')d_{\rmB}(x,x')},
	\end{align}
	or \eqref{tfact1} holds.
\subsection{Proof of Lemma \ref{lem:Vtypi}}
\label{ap:proof6}
Observe that
		\begin{align}
		\bbP[\calV_n^c]
		&=\bbP\bigg[\sum_{P_{XX'} \in \calP_n(\calX \times \calX): D(P_{XX'}\|Q_X Q_X')>2R} \calN(P_{XX'})\geq 1\bigg]\\
		&\leq \bbE\bigg[\sum_{P_{XX'}\in \calP_n(\calX \times \calX): D(P_{XX'}\|Q_X Q_X')>2R} \calN(P_{XX'})\bigg]\\
		&=\bbE\bigg[\sum_{P_{XX'}\in \calP_n(\calX \times \calX): D(P_{XX'}\|Q_X Q_X')>2R} \sum_{i=1}^{M_n}\sum_{j\neq i}\bone\{(\bX_i,\bX_j)\in \calT(P_{XX'})\}\bigg]   \label{exposet1}\\
		&=\sum_{P_{XX'}\in \calP_n(\calX \times \calX): D(P_{XX'}\|Q_X Q_X')>2R} \sum_{i=1}^{M_n}\sum_{j\neq i}\bbP\big[(\bX_i,\bX_j)\in \calT(P_{XX'})\big]\\
		&\leq \sum_{P_{XX'}\in \calP_n(\calX \times \calX): D(P_{XX'}\|Q_X Q_X')>2R} \sum_{i=1}^{M_n}\sum_{j\neq i}2^{-n D(P_{XX'}||Q_X Q_X')}\\
		&\leq \sum_{P_{XX'}\in \calP(\calX \times \calX): D(P_{XX'}\|Q_X Q_X')>2R} \sum_{i=1}^{M_n}\sum_{j\neq i}2^{-n D(P_{XX'}||Q_X Q_X')}
		\label{tm1}\\
		&= M_n(M_n-1) \sum_{P_{XX'}\in \calP(\calX \times \calX): D(P_{XX'}\|Q_X Q_X')>2R}2^{-n D(P_{XX'}||Q_X Q_X')}\\
		&\dotleq  2^{2nR} \sum_{P_{XX'}\in \calP(\calX \times \calX): D(P_{XX'}\|Q_X Q_X')>2R}2^{-n D(P_{XX'}||Q_X Q_X')}\\
		&\leq 2^{2nR} \sum_{P_{XX'}\in \calP(\calX \times \calX): D(P_{XX'}\|Q_X Q_X')>2R} 2^{-n(2R+\alpha(R))} \label{eq132}\\
		&\dotleq  2^{-n\alpha(R)}
		\end{align} for some $\alpha(R)>0$, where \eqref{tm1} follows from \cite{Csis00}, \eqref{eq132} follows from the fact that the number of possible joint types on $\calX \times \calX$ is sub-exponential in $n$.
\subsection{Proof of Lemma \ref{lem:aux}}
\label{ap:proof7}
	Define
	\begin{align}
	\tilV_{ij}&=\sum_{P_{XX'}\in \calP_n(\calX \times \calX): D(P_{XX'}\|Q_X Q_X')\leq 2R-\nu}\bone\{(\bX_i,\bX_j)\in \calT(P_{XX'})\} g_n(P_{XX'}).
	\end{align}
	
	Then, we have
	\begin{align}
	D_n&=\frac{1}{M_n}\sum_{P_{XX'}\in \calP_n(\calX \times \calX): D(P_{XX'}\|Q_X Q_X') \leq 2R-\nu} \calN(P_{XX'}) g_n(P_{XX'})\\
	&= \frac{1}{M_n}\sum_{i=1}^{M_n} \sum_{j\neq i}\sum_{P_{XX'}\in \calP_n(\calX \times \calX): D(P_{XX'}\|Q_X Q_X') \leq 2R-\nu}  \bone\{(\bX_i,\bX_j)\in \calT(P_{XX'})\}g_n(P_{XX'})\\
	&= \frac{1}{M_n}\sum_{i=1}^{M_n} \sum_{j\neq i}\tilV_{ij} \label{betc}.
	\end{align}
	It is easy to see that $\{\tilV_{ij}\}_{i,j=1}^{M_n}$ are pairwise independent. Hence, from \eqref{betc}, we have
	\begin{align}
	\var(D_n)=\frac{1}{M_n^2}\sum_{i=1}^{M_n} \sum_{j\neq i} \var(\tilV_{ij}) \label{b1}.
	\end{align}
	Observe that
	\begin{align}
	&\var(\tilV_{ij})=\var\bigg( \sum_{P_{XX'}\in \calP_n(\calX \times \calX): D(P_{XX'}\|Q_X Q_X')\leq 2R-\nu}\bone\{(\bX_i,\bX_j)\in \calT(P_{XX'})\} g_n(P_{XX'}) \bigg)	\\
	&\qquad=\bbE\bigg[ \sum_{P_{XX'}\in \calP_n(\calX \times \calX): D(P_{XX'}\|Q_X Q_X')\leq 2R-\nu}\bigg(\bone\{(\bX_i,\bX_j)\in \calT(P_{XX'})\}\nn\\
	&\qquad \qquad \qquad -\bbE\bigg[\bone\{(\bX_i,\bX_j)\in \calT(P_{XX'})\}\bigg]\bigg)^2 g_n^2(P_{XX'})  \bigg] \label{pairwise} \\
	&\qquad=  \sum_{P_{XX'}\in \calP_n(\calX \times \calX): D(P_{XX'}\|Q_X Q_X')\leq 2R-\nu}\bigg(\bbE\bigg[\bigg(\bone\{(\bX_i,\bX_j)\in \calT(P_{XX'})\}\bigg)^2\bigg]\nn\\
	&\qquad \qquad \qquad -\bigg(\bbE\bigg[\bone\{(\bX_i,\bX_j)\in \calT(P_{XX'})\}\bigg]\bigg)^2\bigg) g_n^2(P_{XX'}) \\
	&\qquad\leq \sum_{P_{XX'}\in \calP(\calX \times \calX): D(P_{XX'}\|Q_X Q_X')\leq 2R-\nu}\bigg(\bbE\bigg[\bigg(\bone\{(\bX_i,\bX_j)\in \calT(P_{XX'})\}\bigg)^2\bigg]\nn\\
	&\qquad \qquad \qquad -\bigg(\bbE\bigg[\bone\{(\bX_i,\bX_j)\in \calT(P_{XX'})\}\bigg]\bigg)^2\bigg) g_n^2(P_{XX'}) \\
	&\qquad=  \sum_{P_{XX'}\in \calP(\calX \times \calX): D(P_{XX'}\|Q_X Q_X')\leq 2R-\nu}\bbP\bigg[(\bX_i,\bX_j)\in \calT(P_{XX'})\bigg]\bigg(1-\bbP\bigg[(\bX_i,\bX_j)\in \calT(P_{XX'})\bigg]\bigg)  g_n^2(P_{XX'})  \label{vfact}\\
	&\qquad\leq 2^{n \max_{P_{XX'}\in \calP(\calX \times \calX): D(P_{XX'}\|Q_X Q_X')\leq 2R-\nu}-\sum_{x,x'}d_{\rmB}(x,x')P_{XX'}(x,x')}\nn\\
	&\qquad \qquad \times \sum_{P_{XX'}\in \calP(\calX \times \calX): D(P_{XX'}\|Q_X Q_X')\leq 2R-\nu} \bbP\bigg[(\bX_i,\bX_j)\in \calT(P_{XX'})\bigg] g_n(P_{XX'}) \label{h10},
	\end{align} where \eqref{pairwise} follows from the pairwise independence of $\{(\bX_i,\bX_j)\}_{i\neq j}$, and \eqref{h10} follows from \eqref{laday1}. 
	
	Hence, from \eqref{b1} and \eqref{h10}, we obtain
	\begin{align}
	\var(D_n)&=\frac{1}{M_n^2}\sum_{i=1}^{M_n} \sum_{j\neq i} \var(\tilV_{ij})\nn\\
	&\leq 2^{-2nR} \times 2^{n \max_{P_{XX'}\in \calP(\calX \times \calX): D(P_{XX'}\|Q_X Q_X')\leq 2R-\nu}-\sum_{x,x'}d_{\rmB}(x,x')P_{XX'}(x,x')}\nn\\
	&\qquad \times \sum_{P_{XX'}\in \calP(\calX \times \calX): D(P_{XX'}\|Q_X Q_X')\leq 2R-\nu}\sum_{i=1}^{M_n} \sum_{j\neq i}   \bbP\bigg[(\bX_i,\bX_j)\in \calT(P_{XX'})\bigg] g_n(P_{XX'}) \\
	&= 2^{-2nR} \times 2^{n \max_{P_{XX'}\in \calP(\calX \times \calX): D(P_{XX'}\|Q_X Q_X')\leq 2R-\nu}-\sum_{x,x'}d_{\rmB}(x,x')P_{XX'}(x,x')}\nn\\
	&\qquad \times \sum_{P_{XX'}\in \calP(\calX \times \calX): D(P_{XX'}\|Q_X Q_X')\leq 2R-\nu}\bbE[\calN(P_{XX'})] g_n(P_{XX'}) \\
	&= 2^{-2nR} \times 2^{n \max_{P_{XX'}\in \calP(\calX \times \calX): D(P_{XX'}\|Q_X Q_X')\leq 2R-\nu}-\sum_{x,x'}d_{\rmB}(x,x')P_{XX'}(x,x')}\nn\\
	&\qquad \qquad \times \bigg(\sum_{P_{XX'}\in \calP(\calX \times \calX): D(P_{XX'}\|Q_X Q_X')\leq 2R-\nu}\bbE[\calN(P_{XX'})] g_n(P_{XX'}) \bigg) \label{bug3}\\
	&=M_n 2^{-2nR} \times 2^{-n \min_{P_{XX'}\in \calP(\calX \times \calX): D(P_{XX'}\|Q_X Q_X')\leq 2R-\nu}\big(\sum_{x,x'}d_{\rmB}(x,x')P_{XX'}(x,x')\big)} \bbE[D_n]\\
	&=2^{-nR} \times 2^{-n \min_{P_{XX'}\in \calP(\calX \times \calX): D(P_{XX'}\|Q_X Q_X')= 2R-\nu}\big(\sum_{x,x'}d_{\rmB}(x,x')P_{XX'}(x,x')\big)} \bbE[D_n]
	\label{buchi1}, 
	\end{align} where \eqref{buchi1} follows from the fact that the optimizer of the linear programming is in the boundary of the convex constraint set \cite{Boyd04}.
	
	On the other hand, from \eqref{betc}, we have
	\begin{align}
	\bbE[D_n]&=\frac{1}{M_n}\sum_{i=1}^{M_n} \sum_{j\neq i}\bbE[\tilV_{ij}] \\
	&=\frac{1}{M_n}\sum_{i=1}^{M_n} \sum_{j\neq i}\sum_{P_{XX'}\in \calP_n(\calX \times \calX): D(P_{XX'}\|Q_X Q_X')\leq 2R-\nu} \bbP\big[(\bX_i,\bX_j) \in \calT(P_{XX'})\big] g_n(P_{XX'})\\
	&=\frac{1}{M_n}\sum_{i=1}^{M_n} \sum_{j\neq i}\sum_{P_{XX'}\in \calP_n(\calX \times \calX): D(P_{XX'}\|Q_X Q_X')\leq 2R-\nu} 2^{-nD(P_{XX'}\|Q_X Q_X')} g_n(P_{XX'}) \label{bet0}\\
	&=(M_n-1)\sum_{P_{XX'}\in \calP_n(\calX \times \calX): D(P_{XX'}\|Q_X Q_X')\leq 2R-\nu} 2^{-nD(P_{XX'}\|Q_X Q_X')} g_n(P_{XX'})\\
	&\doteq (M-1)\sum_{P_{XX'}\in \calP_n(\calX \times \calX): D(P_{XX'}\|Q_X Q_X')\leq 2R-\nu} 2^{-nD(P_{XX'}\|Q_X Q_X')} 2^{-n \sum_{x,x'} d_{\rmB}(x,x') P_{XX'}(x,x')} \label{bet1}\\
	&\doteq (M_n-1) 2^{-n\min_{P_{XX'} \in \calP_n(\calX \times \calX): D(P_{XX'}\|Q_XQ_X')\leq 2R-\nu}\big( D(P_{XX'}\|Q_X Q_X') +\sum_{x,x'} d_{\rmB}(x,x') P_{XX'}(x,x')\big)}\label{bag2a}\\
	&\doteq (M_n-1) 2^{-n\min_{P_{XX'} \in \calP(\calX \times \calX): D(P_{XX'}\|Q_XQ_X')\leq 2R-\nu}\big( D(P_{XX'}\|Q_X Q_X') +\sum_{x,x'} d_{\rmB}(x,x') P_{XX'}(x,x')\big)}\label{bag2},
	\end{align} where \eqref{bet0} follows from \cite{Csis00}, \eqref{bet1} follows from \eqref{laday1}, and \eqref{bag2} follows from the fact that $\calP_n(\calX \times \calX)$ is dense in $\calP(\calX \times \calX)$.
	
	Similarly, observe that
	\begin{align}
	\bbE[P_{\rme}^{\rm ub}(\Cn)]&=\frac{1}{M_n}\sum_{i=1}^{M_n}\sum_{j\neq i} \bbE\big[\Pro[\bX_i \to \bX_j]\big]\\
	&=(M_n-1)\sum_{P_{XX'} \in \calP_n(\calX \times \calX)} \bbE[\calN(P_{XX'}) ] g_n(P_{XX'})\\
	&= (M_n-1)\sum_{P_{XX'} \in \calP_n(\calX \times \calX)} 2^{-n D(P_{XX'}\|Q_X Q_X')} 2^{-n\sum_{x,x'} d_{\rmB}(x,x') P_{XX'}(x,x')}\\
	&\doteq 2^{-N\min_{P_{XX'} \in \calP_n(\calX \times \calX)}\big(D(P_{XX'}\|Q_X Q_X')+\sum_{x,x'} d_{\rmB}(x,x')P_{XX'}(x,x')-R\big)}\\
	&\doteq 2^{-n\min_{P_{XX'} \in \calP(\calX \times \calX)}\big(D(P_{XX'}\|Q_X Q_X')+\sum_{x,x'} d_{\rmB}(x,x')P_{XX'}(x,x')-R\big)} \label{bag3},
	\end{align} where \eqref{bag3} follows from the fact $\calP_n(\calX \times \calX)$ is dense in $\calP(\calX\times \calX)$.
	
	It follows from \eqref{bag3} that
	\begin{align}
	\ErceQ+R=\min_{P_{XX'} \in \calP(\calX \times \calX)} D(P_{XX'}\|Q_X Q_X')+\sum_{x,x'} d_{\rmB}(x,x') P_{XX'}(x,x') \label{zerovat}
	\end{align}
	Now, we have
	\begin{align}
	&\min_{P_{XX'} \in \calP(\calX \times \calX): D(P_{XX'}\|Q_X Q_X')\geq 2R-\nu} D(P_{XX'}\|Q_X Q_X')+\sum_{x,x'} d_{\rmB}(x,x')P_{XX'}(x,x')\nn\\
	&\qquad \geq \min_{P_{XX'} \in \calP(\calX \times \calX): D(P_{XX'}\|Q_X Q_X')\geq 2R-\nu} 2R-\nu+\sum_{x,x'} d_{\rmB}(x,x')P_{XX'}(x,x')\\
	&\qquad= \min_{P_{XX'} \in \calP(\calX \times \calX): D(P_{XX'}\|Q_X Q_X')=2R-\nu} 2R-\nu+\sum_{x,x'} d_{\rmB}(x,x')P_{XX'}(x,x')\label{fo1}
	\end{align} where \eqref{fo1} follows from the fact that the optimizer of the linear programming is on the boundary of the convex constraint set \cite{Boyd04}. 
	
	Now, following the standard approach to solve the linear programming in \eqref{fo1} we have the following Lagrangian: 
	\begin{align}
	\calL(P_{XX'},\lambda)=2R-\nu+\sum_{x,x'} d_{\rmB}(x,x')P_{XX'}(x,x')+\lambda \big(D(P_{XX'}\|Q_X Q_X')-2R+\nu\big).
	\end{align}
	By setting,
	\begin{align}
	0=\frac{\partial \calL(P_{XX'},\lambda)}{\partial P_{XX'}}(x,x')=d_{\rmB}(x,x')+\lambda\bigg(\log \frac{P_{XX'}(x,x')}{Q_X  Q_X'} +\log e\bigg),
	\end{align} we have
	\begin{align}
	P_{XX'}(x,x')&=\frac{1}{e}Q_X(x) Q_{X'}(x') 2^{-\frac{d_{\rmB}(x,x')}{\lambda}} \label{tag1}.
	\end{align}
	Since
	\begin{align}
	\sum_{x,x'} P_{XX'}(x,x')=1,
	\end{align} we must choose $\lambda$ such that
	\begin{align}
	\frac{1}{e}\sum_{x,x'} Q_X(x) Q_{X'}(x') 2^{-\frac{d_{\rmB}(x,x')}{\lambda}}=1 \label{cubat1}.
	\end{align}
	Since $d_{\rmB}(x,x')=-\log \sum_{y \in \calY} \sqrt{W(y|x)W(y|x')} \geq 0$, from \eqref{cubat1} we must have
	$\lambda<0$, otherwise the LHS of \eqref{cubat1} is less than or equal $1/e<1$.
	
	With the choice $P_{XX'}$ in \eqref{tag1}, we have
	\begin{align}
	&2R-\nu+\sum_{x,x'} d_{\rmB}(x,x') P_{X,X'}(x,x')\nn\\
	&\qquad =2R-\nu+\frac{1}{e}\sum_{x,x'}d_{\rmB}(x,x')Q(x) Q(x')2^{-\frac{d_{\rmB}(x,x')}{\lambda}}\\
	&\qquad = 2R-\nu-\frac{1}{e}\sum_{x,x'} Q(x) Q(x')\log \bigg(\sum_{y \in \calY} \sqrt{W(y|x) W(y|x')}\bigg)\bigg(\sum_{y \in \calY} \sqrt{W(y|x) W(y|x')}\bigg)^{\frac{1}{\lambda}}\label{ca1}.
	\end{align}

	Now, from \eqref{cubat1}, we have
	\begin{align}
	1&=\frac{1}{e}\sum_{x,x'} Q_X(x) Q_{X'}(x') 2^{-\frac{d_{\rmB}(x,x')}{\lambda}}\\
	&= \frac{1}{e}\sum_{x,x'} Q_X(x) Q_{X'}(x') \bigg(\sum_{y \in \calY} \sqrt{W(y|x)W(y|x')}\bigg)^{\frac{1}{\lambda}}\\
	&\leq \frac{1}{e}\bigg(\sum_{x,x'} Q(x) Q(x')\sum_{y \in \calY} \sqrt{W(y|x)W(y|x')}\bigg)^{\frac{1}{\lambda}} \label{taco},
	\end{align} where \eqref{taco} follows from the concavity of the function $x^{1/\lambda}$ for $\lambda<0$ and $x \in [0,1]$. 
	
	On the other hand, since $\lambda<0$, the function $-(\log x) x^{1/\lambda}$ is convex in $x \in [0,1]$. Hence, by Jensen's inequality, from \eqref{ca1}, we have
	\begin{align}
	&2R-\nu+\sum_{x,x'} d_{\rmB}(x,x') P_{X,X'}(x,x')\nn\\
	&\qquad > 2R-\nu- \frac{1}{e}\log \bigg(\sum_{x,x'} Q(x)Q(x')\sum_{y \in \calY} \sqrt{W(y|x) W(y|x')}\bigg)\bigg(\sum_{x,x'} \sum_{y \in \calY}Q(x)Q(x') \sqrt{W(y|x) W(y|x')}\bigg)^{\frac{1}{\lambda}} \label{stre1}\\
	&\qquad= 2R-\nu+\bigg[-\log \bigg(\sum_{y \in \calY} \bigg(\sum_{x \in \calX} Q(x)\sqrt{W(y|x)}\bigg)^2\bigg)\bigg]\bigg[\frac{1}{e} \bigg(\sum_{x,x'} \sum_{y \in \calY}Q(x)Q(x') \sqrt{W(y|x) W(y|x')}\bigg)^{\frac{1}{\lambda}}\bigg]\\
	&\qquad= 2R-\nu+ R_0(Q) \bigg[\frac{1}{e} \bigg(\sum_{x,x'} \sum_{y \in \calY}Q(x)Q(x') \sqrt{W(y|x) W(y|x')}\bigg)^{\frac{1}{\lambda}}\bigg] \label{vat}\\
	&\qquad \geq 2R-\nu+ R_0(Q) \label{vat2}\\
	&\qquad\geq R+ \ErceQ \label{vat3}
	\end{align} for $\nu\leq 2R$, where in \eqref{stre1}, the equality does not hold by the condition \eqref{cond0}, \eqref{vat} follows from \cite{MoserBook} where $R_0(Q)$ is the cut-off rate of the DMC under the underlying distribution $Q$, \eqref{vat2} follows from \eqref{taco}, and \eqref{vat3} follows from \cite{MoserBook}.
	
	From \eqref{fo1} and \eqref{vat3}, we have 
	\begin{align}
	&\min_{P_{XX'} \in \calP(\calX \times \calX): D(P_{XX'}\|Q_X Q_X')\geq 2R-\nu}D(P_{XX'}\|Q_X Q_X')+\sum_{x,x'} d_{\rmB}(x,x')P_{XX'}(x,x') \nn\\
	&\qquad \geq \min_{P_{XX'} \in \calP(\calX \times \calX): D(P_{XX'}\|Q_X Q_X')=2R-\nu}2R-\nu+\sum_{x,x'} d_{\rmB}(x,x')P_{XX'}(x,x') \nn\\
	&\qquad > \ErceQ+R \label{vat4}
	\end{align} if $0\leq \nu\leq 2R$.
	
	Therefore, from \eqref{zerovat} and \eqref{vat4}, we must have
	\begin{align}
	\min_{P_{XX'} \in \calP(\calX \times \calX): D(P_{XX'}\|Q_X Q_X')<2R-\nu} D(P_{XX'}\|Q_X Q_X')+\sum_{x,x'}d_{\rmB}(x,x')P_{XX'}(x,x')= \ErceQ+R \label{keybu}.  
	\end{align}
	It follows that
	\begin{align}
	&\min_{P_{XX'} \in \calP(\calX \times \calX): D(P_{XX'}\|Q_X Q_X')\leq 2R-\nu} D(P_{XX'}\|Q_X Q_X') +\sum_{x,x'} d_{\rmB}(x,x') P_{XX'}(x,x')\nn\\
	&\qquad = \begin{cases}  2R-\nu+\sum_{x,x'} d_{\rmB}(x,x') P_{XX'}^*(x,x'), &\qquad \mbox{if} \qquad D(P_{XX'}^*\|Q_X Q_X')=2R-\nu,\\
	R+\ErceQ,& \qquad \mbox{otherwise} \label{bechao}, 
	\end{cases}
	\end{align} where \eqref{bechao} follows from \eqref{keybu}. Here, $P_{XX'}^*$ is an optimizer of $\min_{P_{XX'}\in \calP} D(P_{XX'}\|Q_X Q_{X'})
	+\sum_{x,x'}d_{\rmB}(x,x')P_{XX'}(x,x')-R$.
	
	From \eqref{bag2} and \eqref{bechao},  we obtain \eqref{eq167a}.
	
	Furthermore, from \eqref{buchi1} and \eqref{eq167a}, we obtain
	\begin{align}
	&\frac{\var(D_n)}{\big(\bbE[D_n]\big)^2} \dotleq \begin{cases} 2^{-n\nu}, \qquad D(P_{XX'}^*|Q_X Q_X')=2R-\nu\\ 2^{-n \big(\min_{P_{XX'} \in \calP(\calX \times \calX): D(P_{XX'}\|Q_X Q_X')= 2R-\nu} R-\nu+\sum_{x,x'} d_{\rmB}(x,x') P_{XX'}(x,x')-\ErceQ\big)}, \qquad \mbox{otherwise}  \end{cases}\\
	&\doteq 2^{-\zeta(\nu,R) n}
	\end{align} for
	\begin{align}
	\zeta(\nu,R)=\begin{cases} \nu, \qquad \mbox{if}\qquad  D(P_{XX'}^*|Q_X Q_X')=2R-\nu\\ \min_{P_{XX'} \in \calP(\calX \times \calX): D(P_{XX'}\|Q_X Q_X')= 2R} R+\sum_{x,x'} d_{\rmB}(x,x') P_{XX'}(x,x')-\ErceQ, \qquad \mbox{otherwise}\end{cases}>0,
	\end{align} which is followed from \eqref{vat4} and $\nu>0$.
	
	This concludes our proof of Lemma \ref{lem:aux}.
	
\subsection{Proof of Lemma \ref{lem:tmc}} 
\label{ap:proof8}

Recall the definition of $\calN(P_{XX'})$ in Lemma \ref{lem:Vtypi}. For a pair of codewords with $(\bX_i,\bX_j) \in \calT(P_{XX'})$, it holds that
	\begin{align}
	\bbP\big(\bX_i \to \bX_j\big)=g_n(P_{XX'}) \label{v1}.
	\end{align}
	\begin{align}
	P_{\rme}^{\rm ub}(\Cn)&=\frac{1}{M_n}\sum_{i=1}^{M_n}\sum_{j\neq i} \bbP(\bX_i \to \bX_j)\\
	&=\frac{1}{M_n}\sum_{P_{XX'}} \calN(P_{XX'}) g_n(P_{XX'}) \label{v2}
	\end{align} where \eqref{v2} follows from \eqref{v1}.
	
	First, we consider the case $R>0$. Take an arbitrary $\nu$ such that $0<\nu\leq 2R$. Let
	\begin{align}
	B_n& =  \frac{1}{M_n}\sum_{P_{XX'}\in  \calP_n(\calX \times \calX)} \calN(P_{XX'}) g_n(P_{XX'}),\\
	\tilD_n& =  \frac{1}{M_n}\sum_{P_{XX'} \in \calP_n(\calX \times \calX): D(P_{XX'}\|Q_X Q_X') \leq 2R} \calN(P_{XX'}) g_n(P_{XX'}),\\
	D_n& =  \frac{1}{M_n}\sum_{P_{XX'} \in \calP_n(\calX \times \calX): D(P_{XX'}\|Q_X Q_X') \leq 2R-\nu} \calN(P_{XX'}) g_n(P_{XX'}).
	\end{align} 
	Since $\bbE[\tilD_n] - \bbE[D_n]\to 0$ as $\nu \to 0$ and that $\bbE[\tilD_n]$ and  $\bbE[D_n]$ are exponentially decaying in $n$, for $\nu$ small enough, it holds that
	\begin{align}
	\bbE[D_n]\leq \bbE[\tilD_n]\leq \bbE[D_n] 2^{\eps n/2} \label{ufact1}.
	\end{align}
	
	Recall the typical set defined in Lemma \ref{lem:Vtypi}. For any given $\eps>0$, observe that
	\begin{align}
	&\bbP\bigg[\bigg|- \frac{\log P_{\rme}^{\rm ub}(\Cn)}{n}+\frac{1}{n}\log\bigg(\frac{1}{M_n}\sum_{P_{XX'}\in  \calP_n(\calX \times \calX): D(P_{XX'}\|Q_X Q_X') \leq 2R} \bbE\big[\calN(P_{XX'})\big] g_n(P_{XX'})\bigg)\bigg|>\eps\bigg]\\
	&\qquad=\bbP\bigg[\bigg|\log \frac{B_n}{\bbE[\tilD_n]}\bigg|>\eps n \bigg]\\
	&\qquad\leq \bbP\bigg[\bigg|\log \frac{B_n}{\bbE[\tilD_n]}\bigg|>\eps n\bigg|\calV_n \bigg]\bbP[\calV_n] +\bbP[\calV_n^c] \label{a1}\\
	&\qquad= \bbP\bigg[\bigg|\log \frac{\tilD_n}{\bbE[\tilD_n]}\bigg|>\eps n\bigg|\calV_n \bigg]\bbP[\calV_n] +\bbP[\calV_n^c] \label{a2}\\
	&\qquad\leq  \bbP\bigg[\bigg|\log \frac{\tilD_n}{\bbE[\tilD_n]}\bigg|>\eps n \bigg] +\bbP[\calV_n^c] \label{a3}\\
	&\qquad=\bbP\big[\tilD_n>\bbE[\tilD_n] 2^{\eps n} \big] +\bbP\big[\tilD_n<\bbE[\tilD_n] 2^{-\eps n} \big]+\bbP[\calV_n^c] \label{a4}\\
	&\qquad\leq \frac{\bbE[\tilD_n]}{2^{\eps n} \bbE[\tilD_n]} +\bbP\big[ D_n<\bbE[D_n] 2^{-\eps/2 n} \big]+\bbP[\calV_N^c] \label{a6}\\
	&\qquad= 2^{-\eps n} +\bbP\big[ D_n<\bbE[\tilD_n] 2^{-\eps n} \big]+2^{-n\alpha(R)} \label{a7a}\\
	&\qquad\leq 2^{-\eps n} +\bbP\big[ D_n<\bbE[D_n] 2^{-(\eps/2) n} \big]+2^{-n \alpha(R)} \label{a7}
	\end{align} where \eqref{a1} follows from $\bbP(A)=\bbP(A|B)\bbP(B)+\bbP(A|B^c)\bbP(B^c) \leq \bbP(A|B)\bbP(B)+\bbP(B^c)$, \eqref{a2} follows from the fact that given $\calV_n$, it holds that $B_n=D_n$, \eqref{a3} follows from $\bbP(A|B)\bbP(B) \leq \bbP(A|B)\bbP(B)+\bbP(A|B^c)\bbP(B^c)=\bbP(A)$, \eqref{a6} follows from Markov's inequality, \eqref{a7a} follows from $D_n \leq \tilD_n$ and from Lemma \ref{lem:Vtypi}, \eqref{a7} follows from  $\bbE[\tilD_n]\leq \bbE[D_n] 2^{\eps n/2}$ for $\nu$ sufficiently small by  \eqref{ufact1}.
	
	Now, we have
	\begin{align}
	\bbP\big[ D_n<\bbE[D_n] 2^{-(\eps/2) n} \big]&=\bbP\big[ D_n-\bbE[D_n]<\bbE[D_n]\big(2^{-(\eps/2) n}-1\big) \big]\\
	&\leq \bbP\big[ \big|D_n-\bbE[D_n]\big|> \bbE[D_n] \big(1-2^{-(\eps/2) n}\big) \big]\\
	&\dotleq \frac{\var(D_n)}{\big(\bbE[D_n]\big)^2} \label{a8}\\
	&\dotleq 2^{-n\beta(\nu,R)} \label{eq191},
	\end{align} where \eqref{eq191} follows from Lemma \ref{lem:aux}. 
	
	From \eqref{a7} and \eqref{eq191}, for any $\eps>0$ and $R> 0$, we have
	\begin{align}
	&\bbP\bigg[\bigg|- \frac{\log P_{\rme}^{\rm ub}(\Cn)}{n}+\frac{1}{n}\log\bigg(\frac{1}{M_n}\sum_{P_{XX'}: D(P_{XX'}\|Q_X Q_X') \leq 2R} \bbE\big[\calN(P_{XX'})\big] g_n(P_{XX'})\bigg)\bigg|>\eps\bigg]\\
	&\qquad \dotleq 2^{-\eps n} + 2^{-n\beta(\nu,R)}+ 2^{-n\alpha(R)} \label{eq194}. 
	\end{align}
	It follows from \eqref{eq194} that
	\begin{align}
	\sum_{n=1}^{\infty} \bbP\bigg[\bigg|- \frac{\log P_{\rme}^{\rm ub}(\Cn)}{n}+\frac{1}{n}\log\bigg(\frac{1}{M_n}\sum_{P_{XX'}: D(P_{XX'}\|Q_X Q_X') \leq 2R} \bbE\big[\calN(P_{XX'})\big]g_n(P_{XX'})\bigg)<\infty.
	\end{align} Hence, by Borel-Cantelli's lemma \cite{Billingsley}, we have
	\begin{align}
	-\frac{\log P_{\rme}^{\rm ub}(\Cn)}{n}+\frac{1}{n}\log\bigg(\frac{1}{M_n}\sum_{P_{XX'}: D(P_{XX'}\|Q_X Q_X') \leq 2R}\bbE\big[\calN(P_{XX'})\big]g_N(P_{XX'})\bigg)\asto 0  \label{tachy1}
	\end{align}
	On the other hand, we have
	\begin{align}
	&\bigg|-\frac{\log P_{\rme}^{\rm ub}(\Cn))}{n}+\frac{1}{n}\log\bigg(\frac{1}{M_n}\sum_{P_{XX'}: D(P_{XX'}\|Q_X Q_X') \leq 2R}\bbE\big[\calN(P_{XX'})\big]g_n(P_{XX'})\bigg)\bigg|\nn\\
	&\qquad \leq -\frac{\log P_{\rme}^{\rm ub}(\Cn)}{n} + \bigg|-\frac{1}{n}\log\bigg(\frac{1}{M_n}\sum_{P_{XX'}: D(P_{XX'}\|Q_X Q_X') \leq 2R}\bbE\big[\calN(P_{XX'})\big]g_n(P_{XX'})\bigg)\bigg|\\
	&\qquad \leq -\frac1n\log \PeCn + \bigg|-\frac{1}{n}\log\bigg(\frac{1}{M_n}\sum_{P_{XX'}: D(P_{XX'}\|Q_X Q_X') \leq 2R}\bbE\big[\calN(P_{XX'})\big]g_n(P_{XX'})\bigg)\bigg|\\
	&\qquad \leq E_{\rm{sp}}(R) + \bigg|-\frac{1}{n}\log\bigg(\frac{1}{M_n}\sum_{P_{XX'}: D(P_{XX'}\|Q_X Q_X') \leq 2R}\bbE\big[\calN(P_{XX'})\big]g_n(P_{XX'})\bigg)\bigg|,  \label{c1}\\
	&\qquad \leq E_{\rm{sp}}(R) + \bigg|-\frac{1}{n}\log\bigg( \frac{\bbE[D_n]}{M_n}\bigg)\bigg|,\label{c2}\\
	&\qquad\leq   E_{\rm{sp}}(R) +R+ \min_{P_{XX'} \in \calP_n(\calX \times \calX): D(P_{XX'}\|Q_XQ_X')= 2R}\sum_{x,x'} d_{\rmB}(x,x') P_{XX'}(x,x') \label{c3}\\
	&\qquad \leq  E_{\rm{sp}}(R) +R+ D_{\rmb}<\infty \label{bd},
	\end{align} where \eqref{c1} follows from \cite{MoserBook}, \eqref{c2}, and \eqref{c3} follows from Lemma \ref{lem:aux}, where \eqref{bd} follows with the fact that $d_{\rmB}(x,x') \leq D_{\rmb}<\infty$ for all $x,x'$ by the condition \eqref{cond0}.
	
	From \eqref{tachy1}, \eqref{bd}, and the bounded convergence theorem \cite{Billingsley}, we have
	\begin{align}
	\lim_{n\to \infty} \bbE\bigg[ -\frac{\log P_{\rme}^{\rm ub}(\Cn)}{n}+\frac{1}{n}\log\bigg(\frac{1}{M_n}\sum_{P_{XX'} \in \calP_n(\calX \times \calX): D(P_{XX'}\|Q_X Q_X') \leq 2R}\bbE\big[\calN(P_{XX'})\big]g_n(P_{XX'})\bigg)\bigg]=0 \label{gac1}.
	\end{align}
	Now, by Lemma \ref{lem:aux}, we have
	\begin{align}
	&\lim_{n\to \infty} -\frac{1}{n}\log\bigg(\frac{1}{M_n}\sum_{P_{XX'}\in \calP_n(\calX \times \calX): D(P_{XX'}\|Q_X Q_X') \leq 2R}\bbE\big[\calN(P_{XX'})\big]g_n(P_{XX'})\bigg)\nn\\
	&\qquad=\min_{P_{XX'} \in \calP(\calX \times \calX): D(P_{XX'}\|Q_XQ_X')\leq 2R}\big( D(P_{XX'}\|Q_X Q_X') +\sum_{x,x'} d_{\rmB}(x,x') P_{XX'}(x,x')-R\big) \label{gac2}.
	\end{align}
	Hence, we obtain \eqref{eqn:key0} from \eqref{gac1} and \eqref{gac2}. Note that \eqref{eqn:key1} can be achieved from \eqref{eqn:key0} by using \eqref{bechao} with $\nu=0$.
\subsection{Proof of Lemma \ref{lem:sup}}

\label{ap:proof9}

Let $g_n(P_{XX'})$ be the pairwise error probability given the joint type $P_{XX'}$ for $P_{XX'} \in \calP_n(\calX \times \calX)$. By Lemma \ref{lem:GLem}, the pairwise error probability can be expressed as
	\begin{align}
	g_n(P_{XX'}) \doteq 2^{-n\sum_{x,x'}d_{\rmB}(x,x') P_{XX'}(x,x')} \label{laday1},
	\end{align}
	where
	\begin{align}
	d_{\rmB}(x,x')= -\log\bigg(\sum_{y \in \calY} \sqrt{W(y|x)W(y|x')}\bigg).
	\end{align}
	
	Now, let
	\begin{align}
	\tilV_{ij} =  \sum_{P_{XX'}: D(P_{XX'}\|Q_X Q_X')\leq 2R}\bone\{(\bX_i,\bX_j)\in \calT(P_{XX'})\} g_n(P_{XX'}) \label{deftilVij}.
	\end{align}
	
	Now, let
	\begin{align}
	\tilde{P}_{\rme}^{\rm ub}(\Cn)& =  \frac{1}{M_n}\sum_{P_{XX'}: D(P_{XX'}\|Q_X Q_X')\leq 2R} \calN(P_{XX'})g_N(P_{XX'})\\
	&=\frac{1}{M_n}\sum_{i=1}^{M_n} \sum_{j\neq i}\sum_{P_{XX'}: D(P_{XX'}\|Q_X Q_X')\leq 2R} \bone\{(\bX_i,\bX_j) \in \calT(P_{XX'})\} g_n(P_{XX'})\\
	&= \frac{1}{M_n}\sum_{i=1}^{M_n} \sum_{j\neq i} \tilV_{ij} \label{betchao},\\ 
	A_n& =  \bbE[\tilde{P}_{\rme}^{\rm ub}(\Cn)] \label{A2}.
	\end{align}
	
	From Lemma \ref{lem:aux} and Lemma \ref{lem:tmc}, we have
	\begin{align}
	A_n=2^{-n \Etrcub(R,Q)} \label{bag2b}.
	\end{align} 
	
	Now, recall the definition of the typical set $\calV_n$ in Lemma \ref{lem:Vtypi}. Observe that
	\begin{align}
	&\bbP\bigg[P_{\rme}^{\rm ub}(\Cn)>\frac{1}{2}2^{-n(\Etrcub(R,Q)-\eps)}\bigg]+ \bbP\bigg[P_{\rme}^{\rm ub}(\Cn)< 2^{-n(\Etrcub(R,Q)+\eps)}\bigg]\nn\\
	&\qquad=\bbP\bigg[P_{\rme}^{\rm ub}(\Cn)-A_n>\frac{1}{2}2^{-n(\Etrcub(R,Q)-\eps)}-A_n]\bigg]\nn\\
	&\qquad \qquad + \bbP\bigg[P_{\rme}^{\rm ub}(\Cn)-A_n< 2^{-n(\Etrcub(R,Q)+\eps)}-A_n\bigg]\\
	&\qquad \leq \bbP\bigg[\big|P_{\rme}^{\rm ub}(\Cn)-A_n\big|>\frac{1}{2}2^{-n(\Etrcub(R,Q)-\eps)}-A_n] \nn\\
	&\qquad \qquad + \bbP\bigg[\big|P_{\rme}^{\rm ub}(\Cn)-A_n\big|> A_n-2^{-n(\Etrcub(R,Q)+\eps)}\bigg]\\
	&\qquad \leq \bbP\bigg[\big|P_{\rme}^{\rm ub}(\Cn)-A_n\big|>\frac{1}{2}2^{-N(\Etrcub(R,Q)-\eps)}-A_n|\calV_n]\bbP[\calV_n] +\bbP(\calV_n^c)\nn\\
	&\qquad \qquad + \bbP\bigg[\big|P_{\rme}^{\rm ub}(\Cn)-A_n\big|> A_n-2^{-n(\Etrcub(R,Q)+\eps)}|\calV_n\bigg]\bbP[\calV_n]+ \bbP(\calV_n^c) \label{xep1}\\
	&\qquad \leq \bbP\bigg[\big|\tilde{P}_{\rme}^{\rm ub}(\Cn)-A_n\big|>\frac{1}{2}2^{-n(\Etrcub(R,Q)-\eps)}-A_n|\calV_n]\bbP[\calV_n] +\bbP(\calV_n^c)\nn\\
	&\qquad \qquad + \bbP\bigg[\big|\tilde{P}_{\rme}^{\rm ub}(\Cn)-A_n\big|> A_n-2^{-n(\Etrcub(R,Q)+\eps)}|\calV_n\bigg]\bbP[\calV_n]+ \bbP(\calV_n^c)\\
	&\qquad \leq \bbP\bigg[\big|\tilde{P}_{\rme}^{\rm ub}(\Cn)-A_n\big|>\frac{1}{2}2^{-n(\Etrcub(R,Q)-\eps)}-A_n] +\bbP(\calV_n^c)\nn\\
	&\qquad \qquad + \bbP\bigg[\big|\tilde{P}_{\rme}^{\rm ub}(\Cn)-A_n\big|> A_n-2^{-n(\Etrcub(R,Q)+\eps)}\bigg]+ \bbP(\calV_n^c)\\
	& \qquad \dotleq \frac{\var(\tilde{P}_{\rme}^{\rm ub} (\Cn))}{2^{-2n\Etrcub(R,Q)}}+ 2^{-\alpha(R)n} \label{tex2}\\
	&\qquad = 2^{2n\Etrcub(R,Q)} \frac{1}{M_n^2}\sum_{i=1}^{M_n} \sum_{j\neq i} \var(\tilV_{ij}) + 2^{-\alpha(R)n} \label{tex3}\\
	&\qquad =2^{2n\Etrcub(R,Q)} 2^{-2nR} 2^{n \max_{P_{XX'}: D(P_{XX'}\|Q_X Q_X')\leq 2R}-\sum_{x,x'}d_{\rmB}(x,x')P_{XX'}(x,x')}2^{nR}2^{-n\Etrcub(R,Q)}  + 2^{-\alpha(R)N} \label{te3b} \\
	&\qquad=2^{-n(R-\Etrcub(R,Q)+\min_{P_{XX'}: D(P_{XX'}\|Q_X Q_X')\leq 2R}\sum_{x,x'}d_{\rmB}(x,x')P_{XX'}(x,x')}+ 2^{-\alpha(R)n} \label{tex4},
	\end{align} where \eqref{xep1} follows from $\bbP(A) =\bbP(A|B)\bbP(B)+\bbP(A|B^c)\bbP(B^c) \leq \bbP(A|B)\bbP(B)+ \bbP(B^c)$, \eqref{tex2} follows from Chebyshev's inequality, \eqref{bag2}, and Lemma \ref{lem:Vtypi}, \eqref{tex3} follows from the pairwise independence of $\tilV_{ij}$, and \eqref{te3b} follows from \eqref{buchi1} and $M_n\doteq 2^{nR}$.
	
	Now, for the case $\Etrcub(R,Q)=\ErceQ$, we must have 
	\begin{align}
	R+\min_{P_{XX'}: D(P_{XX'}\|Q_X Q_X')\leq  2R}d_{\rmB}(x,x')P_{XX'}(x,x')&= \min_{P_{XX'}: D(P_{XX'}\|Q_X Q_X')= 2R} R+\sum_{x,x'} d_{\rmB}(x,x')P_{XX'}(x,x')\\
	&\geq \min_{P_{XX'}: D(P_{XX'}\|Q_X Q_X')\geq 2R} R+\sum_{x,x'} d_{\rmB}(x,x')P_{XX'}(x,x')\\
	&> \ErceQ \label{bf}\\
	&=\Etrcub(R,Q) \label{bf2},
	\end{align} where \eqref{bf} follows from \eqref{vat4} by setting $\nu=0$.

Hence, for this case, from \eqref{tex4}, we have
	\begin{align}
\bbP\bigg[P_{\rme}^{\rm ub}(\Cn)>\frac{1}{2}2^{-n(\Etrcub(R,Q)-\eps)}\bigg]+ \bbP\bigg[P_{\rme}^{\rm ub}(\Cn)< 2^{-n(\Etrcub(R,Q)+\eps)}\bigg] \leq 2^{-n\Delta} \label{text5}
\end{align} for some $\Delta>0$.
	
Now, we consider the case $\Etrcub(R,Q)>\ErceQ$. First of all we need the following lemma, which extends \cite[Th.~2.1]{Barg2002a} to the \ac{DMC} case.
\begin{lemma}\label{teo:mean_exp}
In a DMC, for all rates for which $\Etrcub(R,Q) = \ErceQ+\Delta E$,  $\Delta E>0$ the probability that
a code of length $n$ and rate $R$ from the RCE has a codeword pair with empirical joint type $P_{XX'}$ such that 
 $D(P_{XX'}||Q_X Q_X')>2R$ goes to $0$ exponentially fast as $n\rightarrow\infty$.

Furthermore, if $P_{XX'}$ is such that $D(P_{XX'}||Q_X Q_X')\leq 2R$, the probability that the number of codeword pairs with joint type $P_{XX'}$ satisfies $\calN(P_{XX'})\doteq 2^{-n(2R-D(P_{XX'}||Q_X Q_X'))}$ goes to one exponentially fast as $n\rightarrow\infty$.
\end{lemma}
\begin{IEEEproof}
Let $\calN(P_{XX'})$ be the number of codeword pairs with joint type $P_{XX'}$. Then, we have
\begin{align}
\bbP\big[\calN(P_{XX'})\geq 1\big] &\leq \bbE\big[\calN(P_{XX'})\big] \label{ten}\\
&=\frac{M(M-1)}{2}\text{Pr}\left[P_{XX'}\right]\label{ten1}\\
&\doteq e^{n2R}e^{-nD(P_{XX'}\|Q_X Q_X')}\label{eqn:theo1_1}
\\ & = 2^{-n\left(D(P_{XX'}\|Q_X Q_X')-2R\right)}\\
&\doteq 2^{-n\big(D(P_{XX'}\|Q_X Q_X')-2R\big)} \to 0 \label{tenb},
\end{align} 
where \eqref{ten} follows from the Markov's inequality, $\Pro\{P_{XX'}\}$ in \eqref{ten1} indicates the probability to find a codeword pair with joint type $P_{XX'}$ and \eqref{eqn:theo1_1} follows from  \cite[Th.~11.1.4]{coverThomas}

Similarly, it is not hard to see that
\begin{align}
\var(\calN(P_{XX'}))\doteq 2^{-n\big(D(P_{XX'}\|Q_X Q_X')-2R\big)} \label{afact}.
\end{align}
The second statement can be proven by observing that by Chebyshev's inequality and \eqref{afact}, for any positive number $\Delta$:
\begin{align}
\Pro\left[\left|\calN(P_{XX'}) - E\left[{\mathcal{N}}_n(P_{XX'})\right]\right|>2^{-n(D(P_{XX'}||Q_X Q_X')-2R+\Delta)}\right]\dotleq 2^{-n(2R-D(P_{XX'}||Q_X Q_X')-2\Delta)} \label{mat2}.
\end{align}
Furthermore, since $\bbE[\calN(P_{XX'})]\doteq 2^{-n(2R-D(P_{XX'}||Q_X Q_X'))}$, from \eqref{mat2} the probability that the number of codeword pairs with joint type $P_{XX'}$ satisfies $\calN(P_{XX'})\doteq 2^{-n(2R-D(P_{XX'}||Q_X Q_X'))}$ goes to one exponentially fast as $n\rightarrow\infty$. As a closing remark, note that the region defined by the inequality $D(P_{XX'}||Q_X Q_X')\leq 2R$ is the equivalent for general \ac{DMC} to the Gilbert-Varshamov region in \cite{Barg2002a}.
\end{IEEEproof}
Now, for the case $\EtrcubQ>\ErceQ$, observe that
\begin{align}
\Pro\left[\Peubn\geq \frac{1}{2} e^{-n[\Etrcub(R,Q)-\epsilon]}\right] + \Pro\left[\Peubn\leq e^{-n{[\Etrcub(R,Q)+\epsilon]}}\right]\label{eqn:mean_0_step1_3}\\
\leq \frac{1}{n^{1+\kappa}} + \Pro\left[\Peubn\leq e^{-n{[\Etrcub(R,Q)+\epsilon]}}\right]\label{eqn:mean_0_step1_4}
\end{align}
where \eqref{eqn:mean_0_step1_4} follows from \cite[Eq.~(22)]{Giusseppe2021e} with $\gamma_n = n^{1+\kappa'}$ for some $\kappa'>0$. Next, we bound the second term in \eqref{eqn:mean_0_step1_4} for large values of $n$.

Let $\Delta>0$ small enough and define
\begin{align}
P_{XX'}^*& = \argmin_{P_{XX'}: D(P_{XX'}\|Q_X Q_X')\leq 2R} R+\sum_{x,x'}d_{\rmB}(x,x')P_{XX'}(x,x'), \\
\calA& = \left[\left|\calN(P_{XX'}^*) - E\left[\calN(P_{XX'}^*)\right]\right|>2^{-n(D(P_{XX'}^*||Q_X Q_X')-2R+\Delta)}\right].
\end{align}
Then, on $\calA^c$, by Lemma \ref{teo:mean_exp}, we have
\begin{align}
2\calN(P_{XX'}^*) g_N(P_{XX'}^*) &\doteq 2^{-n(D(P_{XX'}^*||Q_X Q_X')-2R)}  g_N(P_{XX'}^*) \label{amo} \\
&\doteq  2^{-n \sum_{x,x'} d_{\rm B}(x,x') P_{XX'}^*(x,x')} \label{amo2}, 
\end{align} where \eqref{amo} follows from Lemma \ref{lem:GLem}, and \eqref{amo2} follows from Lemma \ref{lem:tmc} and Lemma \ref{lem:aux} which proves that $$D(P_{XX'}^*\|Q_XQ_{X'})=2R$$ for the case $\EtrcQ>\ErceQ$.

Then, we have
\begin{align}
&\Pro\left[\Peubn\leq 2^{-n{(\Etrcub(R,Q)+\epsilon)}}\right]\nn\\
&\qquad =\Pro\left[\frac{2}{M}\sum_{P_{XX'}}\calN(P_{XX'})g_N{(P_{XX'})} \leq 2^{-n{(\Etrcub(R,Q)+\epsilon)}}\right]\label{eqn:mean_0_step1_5}\\
&\qquad \leq \Pro\left[\frac{2}{M}\calN(P_{XX'}^*)g_N{(P_{XX'}^*)}\leq 2^{-n{(\Etrcub(R,Q)+\epsilon)}}\right]\label{eqn:mean_0_step1_6}\\
&\qquad \leq \Pro\left[\frac{2}{M}\calN(P_{XX'}^*)g_N{(P_{XX'}^*)}\leq 2^{-n{(R +\sum_{x,x'}d_{\rmB}(x,x')P_{XX'}^*(x,x') +\epsilon)}}\right]\label{eqn:mean_0_step1_7}\\
&\qquad \leq \Pro\left[2\calN(P_{XX'}^*)g_N{(P_{XX'}^*)}\leq 2^{-n{(\sum_{x,x'}d_{\rmB}(x,x')P_{XX'}^*(x,x') +\epsilon)}}\right]\label{eqn:mean_0_step1_8}\\
&\qquad = \Pro\left[2\calN(P_{XX'}^*)g_N{(P_{XX'}^*)}\leq 2^{-n{(\sum_{x,x'}d_{\rmB}(x,x')P_{XX'}^*(x,x') +\epsilon)}}\bigg|\calA\right] \Pro(\calA)\nn\\ &
\qquad + \Pro\left[2\calN(P_{XX'}^*)g_N{(P_{XX'}^*)}
\leq 2^{-n{(\sum_{x,x'}d_{\rmB}(x,x')P_{XX'}^*(x,x') +\epsilon)}}\bigg|\calA^c\right] \Pro(\calA^c)\label{eqn:mean_0_step1_9}\\
&\qquad \leq  \Pro(\calA)  + \Pro\left[2\calN(P_{XX'}^*)g_N{(P_{XX'}^*)}
\leq 2^{-n{(\sum_{x,x'}d_{\rmB}(x,x')P_{XX'}^*(x,x') +\epsilon)}}\bigg|\calA^c\right]\label{eqn:mean_0_step1_10}\\
&\qquad \leq 2^{-n\Delta'}+ 0 \label{eqn:mean_0_step1_13}\\
&\qquad=2^{-n\Delta'},
\end{align} for $n$ sufficiently large, where \eqref{eqn:mean_0_step1_7}  follows from Lemma \ref{lem:tmc} and $\EtrcQ>\ErceQ$, \eqref{eqn:mean_0_step1_13} follows from Lemma \ref{teo:mean_exp} and \eqref{amo2}, respectively.

Hence, for this case, we have
\begin{align}
\bbP\bigg[P_{\rme}^{\rm ub}(\Cn)>\frac{1}{2}2^{-n(\Etrcub(R,Q)-\eps)}\bigg]+ \bbP\bigg[P_{\rme}^{\rm ub}(\Cn)< 2^{-n(\Etrcub(R,Q)+\eps)}\bigg] \leq \frac{1}{n^{1+\kappa'}} + 2^{-n\Delta'}\label{text6}
\end{align} for some $\kappa'>0$ and $\Delta'>0$.

Finally, from \eqref{text5} and \eqref{text6}, our proof is concluded.
\subsection{Proof of Lemma \ref{caenlem}}
\label{ap:proof10}

	Observe that
	\begin{align}
	\bbE[P_{\rme}^{\rm ub}(\Cn)]&\geq \bbE[\PeCn]\\
	&=\frac{1}{M_n}\sum_{i=1}^{M_n}\bbE\bigg[\bbP\bigg(\bigcup_{j\neq i}\{\bX_i \to \bX_j\}\bigg)\bigg]\\
	&=\frac{1}{M_n}\sum_{i=1}^{M_n}\bbE\bigg[\bbE\bigg[\bone\bigg\{\bigcup_{j\neq i}\{\bX_i \to \bX_j\}\bigg\}\bigg]\bigg]\\
	&\geq \frac{1}{M_n}\sum_{i=1}^{M_n}\sum_{j\neq i}\frac{\big(\bbE\big[ \bbE[\bone\{\bX_i \to \bX_j\}]\big]\big)^2} {\bbE[ \bbE[\bone\{\bX_i \to \bX_j\}]] + \sum_{k\neq i, j}\bbE[ \bbE[\bone\{\{\bX_i \to \bX_j\} \cap \{\bX_i \to \bX_k\}\}]]} \label{eq68}\\
	&= \frac{1}{M_n}\sum_{i=1}^{M_n}\sum_{j\neq i}\frac{\big(\bbE\big[ \Pro[\bX_i \to \bX_j]\big]\big)^2} {\bbE[ \Pro[\bX_i \to \bX_j]] + \sum_{k\neq i, j}\bbE[ \Pro[\{\bX_i \to \bX_j\} \cap \{\bX_i \to \bX_k\}]} \label{eq68b}\\
	&= \frac{(M_n-1)\big(\bbE\big[ \Pro[\bX_1 \to \bX_2]\big]\big)^2}{\bbE[\Pro[\bX_1 \to \bX_2]] + (M_n-2)\bbE[\Pro[\{\bX_1\to \bX_2\}\cap \{\bX_1 \to \bX_3\}]] }\\
	&=\frac{P_{\rme}^{\rm ub}(\Cn)\bbE\big[\Pro[\bX_1 \to \bX_2]\big] }{\bbE\big[\Pro[\bX_1 \to \bX_2]\big] + (M_n-2)\bbE\big[\Pro[\{\bX_1\to \bX_2\}\cap \{\bX_1 \to \bX_3\}] \big]} \label{eq69},
	\end{align} where \eqref{eq68} follows from Caen's inequality in Lemma \ref{DECAENLB} by, for each fixed $i$, setting $\calI_i=\{j \in [M]\setminus \{i\}: j\neq i\}, A_j^{(i)} =  \{\bX_i\to \bX_j\}$ with the probability measure defined as
	$
	\Pro(A_j^{(i)}) =  \bbE[\bbE[\bone\{\bX_i \to \bX_j\}]]=\bbE[\Pro[\bX_i\to \bX_j]],
	$ where the inner expectation is over the BSC channel randomness and the outer one is over the random codebook ensemble. This is the probability of event $\{\bX_i\to \bX_j\}$ on the a product probability space generated from channel statistics and random codebook generations.  By the symmetry of the codebook generation, it is easy to see that $\Pro(A_j^{(i)})=\bbE[\Pro[\bX_i\to \bX_j]]=\bbE[\Pro[\bX_1\to \bX_2]]=\Pro(A_2^{(1)})$ for all $j\neq i$.
	
	From \eqref{eq69}, it holds that
	\begin{align}
	1&\leq \frac{\bbE[P_{\rme}^{\rm ub}(\Cn)]}{\bbE[\PeCn]}\\
	&\leq 1+ (M_n-2)\frac{\bbE[\Pro[\{\bX_1 \to \bX_2\} \cap \{\bX_1 \to \bX_3\}] ]}{\bbE[\Pro[\bX_1 \to \bX_2]]}\label{eq70a1}.
	\end{align}
	
	Recall the definition of $d_{\rmB}(x,x')$ in \eqref{laday1}. Assume that $\bx_1 \in \calT(P_X)$ for some $P_X \in \calP_n(\calX)$, which is a fixed vector. Then, given $(\bx_1,\bx_2) \in \calT(P_{XX'})$ and $(\bx_1,\bx_3) \in  \calT(P_{XX^"})$ where $P_{XX'} \in \calP_n(\calX \times \calX)$ and $P_{XX^"}\in \calP_n(\calX \times \calX)$, it holds that
	\begin{align}
	&\bbP\{\{\bx_1 \to \bx_2\} \cap \{\bx_1 \to \bx_3\} |(\bx_1,\bx_2) \in \calT(P_{XX'}),(\bx_1,\bx_3) \in  \calT(P_{XX^"})\}\nn\\
	&\qquad \leq \min\bigg\{\bbP\{\bx_1 \to \bx_2|(\bx_1,\bx_2) \in \calT(P_{XX'})\},\bbP\{\bx_1 \to \bx_3|(\bx_1,\bx_3) \in \calT(P_{XX^"})\} \bigg\}\\
	&\qquad \leq  \min\bigg\{2^{-n \sum_{x,x'} d_{\rmB}(x,x') P_{XX'}(x,x') },2^{-n \sum_{x,x"} d_{\rmB}(x,x") P_{XX"}(x,x") }\bigg\} \label{intestinemod} \\
	&\qquad= 2^{-n \max\big\{\sum_{x,x'} d_{\rmB}(x,x') P_{XX'}(x,x'), \sum_{x,x"} d_{\rmB}(x,x") P_{XX"}(x,x")\big\}} \label{T1a},
	\end{align} which does not depend on $\bx_1,\bx_2,\bx_3$, where \eqref{intestinemod} follows from Lemma \ref{lem:GLem}.
	
	In addition, we have
	\begin{align}
	\bbP\bigg[(\bx_1,\bX_2)\in \calT(P_{XX'})\bigg]&=\sum_{\bx_2} \bbP(\bx_2)\bone\{(\bx_1,\bx_2)\in \calT(P_{XX'}) \}\\
	&=\sum_{\bx_2} 2^{-n \big(H(P_X')+D(P_{X'}\|Q)\big)}\bone\{(\bx_1,\bx_2)\in \calT(P_{XX'}) \} \label{g1}\\
	&=2^{-n \big(H(P_X')+D(P_{X'}\|Q)\big)}\sum_{\bx_2} \bone\{(\bx_1,\bx_2)\in \calT(P_{XX'}) \}\\
	&=2^{-n \big(H(P_X')+D(P_{X'}\|Q)\big)}\frac{|\calT(P_{XX'}|}{|\calT(P_X)|}\\
	&\doteq 2^{-n \big(H(P_X')+D(P_{X'}\|Q)\big)} \frac{2^{N H(P_{XX'})|}}{2^{N H(P_X)|}} \label{g2}\\
	&=2^{-n \big(I_P(X;X')+D(P_X'\|Q)\big)} \label{smao2a},
	\end{align} where \eqref{g1} and \eqref{g2} follow from
	\cite{Csis00}.
	
	Similarly, we also have
	\begin{align}
	\bbP\bigg[(\bx_1,\bX_3)\in \calT(P_{XX^{"}})\bigg]
	=2^{-n \big(I_P(X;X^{"})+D(P_X^{"}\|Q)\big)} \label{smao4a}.
	\end{align}

	Hence, we have
	\begin{align}
	&\bbP\bigg[\{(\bx_1,\bX_2) \in \calT(P_{XX'})\} \cap \{(\bx_1,\bX_3)\in \calT(P_{XX^{"}}) \}\bigg|\bX_1=\bx_1 \bigg]\nn\\
	&\qquad=\bbP\bigg[\{(\bx_1,\bX_2) \in \calT(P_{XX'})\} \cap \{(\bx_1,\bX_3) \in \calT(P_{XX^{"}})\}\bigg]\\
	&\qquad=\bbP\bigg[(\bx_1,\bX_2) \in \calT(P_{XX'})\bigg]\bbP\bigg[(\bx_1,\bX_3) \in \calT(P_{XX"})\bigg]\\
	&\qquad=2^{-n \big(I_P(X;X')+D(P_X'\|Q)\big)} 2^{-n \big(I_P(X;X^{"})+D(P_X^{"}\|Q)\big)} \label{eq197}\\
	&\qquad=2^{-n (I_P(X;X')+I_P(X;X^{"})+D(P_X'\|Q)+ D(P_{X^{"}}\|Q))} \label{T2a},  
	\end{align} where \eqref{eq197} follows from \eqref{smao2a} and \eqref{smao4a}.
	
	It follows from \eqref{T1a} and \eqref{T2a} that
	\begin{align}
	&\bbE_{\bX}[\Pro[\{\bx_1 \to \bX_2\} \cap \{\bx_1 \to \bX_3\}]]\nn\\
	&\quad = \sum_{P_{X'|X}} \sum_{P_{X^{"}|X}} \bbE\bigg[\bbP\{\{\bx_1 \to \bX_2\} \cap \{\bx_1 \to \bX_3\} \bigg|(\bx_1,\bX_2) \in \calT_{P_{XX'}},(\bx_1,\bX_3)\in \calT_{P_{XX^{"}}}\}\nn\\
	&\qquad \qquad \times \bbP\bigg[\{(\bx_1,\bX_2)\in \calT(P_{XX'})\} \cap \{(\bx_1,\bX_3)\in \calT(P_{XX^{"}})\}\bigg|\bX_1=\bx_1 \bigg]\bigg]\\
	&\quad \leq \sum_{P_{X'|X}} \sum_{P_{X^{"}|X}}2^{-n \max\{\sum_{x,x'} d_{\rmB}(x,x') P_{XX'}(x,x'), \sum_{x,x"} d_{\rmB}(x,x") P_{XX^{"}}(x,x")\}}\nn\\
	&\qquad \qquad \times 2^{-n (I_P(X;X')+I_P(X;X^{"})+D(P_X'\|Q)+ D(P_X^{"}\|Q))}  \label{eq197a}\\
	&\qquad \leq \sum_{P_{X'|X}} \sum_{P_{X^{"}|X}}2^{-\frac{n}{2} \big(\sum_{x,x'} d_{\rmB}(x,x') P_{XX'}(x,x') +\sum_{x,x"} d_{\rmB}(x,x") P_{XX"}(x,x")\big)}\nn\\
	&\qquad \qquad \times 2^{-n (I_P(X;X')+I_P(X;X^{"})+D(P_X'\|Q)+ D(P_{X^{"}}\|Q))}  \label{tag2}\\
	&=\bigg(\sum_{P_{X'|X}}2^{-\frac{n}{2} \big(\sum_{x,x'} d_{\rmB}(x,x') P_{XX'}(x,x')\big)} 2^{-n \big(I_P(X;X')+D(P_X'\|Q)\big)} \bigg)\nn\\
	&\qquad \times \bigg(\sum_{P_{X^{"}|X}}2^{-\frac{n}{2} \big(\sum_{x,x"} d_{\rmB}(x,x") P_{XX^{"}}(x,x")\big)} 2^{-n \big(I_P(X;X^{"})+D(P_{X^{"}}\|Q)\big)} \bigg)\\
	&= \bigg(\sum_{P_{X'|X}}2^{-\frac{n}{2} \big(\sum_{x,x'} d_{\rmB}(x,x') P_{XX'}(x,x')\big)} 2^{-n \big(I_P(X;X')+D(P_X'\|Q)\big)} \bigg)^2 \label{k1},
	\end{align} where \eqref{tag2} follows from $\max\{a,b\}\geq \frac{a+b}{2}$.
	
	It follows from \eqref{k1} that
	\begin{align}
	&\bbE[\Pro[\{\bX_1 \to \bX_2\}\cap \{\bX_1\to \bX_3\}]]\nn\\
	&\qquad =\sum_{\bx_1}\Pro(\bx_1) \bbE[\Pro[\{\bx_1 \to \bX_2\}\cap \{\bx_1\to \bX_3\}]\big|\bX_1=\bx_1]\\
	&\qquad =\sum_{\bx_1}\Pro(\bx_1) \bbE[\Pro[\{\bx_1 \to \bX_2\}\cap \{\bx_1\to \bX_3\}]] \label{f1}\\
	&\qquad =\sum_{P_X}\sum_{\bx_1 \in \calT(P_X)} \Pro(\bx_1) \bbE[\Pro[\{\bx_1 \to \bX_2\}\cap \{\bx_1\to \bX_3\}]] \label{f2}\\
	&\qquad=\sum_{P_X} \sum_{\bx_1 \in \calT(P_X)} 2^{-n(D(P_X\|Q)+ H(P_X))}\bbE[\Pro[\{\bx_1 \to \bX_2\}\cap \{\bx_1\to \bX_3\}]] \label{f3}\\
	&\qquad \leq \sum_{P_X} \sum_{\bx_1 \in \calT(P_X)} 2^{-n(D(P_X\|Q)+ H(P_X))}\bigg(\sum_{P_{X'|X}}2^{-\frac{n}{2} \big(\sum_{x,x'} d_{\rmB}(x,x') P_{XX'}(x,x')\big)} 2^{-n \big(I_P(X;X')+D(P_X'\|Q)\big)} \bigg)^2 \\
	&\qquad \leq \sum_{P_X}  2^{-n D(P_X\|Q)}\bigg(\sum_{P_{X'|X}}2^{-\frac{n}{2} \big(\sum_{x,x'} d_{\rmB}(x,x') P_{XX'}(x,x')\big)} 2^{-n \big(I_P(X;X')+D(P_X'\|Q)\big)} \bigg)^2 \label{f4},
	\end{align} where \eqref{f1} follows from the independence of codewords in the random codebook ensemble, \eqref{f3} follows from \cite{Csis00}.
	
Now, from \eqref{tenb} in the proof of Lemma \ref{teo:mean_exp}, for all joint type $P_{XX'}$ such that $D(P_{XX'}\|Q_X Q_{X'})>2R$, it holds that
	\begin{align}
	\sum_{n=1}^{\infty}\bbP\big[\calN(P_{XX'})\geq 1\big]\leq \sum_{n=1}^{\infty} 2^{-n\big(D(P_{XX'}\|Q_X Q_X')-2R\big)}<\infty \label{taba}.
	\end{align}
	From \eqref{taba} and Borel-Cantelli's lemma \cite{Billingsley}, it holds almost surely that
	$
	\calN(P_{XX'})= 0
	$ for all joint type $P_{XX'}$ such that $D(Q_{XX'}\|Q_X Q_{X'})>2R$.
	
	Hence, from \eqref{k1} and the above fact with noting the number of types or conditional types are sub-exponential in $N$, we have
	\begin{align}
	&\bbE[\Pro[\{\bX_1 \to \bX_2\} \cap \{\bX_1 \to \bX_3\}]]\nn\\
	&\doteq 2^{-n \big(\min_{P_{XX'} \in \calP_n(\calX \times \calX):D(P_{XX'}\|Q_X Q_X')\leq 2R } D(P_X\|Q)+ 2 \big(I_P(X;X')+D(P_X'\|Q)\big)+\sum_{x,x'} d_{\rmB}(x,x') P_{XX'}(x,x')  \big) } \label{eqbesa}\\
	&\doteq 2^{-n \big(\min_{P_{XX'} \in \calP(\calX \times \calX):D(P_{XX'}\|Q_X Q_X')\leq 2R } D(P_X\|Q)+ 2 \big(I_P(X;X')+D(P_X'\|Q)\big)+\sum_{x,x'} d_{\rmB}(x,x') P_{XX'}(x,x')  \big) } \label{eqbes},
	\end{align} where \eqref{eqbesa} follows from the sub-exponential number of possible $n$-types in $\calX \times \calX$ \cite{Csis00}, and \eqref{eqbes} follows from the fact that $\calP_n(\calX \times \calX)$ is dense in $\calP(\calX \times \calX)$.
	
	Now, note that $Q_X=Q_X'=Q$, so we have
	\begin{align}
	I_P(X;X')&=D(P_{XX'}\|P_X P_X')\\
	&=D(P_{XX'}\|Q_X Q_X')-D(P_X\|Q)-D(P_X'\|Q).
	\end{align}
	It follows that
	\begin{align}
	&D(P_X\|Q)+ 2 \big(I_P(X;X')+D(P_X'\|Q)\big)\nn\\
	&\qquad= D(P_X\|Q)+2 \big(D(P_{XX'}\|Q_X Q_X')-D(P_X\|Q)\big)\\
	&\qquad =2 D(P_{XX'}\|Q_X Q_X')- D(P_X\|Q)\\
	&\qquad \geq D(P_{XX'}\|Q_X Q_X') \label{ta},
	\end{align} where \eqref{ta} follows from the data processing for KL divergence (or log-sum inequality \cite{cover}).
	
	Hence, we have
	\begin{align}
	&\min_{P_{XX'} \in \calP(\calX \times \calX):D(P_{XX'}\|Q_X Q_X')\leq 2R} D(P_X\|Q)+ 2 \big(I_P(X;X')+D(P_X'\|Q)\big) +\sum_{x,x'} d_{\rmB}(x,x') P_{XX'}(x,x')\nn\\
	&\qquad\geq \min_{P_{XX'} \in \calP(\calX \times \calX):D(P_{XX'}\|Q_X Q_X')\leq 2R} D(P_{XX'}\|Q_X Q_X') +\sum_{x,x'} d_{\rmB}(x,x') P_{XX'}(x,x')
	\label{tatc}\\
	&\qquad =\EtrcQ+R \label{botat},
	\end{align} where \eqref{tatc} follows from \eqref{ta}, and \eqref{botat} follows from Lemma \ref{lem:tmc}. Note that \eqref{tatc} becomes equality if and only if $P_{XX'}(x,x')=Q(x) Q(x')$ for all $x,x' \in \calX \times \calX$. However, at $P_{XX'}=Q_X Q_X'$, we have
	\begin{align}
	&\min_{P_{XX'} \in \calP(\calX \times \calX):D(P_{XX'}\|Q_X Q_X')\leq 2R} D(P_X\|Q)+ 2 \big(I_P(X;X')+D(P_X'\|Q)\big) +\sum_{x,x'} d_{\rmB}(x,x') P_{XX'}(x,x')\nn\\
	&\qquad= \sum_{x,x'} d_{\rmB}(x,x') Q(x) Q(x')\\
	&\qquad=-\sum_{x,x'} \log\bigg(\sum_{y \in \calY} \sqrt{W(y|x)W(y|x')}\bigg) Q(x) Q(x')\\
	&\qquad > -\log \bigg( \sum_{x,x'}\sum_{y \in \calY} \sqrt{W(y|x)W(y|x')}Q(x) Q(x')\bigg) \label{bet10}\\
	&\qquad=-\log \bigg( \sum_{y \in \calY} \sum_{x,x'} \sqrt{W(y|x)W(y|x')}Q(x) Q(x')\bigg)\\
	&\qquad=-\log \bigg( \sum_{y \in \calY} \bigg(\sum_x \sqrt{W(y|x)}Q(x)\bigg)^2\bigg)\\
	&\qquad= -\log \bigg( \sum_{y \in \calY} \bigg(\sum_x \sqrt{W(y|x)}Q(x)\bigg)^2\bigg)\\
	&\qquad=R_0(Q) \label{eq}\\
	&\qquad=\ErceQ+R \label{eq11},
	\end{align} where \eqref{bet10} follows from the convexity of the function $-\log x$ with noting that the equality does not happen by the condition \eqref{bet10}, and \eqref{eq} follows from \cite[Eq. (8.45)]{MoserBook} with $R_0(Q)$ is the cut-off rate of the DMC at the distribution $Q$. 
	
	Therefore, from \eqref{botat} and \eqref{eq11}, we have
	\begin{align}
	\min_{P_{XX'} \in \calP(\calX \times \calX):D(P_{XX'}\|Q_X Q_X')\leq 2R} D(P_{XX'}\|Q_X Q_X') +\sum_{x,x'} d_{\rmB}(x,x') P_{XX'}(x,x')> \EtrcQ+R
	\end{align} for the case $\Etrcub(R,Q)=\ErceQ$.
	
	Now, for the case $\Etrcub(R,Q)>\ErceQ$, \eqref{botat} happens at the optimizer $P_{XX'}^*$ satisfying $D(P_{XX'}^*\|Q_X Q_X')=2R$, which leads to $P_{XX'}^* \neq Q_X Q_X'$ if $R>0$, so the equality can not happen in \eqref{ta}. 
	
	In summary, at $R>0$ and a fixed underlying distribution $Q$, it holds that
	\begin{align}
	\min_{P_{XX'} \in \calP(\calX \times \calX):D(P_{XX'}\|Q_X Q_X')\leq 2R} D(P_{XX'}\|Q_X Q_X') +\sum_{x,x'}d_{\rmB}(x,x') P_{XX'}(x,x')> \Etrcub(R,Q)+R \label{cond3}.
	\end{align}
	
	Hence, it holds from \eqref{eqbes} and \eqref{cond0} that
	\begin{align}
	\bbE\bigg[\bbP\big[\{\bX_1 \to \bX_2\} \cap \{\bX_1 \to \bX_3\}\bigg] \leq 2\times 2^{-\big(R+\Etrcub(R,Q)\big) n} 2^{-\delta(R) n} \label{eq176}
	\end{align} for some constant $\delta(R)>0$ .
	
	Now, on the other hand, we know that
	\begin{align}
	\bbE[\Pro[\bX_1 \to \bX_2]]&=\frac{1}{M_n(M_n-1)}\sum_{i=0}^{M_n-1}\sum_{j\neq i}\bbE[\Pro[\bX_i \to \bX_j]]\\
	& =\frac{1}{M_n-1}\bbE\bigg[\frac{1}{M_n}\sum_{i=0}^{M_n-1}\sum_{j\neq i}\Pro[\bX_i \to \bX_j]\bigg] \label{H0a}\\
	& = \frac{\bbE[P_{\rme}^{\rm ub}(\Cn)] }{M_n-1}\\
	&\geq \frac{\bbE[\PeCn] }{M_n-1}\\
	&\doteq 2^{-n(\ErceQ-R)}
	\label{eqH1a}.
	\end{align}
	
	From \eqref{eq70a1}, \eqref{eq176}, and \eqref{eqH1a}, we obtain
	\begin{align}
	0&\leq \frac{\bbE[P_{\rme}^{\rm ub}(\Cn)]}{\bbE[P_\rme(\Cn)]}-1\\
	&\doteq 2^{-N(\delta(R)+\Etrcub(R,Q)-\ErceQ)} 
	\label{compass}.
	\end{align}

\subsection{Proof of Lemma~\ref{steinlemmu}}
\label{prooflemma10}
Observe that
\begin{align}
\bbE\big[2^{-tS_n/n}\big]&=\sum_{x_1,x_2,\cdots,x_n} 2^{-\frac{t(x_1+x_2+\cdots+x_n)}{n}} \Pro[X_1=x_1,X_2=x_2,\cdots,X_n=x_n]\\
&=\sum_{x_1,x_2,\cdots,x_n} 2^{-\frac{t(x_1+x_2+\cdots+x_n)}{n}} \Pro[X_1=x_1,X_2=x_2,\cdots,X_n=x_n]\bone\{(x_1,x_2,\cdots,x_n) \in \calV\}\nn\\
&\qquad + \sum_{x_1,x_2,\cdots,x_n} 2^{-\frac{t(x_1+x_2+\cdots+x_n)}{n}} \Pro[X_1=x_1,X_2=x_2,\cdots,X_n=x_n]\bone\{(x_1,x_2,\cdots,x_n) \in \calV^c\}\\
&\leq \sum_{x_1,x_2,\cdots,x_n} 2^{-\frac{t(x_1+x_2+\cdots+x_n)}{n}}\prod_{k=1}^n P(x_k) \bone\{(x_1,x_2,\cdots,x_n) \in \calV\}\nn\\
&\qquad + \sum_{x_1,x_2,\cdots,x_n} 2^{-\frac{t(x_1+x_2+\cdots+x_n)}{n}}\bone\{(x_1,x_2,\cdots,x_n) \in \calV^c\}\\
&\leq \sum_{x_1,x_2,\cdots,x_n} 2^{-\frac{t(x_1+x_2+\cdots+x_n)}{n}} \prod_{k=1}^n P(x_k) + \sum_{x_1,x_2,\cdots,x_n} 2^{-n \zeta}\bone\{(x_1,x_2,\cdots,x_n) \in \calV^c\}\\
&=\bbE_{\prod_{i=1}^n P(x_i)}\big[2^{-tS_n/n}\big]+ 2^{-n \zeta} |\calV^c|.
 \end{align}

 \subsection{Proof of Lemma~\ref{lemma:mean_joint_type_prod}}
 \label{prooflemma11}
 
 We have that

	\begin{align}\label{eqn:crosscorr1}
	\bbE[\calI\{i,j\}\calI\{i,k\}]&=\Pro\{(X_i,X_j)\in \calT(P_{XX'}),(X_i,X_k)\in \calT(P_{XX'})\}\\\label{eqn:crosscorr2}
	&=\sum_{\bx_i}\Pro\{X_i=\bx_i\}\Pro\{(\bx_i,X_j)\in \calT(P_{XX'}),(\bx_i,X_k)\in \calT(P_{XX'})\}\\\label{eqn:crosscorr3}
	&=\sum_{\bx_i}\Pro\{X_i=\bx_i\}\Pro\{(\bx_i,X_j)\in \calT(P_{XX'})\}\Pro\{(\bx_i,X_k)\in \calT(P_{XX'})\}\\\label{eqn:crosscorr4}
	&=\sum_{\bx_i}\Pro\{X_i=\bx_i\}\Pro\{(\bx_i,X_j)\in \calT(P_{XX'})\}^2 \\\label{eqn:crosscorr5}
	&=\sum_{P_X}\calN(P_X)\Pro\{\bx \in\calT(P_X)\}\Pro\{(\bx,X_j)\in \calT(P_{XX'})|\bx\in \calT(P_X)\}^2 \\
	&\dot{=}\max_{P_X}\calN(P_X)\Pro\{\bx\in\calT(P_X)\}\Pro\{(\bx,X_j)\in \calT(P_{XX'})|\bx \in \calT(P_X)\}^2 \label{eqn:crosscorr6}
	\end{align}
	where in \eqref{eqn:crosscorr2} we conditioned to codeword $X_i$ being equal to a given realization $\x_i$, \eqref{eqn:crosscorr3} is because $X_j$ and $X_k$ are independent, \eqref{eqn:crosscorr4} is because they are also identically distributed, in \eqref{eqn:crosscorr5} we grouped codewords $X_i$ according to their type $P_X$ and used the fact that $\Pro\{(\bx_i,X_j)\in \calT(P_{XX'})\}$ takes the same value when $\bx_i$ has the same type. Expression \eqref{eqn:crosscorr6} is hard to calculate because of the term $\Pro\{(\bx,X_j)\in \calT(P_{XX'})|\bx\in \calT(P_X)\}$. Therefore we find a lower bound and an upper bound on Eqn. \eqref{eqn:crosscorr5}. The lower bound is:
	\begin{align}
	&\sum_{P_X}\calN(P_X)\Pro\{\bx\in\calT(P_X)\}\Pro\{(\bx,X_j)\in \calT(P_{XX'})|\bx \in\calT(P_X)\}^2 \nn\\
	&\qquad \geq \left(\sum_{P_X}\calN(P_X)\Pro\{\bx\in\calT(P_X)\}\Pro\{(\bx,X_j)\in \calT(P_{XX'})|\bx\in\calT(P_X)\}\right)^2\\
	&\qquad =\Pro\{(X_i,X_j)\in \calT(P_{XX'})\}^2\\
	&\qquad \dot{=} 2^{-n2D(P_{XX'}||Q_X Q_X')}\label{eqn:lb_joint_indic}
	\end{align}
	while the upper bound is: 
	\begin{align}
	&\sum_{P_X}\calN(P_X)\Pro\{\bx\in\calT(P_X)\}\Pro\{(\bx,X_j)\in \calT(P_{XX'})|\bx \in \calT(P_X)\}^2\nn\\
	&\qquad \leq \sum_{P_X}\calN(P_X)\Pro\{\bx\in\calT(P_X)\}\Pro\{(\bx,X_j)\in \calT(P_{XX'})|\bx \in\calT(P_X)\}\nn\\
	&\qquad \qquad \times \cdot \max_{P_X}\Pro\{(\bx,X_j)\in \calT(P_{XX'})|\bx \in \calT(P_X)\}
	\\
	&\qquad =\Pro\{(X_i,X_j)\in \calT(P_{XX'})\}\max_{P_X}\Pro\{(\bx,X_j)\in \calT(P_{XX'})|\bx \in \calT(P_X)\}\\
	&\qquad \dot{=} 2^{-n[D(P_{XX'}||Q_X Q_X')+\eta]}
	\end{align}
	where 
	$$\eta=-\frac{1}{n}\log \max_{P_X}\Pro\{(\bx,X_j)\in \calT(P_{XX'})|\bx \in\calT(P_X)\}\leq D(P_{XX'}||Q_X Q_X'),$$
	and the inequality follows from \eqref{eqn:lb_joint_indic}. 

\subsection{Proof of Lemma~\ref{lemma:type_enum_double}}
\label{prooflemma12}

	The proof is based on  \cite[Th.~10]{suen_new_JansonRSA1998}. A similar proof of an equivalent result is presented  for the case of constant composition codes in \cite{Tamir2020a}. However, there are several differences with our case. First of all our Lemma \ref{lemma:mean_joint_type_prod} gives a bound rather than a dot equality, which has implications on the minimum exponent starting from which a double exponential decay is found. Other differences with \cite{Tamir2020a} are indicated in the following.
	
	Let us define the quantities:
	\begin{align}
	a=2^{-n\epsilon}
	\end{align}
	\begin{align}
	\Delta=\bbE[\calN(P_{XX'})]\dot{=}2^{n(2R-D(P_{XX'}||Q_X Q_X'))}.
	\end{align}
	Let us consider a graph in which vertices are indicated with pairs (e.g., $(k,l)$ is a vertex). Two vertices $(k,l)$ and $(i,j)$ are connected if exactly one index in both pairs coincide. Let us indicate with $(k,l)\sim (i,j)$ the case in which vertices $(k,l)$ and $(i,j)$ are connected. Let us also indicate with $[M]_*^2$ the set $\{i,j\in\{1,2,\ldots M_n\}:i\neq j\}$.
	Let us define and bound the following quantity:
	\begin{align}\label{eqn:theta}
	\Theta & =  \frac{1}{2}\sum_{(i,j)\in [M]_*^2} \sum_{(k,l)\in [M]_*^2, (k,l)\sim (i,j)}\mathbb{E}[\calI\{i,j\}\calI\{i,k\}]\\\label{eqn:theta1}
	&\dot{=}\frac{1}{2}2^{2nR}(2^{nr}+2^{nr}-2)\bbE[\calI\{1,2\}\calI\{1,3\}]\\\label{eqn:theta2}
	&\dot{\leq}\frac{1}{2}2^{2nR}(2^{nr}+2^{nr}-2) 2^{-nD(P_{XX'}||Q_X Q_X')}\\\label{eqn:theta3}
	&\dot{=}\frac{1}{2}2^{n(3R-D(P_{XX'}||Q_X Q_X'))}
	\end{align}
	where \eqref{eqn:theta1} and \eqref{eqn:theta2} follow from Lemma \ref{lemma:mean_joint_type_prod} as well as from the fact that codewords are i.i.d. and noticing that there are about $M_n^2$ codeword pairs $(i,j)$ and, for each of them, there are exactly $2^{nR}-1$ connected vertices. Note that in \cite{Tamir2020a} a dot equality rather than an inequality is present.
	

	Now let us define and, where needed, bound the following three quantities:
	\begin{align}\label{eqn:omega}
	\Omega & =  \max_{(i,j)\in[M]^2_*} \sum_{(k,l)\in [M]_*^2, (k,l)\sim (i,j)}\bbE[\calI\{i,k\}]\\\label{eqn:omega1}
	&\dot{=}(2^{nr}+2^{nr}-2) 2^{-nD(P_{XX'}||Q_X Q_X')}\\\label{eqn:omega2}
	&\dot{=}2^{n(R-D(P_{XX'}||Q_X Q_X'))}
	\end{align}
	
	\begin{align}\label{eqn:delta_omega}
	\frac{\Delta}{6\Omega} \dot{=} \frac{2^{n[2R-D(P_{XX'}||Q_X Q_X')]}}{2^{n[R-D(P_{XX'}||Q_X Q_X')]}}=2^{nR}
	\end{align}
	
	\begin{align}\label{eqn:delta_teta}
	\frac{\Delta^2}{8\Theta + 2\Delta} &\dot{\geq} \frac{2^{n[4R-2D(P_{XX'}||Q_X Q_X')]}}{e^{n[3R-D(P_{XX'}||Q_X Q_X')]} + 2^{n[2R-D(P_{XX'}||Q_X Q_X')]}}\\
	&= \frac{2^{n[2R-D(P_{XX'}||Q_X Q_X')]}}{2^{nR} + 1}\\
	&\dot{=} 2^{n[R-D(P_{XX'}||Q_X Q_X')]}\\
	\end{align}
	
	Using the definitions above and the result in  \cite[Th.~10]{suen_new_JansonRSA1998} we obtain:
	\begin{align}\label{eqn:type_enum_double1}
	\bbP\left[\calN(P_{XX'})\leq 2^{-n\epsilon}\bbE[\calN(P_{XX'})]\right]&\dot{=} \bbP\left[\calN(P_{XX'})\leq 2^{n[2R-D(P_{XX'}||Q_X Q_X')-\epsilon]}\right]\\&\dotleq \exp\Bigl\{-\min\left(2^{n[R-D(P_{XX'}||Q_X Q_X')]},2^{nR}\right)\Bigr\}\\
	&=\exp\Bigl\{-2^{n[R-D(P_{XX'}||Q_X Q_X')]}\Bigr\}
	\end{align}
	which concludes the proof of the lemma.

\subsection{Proof of Lemma~\ref{lem:extra1}}
\label{prooflemma14}

We prove by induction. The condition \eqref{ta1} shows that \eqref{ta0} holds for $n=2$, $\forall A_1,A_2 \subset \calX$. Now, assume that \eqref{ta0} holds for some $n\geq 2$. We need to show that \eqref{ta0} holds for $n+1$. 
	Indeed, let $A_1,A_2,\cdots, A_n,A_{n+1} \in \calX$ such that there exists $i, j \in [n]$ with $i\neq j$ such that $A_i \cap A_j =\emptyset$.
	
	Now, by reordering $\{A_k\}_{k=1}^n$, we can assume without loss of generality that $i, j \in [n]$. This leads to $\cap_{k=1}^n A_k=\emptyset$.
	Observe that
	\begin{align}
	\Pro\bigg[\bigcap_{k=1}^{n+1}\big\{X \in \calA_k\big\}\bigg]&=\Pro\bigg[\bigcap_{k=1}^n \big\{X \in \calA_k\big\}\bigg]\Pro\bigg[X\in \calA_{n+1}\bigg|X \in \bigcap_{k=1}^n A_k\bigg]\\
	&\qquad\leq \Pro\bigg[\bigcap_{k=1}^n \big\{X \in \calA_k\big\}\bigg]\Pro\bigg[X\in \calA_{n+1}\bigg]\\
	&\qquad \leq \beta \prod_{k=1}^n \Pro[X\in \calA_k]\Pro\big[X\in \calA_{n+1}\big]\\
	&\qquad= \beta \prod_{k=1}^{n+1} \Pro[X\in \calA_k]. 
	\end{align}	
	This concludes our proof by induction.

\subsection{Proof of Lemma~\ref{lem:extra2}}

\label{prooflemma15}

We consider four cases:
	\begin{itemize}
		\item Case 1: $a=1,b=1$. Then, we have
		\begin{align}
		\Pro\big[\{Z_{Q_{XX'}}=a\} \cap \{Z_{\tilQ_{XX'}}=b\} \big]=0.
		\end{align}
		On the other hand, we have
		\begin{align}
		\Pro\big[ Z_{Q_{XX'}}=1\big]\Pro\big[ Z_{\tilQ_{XX'}}=1\big]&=\Pro\big[(\bX_i,\bX_j) \in \calT(Q_{XX'}) \big]\Pro\big[(\bX_i,\bX_j) \in \calT(\tilQ_{XX'}) \big]\\
		&\doteq 2^{-n I_{Q_{XX'}}(X;X')} 2^{-n I_{\tilQ_{XX'}}(X;X')} \label{asmo},
		\end{align} where \eqref{asmo} follows from \cite{Tamir2020a}.
		Hence, we have
		\begin{align}
		\Pro\big[\{Z_{Q_{XX'}}=1\} \cap \{Z_{\tilQ_{XX'}}=1\} \big]&\leq \Pro\big[ Z_{Q_{XX'}}=1\big]\Pro\big[ Z_{\tilQ_{XX'}}=1\big]\\
		&\leq \frac{1}{1-2^{-n I_{\min}}(Q)}\Pro\big[ Z_{Q_{XX'}}=1\big]\Pro\big[ Z_{\tilQ_{XX'}}=1\big],
		\end{align}
		hence, \eqref{bacha1} holds for this case.  
		\item Case 2: $a=1,b=0$. Then, we have
		\begin{align}
		\Pro\big[\{Z_{Q_{XX'}}=1\} \cap \{Z_{\tilQ_{XX'}}=0\} \big]&=\Pro[\{(\bX_i,\bX_j) \in \calT(Q_{XX'})\} \cap \{\{(\bX_i,\bX_j) \not \in \calT(\tilQ_{XX'})\}\}]\\
		&=\Pro[(\bX_i,\bX_j) \in \calT(Q_{XX'})]\\
		&\doteq 2^{-n I_{Q_{XX'}}(X;X')} \label{bo1}.
		\end{align}
		On the other hand, we have
		\begin{align}
		\Pro\big[ Z_{Q_{XX'}}=1\big]\Pro\big[ Z_{\tilQ_{XX'}}=0\big]&=\Pro\big[(\bX_i,\bX_j) \in \calT(Q_{XX'}) \big]\Pro\big[(\bX_i,\bX_j) \not \in \calT(\tilQ_{XX'}) \big]\\
		&\doteq 2^{-n I_{Q_{XX'}}(X;X')} \big(1- 2^{-n I_{\tilQ_{XX'}}(X;X')}\big) \label{bo2}.
		\end{align}
		Now, there are two sub-cases. If  $I_{\tilQ_{XX'}}(X;X')=0$, from \eqref{bo1} and \eqref{bo2} we have
		\begin{align}
		\Pro\big[\{Z_{Q_{XX'}}=1\} \cap \{Z_{\tilQ_{XX'}}=0\} \big]&=\Pro\big[ Z_{Q_{XX'}}=1\big]\Pro\big[ Z_{\tilQ_{XX'}}=0\big]\\
		&\leq \frac{1}{1-2^{-n I_{\min}}(Q)}\Pro\big[ Z_{Q_{XX'}}=1\big]\Pro\big[ Z_{\tilQ_{XX'}}=0\big].
		\end{align} 
		On the other hand, if $I_{\tilQ_{XX'}}(X;X')>0$, we have
		\begin{align}
		\Pro\big[\{Z_{Q_{XX'}}=1\} \cap \{Z_{\tilQ_{XX'}}=0\} \big]&=\frac{1}{1-2^{-n I_{\tilQ_{XX'}}(X;X')}}\Pro\big[ Z_{Q_{XX'}}=1\big]\Pro\big[ Z_{\tilQ_{XX'}}=0\big]\\
		&\leq \frac{1}{1-2^{-n I_{\min}}(Q)}\Pro\big[ Z_{Q_{XX'}}=1\big]\Pro\big[ Z_{\tilQ_{XX'}}=0\big] \label{asto},
		\end{align} where \eqref{asto} follows from $I_{\tilQ_{XX'}}(X;X')\geq I_{\min}(Q)$.
		\item Case 3: $a=0,b=1$. Similarly as case 2, it holds that
		\begin{align}
		\Pro\big[\{Z_{Q_{XX'}}=1\} \cap \{Z_{\tilQ_{XX'}}=0\} \big]&=\Pro\big[ Z_{Q_{XX'}}=1\big]\Pro\big[ Z_{\tilQ_{XX'}}=0\big]\\
		&\leq \frac{1}{1-2^{-n I_{\min}}(Q)}\Pro\big[ Z_{Q_{XX'}}=1\big]\Pro\big[ Z_{\tilQ_{XX'}}=0\big].
		\end{align}
		\item Case 4: $a=0,b=0$. Then, we have
		\begin{align}
		&\Pro\big[\{Z_{Q_{XX'}}=0\} \cap \{Z_{\tilQ_{XX'}}=0\} \big]\nn\\
		&\qquad =\Pro[\{(\bX_i,\bX_j)\notin  \calT(Q_{XX'})\} \cap \{\{(\bX_i,\bX_j) \notin \calT(\tilQ_{XX'})\}\}]\\
		&\qquad =\Pro\big[ (\bX_i,\bX_j)\notin  \calT(Q_{XX'}) \big]+\Pro\big[ (\bX_i,\bX_j)\notin  \calT(\tilQ_{XX'}) \big]\nn\\
		&\qquad \qquad \qquad  -\Pro\big[\{(\bX_i,\bX_j)\notin  \calT(Q_{XX'})\} \cup \{\{(\bX_i,\bX_j) \notin \calT(\tilQ_{XX'})\}\}\big]\\
		&\qquad =\big(1- 2^{-nI_{Q_{XX'}}(X;X')}\big)+ \big(1- 2^{-n I_{\tilQ_{XX'}}(X;X')}\big)\nn\\
		&\qquad \qquad - (1-\Pro\big[ \{(\bX_i,\bX_j)\in  \calT(Q_{XX'})\} \cap \{(\bX_i,\bX_j)\in  \calT(\tilQ_{XX'})\} \big] )\label{ato}\\
		&\qquad=\big(1- 2^{-n I_{Q_{XX'}}(X;X')}\big)+ \big(1- 2^{-n I_{\tilQ_{XX'}}(X;X')}\big)-1\\
		&\qquad=1-2^{-n I_{Q_{XX'}}(X;X')}- 2^{-n I_{\tilQ_{XX'}}(X;X')} \label{besche}.
		\end{align} 
		On the other hand, we also have
		\begin{align}
		&\Pro\big[ Z_{Q_{XX'}}=0\big]\Pro\big[ Z_{\tilQ_{XX'}}=0\big]\nn\\
		&\qquad =\big(1-2^{-n I_{Q_{XX'}}(X;X')}\big)\big(1-2^{-n I_{\tilQ_{XX'}}(X;X')}\big)\\
		&\qquad=1-2^{-n I_{Q_{XX'}}(X;X')}- 2^{-n I_{\tilQ_{XX'}}(X;X')}+ 2^{-n I_{Q_{XX'}}(X;X')}2^{-n I_{\tilQ_{XX'}}(X;X')}
		\label{besche2}.
		\end{align}
		From \eqref{besche} and \eqref{besche2}, we have
		\begin{align}
		&\Pro\big[\{Z_{Q_{XX'}}=0\} \cap \{Z_{\tilQ_{XX'}}=0\} \big]\leq \Pro\big[ Z_{Q_{XX'}}=0\big]\Pro\big[ Z_{\tilQ_{XX'}}=0\big] \\
		&\qquad \leq \frac{1}{1-2^{-n I_{\min}}(Q)}\Pro\big[ Z_{Q_{XX'}}=1\big]\Pro\big[ Z_{\tilQ_{XX'}}=0\big]
		\label{boncha}.    
		\end{align}
	\end{itemize}
	From the four cases above, we finally obtain \eqref{bacha1}
	
	\begin{align}
	\frac{1}{n^{3/2}}\sum_{i=1}^n \bbE[|X_i^3|] \to 0 \label{conbat1b},\\
	\frac{1}{n^2}\sum_{i=1}^n \bbE[X_i^4] \to 0 \label{conbat2b},
	\end{align}

\subsection{Proof of Lemma~\ref{steinlem}}
\label{prooflemma16}

The proof is based on a modification of a proof based on Stein's method in \cite[Lemma 3.2]{Nathan2011a}. Without loss of generality, (or by scaling), we can assume that $\sum_{i=1}^n \bbE[X_i^2]=n$.
	
	Let
	\begin{align}
	T = \frac{S_n}{\sqrt{n}},
	\end{align}
	and
	\begin{align}
	T_i =  \frac{1}{\sqrt{n}} \sum_{j\neq i} X_j, \quad \forall i \in [n].
	\end{align}	
	Observe that
	\begin{align}
	\bbE[Tf(T)]&=\frac{1}{\sqrt{n}}\bbE\bigg[\sum_{i=1}^n X_i(f(T)-f(T_i))-(T-T_i)f'(T)\bigg]\nn\\
	&\qquad + \frac{1}{\sqrt{n}}\bbE\bigg[\sum_{i=1}^n X_i(T-T_i)f'(T)\bigg]+ \frac{1}{\sqrt{n}}\bbE\bigg[\sum_{i=1}^n X_if(T_i)\bigg].
	\end{align}
	Now, we have
	\begin{align}
	\bbE[Tf(T)-f'(T)]&=\frac{1}{\sqrt{n}}\bbE\bigg[\sum_{i=1}^n X_i\big(f(T)-f(T_i)-(T-T_i)f'(T)\big)\bigg]\nn\\
	&\qquad + \frac{1}{\sqrt{n}}\bbE\bigg[\sum_{i=1}^n X_i(T-T_i)f'(T)\bigg]+ \frac{1}{\sqrt{n}}\bbE\bigg[\sum_{i=1}^n X_if(T_i)\bigg]-\bbE[f'(T)]\\
	&\quad \leq \bigg|\frac{1}{\sqrt{n}}\bbE\bigg[\sum_{i=1}^n X_i\big(f(T)-f(T_i)-(T-T_i)f'(T)\big)\bigg]\bigg|\nn\\
	&\qquad \qquad+ \bigg|\frac{1}{\sqrt{n}}\bbE\bigg[\sum_{i=1}^n X_if(T_i)\bigg]\bigg|+ \bigg|\bbE\bigg[f'(T)\bigg(1-\frac{1}{\sqrt{n}}\sum_{i=1}^n X_i(T-T_i)\bigg)\bigg]\bigg| \label{abo}\\
	&\qquad\leq \frac{\|f^{''}\|}{2\sqrt{n}}\sum_{i=1}^n \bbE|X_i(T-T_i)^2|+ \frac{1}{\sqrt{n}}\bigg|\bbE\bigg[\sum_{i=1}^n X_if(T_i)\bigg]\bigg|+ \frac{\|f'\|}{n}\bbE\bigg|\sum_{i=1}^n (1-X_i^2)\bigg| \label{qbo}\\
	&\qquad\leq \frac{\|f^{''}\|}{2\sqrt{n}}\sum_{i=1}^n \bbE|X_i^3|+ \frac{1}{\sqrt{n}}\bigg|\bbE\bigg[\sum_{i=1}^n X_if(T_i)\bigg]\bigg|+ \frac{\|f'\|}{n}\bbE\bigg|\sum_{i=1}^n (1-X_i^2)\bigg| \label{moly1}.
	\end{align} 
	Now, observe that
	\begin{align}
	\frac{1}{\sqrt{n}}\sum_{i=1}^n \bbE[X_i f(T_i)]&=  \frac{1}{\sqrt{n}}\sum_{x_1,x_2,\cdots,x_n} \Pro[X_1=x_1,X_2=x_2,\cdots,X_n=x_n]\sum_{i=1}^n x_i f(t_i)\\
	&=  \frac{1}{\sqrt{n}}\sum_{x_1,x_2,\cdots,x_n \in \calV} \Pro[X_1=x_1,X_2=x_2,\cdots,X_n=x_n]\sum_{i=1}^n x_i f(t_i)\\
	&\qquad+  \frac{1}{\sqrt{n}}\sum_{x_1,x_2,\cdots,x_n \in \calV^c } \Pro[X_1=x_1,X_2=x_2,\cdots,X_n=x_n]\sum_{i=1}^n x_i f(t_i)\\
	&\leq \bigg(\frac{1}{1-2^{-f(n)}}\bigg) \frac{1}{\sqrt{n}}\sum_{x_1,x_2,\cdots,x_n \in \calV} \Pro[X_1=x_1,X_2=x_2,\cdots,X_n=x_n]\sum_{i=1}^n x_i f(t_i)\\
	&\qquad+  \frac{1}{\sqrt{n}}\sum_{x_1,x_2,\cdots,x_n \in \calV^c } \Pro[X_1=x_1,X_2=x_2,\cdots,X_n=x_n]\sum_{i=1}^n x_i f(t_i)\\
	&= \frac{1}{\sqrt{n}}\sum_{x_1,x_2,\cdots,x_n \in \calV^c } \Pro[X_1=x_1,X_2=x_2,\cdots,X_n=x_n]\sum_{i=1}^n x_i f(t_i) \label{GM1}.
	\end{align}
	From \eqref{GM1}, we obtain
	\begin{align}
	\bigg|\frac{1}{\sqrt{n}}\sum_{i=1}^n \bbE[X_i f(T_i)]\bigg|&\leq \frac{1}{\sqrt{n}}\sum_{x_1,x_2,\cdots,x_n \in \calV^c } \Pro[X_1=x_1,X_2=x_2,\cdots,X_n=x_n]\bigg|\sum_{i=1}^n x_i f(t_i)\bigg|\\
	&\leq  \frac{\|f\|_{\infty}}{\sqrt{n}}\sum_{x_1,x_2,\cdots,x_N \in \calV^c } \Pro[X_1=x_1,X_2=x_2,\cdots,X_n=x_n]\bigg|\sum_{k=1}^n x_k \bigg|\\
	&=  \frac{\|f\|_{\infty}}{\sqrt{n}}|\calV^c| \sup_{(x_1,x_2,\cdots,x_n) \in \calV^c} \bigg|\sum_{i=1}^n x_i \bigg|\\
	&\leq \frac{\|f\|_{\infty}}{\sqrt{n}}|\calV^c| \sup_{(x_1,x_2,\cdots,x_n) \in \calV^c}\sum_{k=1}^n x_k^4\\
	&\leq \frac{\|f\|_{\infty}}{\sqrt{n}}|\calV^c| g(n) \to 0  \label{molyy2}
	\end{align} as $n \to \infty$.
	
	Similarly, we have
	\begin{align}
	\frac{1}{n}\bbE\bigg[\bigg|\sum_{i=1}^n (1-X_i^2)\bigg|\bigg]&\leq \frac{1}{1- 2^{-f(n)}}\bigg(\frac{1}{n}\bigg)\sum_{x_1,x_2,\cdots,x_n } \prod_{i=1}^n \Pro(X_i=x_i)\sum_{i=1}^n (1-x_i^2)\bigg|\nn\\
	&\qquad +\frac{1}{n} |\calV^c| \sup_{(x_1,x_2,\cdots,x_n) \in \calV^c}\bigg|\sum_{i=1}^n (1-x_i^2)\bigg| \\
	&= \frac{1}{1- 2^{-f(n)}}\bigg(\frac{1}{n}\bigg)\sum_{x_1,x_2,\cdots,x_n } \prod_{i=1}^n \Pro(X_i=x_i)\sum_{i=1}^n (1-x_i^2)\bigg|\nn\\
	&\qquad +\frac{1}{n} |\calV^c| \sup_{(x_1,x_2,\cdots,x_n) \in \calV^c}\max\bigg\{\sum_{i=1}^n x_i^2,1\bigg\}\\
	&\leq \frac{1}{1- 2^{-f(n)}}\bigg(\frac{1}{n}\bigg)\sum_{x_1,x_2,\cdots,x_n } \prod_{i=1}^n \Pro(X_i=x_i)\sum_{i=1}^n (1-x_i^2)\bigg|\nn\\
	&\qquad +\frac{1}{n} |\calV^c| \sup_{(x_1,x_2,\cdots,x_n) \in \calV^c}\max\bigg\{\sqrt{n \sum_{i=1}^n x_i^4} ,1\bigg\} \\
	&\leq \frac{1}{1- 2^{-f(n)}}\bigg(\frac{1}{n}\bigg)\sum_{x_1,x_2,\cdots,x_n } \prod_{i=1}^n \Pro(X_i=x_i)\sum_{i=1}^n (1-x_i^2)\bigg|\nn\\
	&\qquad +\frac{1}{n} |\calV^c| \max\bigg\{\sqrt{n g(n)} ,1\bigg\}\\
	&\leq \frac{1}{1- 2^{-f(n)}}\bigg(\frac{1}{n}\bigg)\sum_{x_1,x_2,\cdots,x_n } \prod_{i=1}^n \Pro(X_i=x_i)\sum_{i=1}^n (1-x_i^2)\bigg|\nn\\
	&\qquad +|\calV^c|g(n) \bigg(\frac{1}{ng(n)} \max\bigg\{\sqrt{n g(n)} ,1\bigg\}\bigg)\\
	&= \frac{1}{1- 2^{-f(n)}}\bigg(\frac{1}{n}\bigg)\sum_{x_1,x_2,\cdots,x_n } \prod_{i=1}^n \Pro(X_i=x_i)\sum_{i=1}^n (1-x_i^2)\bigg|+o(1) \label{batche1}\\
	&\leq \bigg(\frac{1}{1-2^{-f(n)}}\bigg)\frac{\|f''\|_{\infty}}{2n^{3/2}}\sum_{k=1}^n \bbE[|X_k|^3] +o(1) \label{batche1b},
	\end{align} where \eqref{batche1b} follows from \cite[Proof of Lemma 3.4]{Nathan2011a}. 
	
	Furthermore, we also have
	\begin{align}
	\frac{1}{n^{3/2}}\sum_{i=1}^n \bbE\big[|X_i|^3\big]&\leq \frac{1}{1-2^{-f(n)}} \bigg(\frac{1}{n^{3/2}}\bigg)\sum_{x_1,x_2,\cdots,x_n} \prod_{i=1}^n \Pro[X_i=x_i] \bigg(\sum_{i=1}^n |x_i|^3 \bigg)\nn\\
	&\qquad + \frac{1}{n^{3/2}} |\calV^c| \sup_{(x_1,x_2,\cdots,x_n)\in \calV^c} \bigg(\sum_{i=1}^n |x_i|^3 \bigg)\\
	&= \frac{1}{1-2^{-f(n)}} \bigg(\frac{1}{n^{3/2}}\bigg)\sum_{x_1,x_2,\cdots,x_n} \prod_{i=1}^n \Pro[X_i=x_i] \bigg(\sum_{i=1}^n |x_i|^3 \bigg)\nn\\
	&\qquad + \frac{1}{n^{1/2}} |\calV^c| \sup_{(x_1,x_2,\cdots,x_n)\in \calV^c} \bigg(\frac{1}{n}\sum_{i=1}^n |x_i|^3 \bigg) \label{asmota0}\\
	&\leq \frac{1}{1-2^{-f(n)}} \bigg(\frac{1}{n^{3/2}}\bigg)\sum_{x_1,x_2,\cdots,x_n} \prod_{i=1}^n \Pro[X_i=x_i] \bigg(\sum_{i=1}^n |x_i|^3 \bigg)\nn\\
	&\qquad + \frac{1}{n^{1/2}} |\calV^c| \sup_{(x_1,x_2,\cdots,x_n)\in \calV^c} \bigg(\frac{1}{n}\sum_{i=1}^n |x_i|^4 \bigg)^{3/4}\label{asmota} \\
	&\leq \frac{1}{1-2^{-f(n)}} \bigg(\frac{1}{n^{3/2}}\bigg)\sum_{x_1,x_2,\cdots,x_n} \prod_{i=1}^n \Pro[X_i=x_i] \bigg(\sum_{i=1}^n |x_i|^3 \bigg)\nn\\
	&\qquad + \frac{1}{n^{1/2}} |\calV^c|  \bigg(\frac{g(n)}{n} \bigg)^{3/4}\label{asmota1} \\
	&\leq \frac{1}{1-2^{-f(n)}} \bigg(\frac{1}{n^{3/2}}\bigg)\sum_{x_1,x_2,\cdots,x_n} \prod_{i=1}^n \Pro[X_i=x_i] \bigg(\sum_{i=1}^n |x_i|^3 \bigg)\nn\\
	&\qquad + \frac{1}{n^{1/2}} |\calV^c|g(n) \frac{1}{(g(n)n^5)^{1/4}}\label{asmota2} \\
	&=\frac{1}{1-2^{-f(n)}} \bigg(\frac{1}{n^{3/2}}\bigg)\sum_{x_1,x_2,\cdots,x_n} \prod_{i=1}^n \Pro[X_i=x_i] \bigg(\sum_{i=1}^n |x_i|^3 \bigg)+o(1) \label{asmota3}\\
	&\leq \bigg(\frac{1}{1-2^{-f(n)}}\bigg)\frac{\|f''\|_{\infty}}{n}\sqrt{\sum_{k=1}^n \bbE[X_k^4]} \label{moly4},
	\end{align} 
	where \eqref{asmota} follows from the concavity of the function $x^{3/4}$ on $(0,\infty)$.
	
	From \eqref{moly1}, \eqref{molyy2}, \eqref{batche1}, and \eqref{moly4}, we obtain
	\begin{align}
	\big|\bbE[f'(T)-Tf(T)]\big| \leq \bigg(\frac{1}{1-2^{-f(n)}}\bigg)\frac{\|f''\|_{\infty}}{2n^{3/2}}\sum_{k=1}^n \bbE[|X_k|^3]+ \bigg(\frac{1}{1-2^{-f(n)}}\bigg)\frac{\|f''\|_{\infty}}{n}\sqrt{\sum_{k=1}^n \bbE[X_k^4]}+o(1) \to 0
	\end{align}	as $n \to \infty$ under the conditions \eqref{conbat1} and \eqref{conbat2}.
	
	Then, by  \cite[Th.~3.1]{Nathan2011a}, we conclude that
	\begin{align}
	T \dto  \calN(0,1),
	\end{align} under the conditions \eqref{conbat1} and \eqref{conbat2}. Now, since $\tilT$ is a scaling of $T$, hence by  Slutsky's theorem, it holds that
	\begin{align}
	\tilT \dto  \bigg(\lim_{n\to \infty} \sqrt{\frac{\var(S_n)}{\sum_{i=1}^n \bbE[X_i^2]}}\bigg)\calN(0,1).
	\end{align} 
	Now, since $\bbE[\tilT]=0$ and $\var(\tilT)=1$, by applying Lemma \ref{lem:aux2021} with $\var(Z)=1$, we must have
	\begin{align}
	\lim_{n\to \infty} \sqrt{\frac{\var(S_n)}{\sum_{i=1}^n \bbE[X_i^2]}}=1,
	\end{align}
	which leads to
	\begin{align}
	\tilT \dto \stackrel{(d)}{\to } \calN(0,1).
	\end{align}

\subsection{Proof of Lemma \ref{lem:boundmea}}

\label{ap:proof43}

This proof is based on the proof of \cite[Prop.~1.2]{Nathan2011a}. Consider the function $h_x(w)=\bone\{w \leq x\}$, and the `smooth' $h_{x,\eps}(w)$ defined to be one for $w \leq x$, zero for $w>x+\eps$, and linear between them. Then, it is clear that $h_{x,\eps} \in \calV$ with $a=x$ and $c=\eps$.

First, observe that $\eps h_{x,\eps}(w)$ is $1$-Lipschitz and
\begin{align}
\big\|\eps h_{x,\eps}\big\|_{\infty} \leq \eps \label{R0}. 
\end{align}
Hence, it holds that
\begin{align}
4 \sqrt{2\pi}h_{x,4 \sqrt{2\pi}} \in \calH =  \{h\in \calV: c \leq 4 \sqrt{2\pi} \},
\end{align} so $\calH$ in the definition of Wasserstein metric (cf. Definition \ref{steindef}) is a non-empty set, and $d_W(T,Z)$ is well-defined.

Furthermore, by definition of $d_{W,\rm{mod}}(T,Z)$, it holds that
\begin{align}
d_{W,\rm{mod}}(T,Z) &\leq \sup_{h \in \calH} \bbE[|h(Z)|]+ \bbE[|h(T)|]\\
&\leq 2 \|h\|_{\infty}\\
&=2c\\
&\leq 8 \sqrt{2\pi} \label{R2}.
\end{align}

Now, by setting $\eps =  (2\pi)^{1/4}\sqrt{2d_{W,\rm{mod}}(T,Z)}$, it holds that 
\begin{align}
\big\|\eps h_{x,\eps}\big\|_{\infty} &\leq (2\pi)^{1/4}\sqrt{2d_{W,\rm{mod}}(T,Z)} \label{R1}\\
&\leq 4 \sqrt{2\pi} \label{R3},
\end{align} where \eqref{R1} follows from \eqref{R0}, and \eqref{R3} follows from \eqref{R2}.
This means that $\eps h_{x,\eps} \in \calH$ since $\eps h_{x,\eps} \in \calV$ as mentioned above.

Then, we have
\begin{align}
\bbE[h_x(T)]-\bbE[h_x(Z)]&=\bbE[h_x(T)]-\bbE[h_{x,\eps}(Z)]+ \bbE[h_{x,\eps}(Z)]-\bbE[h_x(Z)]\\
&\leq \bbE[h_{x,\eps}(T)]-\bbE[h_{x,\eps}(Z)]+\bbE[h_{x,\eps}(Z)]-\bbE[h_x(Z)]\\
&=\frac{1}{\eps} \bigg(\bbE[\eps h_{x,\eps}(T)]-\bbE[\eps h_{x,\eps}(Z)]\bigg)+ \big|\bbE[h_{x,\eps}(Z)]-\bbE[h_x(Z)]\big| \\
&\leq \frac{1}{\eps} \big|\bbE[\eps h_{x,\eps}(T)]-\bbE[\eps h_{x,\eps}(Z)]\big|+ \big|\bbE[h_{x,\eps}(Z)]-\bbE[h_x(Z)]\big|
\label{g11a}.
\end{align}
Similarly, by choosing $h_{x,\eps}(\omega)$ to be $1$ when $\omega \leq x-\eps$, $0$ when $\omega \geq x$,  and linear between them, which is also a function in $\calV$, we can show that
\begin{align}
\bbE[h_x(Z)]-\bbE[h_x(T)]&\leq \frac{1}{\eps} \big|\bbE[\eps h_{x,\eps}(T)]-\bbE[\eps h_{x,\eps}(Z)]\big|+ \big|\bbE[h_{x,\eps}(Z)]-\bbE[h_x(Z)]\big| \label{g11b}\\
&\leq \frac{1}{\eps} \big|\bbE[\eps h_{x,\eps}(T)]-\bbE[\eps h_{x,\eps}(Z)]\big|\nn\\
&\qquad + \int_{x}^{x+\eps} \frac{1}{\sqrt{2\pi}} \exp \bigg(-\frac{z^2}{2}\bigg) \big(h_{x,\eps}(z)-h_x(z)\big)dz\\
&\leq \frac{1}{\eps} \big|\bbE[\eps h_{x,\eps}(T)]-\bbE[\eps h_{x,\eps}(Z)]\big|+ \frac{\eps}{2\sqrt{2\pi}}
\label{Q1a}. 
\end{align}
From \eqref{g11a} and \eqref{g11b}, we obtain
\begin{align}
\big|\bbP(T\leq x)-\bbP(Z\leq x)\big|\leq \frac{1}{\eps} \big|\bbE[\eps h_{x,\eps}(T)]-\bbE[\eps h_{x,\eps}(Z)]\big|+ \frac{\eps}{2\sqrt{2\pi}}
\label{Q1}. 
\end{align}

Similarly, we also have
\begin{align}
\big|\bbP(-T\leq x)-\bbP(Z\leq x)\big|\leq \frac{1}{\eps} \big|\bbE[\eps h_{x,\eps}(-T)]-\bbE[\eps h_{x,\eps}(Z)]\big|+ \frac{\eps}{2\sqrt{2\pi}} \label{Q2}.
\end{align}
It follows from \eqref{Q1} and \eqref{Q2} that
\begin{align}
&\sup_{x\in \bbR}\min\{\big|\bbP(T\leq x)-\bbP(Z\leq x)\big|,\big|\bbP(-T\leq x)-\bbP(Z\leq x)\big|\}\nn\\
&\qquad \leq \sup_{h\in \calH} \min\bigg\{\frac{1}{\eps} \big|\bbE[\eps h_{x,\eps}(T)]-\bbE[\eps h_{x,\eps}(Z)]\big|,\frac{1}{\eps} \big|\bbE[\eps h_{x,\eps}(-T)]-\bbE[\eps h_{x,\eps}(Z)]\big|\bigg\}+ \frac{\eps}{2\sqrt{2\pi}}\\
&\qquad=\frac{1}{\eps} d_{W,\rm{mod}}(T,Z) + \frac{\eps}{2\sqrt{2\pi}} \label{Q5} \\
&\qquad= (8\pi)^{-1/4}\sqrt{d_{W,\rm{mod}}(T,Z)} \label{Q6}, 
\end{align}	 where \eqref{Q5} follows from $\eps h_{x,\eps} \in \calH$, and \eqref{Q6} follows from our setting $\eps=(2\pi)^{1/4}\sqrt{2d_{W,\rm{mod}}(T,Z)}$ above.

Now, for any $x\in \bbR$, we have
\begin{align}
&\sup_{x\in \bbR}\min\bigg\{\big|\bbP(T\leq x)-\bbP(Z\leq x)\big|,\big|\bbP(T\leq -x)-\bbP(Z\leq x)\big|\bigg\}\nn\\
&\qquad \geq \min\bigg\{\big|\bbP(T\leq x)-\bbP(Z\leq x)\big|,\big|\bbP(T\leq -x)-\bbP(Z\leq x)\big|\bigg\}\\
&\qquad \geq \min\bigg\{\big|\bbP(T\leq x)-\bbP(Z\leq x)\big|,\big|\bbP(T\leq x)-\bbP(Z\leq x)\big|-\big|\bbP(T\leq x)-\bbP(T\geq -x)\big|\bigg\} \label{be}\\
&\qquad \geq \min\bigg\{\big|\bbP(T\leq x)-\bbP(Z\leq x)\big|,\big|\bbP(T\leq x)-\bbP(Z\leq x)\big|\bigg\}-\big|\bbP(T\leq x)-\bbP(T\geq -x)\big| \label{be20},
\end{align} where \eqref{be} follows from the triangle inequality.

From \eqref{Q6} and \eqref{be20}, we obtain \eqref{cachua}.

Now, if the distribution of $T$ is tight, then there exists a distribution $\tilY$ such that $T \dto \tilY$  \cite{Billingsley}. Then, if $x$ is a continuous point of $\Pro(\tilY\leq x)$ such that $x\to 0$ as $n\to \infty$, we have
\begin{align}
&\lim_{n\to \infty} \min\bigg\{\big|\bbP(T\leq x)-\bbP(Z\leq x)\big|,\big|\bbP(T\leq -x)-\bbP(Z\leq x)\big|\bigg\}\nn\\
&\qquad=\lim_{n\to \infty} \min\bigg\{\big|\bbP(\tilY\leq x)-\bbP(Z\leq x)\big|,\big|\bbP(\tilY\leq -x)-\bbP(Z\leq x)\big|\bigg\} \label{Ea}\\
&\qquad=\min\bigg\{\big|\bbP(\tilY\leq 0)-\bbP(Z\leq 0)\big|,\big|\bbP(\tilY\leq 0)-\bbP(Z\leq 0)\big|\bigg\}\\
&\qquad=\lim_{n\to \infty} \big|\bbP(T\leq 0)-\bbP(Z\leq 0)\big| \label{Q50},
\end{align} where \eqref{Ea} follows from $\lim_{N\to \infty} \min\{A_n,B_n\}=\min\{\lim_{n\to \infty} A_n, \lim_{n\to \infty} B_n\}$ if both the limits $\lim_{n\to \infty} A_n$ and $\lim_{N\to \infty} B_n$ exist.

Hence, we obtain \eqref{cachuab} from \eqref{Q50} and \eqref{Q6}.

\subsection{Proof of Lemma \ref{lem:co1}}

\label{ap:proof45}

By Lemma \ref{lem:nathan1}, we have
\begin{align}
d_{W,\rm{mod}}(T,Z)\leq \sup_{h \in \calH}\bigg\{ \big|\bbE\big[f_h'(T)-Tf_h(T)\big]\big|,\big|\bbE\big[f_h'(-T)+Tf_h(-T)\big]\big|\bigg\} \label{co1}.
\end{align}

Now, observe that
\begin{align}
\bbE\big[f_h'(T)-Tf_h(T)\big]&=\bbE[h(T)]-\bbE[h(Z)] \label{eqsol}\\
&=\bbE[h(T_1)]-\bbE[h(Z)] +\bbE[h(T)-h(T_1)] \label{eqsol3}\\
&= \bbE[f_h'(T_1)-T_1 f_h(T_1)]  +\bbE[h(T)-h(T_1)] \label{eqsol4},
\end{align} where \eqref{eqsol} and \eqref{eqsol4} follow from \eqref{eqkey}.

It follows that
\begin{align}
\bigg|	\bbE\big[f_h'(T)-Tf_h(T)\big]\bigg|&=\bigg|\bbE[f_h'(T_1)-T_1 f_h(T_1)]  +\bbE[h(T)-h(T_1)]\bigg|\\
&\leq \bigg|\bbE[f_h'(T_1)-T_1 f_h(T_1)] \bigg|+ \bigg|\bbE[h(T)-h(T_1)]\bigg|\\
&= \bigg|\bbE[f_h'(T_1)-T_1 f_h(T_1)] \bigg| + \bbE[h(T)-h(T_1)] \label{bu}
\end{align} where \eqref{bu} follows from $T\leq T_1$ and $h$ is non-increasing.

Similarly, we have
\begin{align}
\bigg|	\bbE\big[f_h'(-T)+Tf_h(-T)\big]\bigg|&=\bigg|\bbE[f_h'(-T_1)+T_1 f_h(-T_1)]  +\bbE[h(-T)-h(-T_1)]\bigg|\\
&\leq \bigg|\bbE[f_h'(-T_1)+T_1 f_h(-T_1)] \bigg| + \bbE[h(-T_1)-h(T)] \label{bumod},
\end{align} where \eqref{bumod} follows from $T\leq T_1$ and $h$ is non-increasing.

From \eqref{bu} and \eqref{bumod}, for all $h\in \calH$, we have
\begin{align}
&\min\bigg\{\big|\bbE\big[f_h'(T)-Tf_h(T)\big]\big|,\big|\bbE\big[f_h'(-T)+Tf_h(-T)\big]\big|\bigg\}\nn\\
&\qquad \leq \max\bigg\{\big|\bbE[f_h'(T_1)-T_1 f_h(T_1)]\big|,\big|\bbE[f_h'(T_1)-T_1 f_h(T_1)]\big|\bigg\}+\min\bigg\{\bbE[h(T)-h(T_1)],\bbE[h(-T_1)-h(-T)]\bigg\} \label{eqs}.
\end{align}
Finally, we obtain \eqref{ba1} from \eqref{eqs}.

\subsection{Proof of Lemma \ref{Wasserthm2}}
\label{Wasserthm2:proof} The proof of this lemma is based on the proof of the  \cite[Th.~3.1]{Nathan2011a}. Given $h \in \calH$, we choose $f_h$ be a solution of the following ODE equation:
	\begin{align}
	f_h'(w)-wf_h(w)=h(w)-\Phi(h) \label{ODE2}
	\end{align} where $\Phi(h) =  \bbE[h(Z)]$ with $Z \sim \calN(0,1)$, then we have
	\begin{align}
	f_h(w)&=e^{\frac{w^2}{2}}\int_w^{\infty} e^{-\frac{t^2}{2}}\big(\Phi(h)-h(t))dt\\
	&=-e^{\frac{w^2}{2}}\int_{-\infty}^{w} e^{-\frac{t^2}{2}}\big(\Phi(h)-h(t))dt \label{cuteb2}.
	\end{align}
	Now, it is easy to prove the following facts from \eqref{cuteb2} (see \cite{Nathan2011a}):
	\begin{align}
	\|f_h\|_{\infty} &\leq 2\|h'\|_{\infty}=2, \label{Ta12} \\
	\|f_h^{'}\|_{\infty} &\leq \sqrt{\frac{2}{\pi}}\|h'\|_{\infty}=\sqrt{\frac{2}{\pi}} \label{Ta22},\\
	\|f_h^{"}\|_{\infty} &\leq 2\|h'\|_{\infty}=2 \label{Ta32}. 
	\end{align}
	Now, assume that $\bbE[T]=0$ and $\bbE[T^2=1]$. 
	
	Furthermore, for any $h \in \calH$, from \eqref{ODE2}, it holds that
	\begin{align}
	|f_h^{'}(T)-Tf_h(T)|&=|h(T)-\Phi(h)|\\
	&=|h(T)-\bbE[h(Z)]|\\
	&\leq 2 \|h\|_{\infty}\\
	&\leq 8 \sqrt{2\pi} \label{R112}.
	\end{align}
	Furthermore, from \eqref{ODE2}, we also have
	\begin{align}
	d_W(T,Z)&=\sup_{h \in \calH} \big|\bbE[h(T)]-\bbE[h(Z)]\big|\\
	&\leq \sup_{f_h: h \in \calH} \big|\bbE[Tf_h(T)-f_h^{'}(T)\big]\big| \label{K1}.
	\end{align}
	
	Now, for all $f_h: h \in \calH$, observe that
	\begin{align}
	& Tf_h(T)-f_h^{'}(T)\nn\\
	&\qquad= T\big(f_h(T)-f_h(0)-T f_h^{'}(0)\big)+Tf_h(0)+\big(T^2-1\big)f_h^{'}(0)+ \big(f_h^{'}(0)-f_h^{'}(T)\big) \label{acqui1}.
	\end{align}
	
	It follows from \eqref{acqui1} that
	\begin{align}
	\bbE\big[ Tf_h(T)-f_h^{'}(T)\big]&= \bbE\big[T\big(f_h(T)-f_h(0)-T f_h^{'}(0)\big) \big] + f_h(0)\bbE[T] + f_h^{'}(0)\bbE\big[T^2-1\big] + \bbE[f_h^{'}(0)-f_h^{'}(T)] \\
	&=\bbE\big[T\big(f_h(T)-f_h(0)-T f_h^{'}(0)\big) \big] + \bbE[f_h^{'}(0)-f_h^{'}(T)]
	\label{acqui2},
	\end{align} where \eqref{acqui2} follows from the fact that $\bbE[T]=0$ and $\bbE[T^2]=1$.
	
	Hence, from \eqref{K1} and \eqref{acqui2}, we have
	\begin{align}
	\tild_{W,\rm{mod}}(T,Z)&\leq \sup_{f_h: h \in \calH} \big|\bbE[Tf_h(T)-f_h^{'}(T)\big]\big| \label{K1a}\\
	&\leq \sup_{f_h: h \in \calH} \bbE\big[\big|T\big(f_h(T)-f_h(0)-T f_h^{'}(0)\big)\big| \big] + \bbE[\big|f_h^{'}(0)-f_h^{'}(T)\big|] \label{K2}.
	\end{align}
	Now, observe that
	\begin{align}
	\big|T\big(f_h(T)-f_h(0)-T f_h^{'}(0)\big)\big|&=\big|Tf_h(T)-f_h^{'}(T)+f_h^{'}(T)-Tf_h(0)-T^2f_h^{'}(0)\big|\\
	&=\big|Tf_h(T)-f_h^{'}(T)+f_h^{'}(T)-Tf_h(0)-f_h^{'}(0)+(1-T^2) f_h^{'}(0)\big|\\
	&\leq 	\big|Tf_h(T)-f_h^{'}(T)\big|+ |f_h^{'}(T)|+ |T f_h(0)|+|f_h^{'}(0)|+|f_h^{'}(0)(T^2-1)|\big)\\
	&\leq 8 \sqrt{2\pi}+ 2\sqrt{\frac{2}{\pi}}+ 2 |T| + \sqrt{\frac{2}{\pi}} |T^2-1|
	\label{R11a}\\
	&= \bigg(8+\frac{2}{\pi}\bigg)\sqrt{2\pi}+ 2 |T| + \sqrt{\frac{2}{\pi}} |T^2-1|,
	\end{align} where \eqref{R11a} follows from \eqref{Ta12}, \eqref{Ta22} and \eqref{R112}.
	
	Hence, we have
	\begin{align}
	&\big|T\big(f_h(T)-f_h(0)-T f_h^{'}(0)\big)\big| \big]\nn\\
	&\qquad = \min\bigg\{\bigg(8+\frac{2}{\pi}\bigg)\sqrt{2\pi}+  2 |T|, \big|T\big(f_h(T)-f_h(0)-T f_h^{'}(0)\big)\big| \bigg\}\\
	&\qquad \leq \min\bigg\{\bigg(8+\frac{2}{\pi}\bigg)\sqrt{2\pi},\big|T\big(f_h(T)-f_h(0)-T f_h^{'}(0)\big)\big|  \bigg\}+2 |T| + \sqrt{\frac{2}{\pi}} |T^2-1| \label{K4},
	\end{align} where \eqref{K4} follows from $\min\{A+B,C\}\leq \min\{A,C\}+B $ for all $A,B,C\geq 0$. It follows from \eqref{K4} that
	\begin{align}
	&\bbE\big[\big|T\big(f_h(T)-f_h(0)-T f_h^{'}(0)\big)\big|\big]\leq \bbE\bigg[\min\bigg\{\bigg(8+\frac{2}{\pi}\bigg)\sqrt{2\pi},\big|T\big(f_h(T)-f_h(0)-T f_h^{'}(0)\big)\big|  \bigg\}\bigg]\nn\\
	&\qquad+ 2\bbE\big[|T|\big] \label{K10}. 
	\end{align}
	Now, by Taylor's expansion, for some $\eta \in (0,-|T|)\cup (0,|T|)$, we have
	\begin{align}
	f_h(T)-f_h(0)-T f_h^{'}(0)=\frac{1}{2} f_h^{"}(\eta) T^2,
	\end{align}
	so
	\begin{align}
	\big|T\big(f_h(T)-f_h(0)-T f_h^{'}(0)\big)\big|&=\frac{1}{2}\big| T^3 f_h^{"}(\eta)\big|\\
	&\leq \frac{1}{2}\|f_h^{"}|_{\infty}|T^3| \\
	&\leq |T^3| \label{facto2}.
	\end{align}
	Hence, from \eqref{K10} and \eqref{facto2}, we obtain
	\begin{align}
	\bbE\big[\big|T\big(f_h(T)-f_h(0)-T f_h^{'}(0)\big)\big|\big]\leq \bbE\bigg[\min\bigg\{\bigg(8+\frac{2}{\pi}\bigg)\sqrt{2\pi},\big|T\big|^3\bigg\}\bigg]+ 2\bbE\big[|T|\big] + \sqrt{\frac{2}{\pi}} |T^2-1| \label{K11}.
	\end{align}
	
	Similarly, by Taylor's expansion, for some $\theta \in (0,-|T|)\cup (0,|T|)$, we have
	\begin{align}
	f_h^{'}(T)-f_h^{'}(0)=f_h^{"}(\theta)T,
	\end{align}
	so
	\begin{align}
	\bbE\big[\big|f_h^{'}(T)-f_h^{'}(0)\big|\big]&=\bbE\big[\big|f_h^{"}(\theta)T\big|\big]\\
	&\leq \bbE\big[|f_h^{"}(\theta)||T|\big]\\
	&\leq \|f_h^{"}\|_{\infty} \bbE\big[|T|\big] \\
	&\leq  2 \bbE[|T|]
	\label{facto1}.
	\end{align}
	
	Finally, from \eqref{K4}, \eqref{K11}, and \eqref{facto1}, we have
	\begin{align}
	\tild_{W,\rm{mod}}(T,Z)&\leq \sup_{f_h: h \in \calH} \bbE\big[\big|T\big(f_h(T)-f_h(0)-T f_h'(0)\big)\big| \big] + \bbE[\big|f_h^{'}(0)-f_h^{'}(T)\big|] \label{K2x}\\
	&\leq \bbE\bigg[\min\bigg\{\bigg(8+\frac{2}{\pi}\bigg)\sqrt{2\pi},\big|T\big|^3\bigg\}\bigg]+ 2\bbE\big[|T|\big] + \sqrt{\frac{2}{\pi}} |T^2-1| \label{keyfacto}.
	\end{align}
	
	Now, observe that
	\begin{align}
	&\bbE\bigg[\min\bigg\{\bigg(8+\frac{2}{\pi}\bigg)\sqrt{2\pi},\big|T\big|^3\bigg\}\bigg]\nn\\
	&\qquad =	\bbE\bigg[\min\bigg\{\bigg(8+\frac{2}{\pi}\bigg)\sqrt{2\pi},\big|T\big|^3\bigg\}\bigg||T|\leq 1\bigg]\Pro[|T|\leq 1] \nn\\
	&\qquad + \bbE\bigg[\min\bigg\{\bigg(8+\frac{2}{\pi}\bigg)\sqrt{2\pi},\big|T\big|^3\bigg\}\big||T|> 1\bigg]\Pro[|T|> 1]\\
	&\qquad \leq \bbE\bigg[\min\bigg\{\bigg(8+\frac{2}{\pi}\bigg)\sqrt{2\pi},\big|T\big|\bigg\}\bigg||T|\leq 1\bigg]\Pro[|T|\leq 1] \nn\\
	&\qquad + \bbE\bigg[\min\bigg\{\bigg(8+\frac{2}{\pi}\bigg)\sqrt{2\pi},\big|T\big|\bigg\}\big||T|> 1\bigg]\Pro[|T|> 1]\label{J1}\\
	&\qquad \leq \bbE\bigg[\big|T\big|\bigg||T|\leq 1\bigg]\Pro[|T|\leq 1] \nn\\
	&\qquad + \bbE\bigg[\min\bigg\{\bigg(8+\frac{2}{\pi}\bigg)\sqrt{2\pi},\big|T\big|^3\bigg\}\big||T|> 1\bigg]\Pro[|T|> 1]\label{J2}\\
	&\qquad \leq \bbE\big[|T|\big]+ \bbE\bigg[\min\bigg\{\bigg(8+\frac{2}{\pi}\bigg)\sqrt{2\pi},\big|T\big|^3\bigg\}\big||T|> 1\bigg]\Pro[|T|> 1]\label{J3}\\
	&\qquad \leq \bbE\big[|T|\big]+\bigg(8+\frac{2}{\pi}\bigg)\sqrt{2\pi} \Pro[|T|> 1]\\
	&\qquad \leq \bbE\big[|T|\big]+\bigg(8+\frac{2}{\pi}\bigg)\sqrt{2\pi}\bbE[|T|] \label{J4}\\
	&\qquad \leq \bigg(10+\frac{1}{\pi}\bigg)\sqrt{2\pi}\bbE[|T|] \label{J5}, 
	\end{align} where \eqref{J1} follows from $|T|^3 \leq |T|$ for all $|T|\leq 1$, \eqref{J3} follows from $\bbE[X]= \bbE[X|A]\Pro(A)+ \bbE[X|A^c]\Pro(A^c)\geq \bbE[X|A]\Pro(A)$ for all non-negative random variable $X$, and \eqref{J4} follows from Markov's inequality.
	
	From \eqref{keyfacto} it follows that
	\begin{align}
	\tild_{W,\rm{mod}}(T,Z)&\leq  \bigg( \bigg(10+\frac{1}{\pi}\bigg)\sqrt{2\pi}+\bigg(4+ \sqrt{\frac{2}{\pi}}\bigg)\bigg)\bbE[|T|]+ \sqrt{\frac{2}{\pi}} \bbE\big[|T^2-1|\big]\\
	&<40\bbE[|T|]+ \sqrt{\frac{2}{\pi}} \bbE\big[|T^2-1|\big] \label{tas2}.
	\end{align}
	
	By combining Lemma \ref{lem:boundmea} and \eqref{tas2}, we have
	\begin{align}
	d_K(T,Z)&<  2(8\pi)^{-1/4}\sqrt{40\bbE[|T|]+ \sqrt{\frac{2}{\pi}}\bbE \big[|T^2-1|\big]}\\
	&\leq 14 (8\pi)^{-1/4}\sqrt{\bbE[|T|]+ \bbE \big[|T^2-1|\big]}.
	\end{align}
	This concludes our last proof.

\bibliographystyle{ieeetr}
\bibliography{IEEEabrv,isitbib}

\begin{thebibliography}{10}

\bibitem{Shannon48}
C.~E. Shannon, ``A mathematical theory of communication,'' {\em Bell System
  Technical Journal}, vol.~27, pp.~379--423, 1948.

\bibitem{Fano}
R.~M. Fano, {\em Transmission of Information}.
\newblock New York: Wiley, 1961.

\bibitem{Gallager1965a}
R.~G. Gallager, ``Simple derivation of the coding theorem and some
  applications,'' {\em {IEEE} Trans. Inf. Theory}, vol.~11, pp.~3--18, Jan
  2008.

\bibitem{sgb}
C.~E. Shannon, R.~G. Gallager, and E.~R. Berlekamp, ``Lower bounds to error
  probability for coding in discrete memoryless channels {I-II},'' {\em
  Information and Control}, vol.~10, pp.~65--103,~522--552, 1967.

\bibitem{Nakiboglu2020}
B.~Nakibo{\u{g}}lu, ``The sphere packing bound for memoryless channels,'' {\em
  Problems of Information Transmission}, vol.~56, pp.~201--244, 2020.

\bibitem{nakibouglu2019augustin}
B.~Nakibo{\u{g}}lu, ``The {A}ugustin capacity and center,'' {\em Problems of
  Information Transmission}, vol.~55, no.~4, pp.~299--342, 2019.

\bibitem{Barg2002a}
A.~{Barg} and G.~D. {Forney}, ``Random codes: minimum distances and error
  exponents,'' {\em {IEEE} Trans. Inf. Theory}, vol.~48, no.~9, pp.~2568--2573,
  2002.

\bibitem{Nazari}
A.~Nazari, A.~Anastasopoulos, and S.~S. Pradhan, ``Error exponent for
  multiple-access channels: Lower bounds,'' {\em {IEEE} Trans. Inf. Theory},
  vol.~60, no.~9, pp.~5095--5115, 2014.

\bibitem{Merhav2018a}
N.~Merhav, ``Error exponents of typical random codes,'' {\em {IEEE} Trans. Inf.
  Theory}, vol.~64, no.~9, pp.~6223--6235, 2018.

\bibitem{merhav2019error}
N.~Merhav, ``Error exponents of typical random codes for the colored {G}aussian
  channel,'' {\em IEEE Trans. Inf. Theory}, vol.~65, no.~12, pp.~8164--8179,
  2019.

\bibitem{merhav2019error2}
N.~Merhav, ``Error exponents of typical random trellis codes,'' {\em IEEE
  Trans. Inf. Theory}, vol.~66, no.~4, pp.~2067--2077, 2019.

\bibitem{Tamir2020a}
R.~Tamir, N.~Merhav, N.~Weinberger, and A.~{Guill\'{e}n i F\`{a}bregas},
  ``Large deviations behavior of the logarithmic error probability of random
  codes,'' {\em {IEEE} Trans. Inf. Theory}, vol.~66, no.~11, pp.~6635--6659,
  2020.

\bibitem{Ahlswede1982}
R.~Ahlswede and G.~Dueck, ``Good codes can be produced by a few permutations,''
  {\em IEEE Trans. Inf. Theory}, vol.~28, no.~3, pp.~430--443, 1982.

\bibitem{tamir2021universal}
R.~T. (Averbuch) and N.~Merhav, ``Universal decoding for the typical random
  code and for the expurgated code,'' {\em IEEE Trans. Inf. Theory}, 2022.

\bibitem{Giusseppe2021e}
G.~Cocco, A.~{Guill\'{e}n i F\`{a}bregas}, and J.~Font-Segura, ``A dual-domain
  achievability of the typical error exponent,'' in {\em IEEE Int. Symp. Inf.
  Theory}, (Melbourne, Australia), 2021.

\bibitem{Scarlett13}
J.~Scarlett, A.~Martinez, and A.~{Guill\'en i F\`abregas}, ``Mismatched
  decoding: {Finite-Length} bounds, error exponents and approximations,'' {\em
  IEEE Trans. Inf. Th.}, 2014.

\bibitem{gallager_fcc_notes}
R.~G. Gallager, ``Fixed composition arguments and lower bounds to the error
  probability, lecture notes {MIT} 6.441,'' 1994.

\bibitem{Durrett}
R.~Durrett, {\em Probability: Theory and Examples}.
\newblock Cambridge Univ. Press, 4th~ed., 2010.

\bibitem{coverThomas}
T.~Cover and J.~Thomas, {\em Elements of Information Theory}.
\newblock John Wiley and Sons, 2006.

\bibitem{feller}
W.~Feller, {\em An Introduction to Probability Theory and Its Applications}.
\newblock John Wiley and Sons, 2nd~ed., 1971.

\bibitem{cohen2004lower}
A.~Cohen and N.~Merhav, ``Lower bounds on the error probability of block codes
  based on improvements on de caen's inequality,'' {\em IEEE Trans. Inf.
  Theory}, vol.~50, no.~2, pp.~290--310, 2004.

\bibitem{ScarlettMF2013aler}
J.~{Scarlett}, A.~{Martinez}, and A.~{Guill{\'e}n i F{\`a}bregas},
  ``Ensemble-tight error exponents for mismatched decoders,'' in {\em Proc. of
  Allerton Conference}, 2012.

\bibitem{Billingsley}
P.~Billingsley, {\em Probability and Measure}.
\newblock Wiley-Interscience, 3rd~ed., 1995.

\bibitem{gallagerBook}
R.~G. Gallager, {\em Information Theory and Reliable Communication}.
\newblock USA: John Wiley \& Sons, Inc., 1968.

\bibitem{Nathan2011a}
N.~Ross, ``{Fundamentals of Stein's method},'' {\em Probability Surveys},
  vol.~8, no.~none, pp.~210 -- 293, 2011.

\bibitem{paley_zyg1932}
R.~Paley and A.~Zygmund, ``On some series of functions, (3),'' {\em
  Mathematical Proceedings of the Cambridge Philosophical Society}, vol.~28,
  no.~2, pp.~190--205, 1932.

\bibitem{Caenlem}
D.~{de Caen}, ``A lower bound on the probability of a union,'' {\em Discrete
  Mathematics}, vol.~169, no.~1, pp.~217--220, 1997.

\bibitem{MoserBook}
S.~M. Moser, {\em Advanced Topics in Information Theory. Lecture Notes}.
\newblock 2019.

\bibitem{Csis00}
I.~Csisz\'{a}r, ``The method of types,'' {\em IEEE Trans. Inf. Th.}, vol.~44,
  no.~6, pp.~2505--23, 1998.

\bibitem{Royden}
H.~Royden and P.~Fitzpatrick, {\em Real Analysis}.
\newblock Pearson, 4th~ed., 2010.

\bibitem{Stewart}
J.~Stewart, {\em Multivariate Calculus}.
\newblock Cengate Learning, 6th~ed., 2008.

\bibitem{Gallager1973a}
R.~Gallager, ``The random coding bound is tight for the average code
  (corresp.),'' {\em {IEEE} Trans. Inf. Theory}, vol.~19, no.~2, pp.~244--246,
  1973.

\bibitem{Boyd04}
S.~Boyd and L.~Vandenberghe, {\em Convex Optimization}.
\newblock Cambridge University Press, 2004.

\bibitem{cover}
T.~Cover, ``Comments on broadcast channels,'' {\em IEEE Trans. Inf. Th.},
  vol.~44, no.~6, pp.~2524--30, 1998.

\bibitem{suen_new_JansonRSA1998}
S.~Janson, ``New versions of {S}uen's correlation inequality,'' {\em Random
  Struct. Algorithms}, vol.~13, pp.~467--483, Oct. 1998.

\end{thebibliography}
\end{document}